\documentclass[preprint,12pt]{elsarticle}
\usepackage{amssymb}

\usepackage{amsthm}

\newtheorem{thm}{Theorem}

\usepackage{amsmath}
\usepackage{lineno}

\usepackage{enumerate}
\usepackage{here}
\usepackage{indentfirst}
\usepackage{bm}
\usepackage{url}
\usepackage{color}
\usepackage{ulem}
\usepackage{here}

\usepackage{mathbbol}
\usepackage{tikz}
\usetikzlibrary{shapes,arrows} 

\usepackage{caption}
\usepackage[subrefformat=parens]{subcaption}
\usepackage{color}

\def\XXint#1#2#3{{\setbox0=\hbox{$#1{#2#3}{\int}$ }
		\vcenter{\hbox{$#2#3$ }}\kern-.6\wd0}}

\def\x{\bm{x}}
\def\d{\text{d}}

\journal{Computer Methods in Applied Mechanics and Engineering,~}

\begin{document}

\begin{frontmatter}



\title{Extended level set method: a multiphase representation \\
	with perfect symmetric property, and its application \\
	to multi material topology optimization}

\author[label1]{Masaki~Noda}
\author[label1,label2]{Yuki~Noguchi}
\author[label1,label2]{Takayuki~Yamada \corref{cor1}}
\ead{t.yamada@mech.t.u-tokyo.ac.jp}
\cortext[cor1]{Corresponding author.
Tel.: +81-3-5841-0294;
Fax: +81-3-5841-0294.}

\address[label1]{Department of Mechanical Engineering, Graduate School of Engineering, The University of Tokyo, Yayoi 2-11-16, Bunkyo-ku, Tokyo 113-8656, Japan.}

\address[label2]{Department of Strategic Studies, Institute of Engineering Innovation, Graduate School of Engineering, The University of Tokyo, Yayoi 2-11-16, Bunkyo-ku, Tokyo 113-8656, Japan.}

\begin{abstract}
	This paper provides an extended level set (X-LS) based topology optimization method for multi material design. In the proposed method, each zero level set of a level set function $\phi_{ij}$ represents the boundary between materials $i$ and $j$. Each increase or decrease of $\phi_{ij}$ corresponds to a material change between the two materials. This approach reduces the dependence of the initial configuration in the optimization calculation and simplifies the sensitivity analysis. First, the topology optimization problem is formulated in the X-LS representation. Next, the reaction-diffusion equation that updates the level set function is introduced, and an optimization algorithm that solves the equilibrium equations and the reaction-diffusion equation using the finite element method is constructed. Finally, the validity and utility of the proposed topology optimization method are confirmed using two- and three-dimensional numerical examples.
\end{abstract}


\end{frontmatter}

%\linenumbers
%\pagewiselinenumbers

\section{Introduction}\label{sec: Intro}
Many problems involve multiple material phases with varying shapes. For instance, image processing \cite{vese2002multiphase} must identify the shape of the region occupied by an object in an image and flow simulations \cite{merriman1994motion} must accurately track the temporal changes in each material phase. Meanwhile, structural optimization finds the structure that optimizes some physical state field that depends on the structure. A method that represents this structure is fundamental and has a significant impact on the results.

A properly designed structure must satisfy not only various performance requirements, such as weight, strength, and heat transfer characteristics but also design requirements such as price and productivity. To satisfy these requirements simultaneously in a single material design, multifunctional high-performance materials might be required, which do not always exist. To solve such design problems, researchers are turning to design methods that integrate materials with different properties in effective combinations. For example, in the structural design of automobile bodies, weight reduction is required to comply with CO2 emission regulations and to extend the driving range of electric vehicles but without degrading the structure's mechanical properties such as rigidity and collision resistance. Currently, the research and development of automobile body structures combines multiple materials such as aluminum alloys and carbon fiber reinforced polymers. In addition, structural designs with negative thermal expansion coefficients have been reported but cannot be achieved in structures formed from general single materials \cite{sigmund1997design}. 
As described above, the use of multiple materials in designing structures is potentially effective. However, the design of structures that fully exploit the properties of multiple materials requires considerable trial and error, and a design solution that highly satisfies multiple design requirements is difficult to obtain.

Structural optimization methods can derive the optimal design solution using mathematical and mechanical theories. Topology optimization has the highest degree of freedom among the structural optimization methods and is attracting attention because its design solution maximizes the material properties of the constituent materials. Topology optimization is usually performed using the homogenized design method \cite{bendsoe1988generating}, density method  \cite{bendsoe1989optimal}, or methods based on the level set method \cite{sethian2000structural}. 
The homogenized design method and density method optimize the structure while allowing for microstructure and intermediate materials, which cannot be manufactured in practice, whereas topology optimization based on the level set method has the advantage of obtaining clear boundaries. Structural optimization methods based on level sets can be roughly classified into two groups that update the level set function with different information: shape sensitivity \cite{allaire2004structural} and topological derivatives \cite{yamada2010topology}.
The shape sensitivity methods obtain the optimal structure by varying the external shape of the structure based on the distribution of shape sensitivity, which exclude topological changes resulting in new boundaries. The optimal structure tends to depend on the initial structure. 
In topological derivative methods, the structure is represented by a level set function      and the optimal structure is obtained by optimizing the material distribution based on the distribution of topological derivatives. 
The optimal structures are less dependent on the initial structure because topological derivative methods allow both disappearance of an existing boundary and generation of a new boundary.
 
Topology optimization is known as an ill-posed problem, which allows infinitely small structures as the optimal solution. Therefore, it requires a regularization process. Some regularization methods, such as sensitivity filters, density filters, and density methods, require trial and error in parameter selection \cite{sigmund2013topology}. 
By contrast, the method proposed by Yamada et al. \cite{yamada2010topology}, which updates the level set function with the reaction-diffusion equation, allows easy control of the geometric complexity of the optimal structure and is applicable to a wide range of design problems. 
In fact, this method has been used to solve 
elasticity \cite{emmendoerfer2016topology,emmendoerfer2018level,emmendoerfer2020stress}, thermal \cite{yamada2011level,jahangiry2019combination}, acoustic \cite{isakari2014topology,isakari2017level,lanznaster2021level}, electromagnetic \cite{otomori2012topology,jung2021multi}, fluid \cite{yaji2014topology}, current control \cite{fujii2019dc}, thermoelectric \cite{fujii2019optimizing}, multiple material  \cite{cui2016level,kishimoto2017optimal,MasakiNODA202120-00412}, and other optimal design problems, confirming its high extendibility.

When extending topology optimization to multi material design problems, the material representation method is important because the design variables depend on the method, and the design sensitivity changes accordingly.
Typical level set methods for multiple materials include the piecewise-constant level set (PCLS) method, the color level set (ColorLS) method, the multi-material level set (MMLS) method, and the vector-valued level set (VVLS) method.

The PCLS method \cite{wei2009piecewise,luo2009design} represents multiple phases with a single level set function that assigns a unique integer value to each phase region. The PCLS provides a simple multiphase representation at low computational cost and low data requirements. The ColorLS \cite{wang2004color} and MMLS \cite{wang2015multi,cui2016level,kishimoto2017optimal} methods compute multiple level set functions and combine their positive and negative values to represent multiple phases. The MMLS method requires more design variables than the ColorLS method but its sensitivity analysis is less complex. The VVLS method \cite{gangl2020multi} represents multiple phases in different value ranges of a vector-valued function. This method is advantaged by low dependence on the initial structure because continuous changes are allowed among all phases. In the authors' previous research \cite{MasakiNODA202120-00412}, all phases were defined symmetrically with respect to the other phases. The details of each method are described in \ref{sec:generalization}.

These methods represent the phase configuration based on a specific value of the level set function inside each phase region. However, the domains of each phase can also be defined as regions surrounded by the boundaries of other phases. Based on this idea, we construct a new multiphase representation method named the extended level set (X-LS) method. For all two-phase combinations, X-LS method represents a boundary between the two phases by an independent level set function. This construct allows a concise correspondence between the level set function and the topological derivatives, as all phases are defined symmetrically with respect to the other phases, and the geometric complexity of the boundary shape between any two materials can be controlled. Furthermore, the proposed method can be considered a generalization of the previous methods described above.

The following sections are organized as follows. First, the X-LS method is formulated. Next, the material representation of the X-LS method is applied to the multi material topology optimization problem. The numerical implementation method is then described. Finally, the proposed method is applied to two- and three-dimensional stiffness maximization problems and the design problem of compliant mechanisms. These numerical applications demonstrate the validity and usefulness of the proposed method.

\section{Extended level set method}\label{sec: xls Formula}
\begin{figure}[H]
	\centering
	\includegraphics[width=9cm]{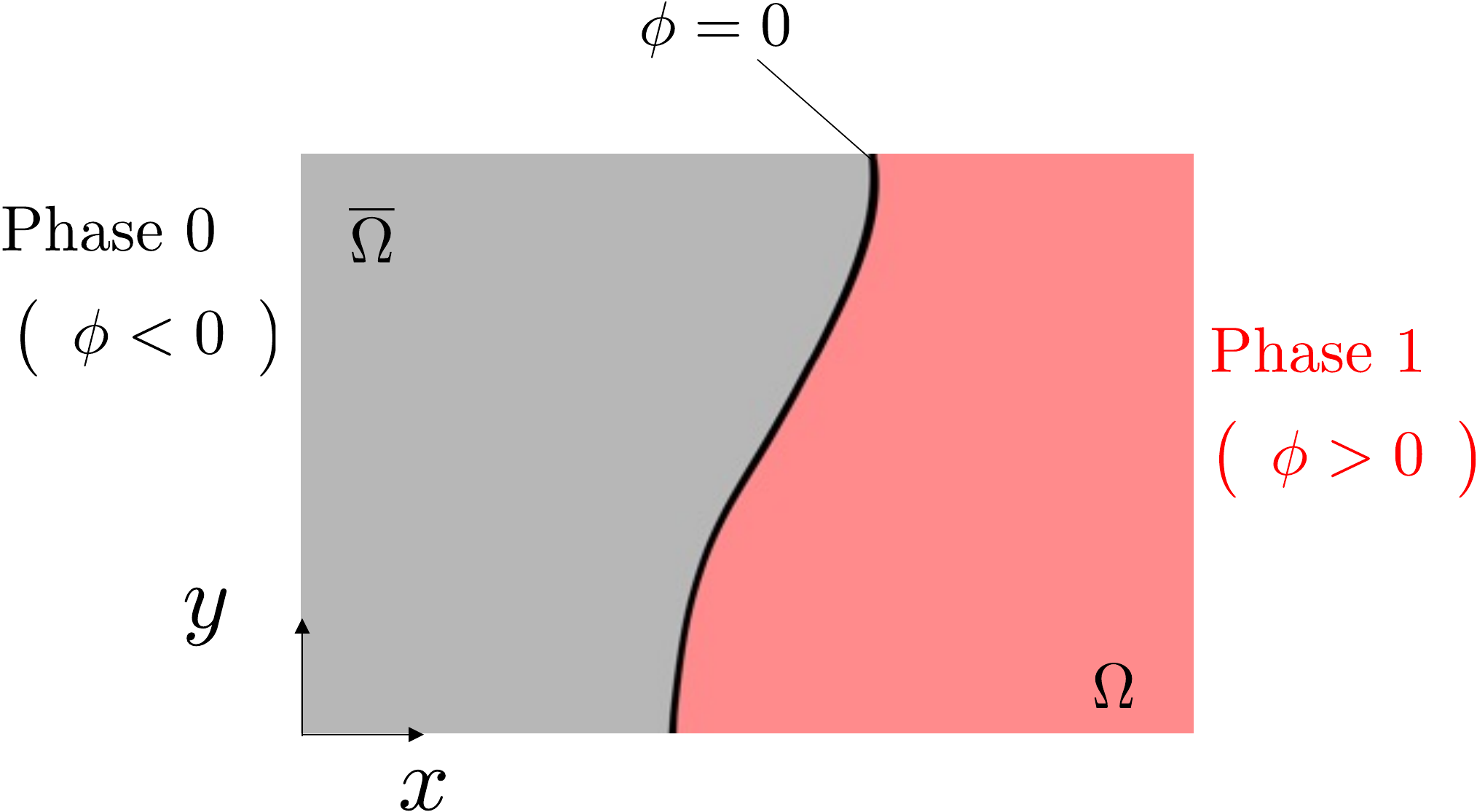}
	\caption{
		Concept of the (original) level set method. The solid line represents the zero isosurface ($\phi=0$) of the level set function. In the regions of phases 1 and 0 (shaded red and gray, respectively), the level set function is positive ($\phi>0$) and negative ($\phi<0$), respectively. Note that the zero isosurface corresponds to the boundary between the two phases.   
	}
	\label{fig: original ls concept}       % Give a unique label
\end{figure}

\begin{figure}[H]
	\centering
	\includegraphics[width=12cm]{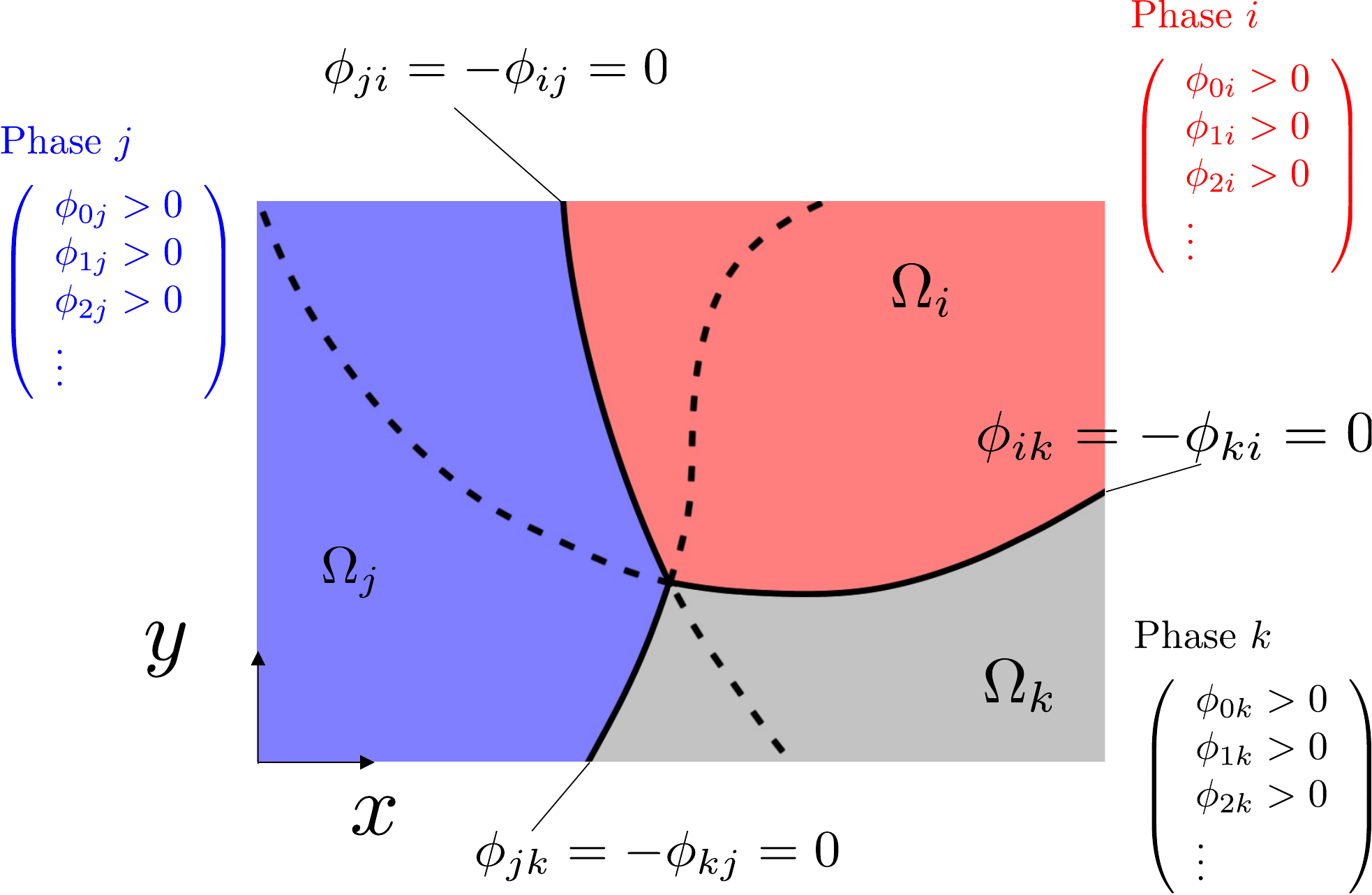}
	\caption{
		Concept of the X-LS method. Different phases in the domain are represented by different X-LS functions. Each zero level set of the X-LS function $\phi_{ij}=0$ corresponds to the boundary between phases $i$ and $j$. The solid and dashed lines are the zero isosurfaces coincident and not coincident with the actual boundaries, respectively. In the domains of phases $i, j$, and $k$ (shaded red, blue, and gray, respectively), all the level set functions $\phi_{li},\phi_{lj}$, and $\phi_{lk}$, respectively, are positive.	($l$ is not equal to $i,j$, and $k$, respectively.)
	}
	\label{fig: xls concept}       % Give a unique label
\end{figure}

In the two-phase case (e.g., cavity + single material), the level set method defines a scalar function called the level set function $\phi(\x)~~(\x\in D)$ in a domain $D$. The two-phase boundary is then cryptically expressed as the zero isosurface of the level set function, i.e., $\phi(\x)=0$. Fig. \ref{fig: original ls concept} shows the concept of the original two-phase level set method. The phases in domains $\Omega$, and $\overline\Omega$, (shaded in different colors) are separated by the zero isosurface (solid line). 
In the present paper, the level set method is extended to $M$ phases. The proposed X-LS method is conceptualized in Fig. \ref{fig: xls concept}. We define $M\times M$ scalar X-LS functions $\phi_{ij}:\mathbb R^{d} \to \mathbb R~~(i,j = 0,1,2,\ldots,M-1)$ in a domain $D \in \mathbb R^{d}$ (where $d$ is the spatial dimension). Each zero level set of the X-LS function  $\phi_{ij}=0$ corresponds to the boundary between phases $i$ and $j$. The dashed and solid lines in Fig. \ref{fig: xls concept} are the zero isosurfaces of the level set functions along the actual boundaries of the phases and within the phases, respectively. In the domain $\Omega_i$ of phase $i$, the level set functions $\phi_{li}$ are positive for all $l$ in $\{0,1,\ldots,i-1,i+1,\ldots,M-1\}$. Similarly, in the domains $\Omega_j$ and $\Omega_k$ of phases $j$ and $k$, respectively, the level set functions $\phi_{lj}$ and $\phi_{lk}$ are positive for all $l$ in $\{0,1,\ldots,j-1,j+1,\ldots,M-1\}$ and $\{0,1,\ldots,k-1,k+1,\ldots,M-1\}$, respectively.

%\textcolor{blue}{（以後もi,j,kが使われていますが，暗に$i\neq j$などが仮定されているように思います．面倒とは思いますが，添字の意味をその都度表すようにしてください．）}

%以下のように特性関数を定義する
Here, we define the characteristic functions $\psi_m(\bm x)$ on domain $\Omega_m\quad(m\in \{0,1,\ldots,M-1\})$ as follows:
\begin{align}
	\psi_m(\bm x) = \begin{cases}
		1\qquad \text{if}\quad&\bm x \in \Omega_m\\
		0\qquad \text{if}\quad&\bm x \in D\setminus\Omega_m
		\end{cases}.
\end{align}
In terms of the X-LS functions $\phi_{ij}$, the characteristic function of each phase is given as
\begin{align}
	\psi_m = \prod_{i\ne m} H(\phi_{im}),\label{eq: psi}
\end{align}
where $H(s)$ is the Heaviside function defined as follows:
\begin{align}
	H(s) =  \begin{cases}
		1\qquad \text{if}\quad&s\ge0\\
		0\qquad \text{if}\quad&s<0
	\end{cases}.\label{eq: heaviside def}
\end{align}
The multiple phases expressed by Eq. (\ref{eq: psi}) are valid only when certain requirements of the X-LS requirements are met. These requirements are described below.

First, from the definition of the X-LS function, $\phi_{ii}~~(i\in \{0,1,\ldots,M-1\})$ are not variables of the boundary surfaces of any two phases and may take any value. 

%Second, 
%%\textcolor{blue}{both} 
%both $\phi_{ij}$ and $\phi_{ji}$ are the level set functions that determine the boundary between phase $i$ and phase $j$.
%%either $\phi_{ij}=0$ iso value surface and $\phi_{ji}=0$  iso value surface corresponds to the boundery suface between material $i$ and material $j$. 
%So we assume X-LS functions satisfy following equations:
%\begin{align}
%	\phi_{ij} = -\phi_{ji}~~(i\ne j).\label{eq: phi ij}
%\end{align}
%Then, the boundary surfaces obtained from $\phi_{ij}=0$ and $\phi_{ji}=0$ are identical.
%By imposing these constraints and excluding the components $\phi_{ii}$, the X-LS function has $M(M-1)/2$ independent components.
Second, the level set functions $\phi_{ij}$ and $\phi_{ji}$ determine the boundary between phases $i$ and $j$. Therefore, the boundary surfaces obtained from $\phi_{ij}=0$ and $\phi_{ji}=0$ should be identical. To satisfy this requirement, we assume the following constraint on the X-LS functions:
\begin{align}
	\phi_{ij} = -\phi_{ji}~~(i\ne j).\label{eq: phi ij}
\end{align}

Third, at each coordinate $\x \in D$, one of the $M$ phases must be assigned, so the X-LS functions should satisfy 
\begin{align}
\sum_m\psi_m =  1.\label{eq: phi sum}
\end{align}%
\section{Multi-material topology optimization} \label{sec: mmto}
We now apply the proposed multiphase representation method to the multi-material topology optimization problem. First, we formulate the following optimization problem, which obtains the configuration of $M$ materials:
\begin{align}
	\min_{\psi_m}\quad&J(\psi_m, U)\nonumber\\
	\mathrm{subject~to}\quad&\mathrm{Governing~equations~for}~U(\psi_m),\nonumber\\
	&g_k(\psi_m, U)\le0.
	\label{eq: optimization problem}
\end{align}%
In these expressions, $U$ is a state variable in the target system, $J$ is the objective function, and $g_k~(k=0,1,2,\ldots)$ are constraint functions. In this research, the domain was assumed as a fixed design domain $D$.

\subsection{Introduction of the reaction-diffusion equation}
The optimization problem in Eq. (\ref{eq: optimization problem}) is difficult to solve directly; therefore, it is usually solved by specifying and updating an appropriate initial value. Following the method of Yamada et al. \cite{yamada2010topology}, we introduce a fictitious time $t$ and replace the topology optimization problem [Eq. (\ref{eq: optimization problem})] with the time evolution problem of level set functions, which is based on reaction-diffusion equations. The time evolution of the level set function $\phi_{ij}$ for materials $i$ and $j$ is given by
\begin{align}
	\frac{\partial \phi_{ij}}{\partial t}=\frac{-\mathcal{D}_{i j} J-C^{\text{ALL}}\sum_k\lambda_k\mathcal{D}_{ij}g_k~}{C_{ij}}+\tau_{ij} L^2\nabla^2 \phi_{ij}&&\mathrm{in} ~D,\nonumber\\
	\phi_{ij}=-1&&\mathrm{on} ~\Gamma_i,\nonumber\\
	\phi_{ij}=1&&\mathrm{on} ~\Gamma_j,\nonumber\\
	\bm n^{(D)} \cdot \nabla\phi_{ij}=0&&\mathrm{on} ~\partial D \backslash(\Gamma_i\cup\Gamma_j),
	\label{eq:reaction-diffusion}
\end{align}
where $\mathcal{D}_{i j} J$ is the extended topological derivative (X-TD) defined in subsec. \ref{sec:Extended topological derivative} and $\partial D$ is the boundary of the fixed design domain $D$. $\Gamma_i$ and $\Gamma_j$ are the boundaries of the fixed design domain for materials $i$ and $j$, respectively, $\bm n^{(D)}$ is the unit normal vector of the boundary of the fixed design domain. Where the material is specified, the boundaries are subjected to Dirichlet boundary conditions. Where the material is not specified or is specified but is neither of $i$ or $j$, the boundaries are subjected to Neumann boundary conditions to prevent influence from outside the fixed design domain. $\lambda_k$ is a control multiplier determined to satisfy the constraint $g_k\le0$. The second term is a regularization term that smooths the boundary shape. $\tau_{ij}>0$ are regularization parameters that control the strength of the regularization. These regularization parameters are set to be symmetric; that is, $\tau_{ij}=\tau_{ji}$. $L$ is the characteristic length of the design domain $D$ and $C^{\text{ALL}}$ and $C_{ij}$ are coefficients for normalizing the sensitivities, respectively defined as follows:
\begin{align}
	C_{ij}=\frac{\int_D  |-\mathcal{D}_{i j} J| \d\Omega} {\int_D \d\Omega~},\label{eq:c ij}\\
	C^{\text{ALL}}=\sum_{i,j}C_{ij}.
\end{align}
To concentrate the effect of regularization near the boundary, following the two-material case \cite{yamada2010topology}, X-LS functions are normalized by the following side constraint:
\begin{align}
	-1 \le \phi_{ij} \le 1.\label{eq: phi limit}
\end{align}
%% to make Eq. (\ref{eq:reaction-diffusion}) a dimensionless equation.
Note that the constraints on the X-LS function [Eqs. (\ref{eq: phi ij}) and (\ref{eq: phi sum})] must be satisfied in the optimal structure. The constraints are discussed in Sec. \ref{sec: xls Formula} and subsec. \ref{sec:Approximation}.

\subsection{Extended topological derivative (X-TD)}\label{sec:Extended topological derivative}
We now construct the extended topological derivatives, the sensitivities that update the X-LS functions described in Sec. \ref{sec: mmto}. In the material representation of the X-LS method, $\phi_{ij}$ is positive and negative in the regions containing materials $j$ and $i$, respectively. In the other regions, the material remains unchanged for any value of $\phi_{ij}$. Therefore, in a structure satisfying Eqs. (\ref{eq: phi ij}) and (\ref{eq: phi sum}), optimality is sufficiently satisfied if the objective function increases when $\phi_{ij}$ increases in material $i$ and when $\phi_{ij}$ decreases in material $j$. When $\phi_{ji}$ ($\phi_{ij}$) decreases in material $i$ ($j$), part of that material will be replaced by material $j$ ($i$). Based on the concept of topological derivatives, we thus introduce X-TD as design sensitivity and update the X-LS functions. Let $\mathcal{D}_{ij} J$ denote the X-TD of the X-LS function $\phi_{ij}$. $\mathcal{D}_{ij} J$ in the multi-material topology optimization problem is then simply expressed as
\begin{align}
	\mathcal{D}_{ij} J
	={D^\text{T}}_{i\to j}J-{D^\text{T}}_{j\to i}J
		\label{eq:definition of X-TD},
\end{align}
%\textcolor{blue}{（X-TDをこのような定義にした動機を軽く説明してください．例えば，X-TDの正負はどんな動きに対応しますか？）}
where ${D^\text{T}}_{i\to j} J$ is the traditional topological derivative representing the rate of change of the objective function when an inclusion of material $j$ with small radius $\varepsilon$ is inserted at coordinate $\x$ in domain $\Omega_i$. It is defined as follows:
\begin{equation}
	{D^\text{T}}_{i\to j}J(\x) =\begin{cases}
		\displaystyle\lim_{\varepsilon \to +0} 
		\frac
		{J({\mathbb C^\varepsilon}_{i\to j}(\bm z))-J(\mathbb C^\text{PRE}(\bm z))}
		%	{J(\Omega_i-\Omega_\varepsilon,\Omega_j+\Omega_\varepsilon)-J(\Omega_i,\Omega_j)}
		{V(\varepsilon)}\qquad&(\x \in \Omega_i)\\
		0\qquad&(\x \notin \Omega_i)
	\end{cases} \label{eq:td},
\end{equation}
where $V(\varepsilon)$ is a function of radius $\varepsilon$ and is proportional to the volume (in a three-dimensional problem) or area (in a two-dimensional problem), and $\mathbb C^\text{PRE}(\bm z)$ and $\mathbb {C^\varepsilon}_{i\to j}(\bm z)$ are functions representing the distribution of material properties in multiple materials with and without the inclusion domain. In $\bm z \in D$, these functions are respectively given as
\begin{align}
		\mathbb C^\text{PRE}(\bm z) &= \mathbb C_i\chi_{\Omega_i}+ \mathbb C(\bm z)(1-\chi_{\Omega_i}),\\
		{\mathbb C^\varepsilon}_{i\to j}(\bm z) &= \left( \mathbb C_j\chi_{\Omega_\varepsilon}+\mathbb C_i(1-\chi_{\Omega_\varepsilon})\right) \chi_{\Omega_i}
		+ \mathbb C(\bm z)(1-\chi_{\Omega_i}),
\end{align}
where $\mathbb C_k$ is the material constant of material $k$
and $\mathbb C(\bm z)$ represents the spatial distribution of material constant in the external region where the inclusion domain is not placed.
$\chi_{\Omega_*}(\bm z)$ is a characteristic function indicating the domain $\Omega_*$, defined as follows:
\begin{align}
		\chi_{\Omega_*}(\bm z)=\begin{cases}
			1\quad&\mathrm{if}\quad\bm z\in \Omega_*\\
			0\quad&\mathrm{if}\quad\bm z\notin \Omega_*
			\end{cases}.
\end{align}

\subsubsection{Behavior of $\phi_{ij}$ when $t\to \infty$}\label{sec: behavior of phi}
Here, we consider the behavior of a fictitious time evolution of the X-LS functions, which follows the reaction-diffusion equation [Eq. (\ref{eq:reaction-diffusion})].
%\textcolor{blue}{（hypotheticalではなくfictitiousがよいと思います．hypotheticalは仮定・仮説の意味合いが強そうです．fictitousには想像上の，架空の〜等の意味があり，どちらかというとこちらの意味合いが良いのではと思います．）}
\begin{thm}\label{thm:ij}
	Assume that Eq. (\ref{eq: phi ij}) is satisfied at fictitious time $t=0$. Eq. (\ref{eq: phi ij}) is satisfied over all fictitious times $t>0$.
\begin{proof}

First, Eq. (\ref{eq:reaction-diffusion}) can be simplified as follows:
\begin{align}
	\frac{\partial \phi_{ij}}{\partial t}
	&=\frac{-\mathcal{D}_{i j} J-C^{\text{ALL}}\sum_k\lambda_k\mathcal{D}_{ij}g_k~}{C_{ij}}+\tau_{ij} L^2\nabla^2 \phi_{ij}\nonumber\\
	&=:D_{ij}\mathcal L +\tau_{ij} L^2\nabla^2 \phi_{ij}\label{eq: organized rde}.
\end{align}
%where $D^{ij}\mathcal L$ is the driving force of $\phi_{ij}$.
From Eq. (\ref{eq:definition of X-TD}), we have;
\begin{align}
	D_{ij}\mathcal L =-D_{ji}\mathcal L.
\end{align}
%\textcolor{blue}{Assuming $\phi_{ij}(t^*)=-\phi_{ji}(t^*)$ on $t=t^*$}
Assuming that $\phi_{ij}(t^*)=-\phi_{ji}(t^*)$ on $t=t^*$, Eq. (\ref{eq:reaction-diffusion}) can be transformed as follows:
\begin{align}
	\frac{\partial \phi_{ij}}{\partial t}(t^*)
%	&=\frac{-\mathcal{D}^{i j} J-C_{\text{ALL}}\sum_k\lambda_k\mathcal{D}^{ij}g_k~}{C_{ij}}+\tau_{ij} L^2\nabla^2 \phi_{ij}\nonumber\\
%	&=\frac{\mathcal{D}^{j i} J+C_{\text{ALL}}\sum_k\lambda_k\mathcal{D}^{ji}g_k~}{C_{ji}}-\tau_{ji} L^2\nabla^2 \phi_{ji}\nonumber\\
	&=-D_{i j} \mathcal L+\tau_{ij} L^2\nabla^2 \phi_{ij}\nonumber\\
	&=D_{j i} \mathcal L-\tau_{ji} L^2\nabla^2 \phi_{ji}\nonumber\\
	&=-	\frac{\partial \phi_{ji}}{\partial t}(t^*).
\end{align}
We then get
\begin{align}
	\phi_{ij}(t^*+\d t)
	&=\phi_{ij}(t^*)+\d t\frac{\partial \phi_{ij}}{\partial t}(t^*)\nonumber\\
	&=-\phi_{ji}(t^*)-\d t\frac{\partial \phi_{ji}}{\partial t}(t^*)\nonumber\\
	&=-\phi_{ji}(t^*+\d t).
\end{align}
Under the assumption  $\phi_{ij}(t=0)=-\phi_{ji}(t=0)$, Theorem \ref{thm:ij} is proven correct by induction.
\end{proof}
\end{thm}

\subsection{Application to linear elasticity problems}
The validity and utility of the proposed method was evaluated on three problems. We consider static systems comprising several linear homogeneous isotropic elastic materials.

\begin{figure}[H]
	\centering
	\includegraphics[width=10cm]{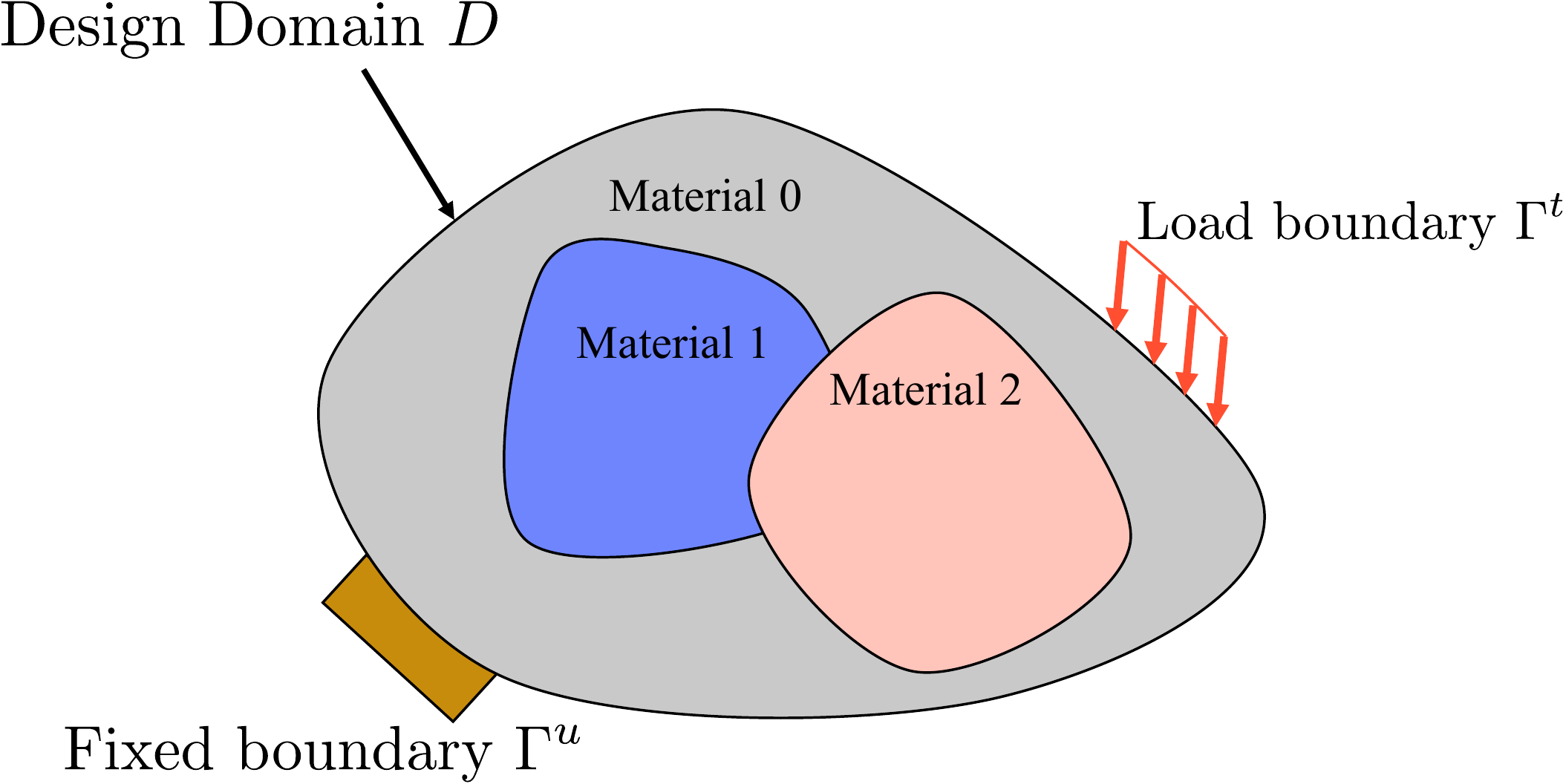}
	\caption{Design domain $D$, material domains, and boundaries in the minimal mean compliance problem}
	\label{fig: stiff}
\end{figure}

First, the proposed method was applied to the minimum mean-compliance problem. A fixed design domain $D$ was composed of domains $\Omega_m$ of materials $m$, with fixed displacement at boundary $\Gamma^u$ and traction $t_i$ imposed at boundary $\Gamma^t$. The displacement field was denoted by $u_i$ in the static equilibrium state. Fig. \ref{fig: stiff} shows the domains and boundary conditions in this problem.
The minimum mean compliance problem was then formulated as follows:
\begin{equation}
	\begin{array}{lll}
		\displaystyle\inf_{\phi_{mn}}\qquad& \displaystyle J_1=\int_{\Gamma^t} t_i u_i \d\Gamma,&\\
		\mathrm{subject~to}\qquad
		& C_{ijkl}u_{k,lj}=0\qquad&\mathrm{in}~D, \\
		&u_{i}=0  &\mathrm{on}~\Gamma^{u}, \\
		&\sigma_{i j} n_{j}=t_{i} &\mathrm{on}~\Gamma^{t},\\
		&g_{\text{V}_m} = \dfrac{~\int_{D} \psi_m \text{d}\Omega~}{\int_{D} \text{d}\Omega} - {V^\text{max}}_{m} \le0&.
	\end{array}\label{eq: J 1}
\end{equation}
Here, the indices $i,j,k$, and $l$ follow the summation convention and the indices after the comma denote the partial derivative of the coordinate components. $\sigma_{ij}$ is the stress tensor, $J_1$ is the objective function, and $g_{\text{V}_m}$ is the volume constraint function of material $m$. ${V^\mathrm{max}}_{m}$ is the maximum volume ratio, defining the upper limit of the ratio of the design area occupied by material $m$. 
$C_{ijkl}(\bm x)$ is the elasticity tensor of multiple materials, defined in terms of the characteristic function $\psi_m$ of each material $m$ and the single material elasticity tensor $C^{(m)}_{ijkl}$ as follows:
\begin{equation}
	C_{ijkl}=\sum^{M-1}_{m=0}\psi_m C^{(m)}_{ijkl}
\end{equation}
The traditional topological derivative of $J_1$ corresponding to the phase change from $a$ to $b$ is obtained as
\begin{align}
	{D^\text{T}}_{a\to b}J_1(\bm x)&=\frac{1}{2} u_{i,j}(\x) \mathcal{A}_{ijkl} u_{k,l}(\x),\label{eq: traditional DT for J1}
\end{align}
where 
the 4-rank tensor $\mathcal{A}_{ijkl}$ is called an elastic moment tensor.

$\mathcal{A}_{ijkl}$ is given by Eq. \ref{eq: 3demt} in a three-dimensional problem \cite{bonnet2013topological}
\begin{align}
	\mathcal{A}_{ijkl}&:=\frac{4 \pi}{3}\left[3 \kappa_a \frac{\Lambda_{1}-1}{1+\zeta_{1}\left(\Lambda_{1}-1\right)} \mathcal{J}_{ijkl}+2 \mu_a \frac{\Lambda_{2}-1}{1+\zeta_{2}\left(\Lambda_{2}-1\right)} \mathcal{K}_{ijkl}\right],\label{eq: 3demt} 
\end{align}
%また，2次元平面応力問題においては以下の様になる(Giusti et al., 2016)．
%以下の式ではC_pqklを除いた部分がpolarization tensor
and by Eq. \ref{eq: 2demt} in a two-dimensional plane stress problem \cite{giusti2016topological}
\begin{align}
	\mathcal{A}_{ijkl}&:=\frac{-1}{\beta\Lambda_3+\eta_1}\nonumber\\
	&\times\left[(1+\beta)(\eta_1-\Lambda_3)\mathcal{I}_{ijpq}+(\alpha-\beta)\frac{\Lambda_3(\Lambda_3-2\eta_3)+\eta_1\eta_2}{\alpha\Lambda_3+\eta_2}\mathcal J_{ijpq} \right]\nonumber\\
	&\times\left(
	\frac{E_a}{1+\nu_a}\mathcal{I}_{pqkl}+\frac{2E_a\nu_a}{1-\nu_a^2}\mathcal{J}_{pqkl}\right),\label{eq: 2demt} 
\end{align}
where the indices $p$ and $q$ follow the summation convention, and the constants $\kappa_m, \alpha,  \beta,$
$\Lambda_{1}, \Lambda_{2},  \Lambda_{3}, $
$\zeta_{1}, \zeta_{2},\eta_{1}, \eta_{2},$ and $\eta_{3}$ are respectively defined as
\begin{align}
	\kappa_m=\frac{E_m}{3(1-2 \nu_m)},~~
	\alpha=\frac{1+\nu_a}{1-\nu_a},~~
	\beta=\frac{3-\nu_a}{1+\nu_a},\nonumber\\
	\Lambda_{1}=\kappa_b / \kappa_a,~~
	\Lambda_{2}=\mu_b/ \mu_a,~~
	\Lambda_{3}=E_b/ E_a,\nonumber\\
	\zeta_1=\frac{1+\nu_a}{3(1-\nu_a)},~~
	\zeta_2=\frac{8-10\nu_a}{15(1-\nu_a)},\nonumber\\
	\eta_1=\frac{1+\nu_b}{1+\nu_a},~~
	\eta_2=\frac{1-\nu_b}{1-\nu_a},~~
	\eta_3=\frac{\nu_b(3\nu_a-4)+1}{\nu_a(3\nu_a-4)+1}.
\end{align}
In these expressions, $d$ is the number of spatial dimensions of the design domain $D$ and $E_m,\nu_m,$ and $\mu_m$ are the Young's modulus, Poisson's ratio, and shear modulus of material $m$, respectively.
%また$\mathcal I_{ijkl},\mathcal J_{ijkl},\mathcal K_{ijkl}$は以下のように定義される4階のテンソルである．
$\mathcal I_{ijkl},\mathcal J_{ijkl}$, and $\mathcal K_{ijkl}$ are respectively defined as follows:
\begin{align}
	\mathcal I_{ijkl}&:=\frac{1}{2}(\delta_{ik}\delta_{jl}+\delta_{il}\delta_{jk}),\\
	\mathcal J_{ijkl}&:=\frac{1}{d}\delta_{ij}\delta_{kl},\\
	\mathcal K_{ijkl}&:=\mathcal I_{ijkl}-\mathcal J_{ijkl},
\end{align}
where $\delta_{ij}$ is the Kronecker delta.

The X-TD of $J_1$ can be estimated by substituting Eq. (\ref{eq: traditional DT for J1}) into Eq. (\ref{eq:definition of X-TD}).
Meanwhile, the X-TDs of the volume constraints $g_{\mathrm{V}_m}$ are derived from the definition of X-TD [Eq. (\ref{eq:definition of X-TD})] as follows:
\begin{align}
	\mathcal{D}_{i j}g_{\text{V}_m}(\x) = -\delta_{im}\psi_i+\delta_{jm}\psi_j.
\end{align}

Next, the proposed method was applied to the optimal design of a compliant mechanism (defining a mechanism with no joints, which converts an applied force into the desired motion through elastic deformation). The fixed design domain $D$ was divided into domains $\Omega_0,\ldots,\Omega_{M-1}$ containing materials $0,\ldots,{M-1}$. The displacement was fixed at boundary $\Gamma^u$ and a traction ${t^\text{in}}_i$ was imposed at boundary $\Gamma^{\text{in}}$. The optimization problem was then formulated as follows:
\begin{align}
	\inf_{\phi_{mn}}~ &J_2 = -\int_{\Gamma^\mathrm{out}} {t^\mathrm{out}}_i \cdot  u_i \d\Omega,\nonumber\\
	\mathrm{subject ~ to}~
	& C_{ijkl}u_{k,lj}=0\qquad&\mathrm{in}~D, \nonumber\\
	&u_{i}=0  &\mathrm{on}~\Gamma^{u}, \nonumber\\
	&\sigma_{i j} n_{j}={k^\mathrm{in}}_{ij} u_{j}+{t^\mathrm{in}}_i &\mathrm{on}~\Gamma^{\mathrm{in}}, \nonumber\\
	&\sigma_{i j} n_{j}={k^\mathrm{out}}_{ij} u_{j} &\mathrm{on}~\Gamma^{\mathrm{out}}, \nonumber\\
	&g_{\text{V}_m} = \dfrac{~\int_{D} \psi_m \text{d}\Omega~}{\int_{D} \text{d}\Omega} - {V^\text{max}}_{m} \le0,\qquad&\label{eq: comp mech}
\end{align}
where ${t^\mathrm{out}}_i$ is a dummy traction vector representing the direction of the specified deformation at output port $\Gamma^\mathrm{out}$. 

Adopting Lazarov's formulation \cite{lazarov2011robust}, 
%\cite{sigmund2009manufacturing}, 
%\cite{sigmund1997design}
the output and input springs with stiffness values ${k^\mathrm{out}}_{ij}$ and ${k^\mathrm{in}}_{ij}$, respectively, were located at boundaries $\Gamma^\mathrm{out}$ and $\Gamma^\mathrm{in}$, respectively. To overcome this spring resistance, the optimal structure will automatically have a certain degree of robustness. The traditional topological derivative of the objective function $J_2$ was computed as 
\begin{align}
	{D^\text{T}}_{a\to b}J_2(\bm x)&=-\frac{1}{2} u_{i,j}(\x) \mathcal{A}_{ijkl} v_{k,l}(\x).
\end{align}
Here, $v_i$ is the adjoint field of $u_i$, which satisfies the following equations:
\begin{align}
 C_{ijkl}v_{k,lj}&=0\qquad&\mathrm{in}~D ,\nonumber\\
v_{i}&=0 \qquad &\mathrm{on}~\Gamma^{u}, \nonumber\\
\sigma_{i j} n_{j}&={k^\mathrm{in}}_{ij} v_{j} \qquad&\mathrm{on}~\Gamma^{\mathrm{in}},\nonumber\\
\sigma_{i j} n_{j}&={k^\mathrm{out}}_{ij} v_{j}-{t^\mathrm{out}}_i \qquad&\mathrm{on}~\Gamma^{\mathrm{out}}.
\end{align}

Finally, the proposed method was applied to the mean compliance and moment of inertia minimization problem. Suppose that the axis of rotation passes through $\bm n^{\mathrm C}$, and let $\bm n$ be a unit vector parallel to the direction of rotation axis. This topology optimization problem can be formulated as follows:
\begin{equation}
	\begin{array}{lll}
		\displaystyle\inf_{\phi_{mn}}\quad& \displaystyle J=\int_{\Gamma_t} t_i u_i \d\Gamma+w J_{\text{I}},&\\
		&J_{\text{I}} = \displaystyle\int_{D}\left( \{
		%		\min_a(||\bm x - \bm x_A+a\bm e_A||^2)
		||\x-\bm n^{\mathrm C}||^2-(\bm x\cdot \bm n)^2
		\} \sum_m\rho_m\psi_m \right)\d\Omega,&\\
		\mathrm{subject~to}\quad
		& C_{ijkl}u_{k,lj}=0\qquad&\mathrm{in}~D, \\
		&u_{i}=0  &\mathrm{on}~\Gamma^{u}, \\
		&\sigma_{i j} n_{j}=t_{i} &\mathrm{on}~\Gamma^{t},\\
		&g_{\text{V}_m} = \dfrac{~\int_{D} \psi_m \text{d}\Omega~}{\int_{D} \text{d}\Omega} - {V^\text{max}}_{m} \le0,\qquad&
	\end{array}\label{eq: J inertia}
\end{equation}
where $w\in \mathbb R_{>0}$ is a weighting factor and
$\rho_m$ is the density of material $m$.

The X-TD of the moment of inertia $J_{\text{I}}$ was derived from definition of X-TD [Eq. (\ref{eq:definition of X-TD})] as follows:
\begin{align}
	\mathcal{D}_{i j}J_{\text{I}}(\x) = \{||\x-\bm n^{\mathrm C}||^2-(\bm x\cdot \bm n)^2\}(\rho_j-\rho_i)(\psi_i+\psi_j).
\end{align}

\section{Numerical implementation}\label{sec:Numerical implementation}

\subsection{Optimization algorithm}
The optimization problems were solved using the following optimization processes: 
\begin{description}
	\item[step 1] Initialize the X-LS functions $\phi_{ij}$
	\item[step 2] Compute the approximated characteristic functions ${\hat\psi}^{'}_m$ from the X-LS function
	\item[step 3] Solve governing and adjoint equations in the FEM simulation
	\item[step 4] Evaluate the objective and constraint functions
	\item[step 5] Calculate the X-TD $\mathcal D_{ij} J$ and control multipliers $\lambda_k$
	\item[step 6] Solve the reaction-diffusion equations using FEM and normalize the X-LS functions to satisfy the side constraints
	\item[step 7] If  the solution has converged, end the optimization calculation. Otherwise, return to Step 2.
\end{description}
\textbf{step~2} approximates the characteristic functions defined in subsec. \ref{sec:ersatz}. FEM was performed in the general purpose finite element analysis software FreeFEM++ \cite{hecht2012new}.

%--------------
%     SECTION
%--------------
\subsection{Approximation of level set functions}\label{sec:Approximation}
%本来式5はトポロジー最適化における制約条件として与えられ、これが最適構造において満たされているように適当な処理が必要であるが、
%節\ref{sec: behavior of phi}で述べたような仮定をおけばほぼ満たされる。
%本研究では、この仮定が満たされるような問題を対象とするため、レベルセット関数に直接制約を与えることはせず、制約を満たす近似的されたレベルセット関数により構造を表現する。

\begin{figure}[H]
	\centering
	\begin{minipage}[t]{0.49\linewidth}
		\centering
		\includegraphics[width=0.86\linewidth]{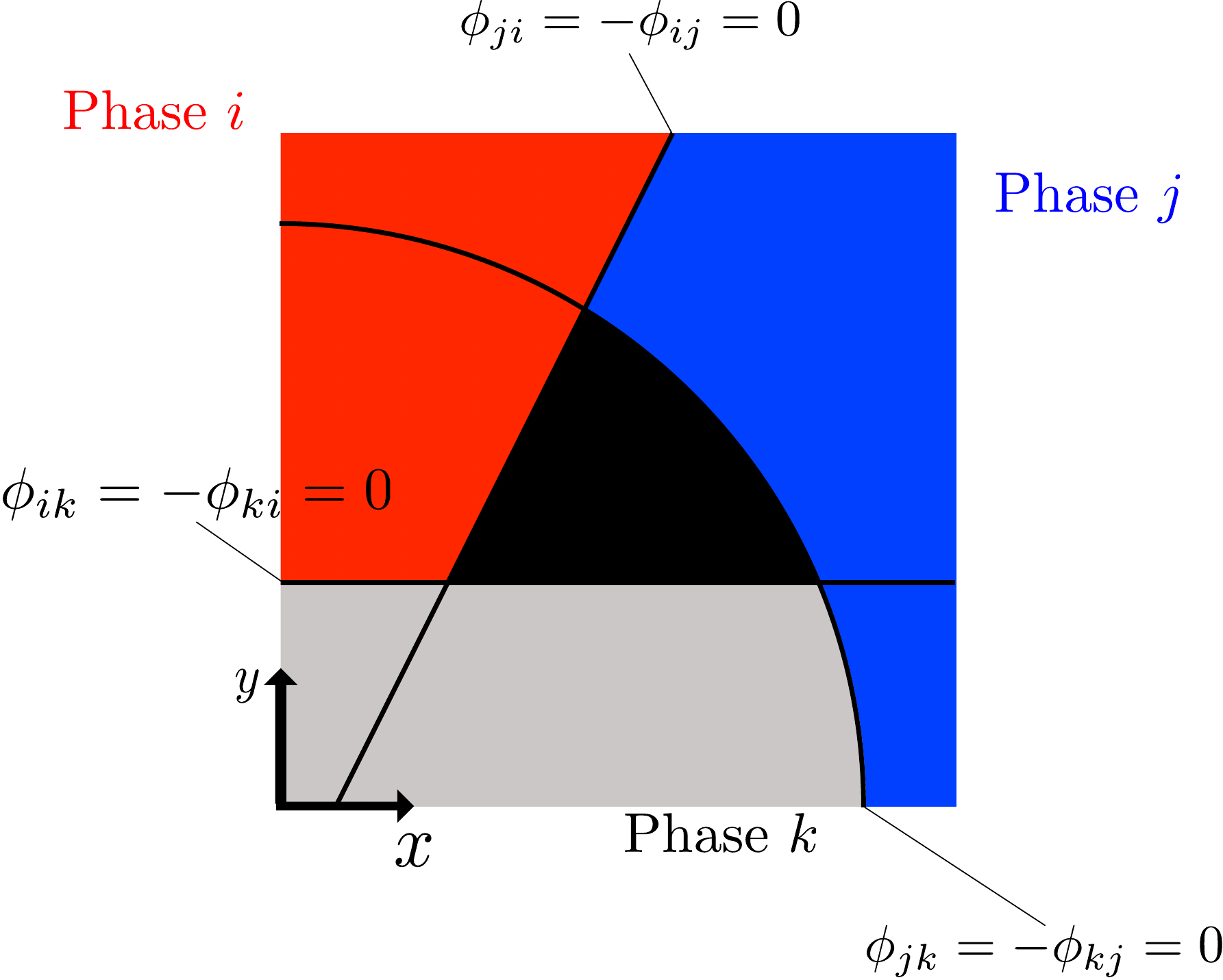}
		\subcaption{before apporoximation}\label{fig: no material}       % Give a unique label
	\end{minipage}
	\begin{minipage}[t]{0.49\linewidth}
		\centering
		\includegraphics[width=0.95\linewidth]{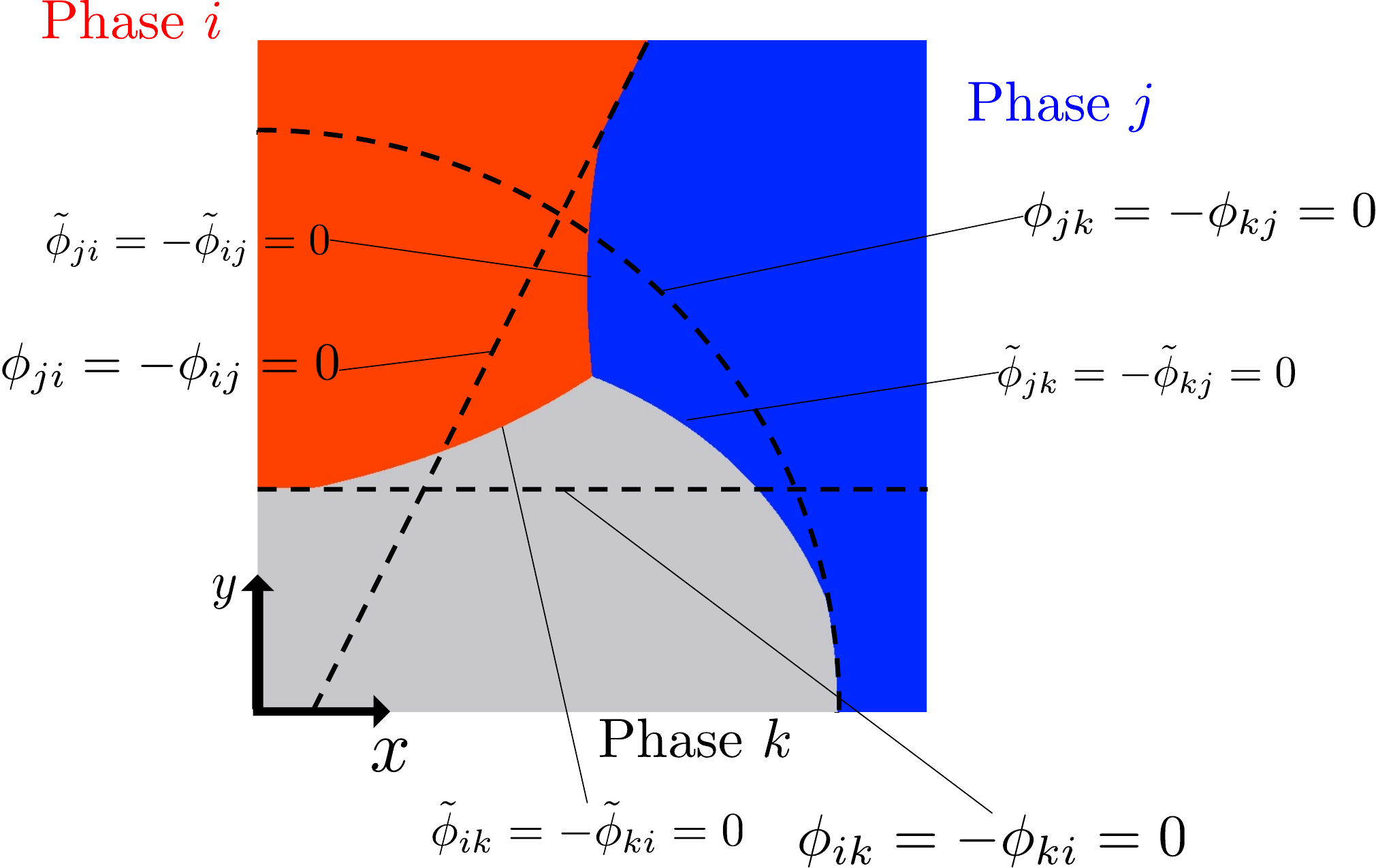}
		\subcaption{after apporoximation}\label{fig: distribution}  
	\end{minipage}
	\caption{
		Example of an area (black region) that cannot be assigned to any phase \subref{fig: no material} and the approximated multiphase distribution \subref{fig: distribution}
	}
	\label{fig: apporoximated distribution}       % Give a unique label
\end{figure}

Eq. (\ref{eq: phi sum}), which constrains the topology optimization problem, must be satisfied in the optimal structure and must be appropriately handled for this purpose. However, in preliminary numerical experiments under the current settings of the optimization problems, this constraint was naturally satisfied almost everywhere over the design domain. Rather than directly constraining the level set functions, we thus represent the material with approximated level set functions that satisfy the constraints in each iteration of the optimization. For example, in Fig. \ref{fig: apporoximated distribution}\subref{fig: no material}, Eq. (\ref{eq: phi sum}) is not perfectly satisfied and one area cannot be assigned to any material. Let the X-LS functions take the following forms:
\begin{align}
	\phi_{ij}=-\phi_{ji}=-3y+6x-1/2,\nonumber\\
	\phi_{jk}=-\phi_{kj}=-4x^2-4y^2+3,\nonumber\\
	\phi_{ki}=-\phi_{ik}=6y-2.
\end{align}
Based on Eq. (\ref{eq: psi}), the red, blue, and gray areas in Fig. \ref{fig: apporoximated distribution}\subref{fig: no material} were assigned to phases $i, j$, and $k$, respectively, and the black area was not assigned to any phase.

To remove the unassigned region, we approximated the characteristic functions defined in Eq. (\ref{eq: psi}) as follows:
\begin{align}
	\hat\psi_m = \prod_{i\ne m} H(\tilde{\phi}_{im}),\label{eq: replaced caracteristic functions}
\end{align}
where $\tilde{\phi}_{im}$ are the approximated X-LS functions, defined as
\begin{align}
	\tilde{\phi}_{im}=\tilde \psi_m - \tilde \psi_i,\label{eq: tilde phi}\\
%	\tilde\psi_m = \prod_{i} \tilde H(\phi_{im}),\label{eq: tilde psi}
	\tilde\psi_m = \prod_{i} (\phi_{im}+1)/2.\label{eq: tilde psi}
\end{align}
In Eq. (\ref{eq: tilde psi}), the functions $\tilde\psi_{m}$ indicate the appearance priority of phase $m$ at coordinate $\x$. The phase $m$ with the largest $\tilde\psi_{m}$ was assigned to coordinate $\x$.

Using Eqs. (\ref{eq: replaced caracteristic functions}), (\ref{eq: tilde psi}), and (\ref{eq: tilde heaviside}), the distribution in Fig. \ref{fig: apporoximated distribution}\subref{fig: no material} was approximated by that in Fig. \ref{fig: apporoximated distribution}\subref{fig: distribution}, where the dotted lines show the boundaries before the approximation. The three approximated boundaries intersected at a single point, and the unassigned region in Fig. \ref{fig: apporoximated distribution}\subref{fig: no material} was divided into the three existing phases. Note that the approximated boundaries of the level set function coincided with the nonapproximated boundaries far from the intersection point.

%\subsection{特性関数の近似}
%本来有限要素法解析では材料境界に沿ったメッシュを生成して解析を行う必要があるが，異種材料の境界で材料定数を滑らかに遷移させることで材料境界を陰的に表現すれば，有限要素を再生成する手間を省略でき，また有限要素解析における数値安定性を確保できる．そのためにまず，特性関数$\psi_m$を以下のように$\tilde\psi_m$に近似する．
\subsection{Ersatz material approach}\label{sec:ersatz}
The ersatz material approach \cite{allaire2004structural} implicitly represents the boundaries of each material as smooth transitions of the material constants at the boundaries. This approach eliminates the computational time and effort of regenerating finite elements and ensures the numerical stability of the finite element analysis. For this purpose, we smoothed out the approximated characteristic function $\hat\psi_m$ as follows:
\begin{align}
	{\hat\psi}^{'}_m = 
	\dfrac
	{\varepsilon^p +\displaystyle\prod_{i\ne m} \tilde{H}((\tilde \phi_{im})/w^p)}{~\displaystyle\sum_{i}\left( \varepsilon^p+ \displaystyle\prod_{j\ne i} \tilde{H}((\tilde\phi_{ji} )/{w^p})\right) ~},
	\label{eq: tilde psi dash}
\end{align}
where $\tilde H$ is the approximated Heaviside function defined by Eq. (\ref{eq: tilde heaviside}), $\varepsilon^p$ is a minimal real number inserted for computational stability, and $w^p$ is a constant that determines the width of the material transition. In this research, $\varepsilon^p$ and $w^p$ were set to $1.0\times10^{-6}$ and $0.2$, respectively.
\begin{align}
	\tilde{H}(s) &=
	\begin{cases}
		0 &(s<-1)\\
		\frac{1}{2}+s
		\left[\frac{15}{16}-s^2 \left(\frac{5}{8}-\frac{3}{16}s^2 \right) \right] \qquad&(-1 \le s \le 1)\\
		1 &(s>1)
	\end{cases}.\label{eq: tilde heaviside}
\end{align}
%式(\ref{eq:mmchi})の分母は，近似された特性関数$\tilde\psi_m$を規格化する役割がある．これは，近似しない場合の特性関数はいずれかが1で他は0であるから，総和は設計領域全体で1であり，それに対応させるためである．

\subsection{Updating scheme of X-LS functions and control multipliers}
Discretizing Eq. (\ref {eq:reaction-diffusion}) over time using the finite difference method, we obtain
\begin{align}
	&\hat\phi_{ij}(\x,t+\Delta t)
	=\phi_{ij}(\x,t)\nonumber\\
	&+{\Delta t} \left(\frac{-\mathcal{D}_{i j} J-C^{\text{ALL}}\sum_k\lambda_k\mathcal{D}_{ij}g_k~}{C_{ij}}+\tau_{ij} L^2\nabla^2 \phi_{ij}(\x,t+\Delta t) \right),
\label{eq:rdenum}
\end{align}
where $\Delta t$ is the time width in one step of the optimization.

Following \cite{tovar2006topology} and other studies, the control multipliers $\lambda_k$ were determined under the proportional-integral-derivative control concept as follows:
\begin{align}
	\lambda_k(t)&= \max\left({K^\text{P}}_k g_k(t),0\right)
	+{g^\text{I}}_k(t) +{K^\text{D}}_k {g^\mathrm{D}}_k(t) ,\label{eq:pid}
\end{align}
where ${g^\text{I}}_k(t)$ and ${g^\mathrm{D}}_k(t)$ are respectively determined as
\begin{align}
	{g^\mathrm{I}}_k(t) &= \max\left({g^\mathrm{I}}_k(t-\Delta t) + \left({K^\text{IP}}_k g_k(t)+{K^\text{ID}}_k {g^\mathrm{D}}_k(t)\right) \Delta t,0\right),\\ \label{eq:pid2}
	{g^\mathrm{D}}_k(t) &=g_k(t)-g_k(t-\Delta t),\\
	{g^\mathrm{I}}_k(-\Delta t)&=0,\\
	g_k(-\Delta t)&=g_k(0).
\end{align}
In Eqs. (\ref{eq:pid}) and (\ref{eq:pid2}), ${K^\text{P}}_k$, ${K^\text{IP}}_k$, and ${K^\text{D}}_k$ are the proportional, integral, and differential gains, respectively, and ${K^\text{ID}}_k$ is a gain that improves the numerical stability.

To satisfy the side constraint of the level set functions given by Eq. (\ref{eq: phi limit}), the X-LS functions were normalized as follows:
\begin{equation}
	\phi_{ij}(\x,t+\Delta t) = \max(-1,\min(1,\hat\phi_{ij}(\x,t+\Delta t))).\label{eq:limnum}
\end{equation}

\subsection{Time-directionally filtered sensitivity}
As is well known, the optimization problem for the compliant mechanism expressed by Eq. (\ref{eq: comp mech}) is prone to numerical instability. Here, we used the following fictitious time-directionally filtered sensitivity
\begin{align}
	\overline{\mathcal D_{ij}J_2(\bm x,t)}&=\int_0^t f(s)\mathcal D_{ij}J_2(\bm x,s)\d s,\label{eq: time-directionally filtered sensitivity}
\end{align}
where $f(s)$ is a weighting function. This method stabilizes the sensitivity of the optimization, which tends to be differently distributed at each iteration, by averaging it over the previous steps. 
To numerically implement the fictitious time-directionally filtered sensitivity, we can simply sum the sensitivities over several previous iterations. The weighted function $f(s)$ is then defined as a simple function that takes a constant value over a certain time interval and 0 at all other times. 
However, to implement this function, we must maintain a large number of topological derivatives $\mathcal D_{ij}J_2(\bm x,s) \quad (s=t, t-\Delta t,  t-2\Delta t, \ldots)$ over a large data volume. 
To avoid this problem, the present research borrows the low-pass filter often used in sensing, where $f(s)$ is defined as
\begin{align}
	f(s)=\frac{\exp(K^{\text{T}}(s-t))}{1-\exp(-K^{\text{T}}\Delta t)}.\label{eq:weighting function}
\end{align}
In Eq. (\ref{eq:weighting function}), the parameter $K^{\text{T}}>0$ controls the amount of past information $\mathcal D_{ij}J_2(\bm x,s) \quad (0<s<t)$ used to update the X-LS functions at fictitious time $t$. The time-directionally filtered sensitivity is discretized over fictitious time as follows:
\begin{align}
	\overline{\mathcal D_{ij}J_2(\bm x,t)}
	=(1-{K^\text{T}}')\overline{\mathcal D_{ij}J_2(\bm x,t-\Delta t)}
	+{K^\text{T}}'\mathcal D_{ij}J_2(\bm x,t),\\
	{K^\text{T}}' = 1-\exp(-K^\text{T}\Delta t).
\end{align}
As seen in Eq. (\ref{eq:weighting function}), when $s$ is sufficiently smaller than $t$, the asymptotic weights are 0. Therefore, the sensitivity can be stabilized by averaging the past sensitivities, while the new sensitivity is kept up to date by ignoring the old sensitivity. Decreasing ${K^\text{T}}'$ enhances the stability of the sensitivity but lengthens the response fictitious time and consequently slows the convergence.

\section{Numerical examples}\label{sec: numerical examples}
This section confirms the utility and validity of the proposed optimization method through numerical examples. In these examples, we consider multi-material optimization with $2 \le M \le 9$ phases of isotropic linear elastic materials. The Young's moduli of the nine materials are listed in Table \ref{tab:materials}. The Poisson ratios were all set to 0.3. In the resulting figures, the areas of the materials are shaded according to their corresponding colors in Table \ref{tab:materials}. The $E_0$ of the voids was assumed to be sufficiently smaller than those of the regions occupied by other materials, following the ersatz material approach \cite{allaire2004structural}.
\begin{table}[h]
	\centering
	\caption{Number assignment, Young's modulus, and corresponding colors of the materials in the numerical examples}
	\begin{tabular}{ccc}\hline
		Material number $m$&Young's modulus $E_m$[GPa]&Color\\ \hline
		0&0.1&gray(2D)/void(3D)\\ 
		1&200&red\\
		2&100&blue\\
		3&150&yellow\\
		4&175&green \\
		5&125&light blue \\
		6&75&orange \\
		7&50&pink \\
		8&25&purple\\ \hline
	\end{tabular}
	\label{tab:materials}
\end{table}

\subsection{Two-dimensional minimum mean compliance problems}\label{sec: 2d comp minim}
\begin{figure}[H]
	\centering
	\includegraphics[width=10cm]{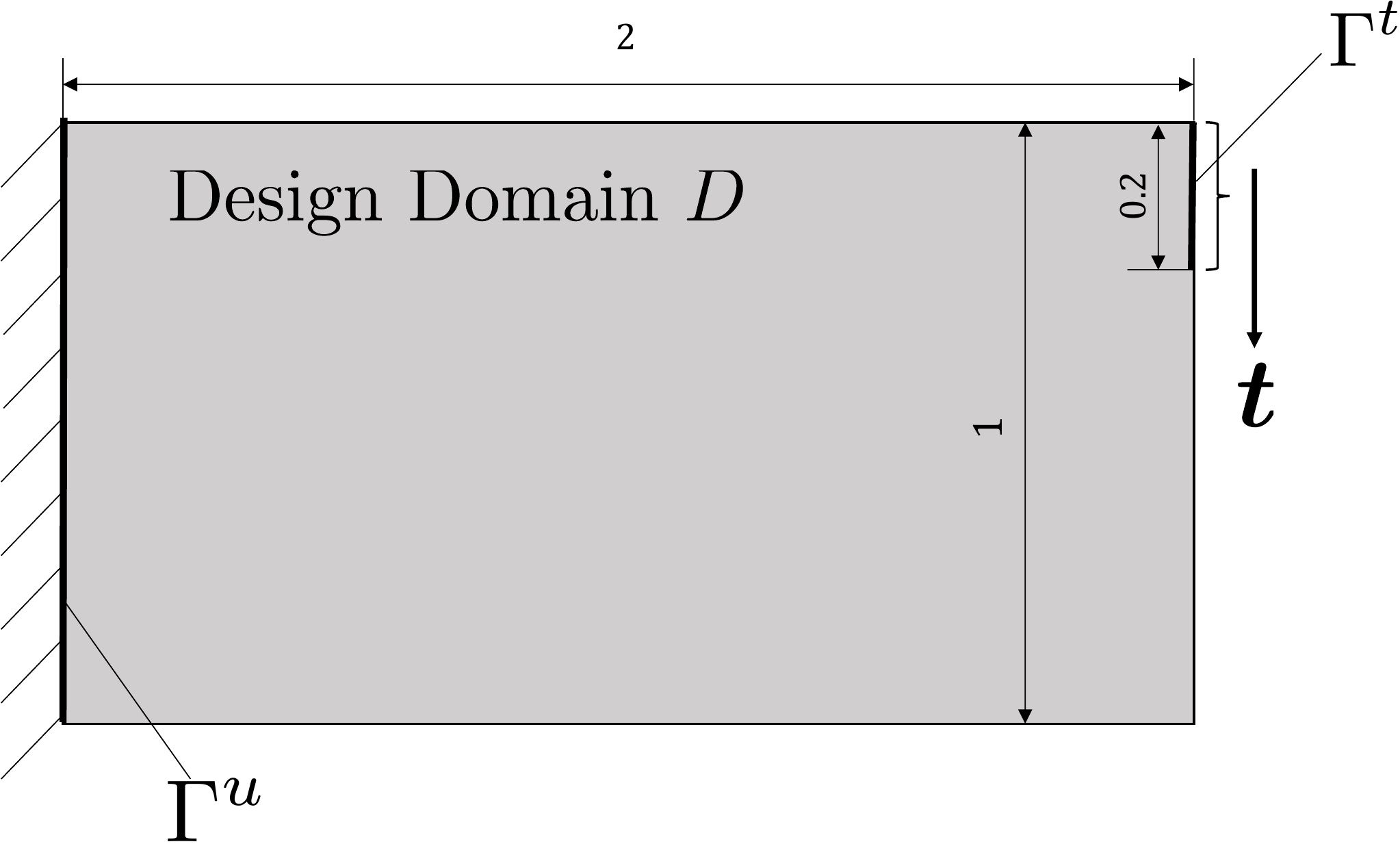}
	\caption{Two-dimensional problem settings}
	\label{fig: problem 2d}
\end{figure}

In this subsection, the proposed optimization method is applied to two-dimensional minimum-compliance problems. Fig. \ref{fig: problem 2d} shows the fixed design domain and the boundary conditions.
No material was specified at the fixed displacement boundary $\Gamma^u$, material 1 was specified at the boundary $\Gamma^t$ subjected to traction force $\bm t$, and material 0 was specified at the other boundaries.
% and the boundary conditions of the level set functions were those in Eq. (\ref{eq:reaction-diffusion}). 
The characteristic length $L$ was set to 1 m.

%\textcolor{blue}{（数値例が多いので，First,  Second,...に対応するようにsubsubsectionを設けるとより分かりやすいと思います．）}
\subsubsection{Examples for various values of $M$}\label{sec:2d M}
\begin{figure}[H]
	\begin{minipage}[b]{0.24\linewidth}
		\centering
		\includegraphics[width=\linewidth]{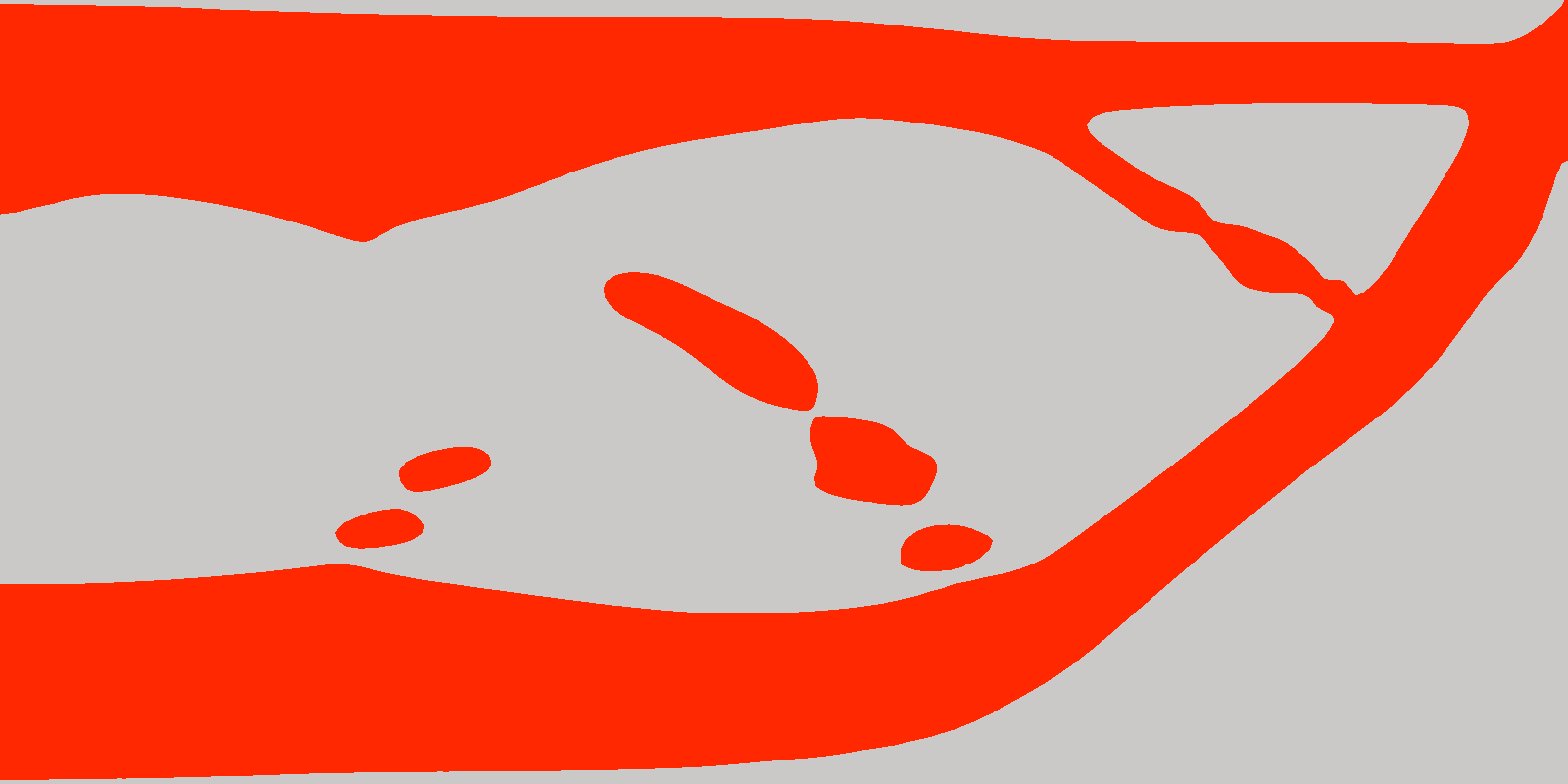}
		\subcaption{Step 10}\label{fig:M2 10}
	\end{minipage}
	\begin{minipage}[b]{0.24\linewidth}
		\centering
		\includegraphics[width=\linewidth]{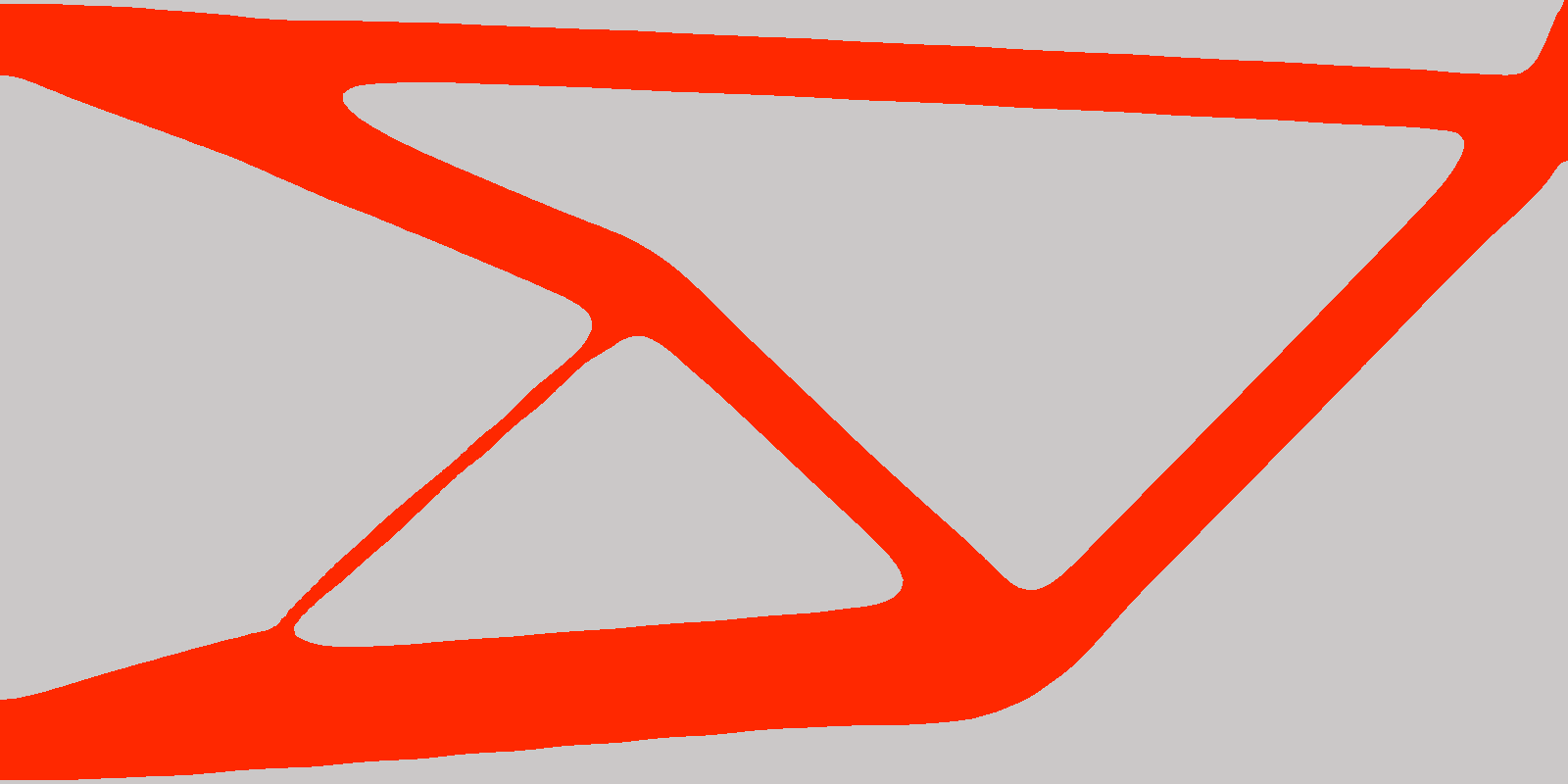}
		\subcaption{Step 50}\label{fig:M2 50}
	\end{minipage}
	\begin{minipage}[b]{0.24\linewidth}
		\centering
		\includegraphics[width=\linewidth]{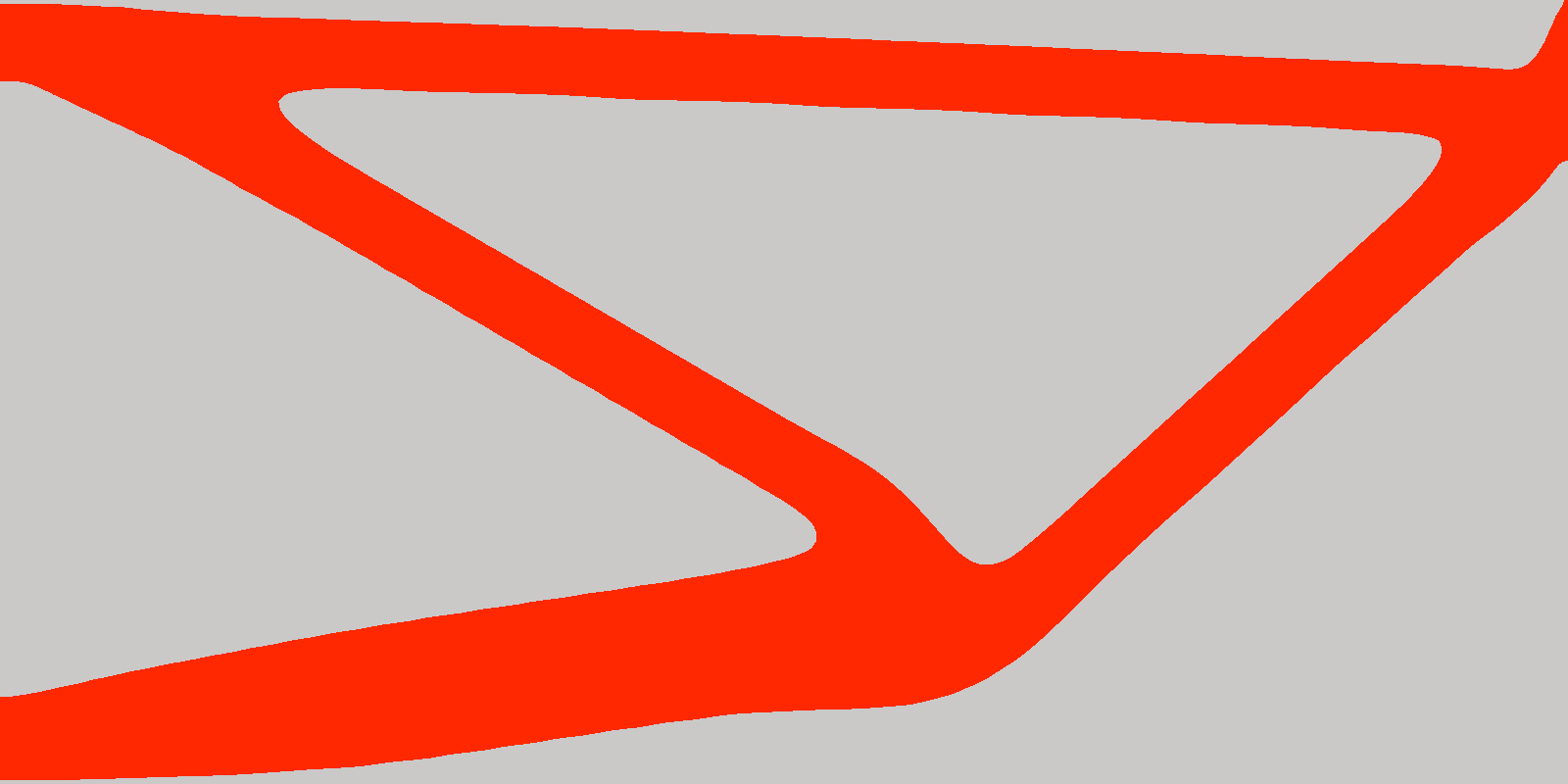}
		\subcaption{Step 200}\label{fig:M2 200}
	\end{minipage}
	\begin{minipage}[b]{0.24\linewidth}
		\centering
		\includegraphics[width=\linewidth]{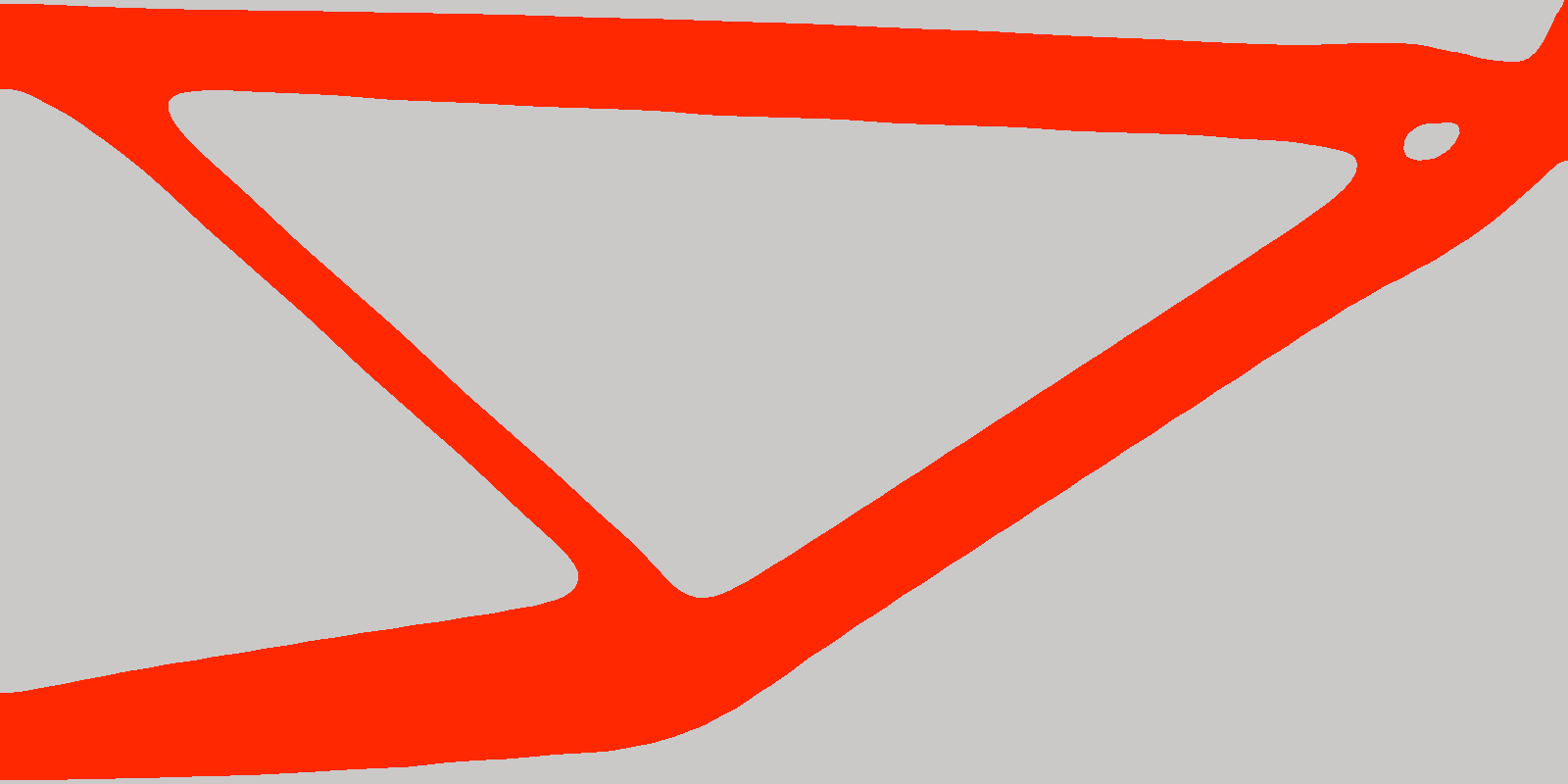}
		\subcaption{Optimal Configration}\label{fig:M2 opt}
	\end{minipage}
	\caption{Intermediate results and optimal configuration of two materials (Case 1)}\label{fig:M2}
\end{figure}

\begin{figure}[H]
	\begin{minipage}[b]{0.24\linewidth}
		\centering
		\includegraphics[width=\linewidth]{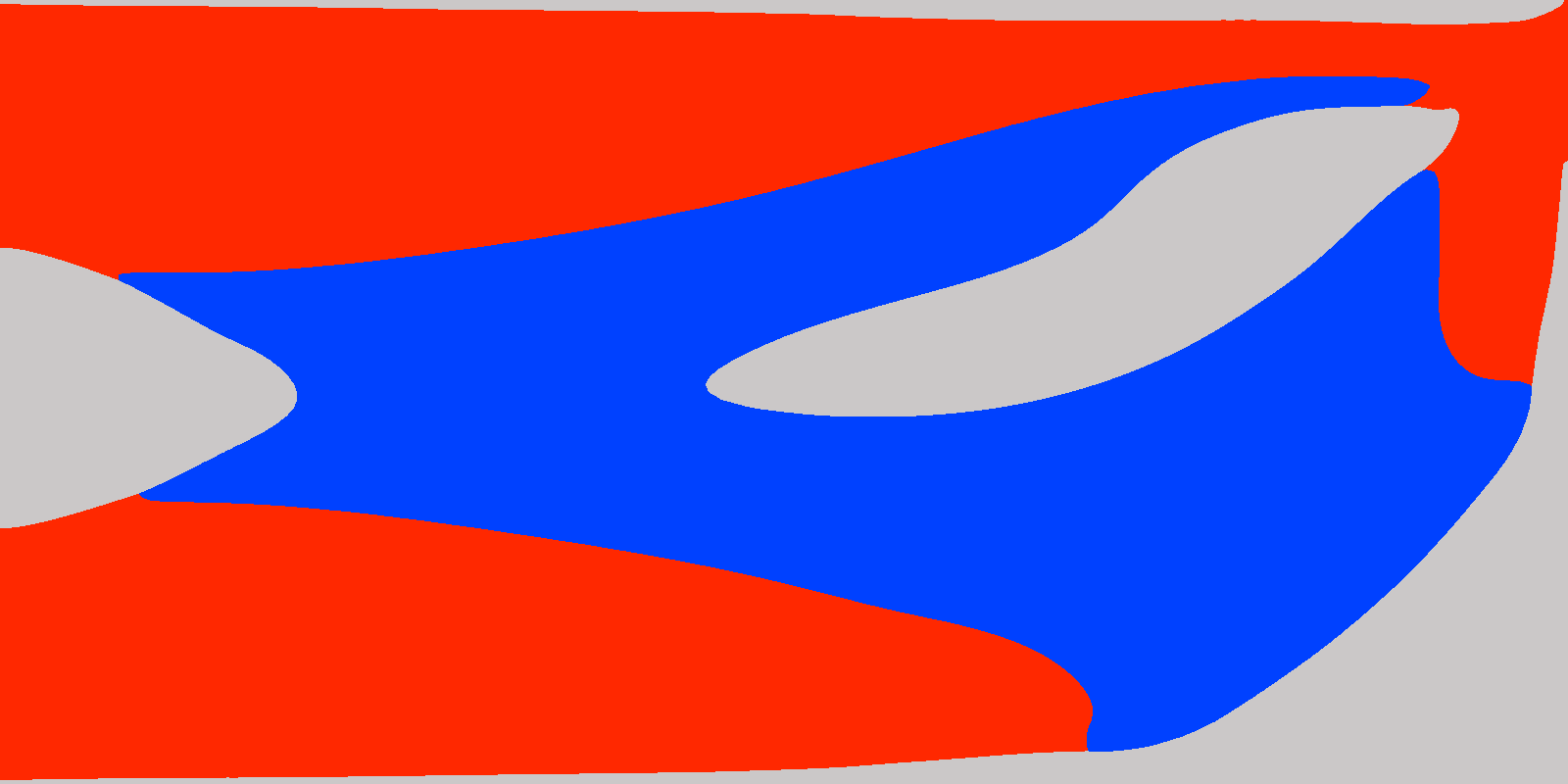}
		\subcaption{Step 10}\label{fig:M3 10}
	\end{minipage}
\begin{minipage}[b]{0.24\linewidth}
	\centering
	\includegraphics[width=\linewidth]{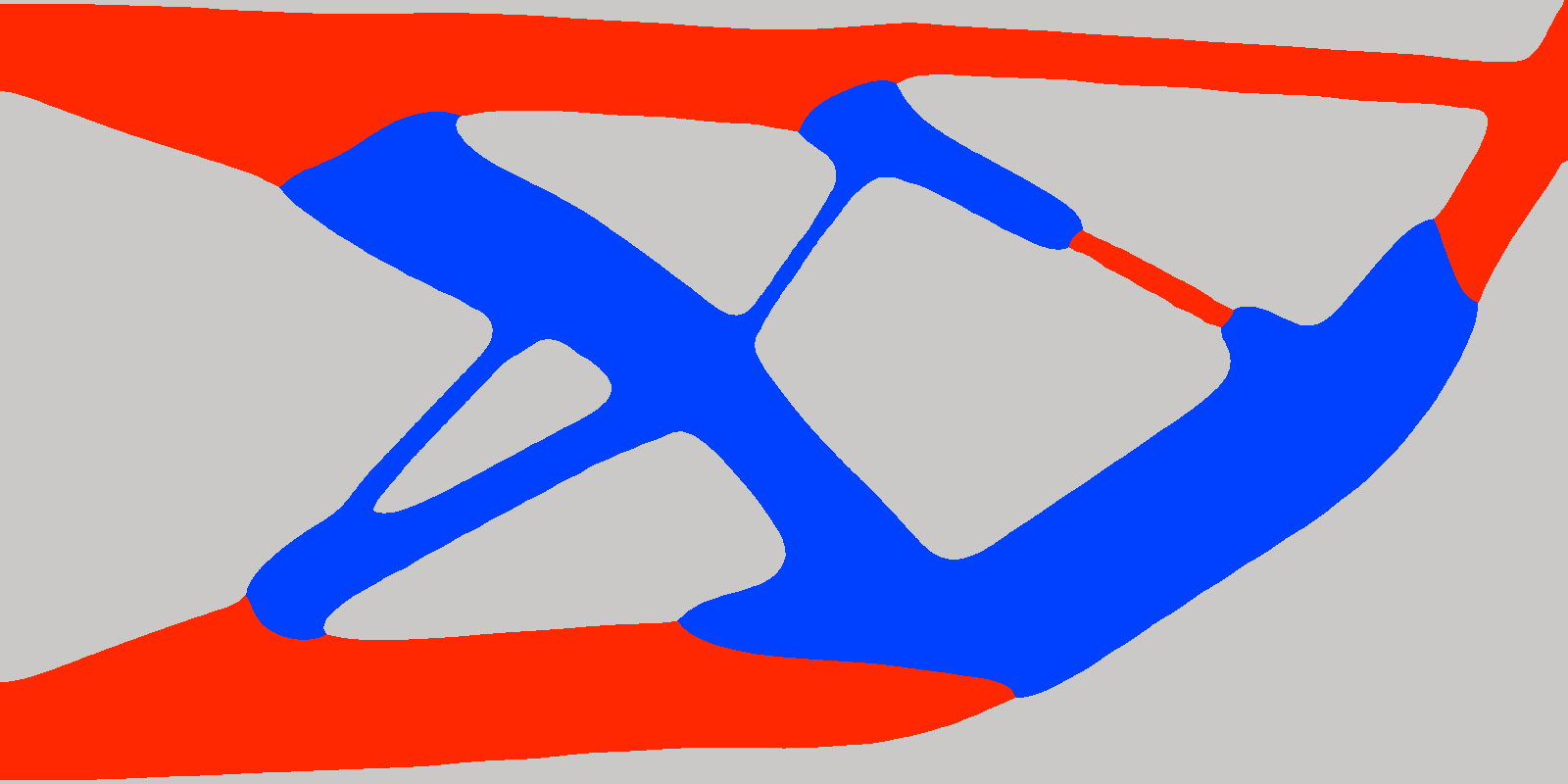}
	\subcaption{Step 50}\label{fig:M3 50}
\end{minipage}
\begin{minipage}[b]{0.24\linewidth}
	\centering
	\includegraphics[width=\linewidth]{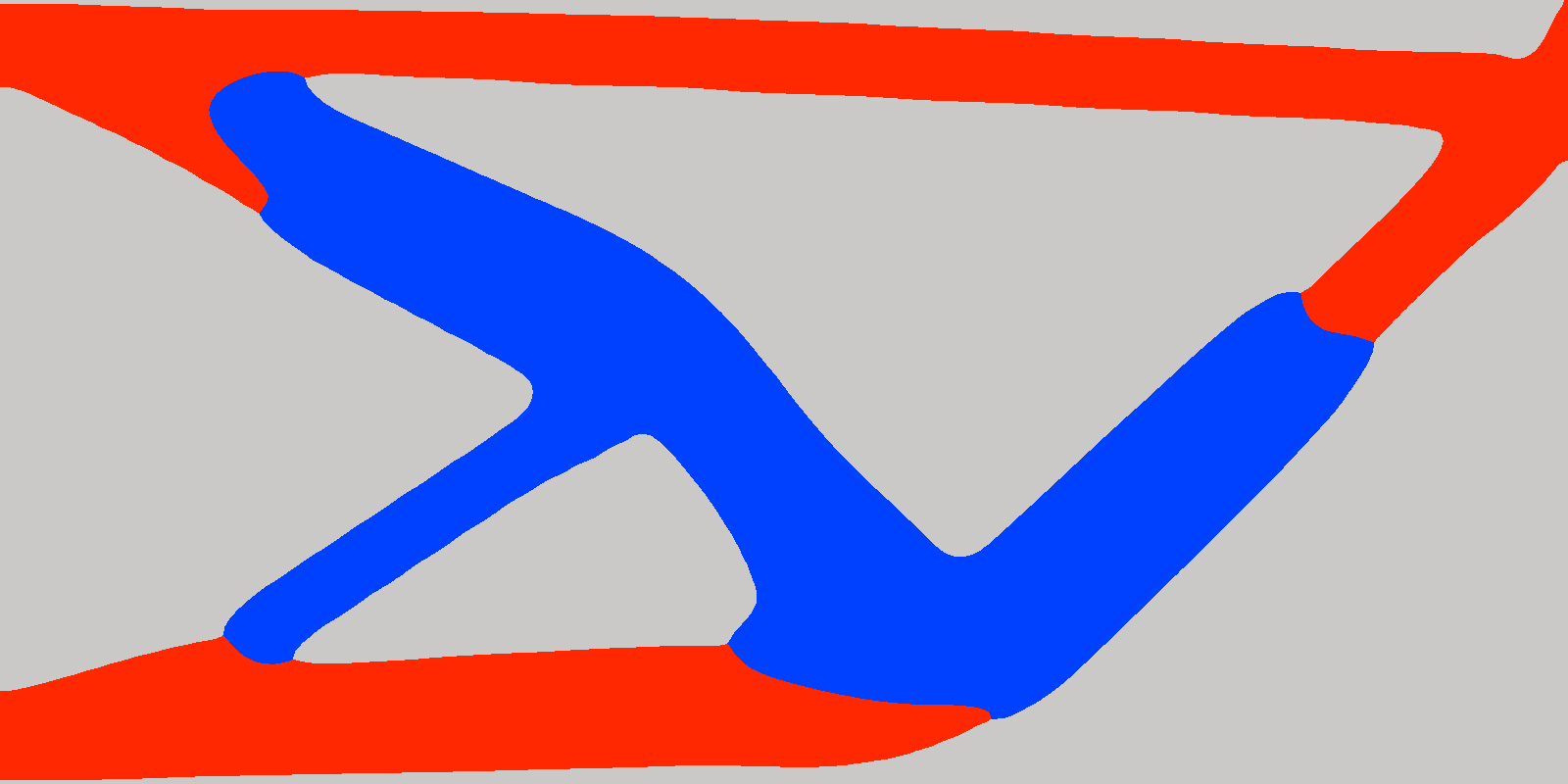}
	\subcaption{Step 200}\label{fig:M3 200}
\end{minipage}
\begin{minipage}[b]{0.24\linewidth}
	\centering
	\includegraphics[width=\linewidth]{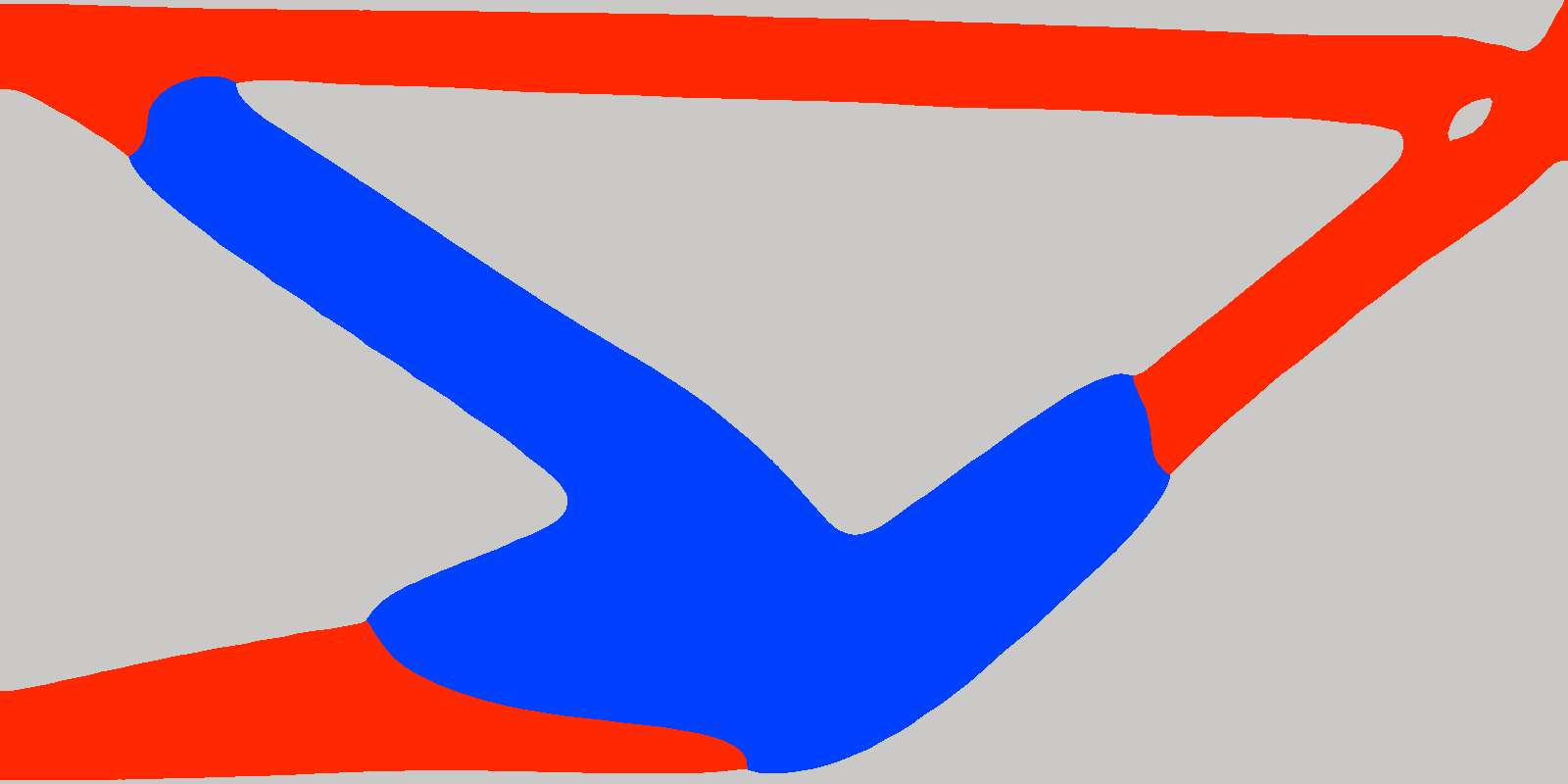}
	\subcaption{Optimal Configration}\label{fig:M3 opt}
\end{minipage}
\caption{Intermediate results and optimal configuration of three materials (Case 2)}\label{fig:M3}
\end{figure}

\begin{figure}[H]
	\begin{minipage}[b]{0.24\linewidth}
		\centering
		\includegraphics[width=\linewidth]{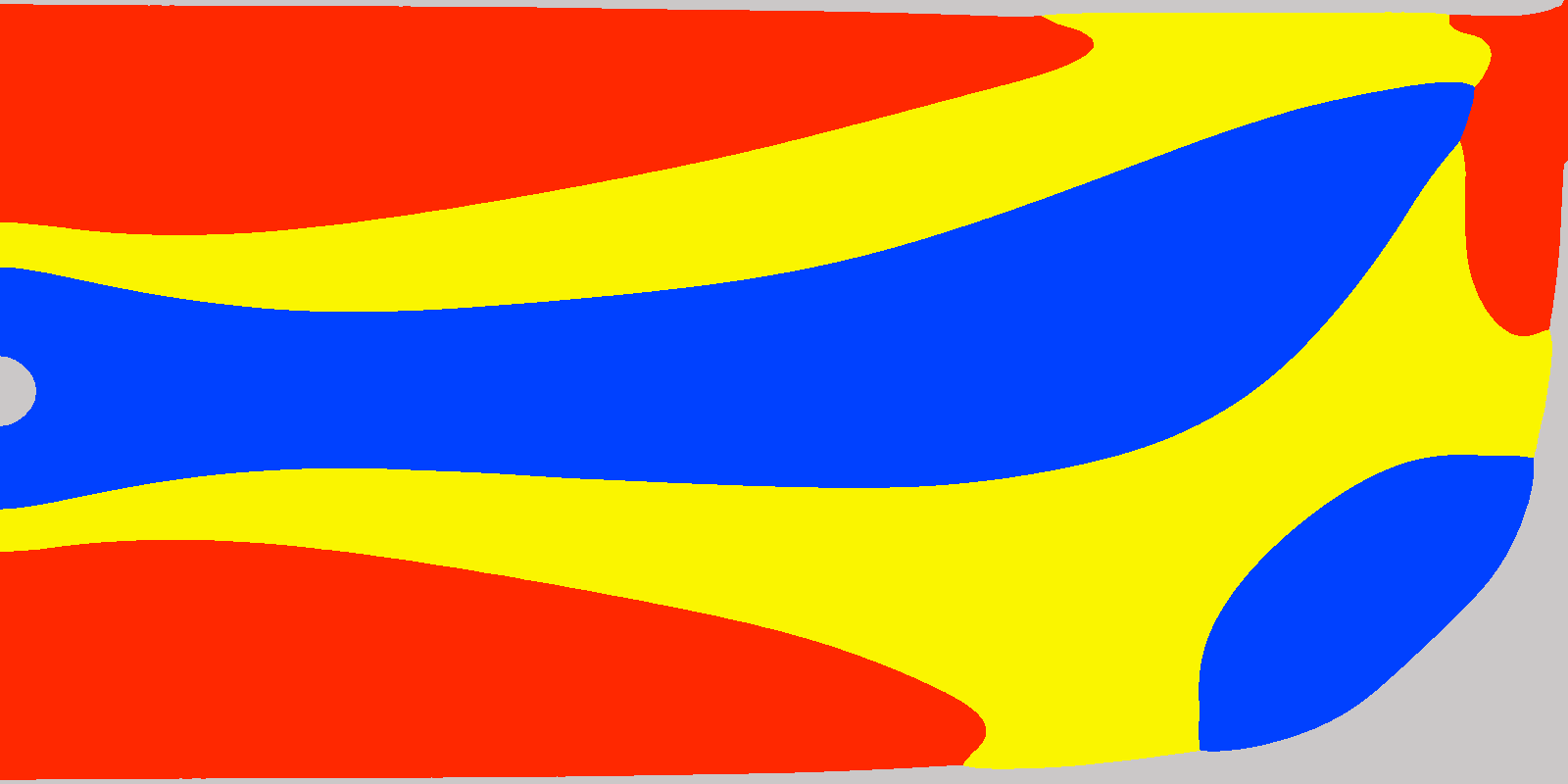}
		\subcaption{Step 10}\label{fig:M4 10}
	\end{minipage}
	\begin{minipage}[b]{0.24\linewidth}
		\centering
		\includegraphics[width=\linewidth]{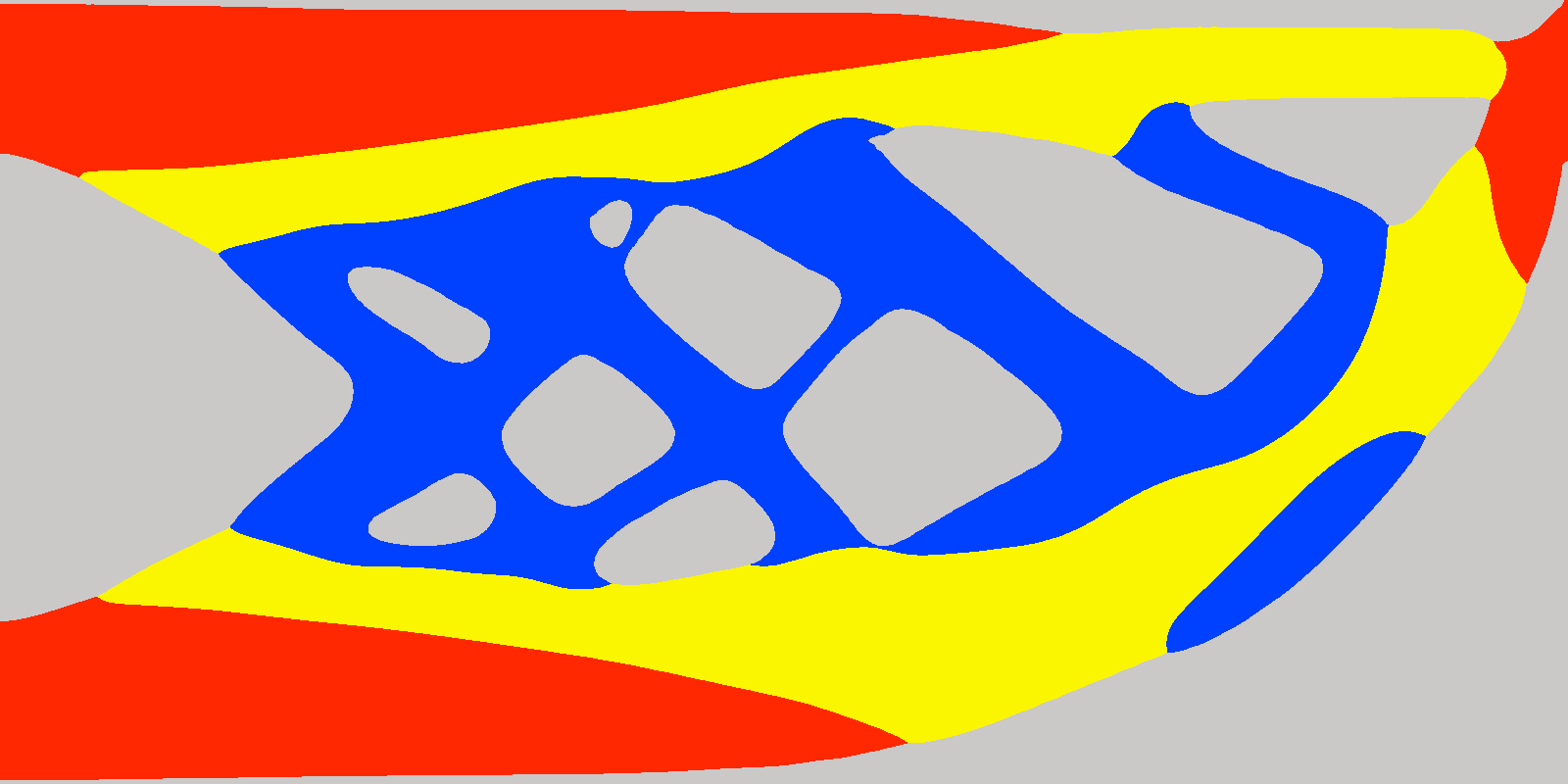}
		\subcaption{Step 50}\label{fig:M4 50}
	\end{minipage}
	\begin{minipage}[b]{0.24\linewidth}
		\centering
		\includegraphics[width=\linewidth]{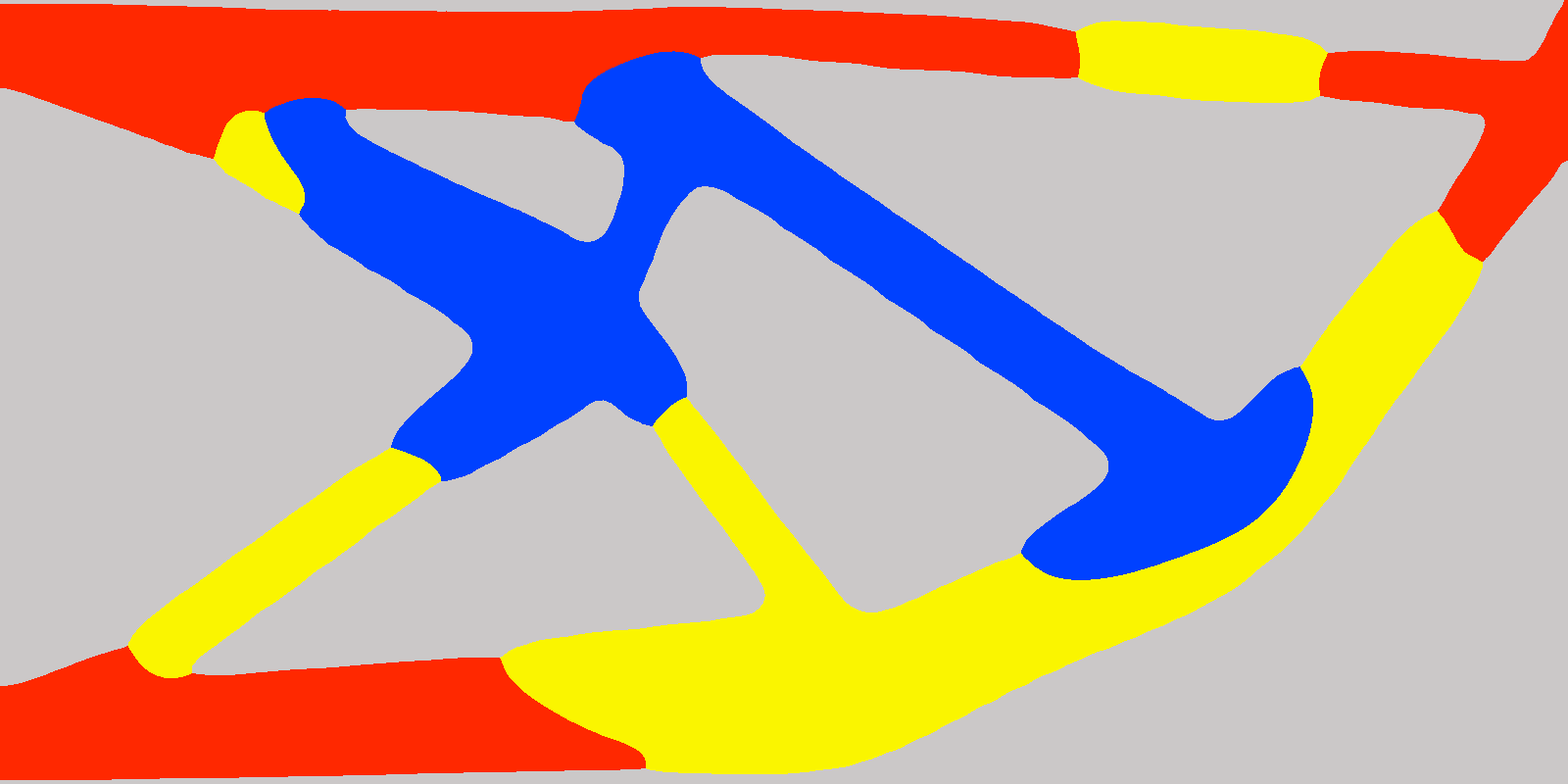}
		\subcaption{Step 200}\label{fig:M4 200}
	\end{minipage}
	\begin{minipage}[b]{0.24\linewidth}
		\centering
		\includegraphics[width=\linewidth]{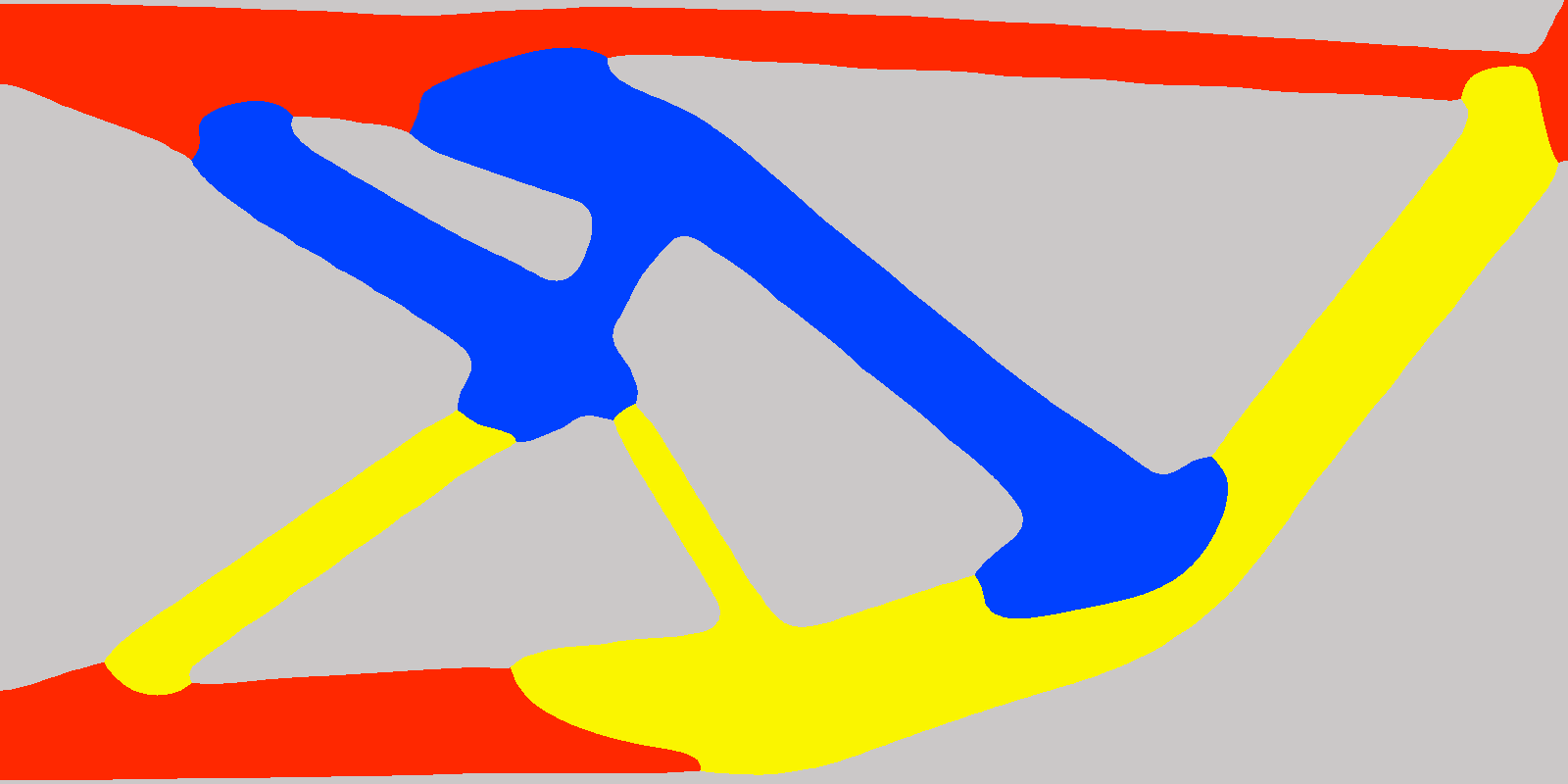}
		\subcaption{Optimal Configration}\label{fig:M4 opt}
	\end{minipage}
	\caption{Intermediate results and optimal configuration of four materials (Case 3)}\label{fig:M4}
\end{figure}

\begin{figure}[H]
	\begin{minipage}[b]{0.24\linewidth}
		\centering
		\includegraphics[width=\linewidth]{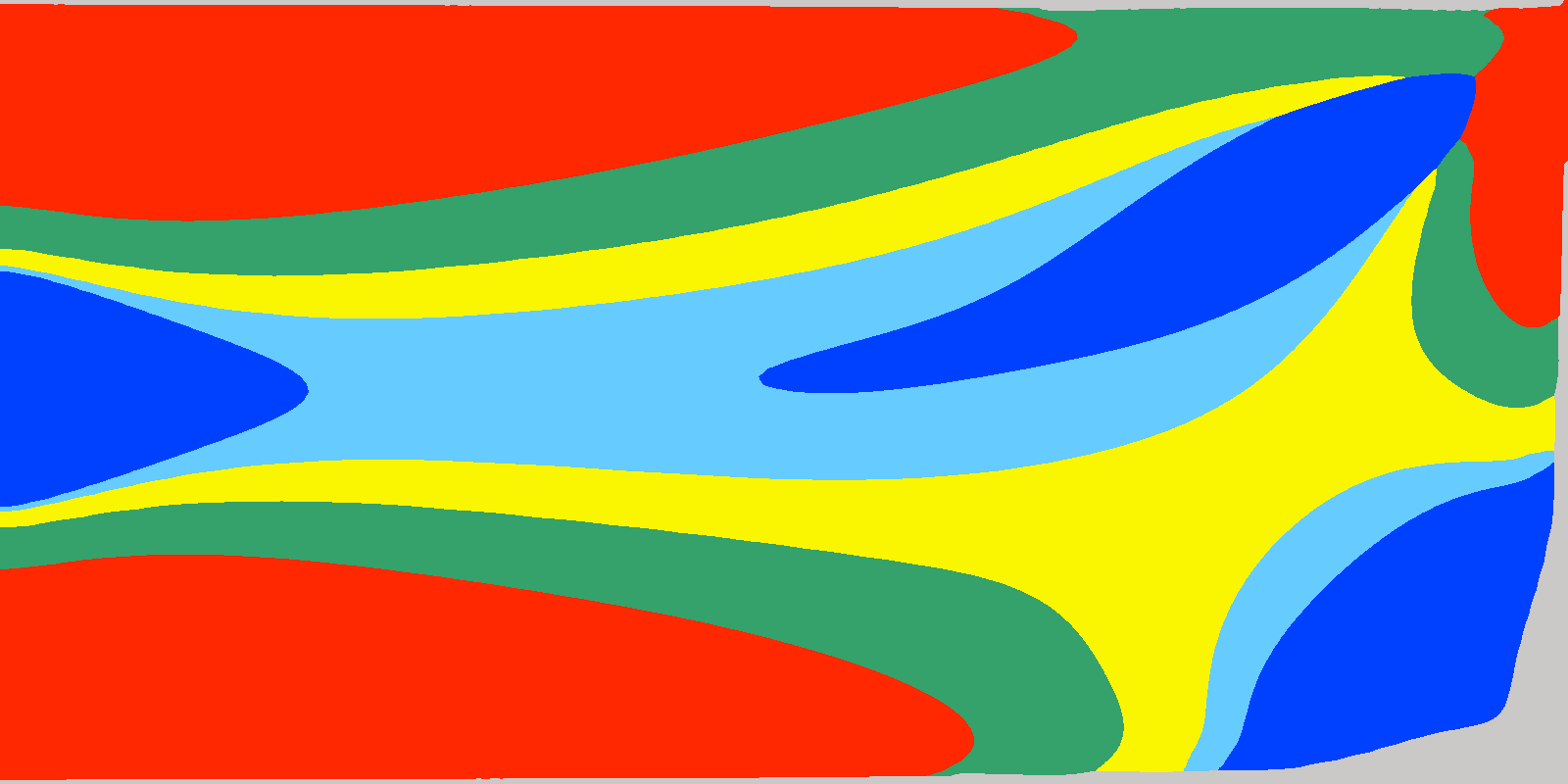}
		\subcaption{Step 10}\label{fig:M9 10}
	\end{minipage}
	\begin{minipage}[b]{0.24\linewidth}
		\centering
		\includegraphics[width=\linewidth]{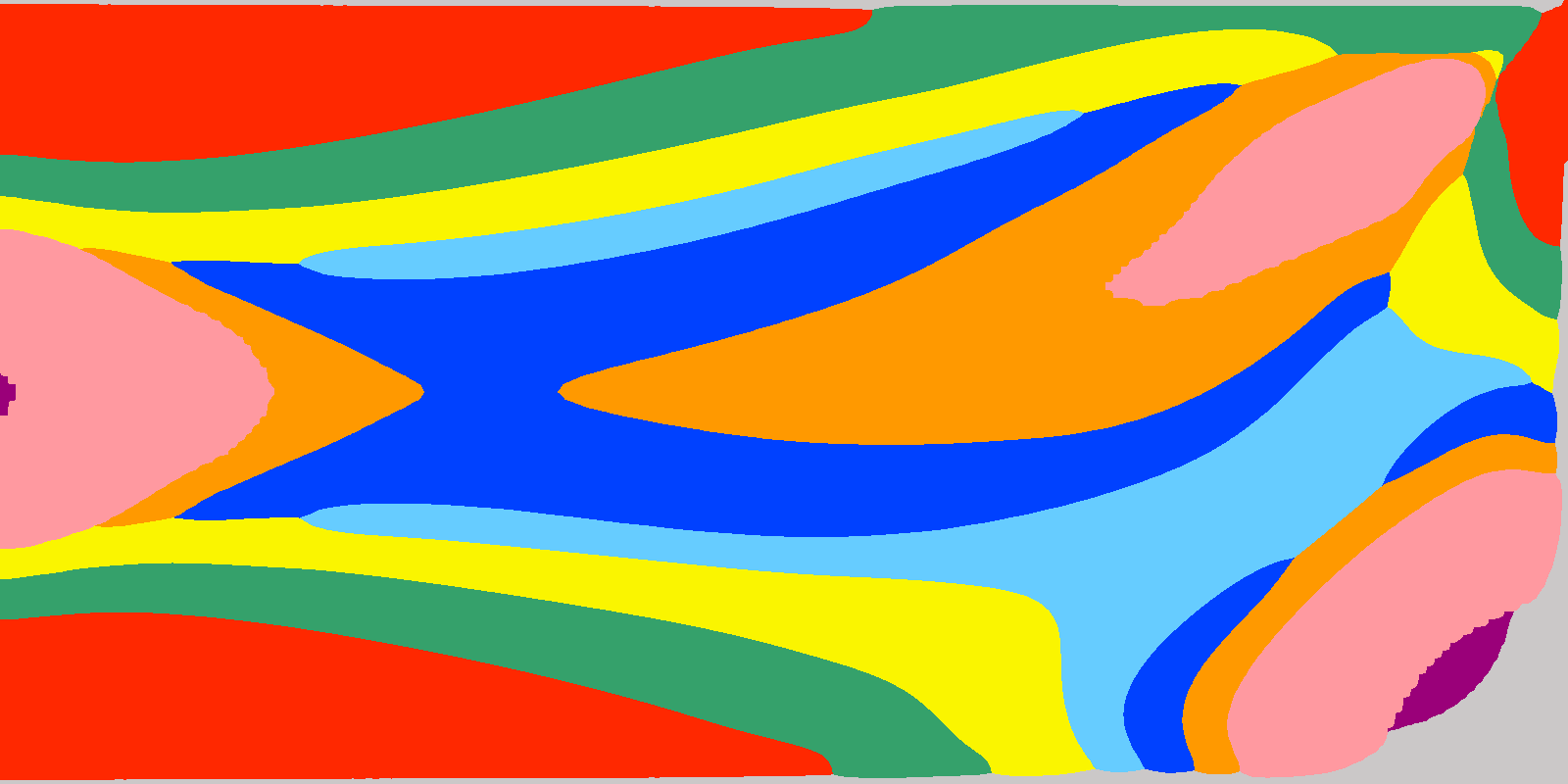}
		\subcaption{Step 50}\label{fig:M9 50}
	\end{minipage}
	\begin{minipage}[b]{0.24\linewidth}
		\centering
		\includegraphics[width=\linewidth]{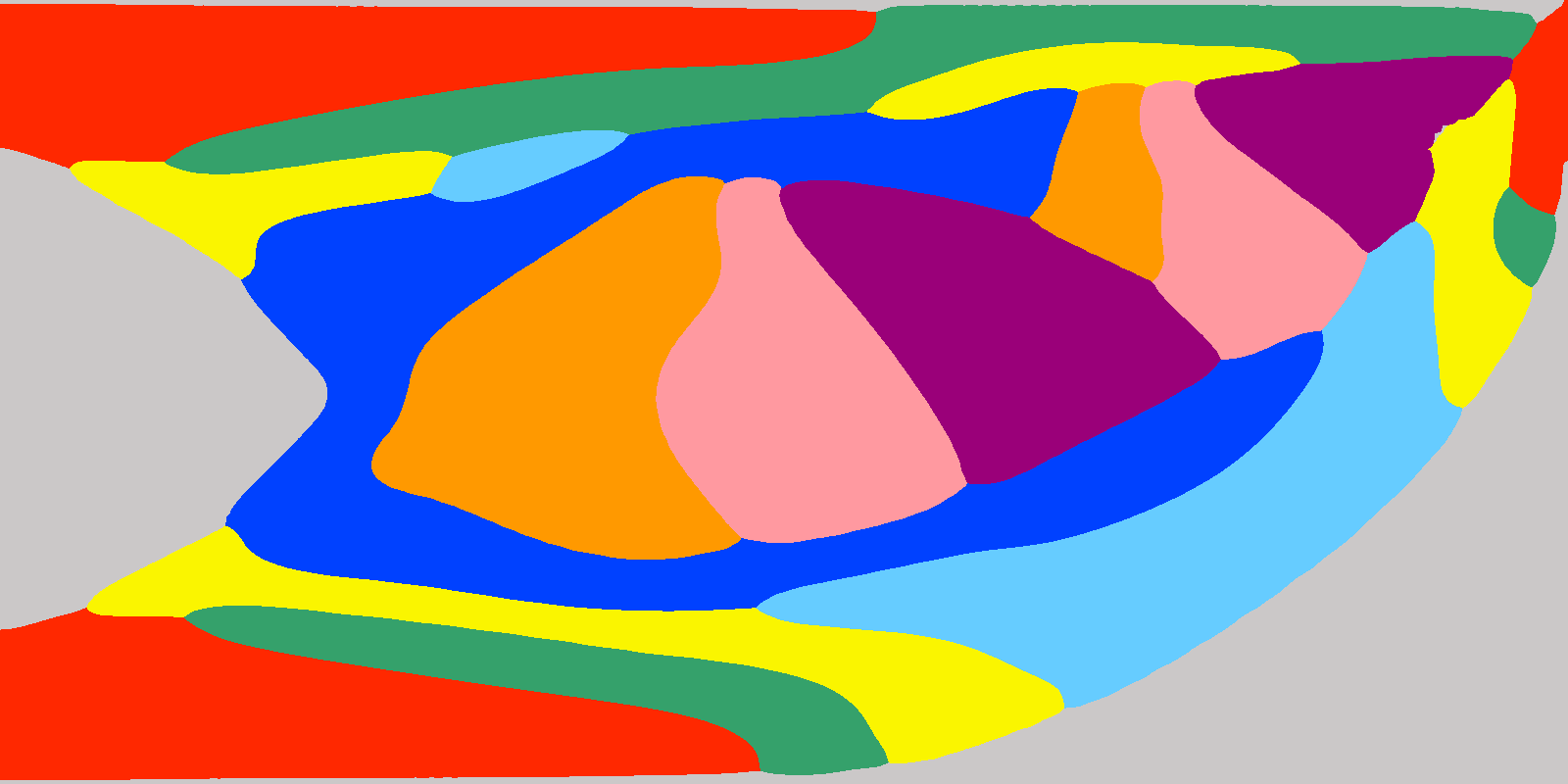}
		\subcaption{Step 200}\label{fig:M9 200}
	\end{minipage}
	\begin{minipage}[b]{0.24\linewidth}
		\centering
		\includegraphics[width=\linewidth]{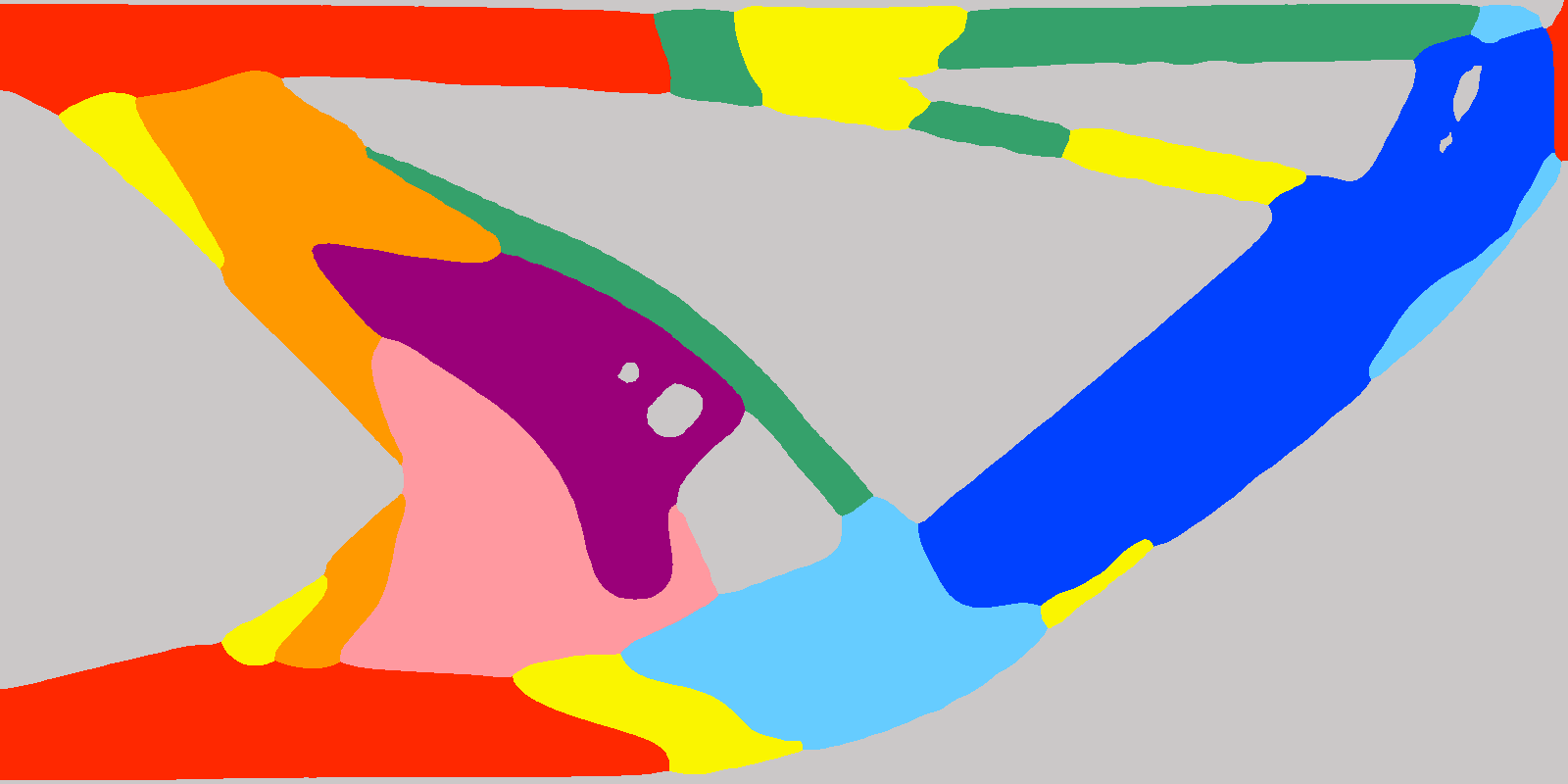}
		\subcaption{Optimal Configration}\label{fig:M9 opt}
	\end{minipage}
	\caption{Intermediate results and optimal configuration of nine materials (Case 4)}\label{fig:M9}
\end{figure}

The proposed method was first applied to the two-, three-, four-, and nine-material cases ($M=2,3,4,9$), which were named Cases 1, 2, 3, and 4, respectively. The maximum volume ratios ${V^\text{max}}_m$ are given in Table \ref{tab:volumes}. 
%なお，本研究の主眼は複数の材料の配置が最適化される最適化法を提案することにあるため，それぞれの材料が配置される割合が明らかであり，得られる結果が比較的容易に推測される問題設定を用いた．
%\textcolor{blue}{（次の文は表で説明されているパラメータの設定理由でしょうか．文章が浮いてみえるので The reason why we set these parameters is ...等とつなげばよいかと思います）}
%\g{あえて体積制約を用いたことについての説明のつもりでした．特に書かなければ気にならないところのように思えてきたので削除します．}
%Since the main focus of this study is to propose an optimization method in which the arrangement of multiple materials is optimized, we used a problem setting in which the proportion of each material arrangement is clear and the results obtained are relatively easy to estimate. results and the optimal configurations.
\begin{table}[h]
	\centering
	\caption{The maximum volume ratios ${V^\text{max}}_m$[\%]}
	\begin{tabular}{lrrrrrrrrr}\hline
		The number of &\multicolumn{9}{c}{Material number $m$} \\
		materials $M$& 
		~~0  & ~~1 &~~2 &~~3 &~~4 &~~5 &~~6 &~~7 &~~8 \\ \hline
		2(Case 1)&100&30&-&- &-&-&-&-&- \\
		3(Case 2)&100&20&20&- &-&-&-&-&- \\
		4(Case 3)&100&13.3&13.3&13.3 &-&-&-&-&- \\
		9(Case 4)&100&10&10&5 &5&5&5&5&5 \\ \hline
	\end{tabular}
	\label{tab:volumes}
\end{table}
The regularization parameters were set to $\tau_{ij}=1\times10^{-3}$ and the initial values of the X-LS functions $\phi_{ij}$ were set to 0. By the definition of the level set method [Eq. (\ref{eq: psi})], the entire design domain is not assigned to any material, but in the ersatz material approach [Eq. (\ref{eq: tilde psi dash})], the entire area is assumed to be occupied by a material with the average Young's modulus of the existing materials. 
Figs. \ref{fig:M2}--\ref{fig:M9} show the obtained intermediate results and the optimal configurations.
In all cases, the optimal configurations were smooth and clear. The material with the highest Young's modulus was placed in the upper-left and lower-left regions, where the stress was concentrated; the lower-right region, which contributed little to the stiffness, was hollow. The obtained structures were therefore mechanically appropriate. The values of objective functions in Cases 1, 2, 3, and 4 were $2.17\times 10^{-11}, 2.10\times 10^{-11}, 2.00\times 10^{-11}$, and $1.97\times 10^{-11}$, respectively.
The objective function improved as the number of material types increased because a wide range of materials with intermediate Young's moduli could be distributed in the region requiring relatively low stiffness. The upper volume constraints were satisfied in Cases 1--3 (volume constraint function $< 10^{-10}$), but in Case 4, the amount of violation was relatively large (volume constraint function $< 10^{-3}$). It was thought that as the number of materials increased, the volumes of the materials changed in conjunction; therefore, the control multipliers were difficult to adjust.

\subsubsection{Examples with different $\tau_{ij}$}
\begin{figure}[H]
	\begin{minipage}[b]{0.32\linewidth}
		\centering
		\includegraphics[width=\linewidth]{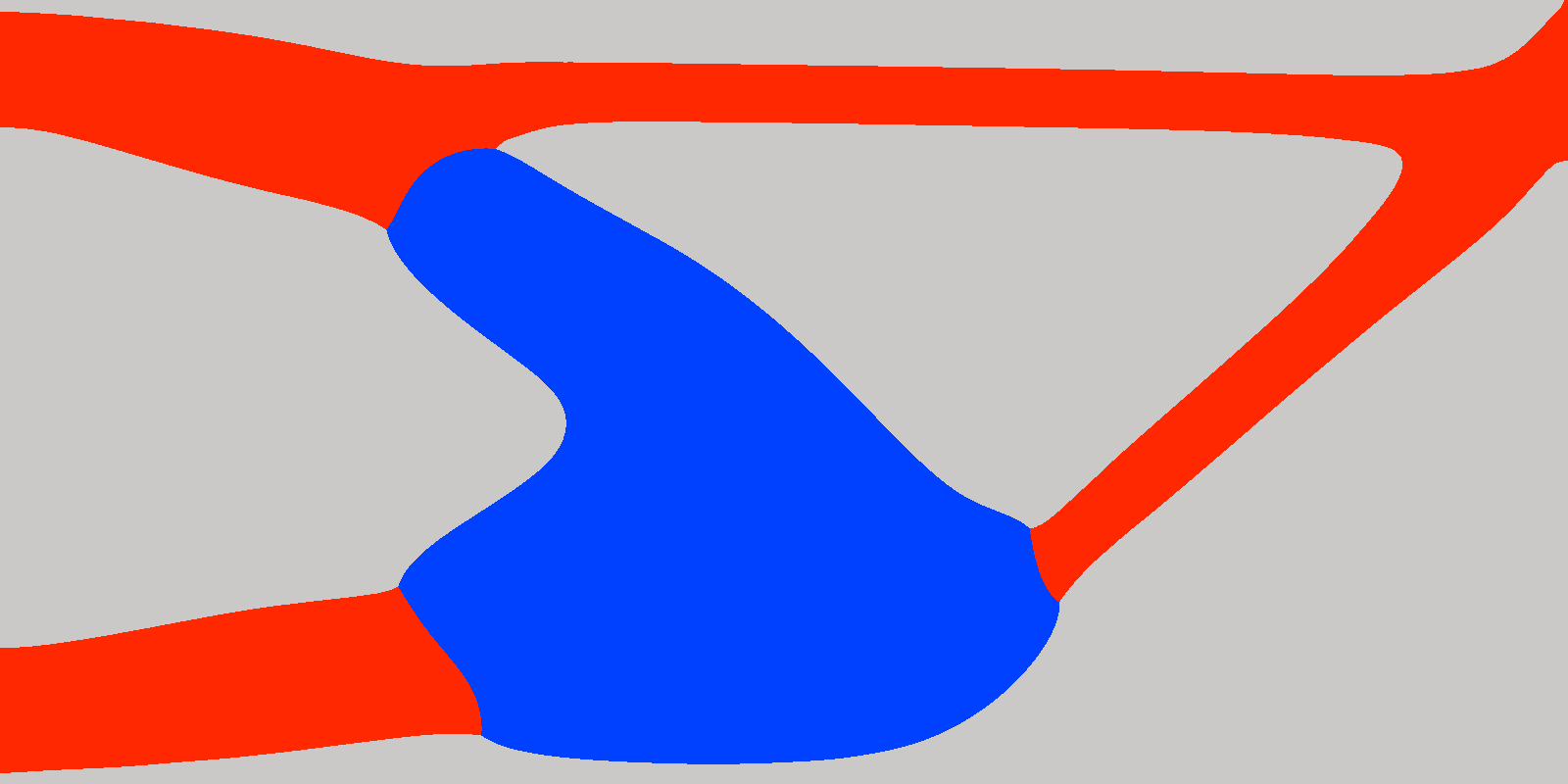}
		\subcaption{$\tau_{ij}=1.0\times10^{-2}$\\(Case 5)}\label{fig:tau2}
	\end{minipage}
	\begin{minipage}[b]{0.32\linewidth}
		\centering
		\includegraphics[width=\linewidth]{fig/2d_results/M/2D_v9_M3_1617978141_10000.png}
		\subcaption{$\tau_{ij}=1.0\times10^{-3}$\\(Case 2)}\label{fig:tau3}
	\end{minipage}
	\begin{minipage}[b]{0.32\linewidth}
		\centering
		\includegraphics[width=\linewidth]{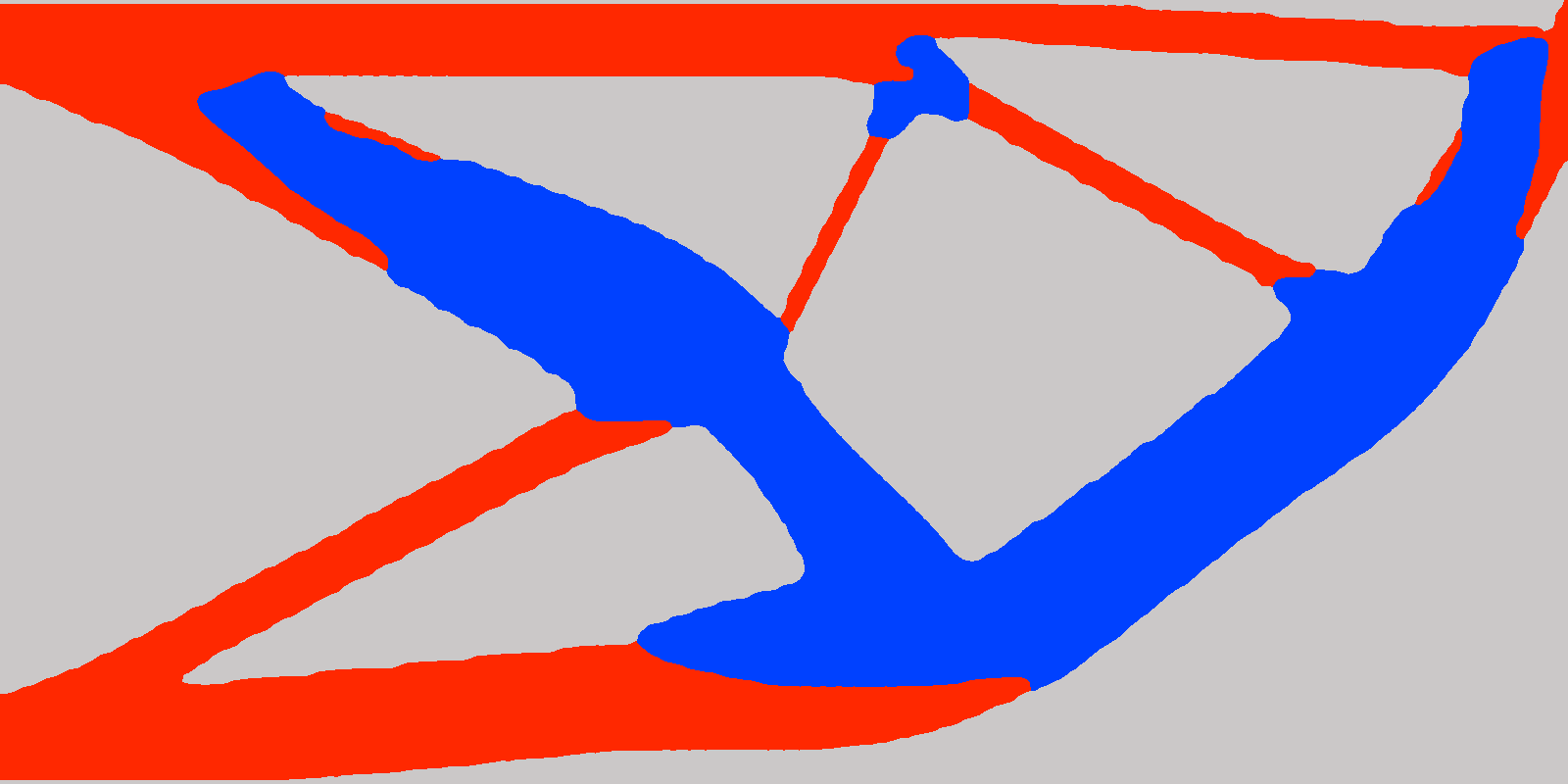}
		\subcaption{$\tau_{ij}=1.0\times10^{-4}$\\(Case 6)}\label{fig:tau4}
	\end{minipage}
	\caption{
	Optimal configurations for different values of regularization parameter $\tau_{ij}$. The smaller the regularization parameter, the more complex is the optimal configuration.}\label{fig:tauall}
\end{figure}

We next examined the effect of regularization parameter $\tau_{ij}$ on the optimal configuration under the conditions of the three-material case described in subsec. \ref{sec:2d M}. Fig. \ref{fig:tauall} shows the optimization results for $\tau_{ij}=1\times10^{-2},1\times10^{-3},1\times10^{-4}$ (Cases 5, 2, and 6, respectively).
The values of the objective functions in Cases 5, 2, and 6 were $2.32\times 10^{-11}, 2.10\times 10^{-11}$, and $2.00\times 10^{-11}$, respectively. The upper volume constraints were mainly satisfied (volume constraint functions $<10^{-10}$). Reducing the regularization parameter improved the value of the objective function. Comparing the obtained optimal structures, we find that when the regularization parameter decreases, the configuration becomes more complex, indicating that the geometric complexity of the structure can be qualitatively controlled as in the two-phase case demonstrated in \cite{yamada2010topology}.
% or in the MMLS method \cite{kishimoto2017topology}

\begin{figure}[H]
	\begin{minipage}[t]{0.32\linewidth}
		\centering
		\includegraphics[width=\linewidth]{fig/2d_results/tau/2D_v9_M3tau4_1617978175_10000.png}
		\subcaption{$\tau_{ij}=1.0\times10^{-4}$ (Case 6)}\label{fig:tau4 2}
	\end{minipage}
\begin{minipage}[t]{0.32\linewidth}
	\centering
	\includegraphics[width=\linewidth]{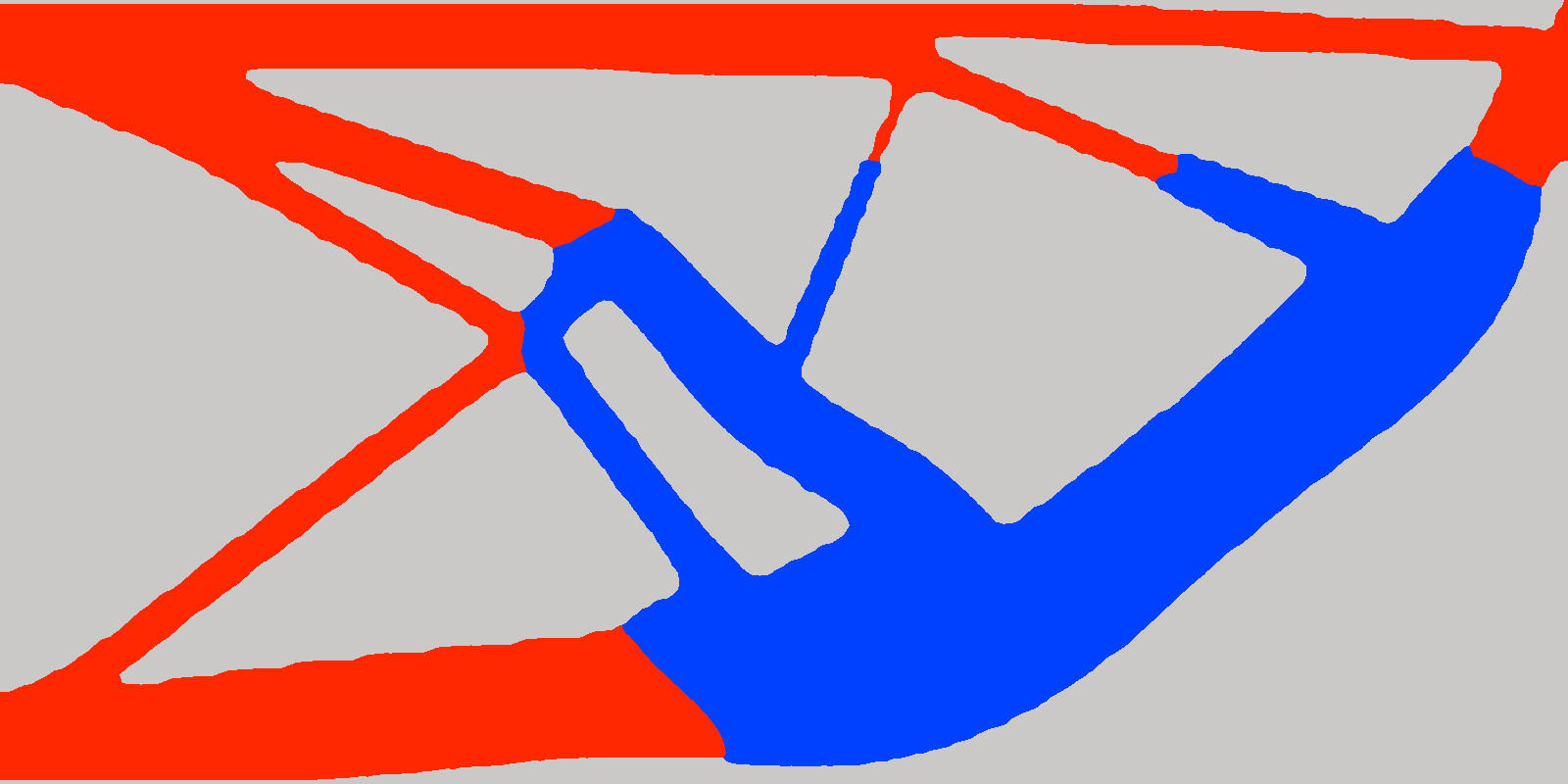}
	\subcaption{$\tau_{12}=1.0\times10^{-2},\tau_{01}=\tau_{02}=1.0\times10^{-4}$ (Case 7)}\label{fig:tau12 2}
\end{minipage}
\begin{minipage}[t]{0.32\linewidth}
	\centering
	\includegraphics[width=\linewidth]{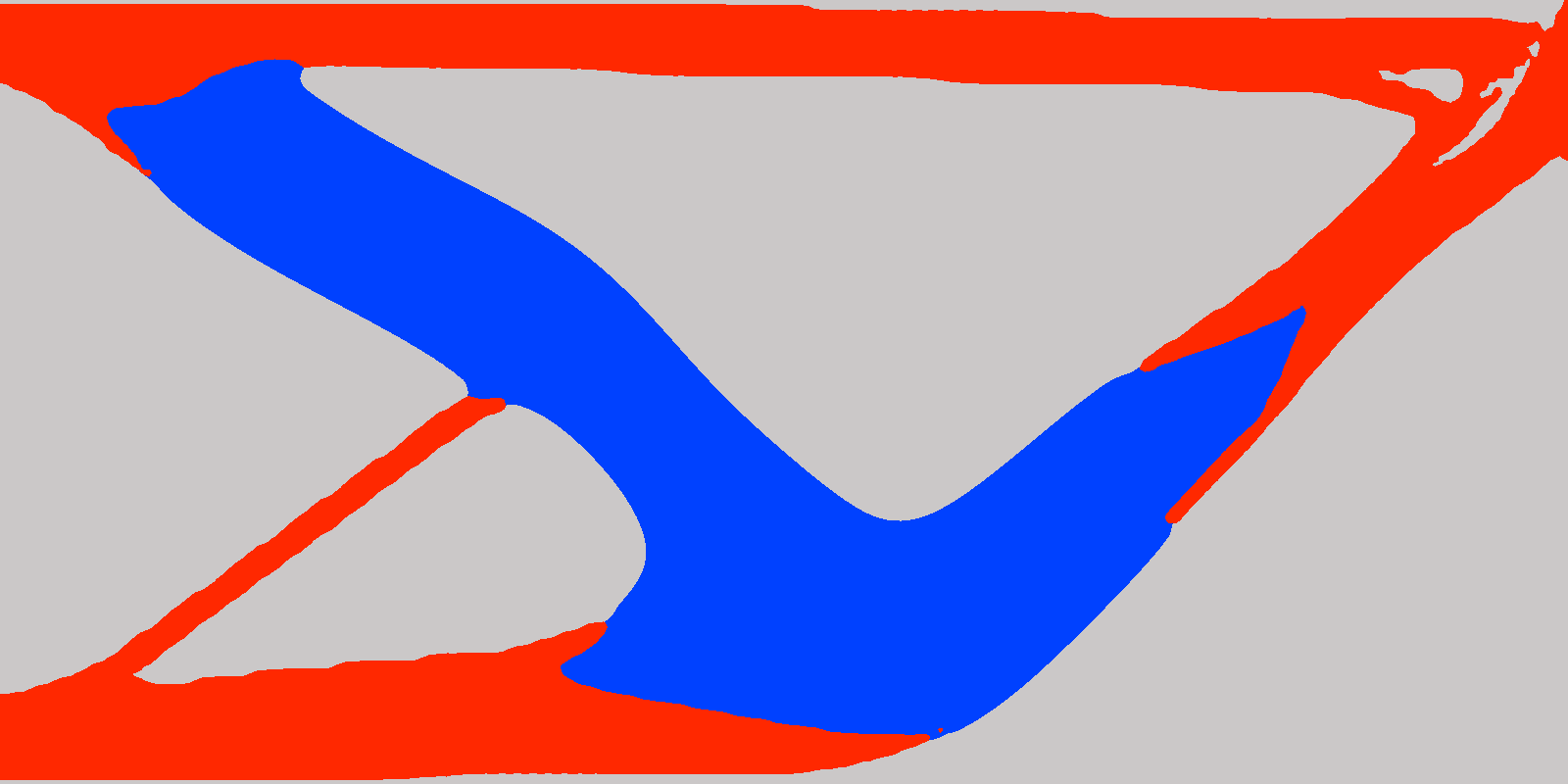}
	\subcaption{$\tau_{02}=1.0\times10^{-2},\tau_{01}=\tau_{12}=1.0\times10^{-4}$ (Case 8)}\label{fig:tau02 2}
\end{minipage}
\caption{
	Optimal configurations for various values of the regularization parameter $\tau_{ij}$
 }\label{fig:taueach}
\end{figure}

Next, we strengthened the regularization of one boundary relative to the other boundaries. Here, the regularization parameters
$(\tau_{01},\tau_{12},\tau_{20})\\ \left[=(\tau_{10},\tau_{21},\tau_{02})\right]$
were set to 
$(1\times10^{-4},1\times10^{-2},1\times10^{-4})$ and 
$(1\times10^{-2},1\times10^{-4},1\times10^{-4})$ in Cases 7 and 8, respectively. 
Fig.\ref{fig:taueach} shows the optimal configurations in Cases 6, 7, and 8.
The values of the objective functions in Cases 6, 7, and 8 were $2.00\times 10^{-11}, 1.98\times 10^{-11}$, and $2.39\times 10^{-11}$, respectively. The upper volume constraints were mostly satisfied (volume constraint functions $<10^{-10}$ in Cases 6, 7 and $<10^{-4}$ in Case 8). The boundaries between Materials 1 and 2 were shorter in the optimal structure in Case 7 (Fig. \ref{fig:taueach}\subref{fig:tau12 2}) than in Case 6 (Fig. \ref{fig:taueach}\subref{fig:tau4 2}), but the overall complexities of the structure were very similar. 
In the optimal structure shown in Fig. \ref{fig:taueach}\subref{fig:tau02 2}, where the regularization parameter was set to a large value ($\tau_{02}=1\times10^{-2}$), 
the low curvatures of the boundary between the blue and gray regions indicate a strong regularization effect. 
In the lower left part of Fig. \ref{fig:taueach}\subref{fig:tau02 2}, where the regularization parameter was small ($\tau_{01}=1\times10^{-4}$), 
there was a thin red area in the gray region.
In the right part of Fig. \ref{fig:taueach}\subref{fig:tau02 2}, where the regularization parameter was also small ($\tau_{12}=1\times10^{-4}$), the boundary between the red and blue regions was intricately shaped, indicating that geometric complexity was allowed for these boundaries. In other words, the geometric complexity of the individual boundaries could be controlled.

\subsubsection{Uniform cross-section surface constraint}\label{sec:ucss 2d}
Next, we considered the uniform cross-section surface constraint, which is important from a manufacturing viewpoint. Under this constraint, all cross-section surfaces of a structure are equal when viewed from a certain direction. If the material boundary satisfies this constraint, its manufacturing process can be simplified.

To implement the uniform cross-section surface constraint, we replaced the second term on the right-hand side of Eq. (\ref{eq:reaction-diffusion}) with the anisotropic regularization factors $\tilde\tau_{ijk}(=\tilde\tau_{jik})~~(k\in\{x,y,z\})$ as follows:
\begin{align}
	\frac{\partial \phi_{ij}}{\partial t}
	&=\frac{-\mathcal{D}_{i j} J-C^{\text{ALL}}\sum_l\lambda_l\mathcal{D}_{ij}g_l~}{C_{ij}}
	+\tau_{ij} L^2\nabla^2 \phi_{ij}\nonumber\\
	&\to\frac{-\mathcal{D}_{i j} J-C^{\text{ALL}}\sum_l\lambda_l\mathcal{D}_{ij}g_l~}{C_{ij}}{K^{\text{ucss}}}_{ij}\nonumber\\
	&+\tau_{ij} L^2 (\tilde\tau_{ijx}\frac{\partial^2}{\partial x^2}\phi_{ij}+\tilde\tau_{ijy}\frac{\partial^2}{\partial y^2}\phi_{ij}+\tilde\tau_{ijz}\frac{\partial^2}{\partial z^2}\phi_{ij}).\label{eq:ucss}
\end{align}

To prevent the level set function from being 0 almost everywhere in the design domain under the uniform cross-section surface constraints, we introduced normalization coefficients ${K^{\text{ucss}}}_{ij}$ satisfying ${K^{\text{ucss}}}_{ij}={K^{\text{ucss}}}_{ji}$. When the uniform cross-section surface constraints are imposed, the distribution of the level set functions is affected more by the diffusion term than by the topological derivatives. Consequently, the spatial distributions of the obtained level set functions will be similar to those updated by the design sensitivity with a small absolute value averaged along the direction of the uniform cross-section surface constraint. Therefore, ${K^{\text{ucss}}}_{ij}$ must be set large to avoid flattening of the level set function.

\begin{figure}[H]
	\begin{minipage}[b]{0.49\linewidth}
		\centering
		\includegraphics[width=\linewidth]{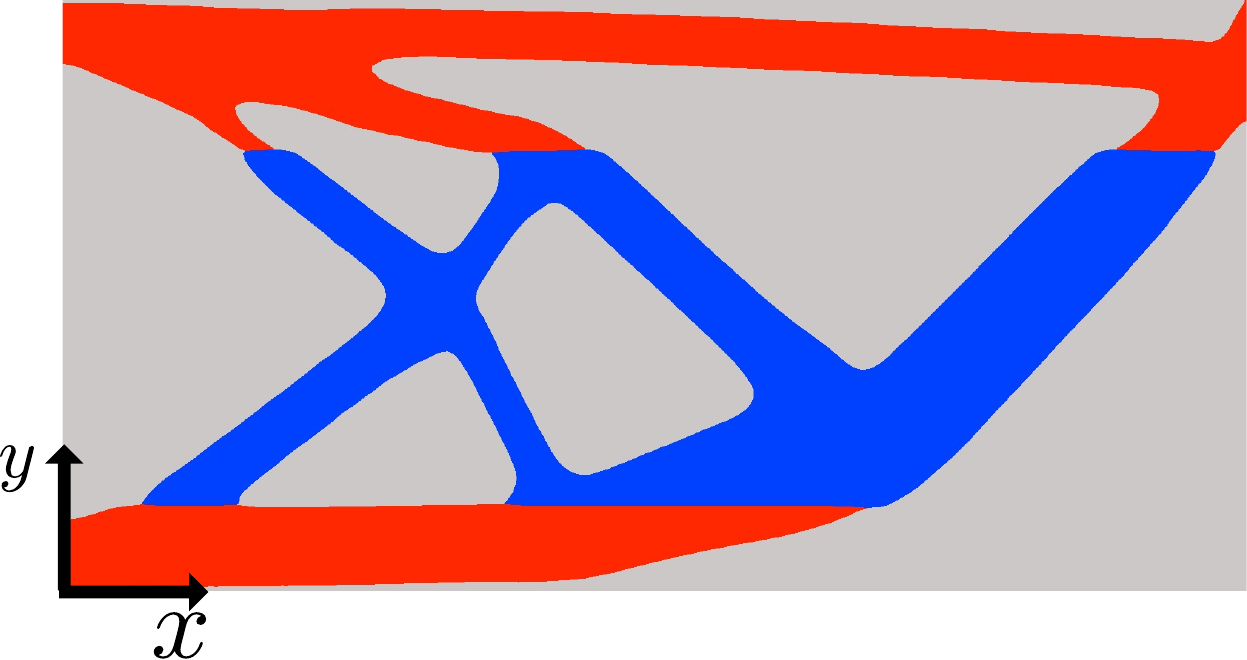}
		\subcaption{Case 9}\label{fig:tau12 x}
	\end{minipage}
	\begin{minipage}[b]{0.49\linewidth}
		\centering
		\includegraphics[width=\linewidth]{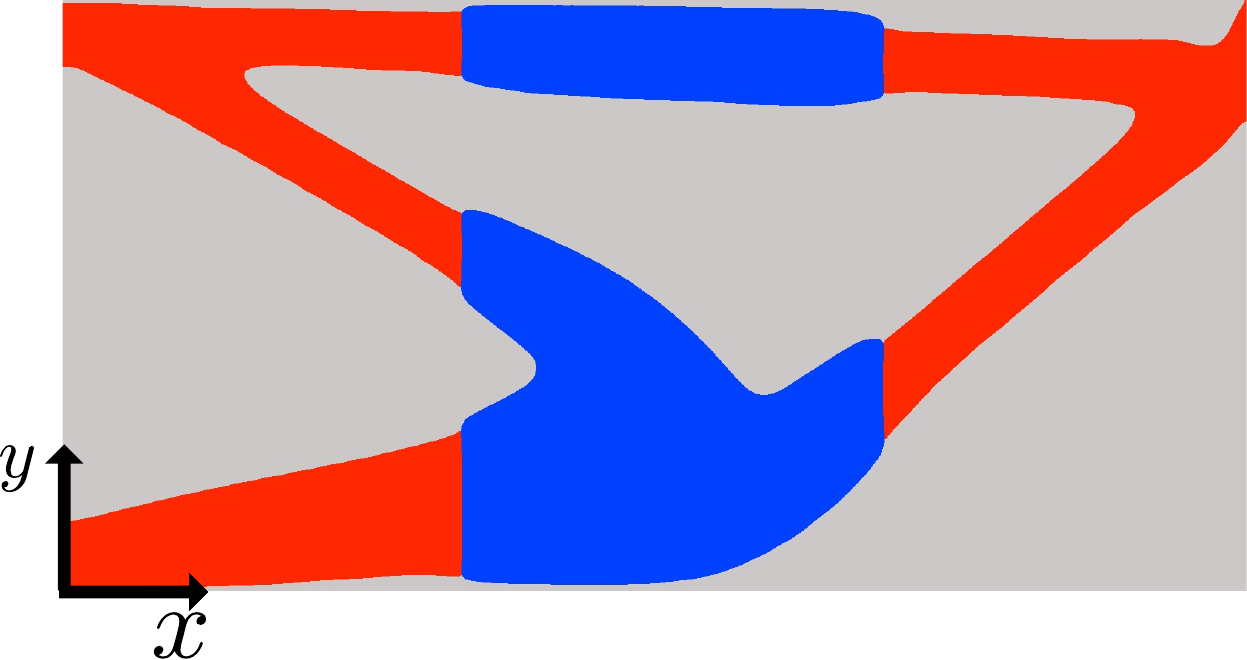}
		\subcaption{Case 10}\label{fig:tau12 y}
	\end{minipage}
	\caption{
	Optimal configurations under the uniform cross-section surface constraint between materials 1 and 2. In \subref{fig:tau12 x} and \subref{fig:tau12 y}, the constraint was imposed so that the cross section was parallel to the x-axis and y-axis, respectively.
}\label{fig:tau12 ucss}
\end{figure}

Here, we show the effect of uniform cross-section surface constraints on the obtained optimal configuration in the two-dimensional case. The regularization parameters are set to $\tau_{ij}=1\times 10^{-3}$ and anisotropic regularization factors $\tilde\tau_{ijx}, \tilde\tau_{ijy}$ are set to 1, except for $\tau_{12x}(=\tau_{21x})=1\times 10^5$ in Case 9  and $\tau_{12y}(=\tau_{21y})=1\times 10^5$ in Case 10. ${K^{\text{ucss}}}_{ij}$ is set to 1. Optimal configurations for Case 9 and 10 are shown in Fig.\ref{fig:tau12 ucss}. 

Here, we show the effect of uniform cross-section surface constraints on the obtained optimal configuration in a two-dimensional case. The regularization parameters $\tau_{ij}$ were set to $1\times 10^{-3}$ and the anisotropic regularization factors $\tilde\tau_{ijx}$ and $ \tilde\tau_{ijy}$ were set to 1, except for $\tau_{12x}(=\tau_{21x})$ in Case 9 (set to $1\times 10^5$) and $\tau_{12y}(=\tau_{21y})$ in Case 10 (set to $1\times 10^5$). ${K^{\text{ucss}}}_{ij}$ was set to 1. The optimal configurations in Cases 9 and 10 are shown in Fig. \ref{fig:tau12 ucss}.
The values of the objective functions in Case 9 and 10 were $2.03\times 10^{-11}$ and $2.12\times 10^{-11}$, respectively. 
The upper volume constraints were mostly satisfied (volume constraint functions $<10^{-2}$). The red and blue regions in Fig. \ref{fig:tau12 ucss}, corresponding to materials 1 and 2 respectively, were divided by straight lines spanning the design domain, indicating that the cross-sectional surfaces were uniform under the constraints.

\subsubsection{Piecewise-linear surface constraint}
%but it still has the benefit that the surface properties can be easily adjusted.
The uniform cross-section surface constraint imposes equal cross-sections over the entire region. Here, we describe another constraint which imposes a piecewise-linear interface in a given direction but allows the interfaces to be stepped over the entire region. The optimal configuration obtained under this constraint is slightly more difficult to manufacture than one constructed under the uniformed cross-section surface constraint, but owing to the relatively high degree of freedom, an optimal configuration with a superior objective function value is easily obtained.

\begin{figure}[H]
	\begin{minipage}[b]{0.49\linewidth}
		\centering
		\includegraphics[width=\linewidth]{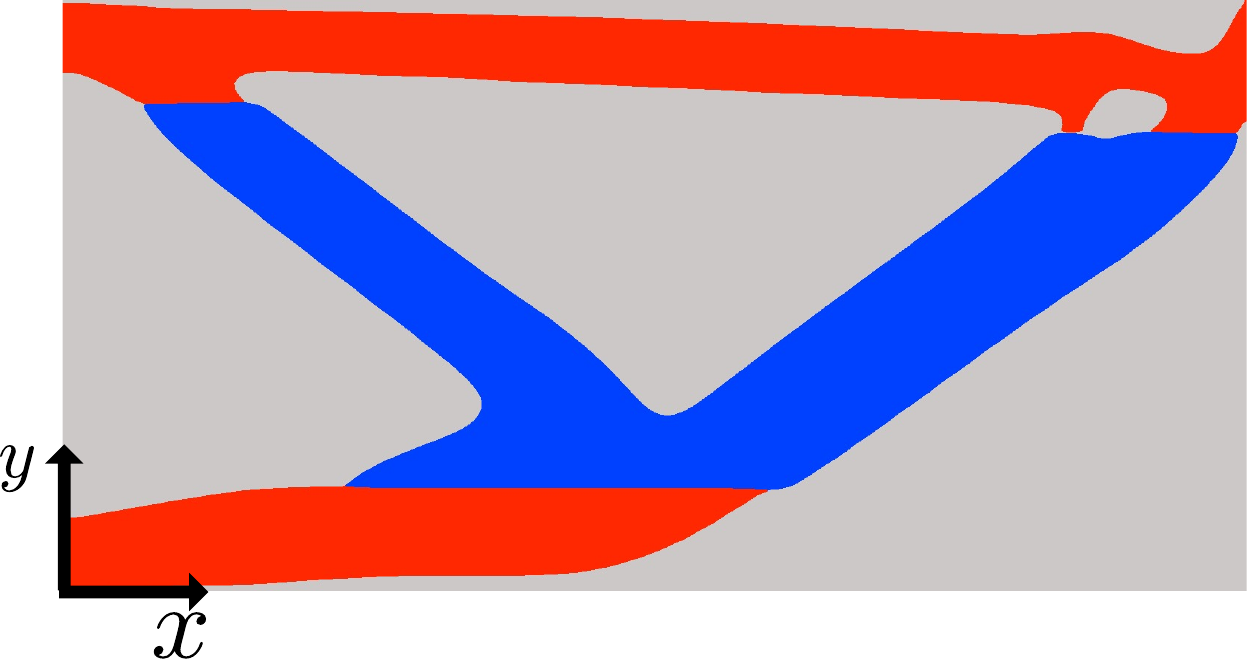}
		\subcaption{Case 11}\label{fig:tau12 lx}
	\end{minipage}
	\begin{minipage}[b]{0.49\linewidth}
		\centering
		\includegraphics[width=\linewidth]{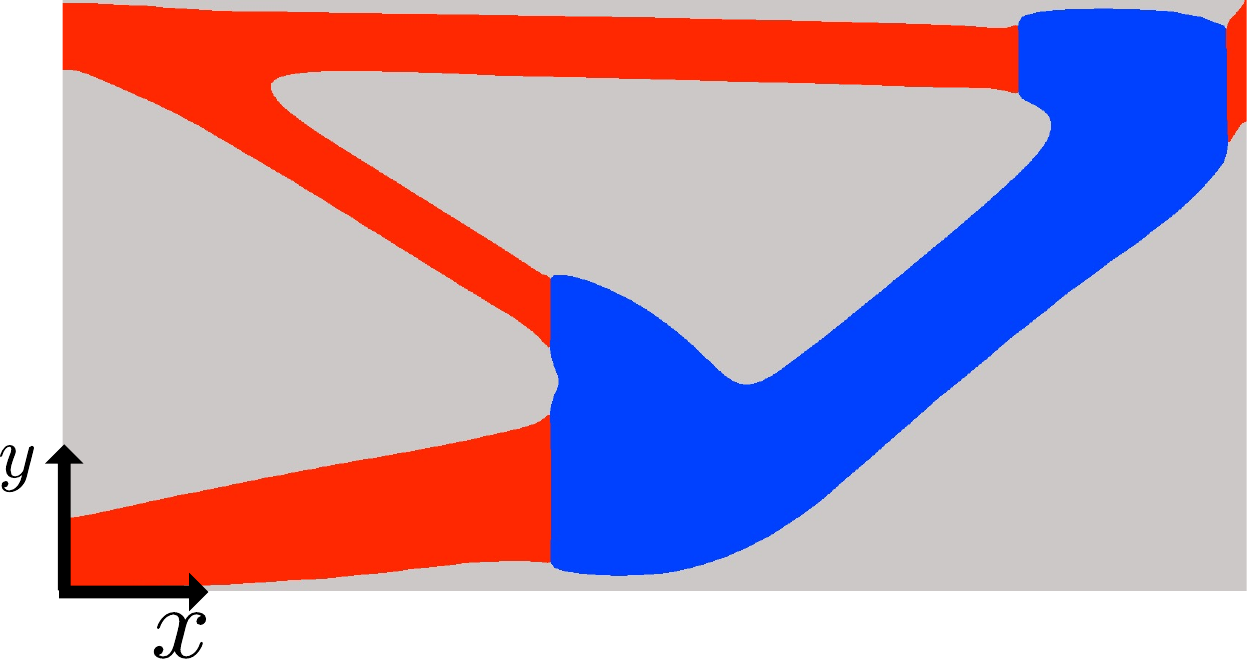}
		\subcaption{Case 12}\label{fig:tau12 ly}
	\end{minipage}
	\caption{
		Optimal configuration under the piecewise-linear surface constraint between materials 1 and 2. In \subref{fig:tau12 lx} and \subref{fig:tau12 ly}, the constraint was imposed so that the interfaces were parallel to the x-axis and y-axis, respectively.}\label{fig:tau12 lucss}
\end{figure}

In particular, depending on the characteristic function, the anisotropic regularization parameter
${\tilde{\tau}}_{ijk}~~(k\in\{x,y,z\})$ 
is changed in a piecewise manner using the piecewise anisotropic regularization parameters 
$\tilde\tau_{ijk}^{'}(=\tilde\tau_{jik}^{'})~~(k\in\{x,y,z\})$ as follows:
\begin{align}
	\tilde\tau_{ijk}=1+\tilde\tau_{ijk}^{'}({\hat\psi}^{'}_i+{\hat\psi}^{'}_j)~~(k\in\{x,y,z\}).
\end{align}
The boundary between two materials $i$ and $j$ is then constructed piecewise linearly along direction $k$. Fig. \ref{fig:tau12 lucss} shows the optimal configurations under this constraint. Here, the regularization parameter $\tau_{ij}$ was set to $1\times 10^{-3}$ and the anisotropic regularization factors $\tilde\tau_{ijx}^{'}, \tilde\tau_{ijy}^{'}$ were set to 1. The exceptions were $\tilde\tau_{12x}^{'}=1\times 10^5$ in Case 11 and $\tilde\tau_{12y}^{'}=1\times 10^5$ in Case 12. ${K^{\text{ucss}}}_{ij}$ was set to 1.
The values of the objective functions in Cases 11 and 12 were $2.16\times 10^{-11}$ and $2.10\times 10^{-11}$, respectively. 
Upper volume constraints were mostly satisfied (the volume constraint functions were less than  $2\times10^{-3}$).
The upper volume constraints were mostly satisfied (volume constraint functions $<2\times10^{-3}$). 
The effect of the piecewise-linear surface constraints was observed on the interfaces between materials 1 and 2, but was nonevident at the interfaces between materials 0 and 1 and between materials 0 and 2.

\subsubsection{Examples of various initial configurations}
\begin{figure}[H]
	\begin{minipage}[t]{0.24\linewidth}
		\centering
		\includegraphics[width=\linewidth]{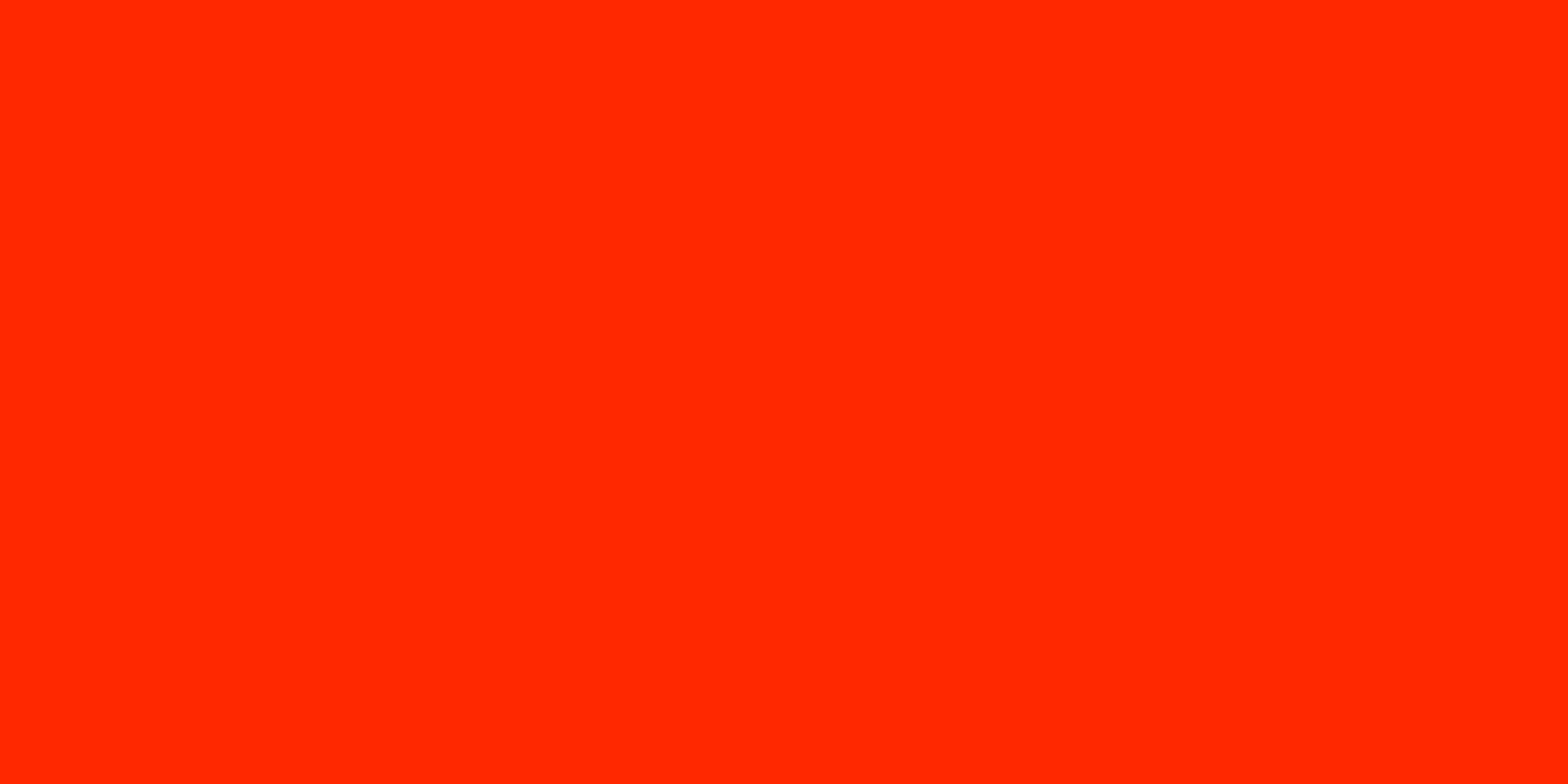}
		\subcaption{Case 13, initial configuration}\label{fig:init1 0}
	\end{minipage}
	\begin{minipage}[t]{0.24\linewidth}
		\centering
		\includegraphics[width=\linewidth]{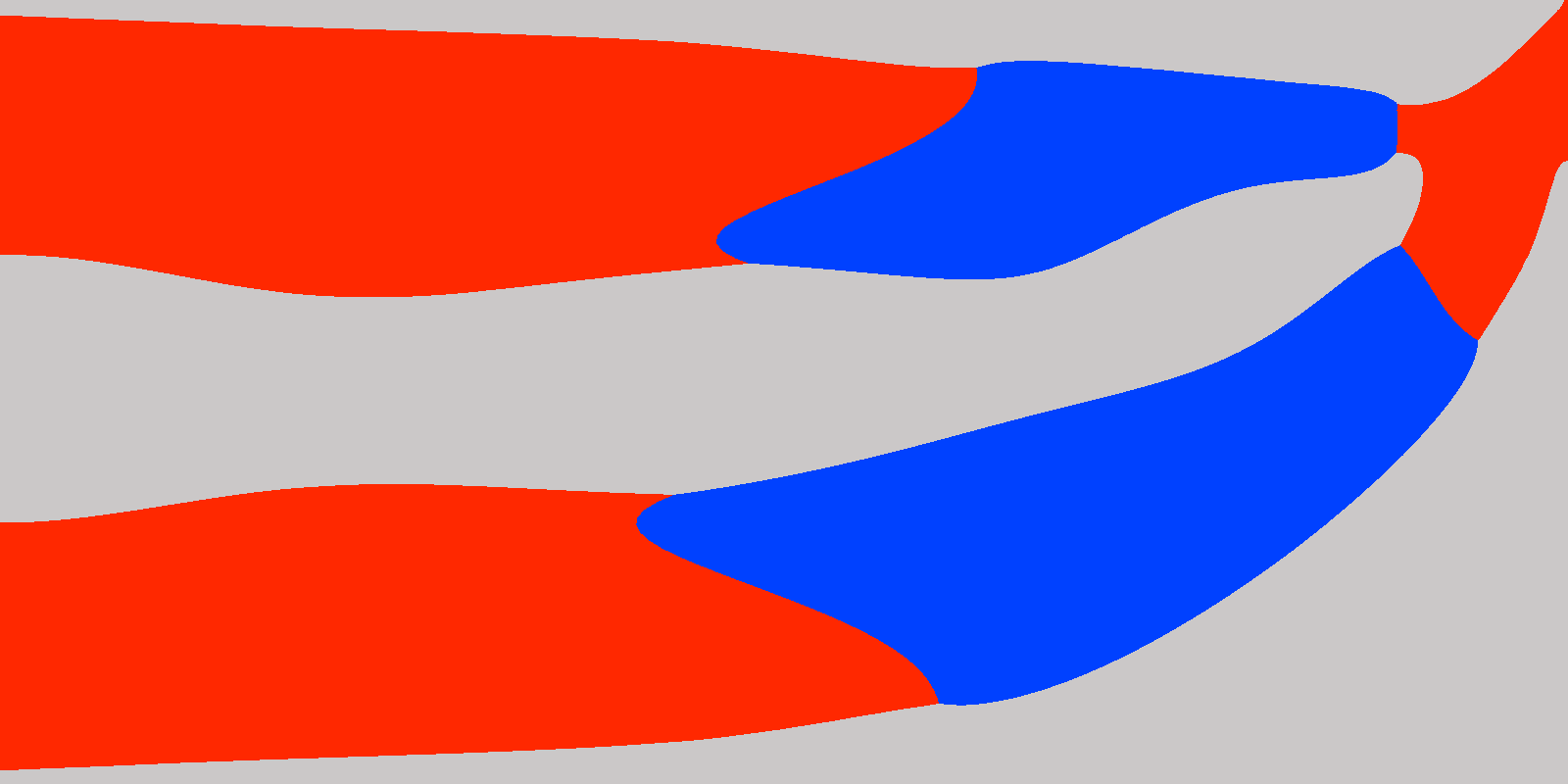}
		\subcaption{Case 13, step 20}\label{fig:init1 20}
	\end{minipage}
	\begin{minipage}[t]{0.24\linewidth}
		\centering
		\includegraphics[width=\linewidth]{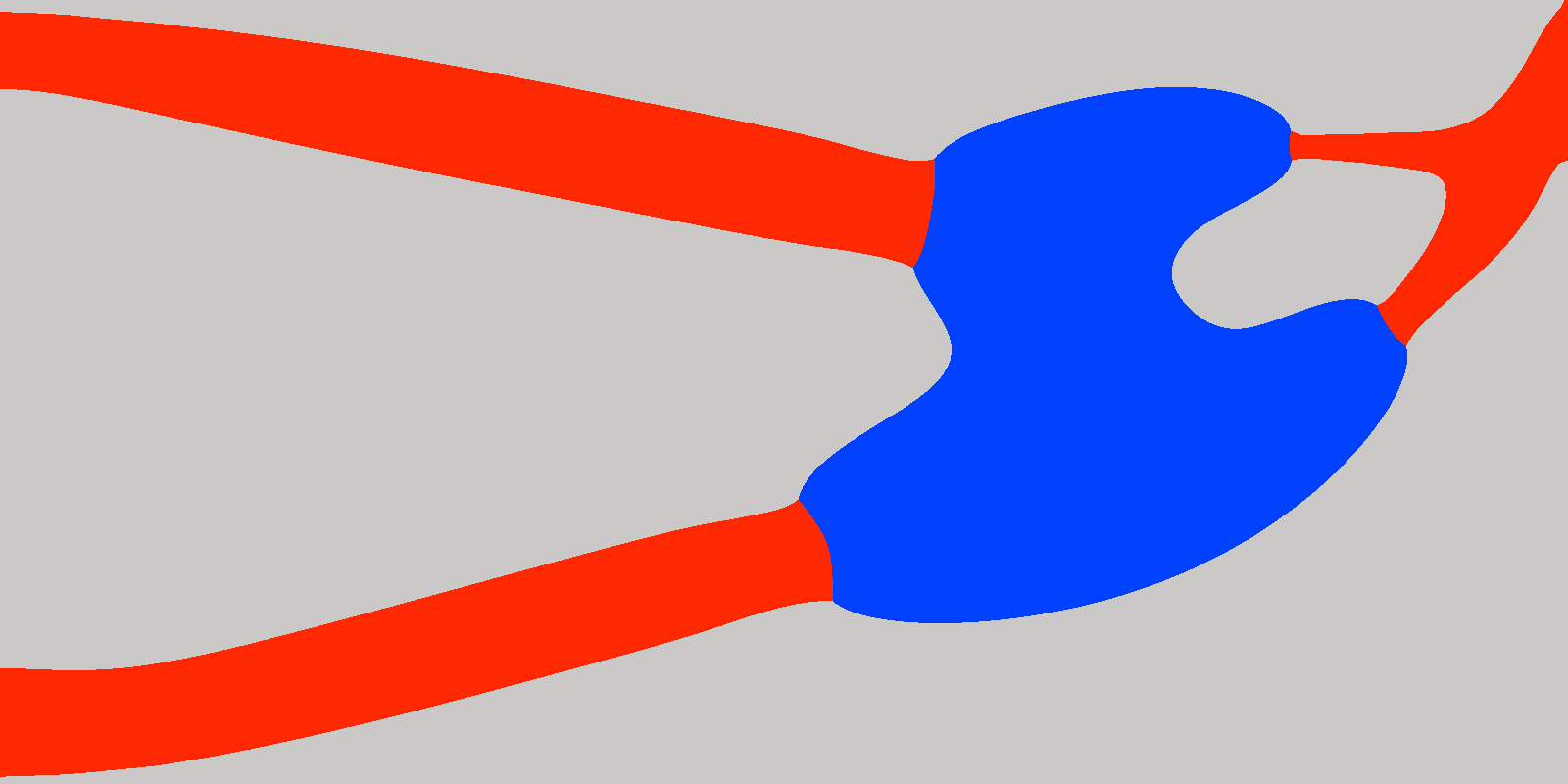}
		\subcaption{Case 13, step 50}\label{fig:int1 50}
	\end{minipage}
	\begin{minipage}[t]{0.24\linewidth}
		\centering
		\includegraphics[width=\linewidth]{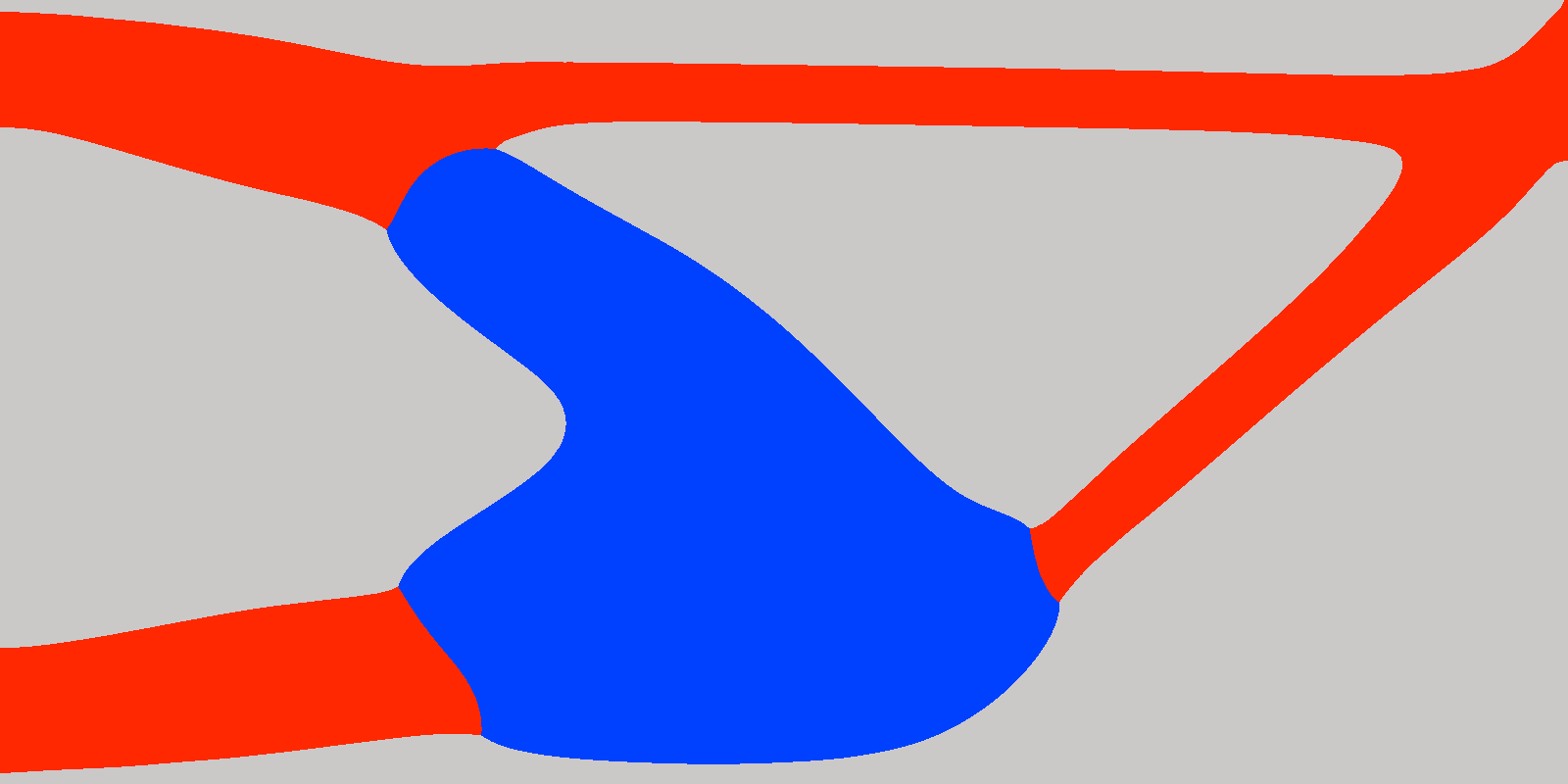}
		\subcaption{Case 13, optimal Configration}\label{fig:init1 opt}
	\end{minipage}\\
	\begin{minipage}[t]{0.24\linewidth}
		\centering
		\includegraphics[width=\linewidth]{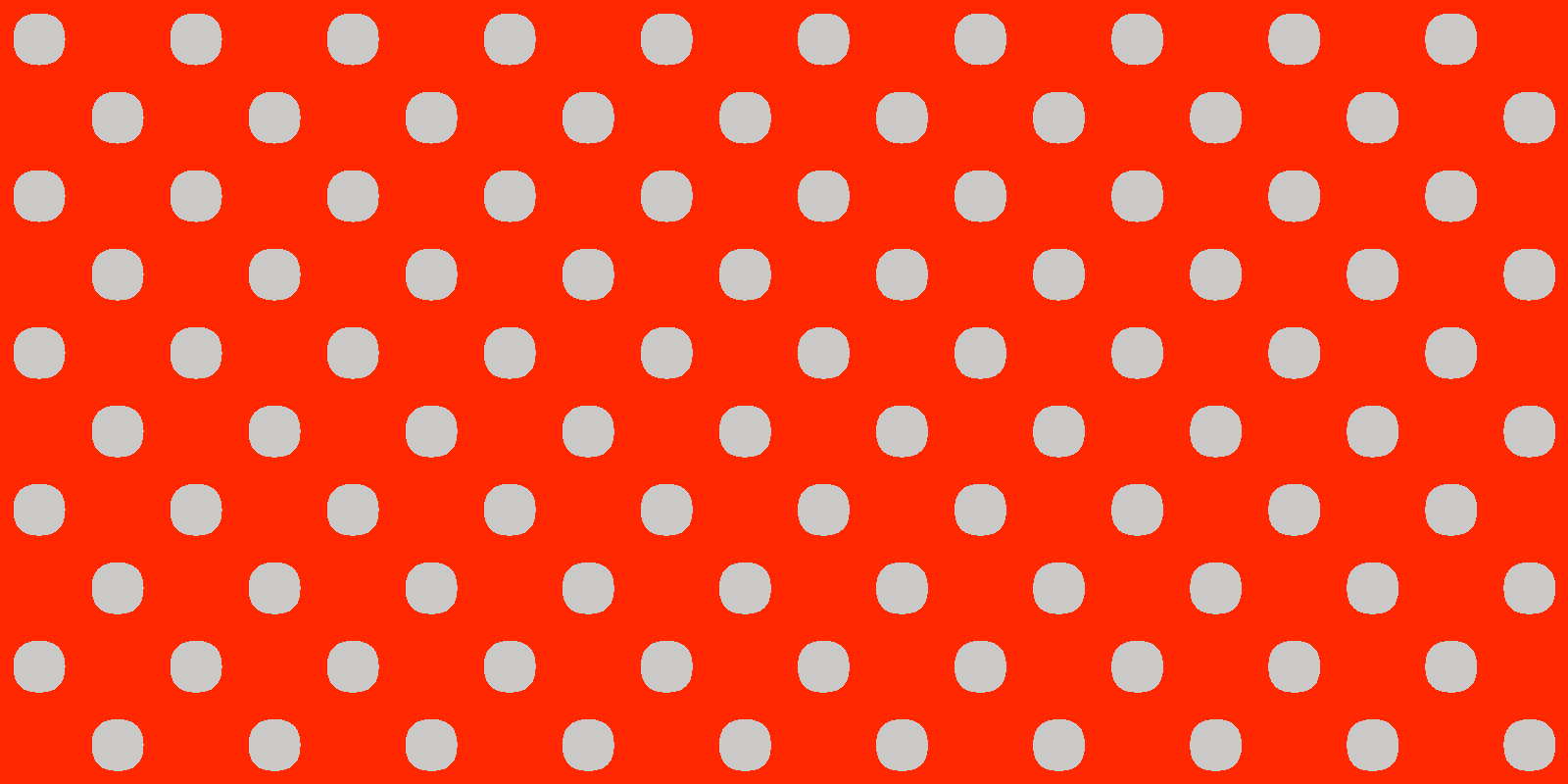}
		\subcaption{Case 14, initial configuration}\label{fig:init2 0}
	\end{minipage}
	\begin{minipage}[t]{0.24\linewidth}
		\centering
		\includegraphics[width=\linewidth]{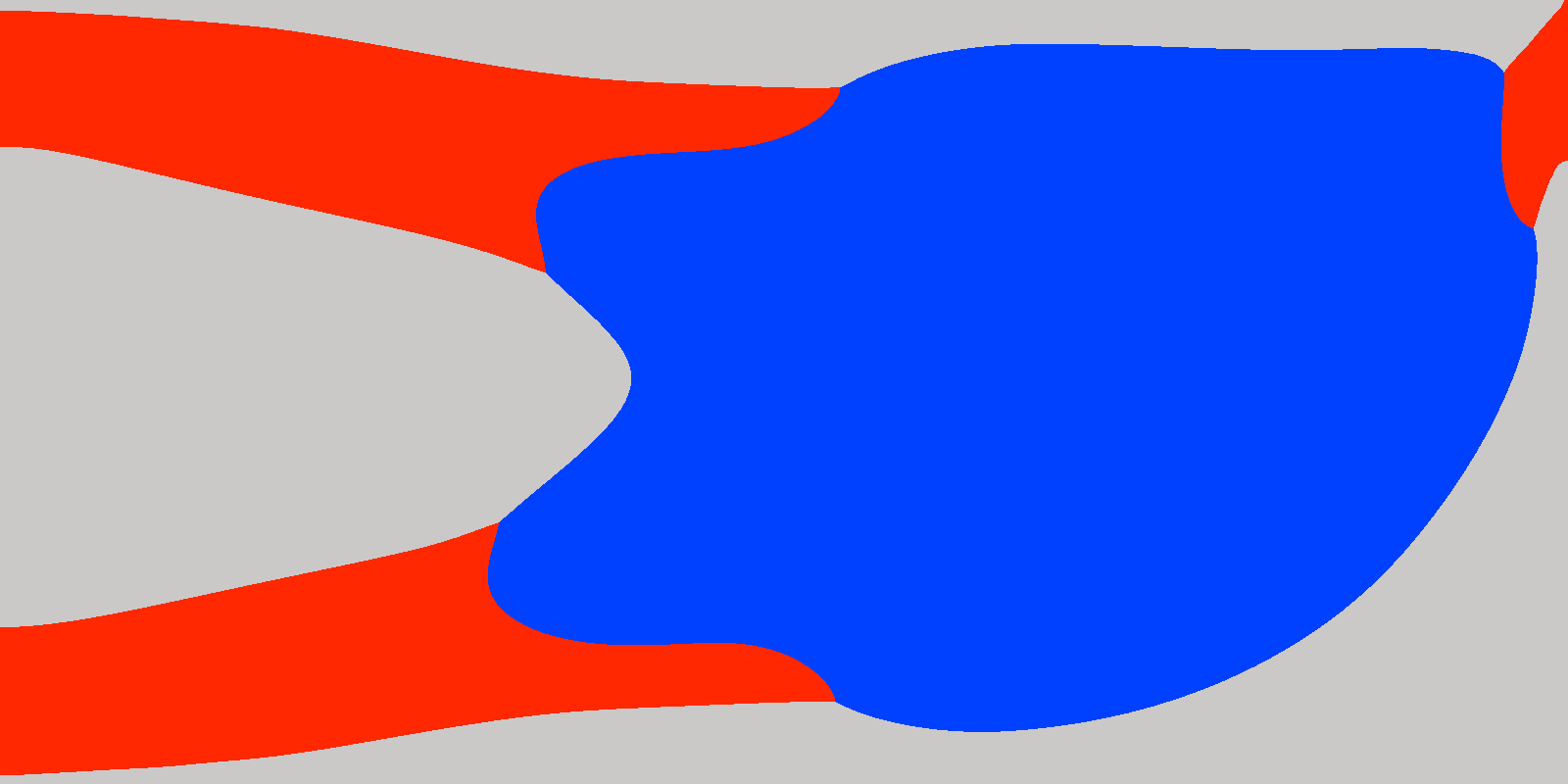}
		\subcaption{Case 14, step 20}\label{fig:init2 20}
	\end{minipage}
	\begin{minipage}[t]{0.24\linewidth}
		\centering
		\includegraphics[width=\linewidth]{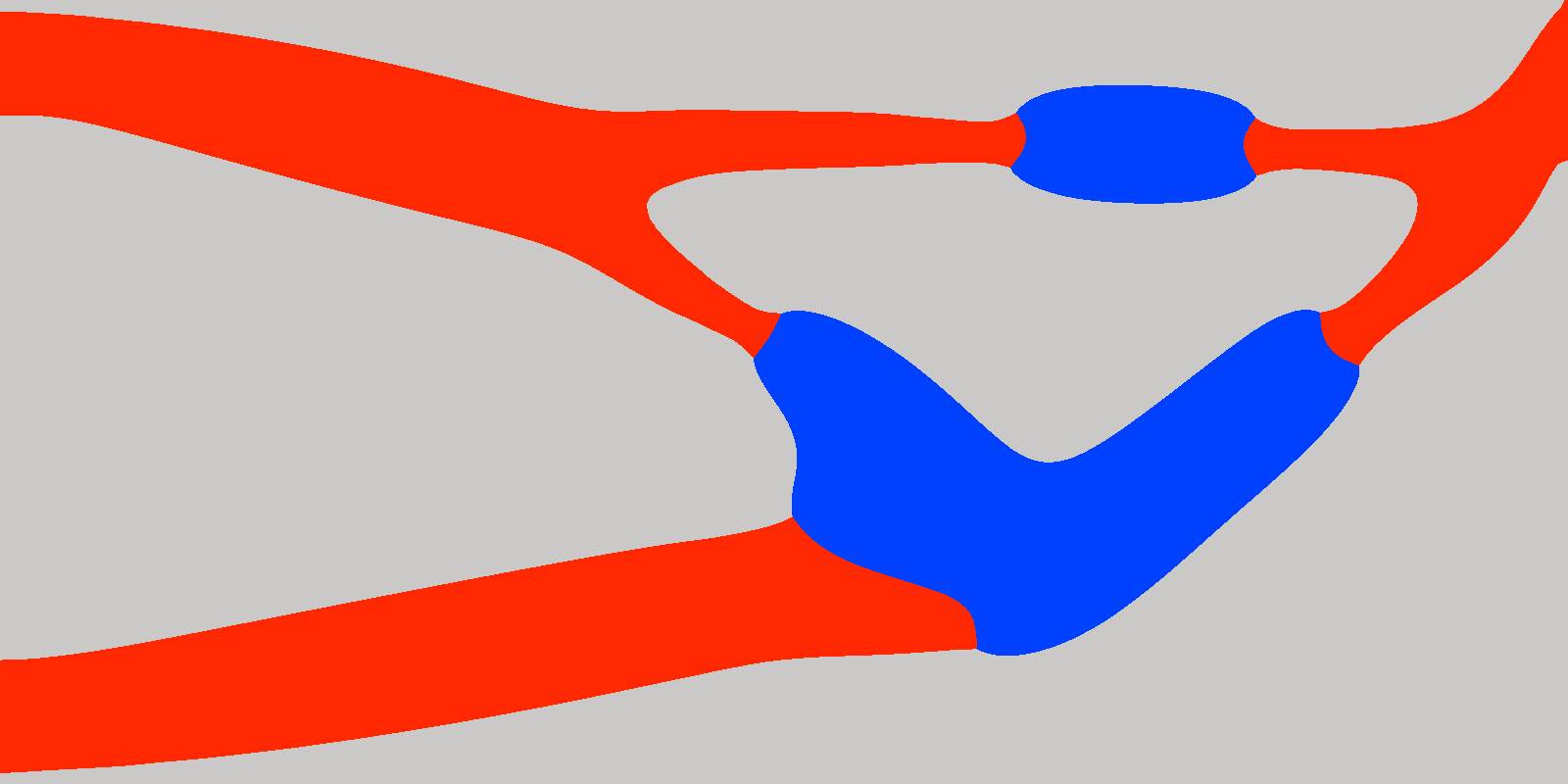}
		\subcaption{Case 14, step 50}\label{fig:int2 50}
	\end{minipage}
	\begin{minipage}[t]{0.24\linewidth}
		\centering
		\includegraphics[width=\linewidth]{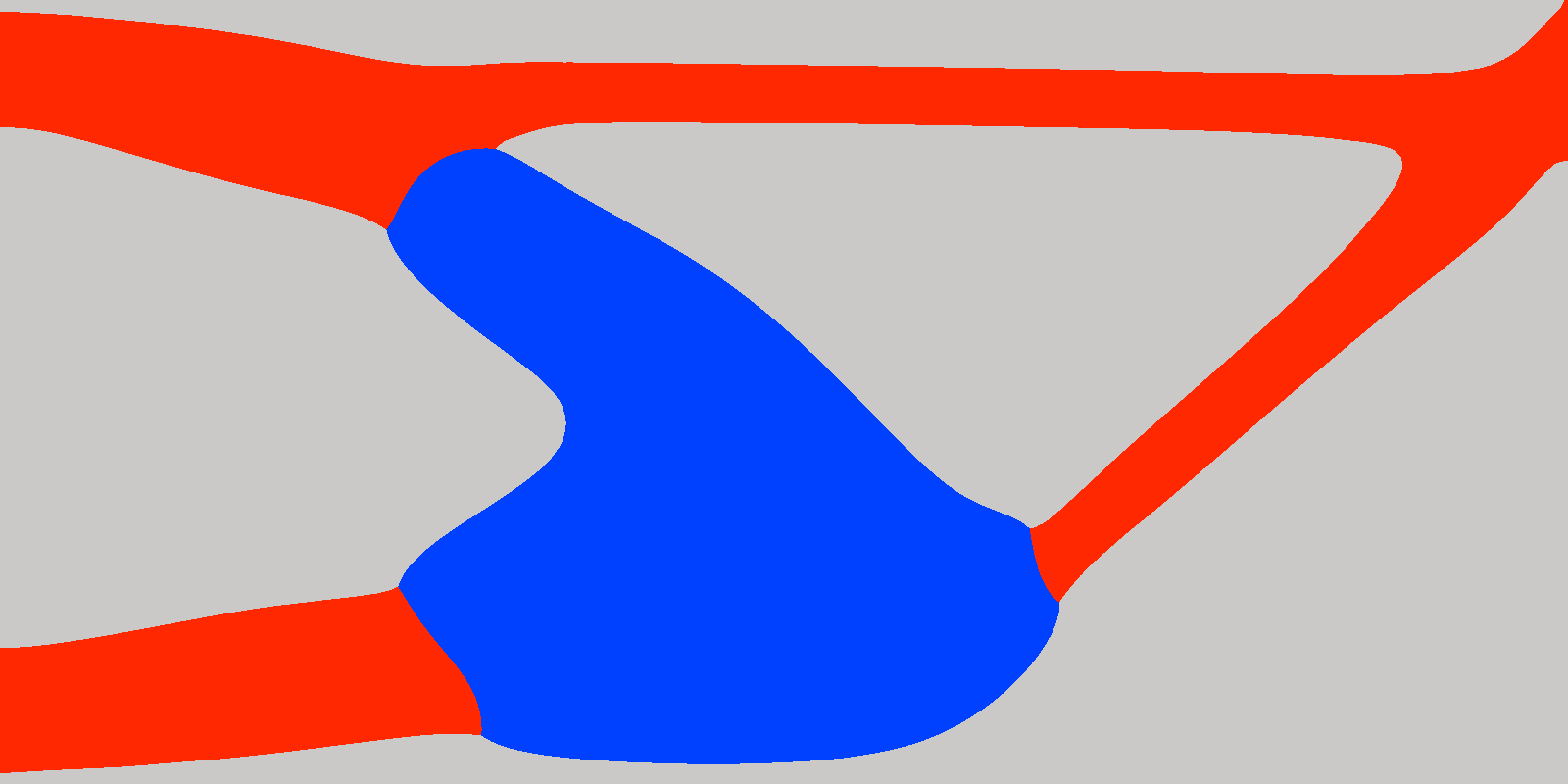}
		\subcaption{Case 14, optimal configration}\label{fig:init2 opt}
	\end{minipage}\\
	\begin{minipage}[t]{0.24\linewidth}
		\centering
		\includegraphics[width=\linewidth]{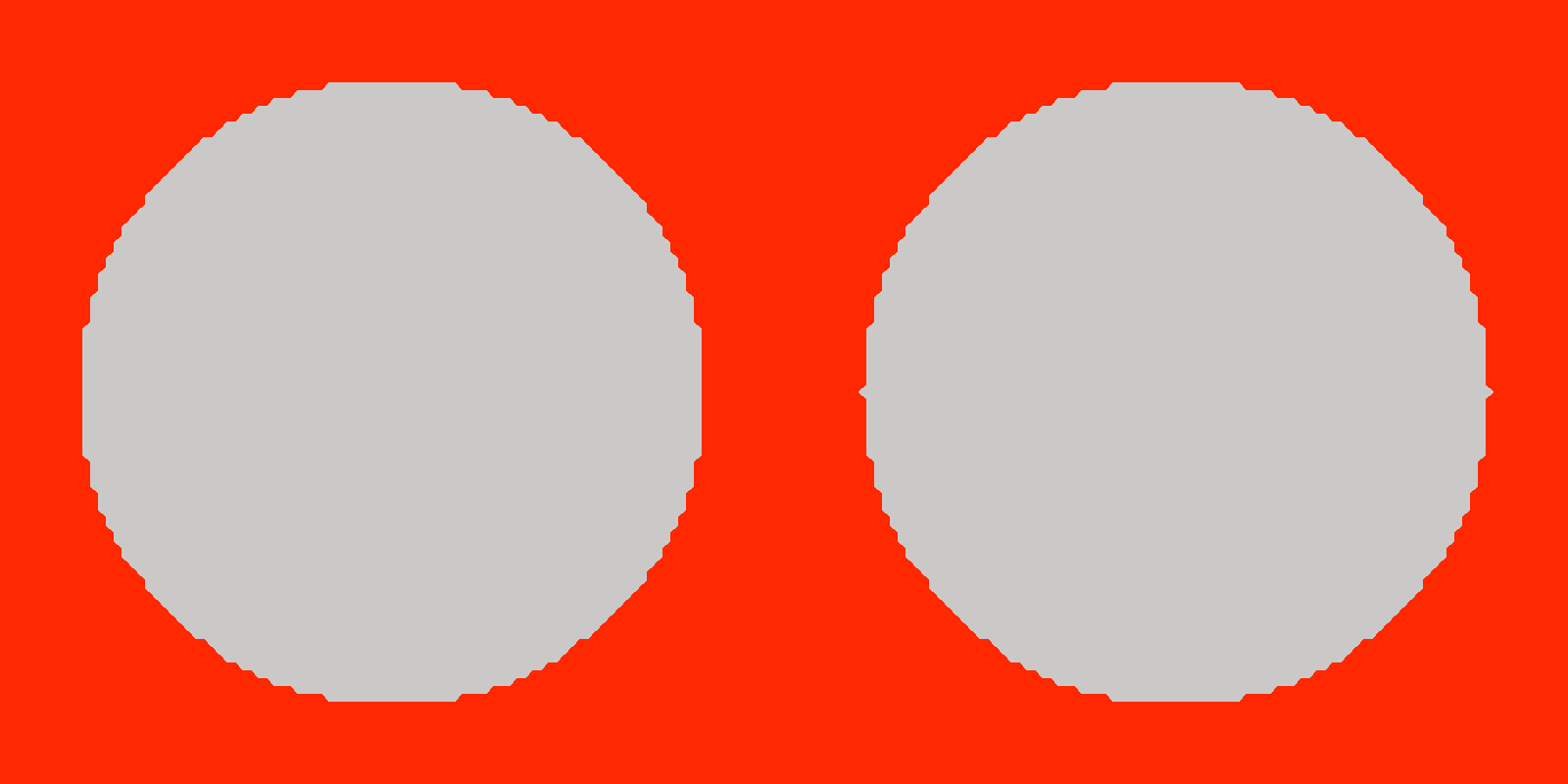}
		\subcaption{Case 15, initial configuration}\label{fig:init3 0}
	\end{minipage}
	\begin{minipage}[t]{0.24\linewidth}
		\centering
		\includegraphics[width=\linewidth]{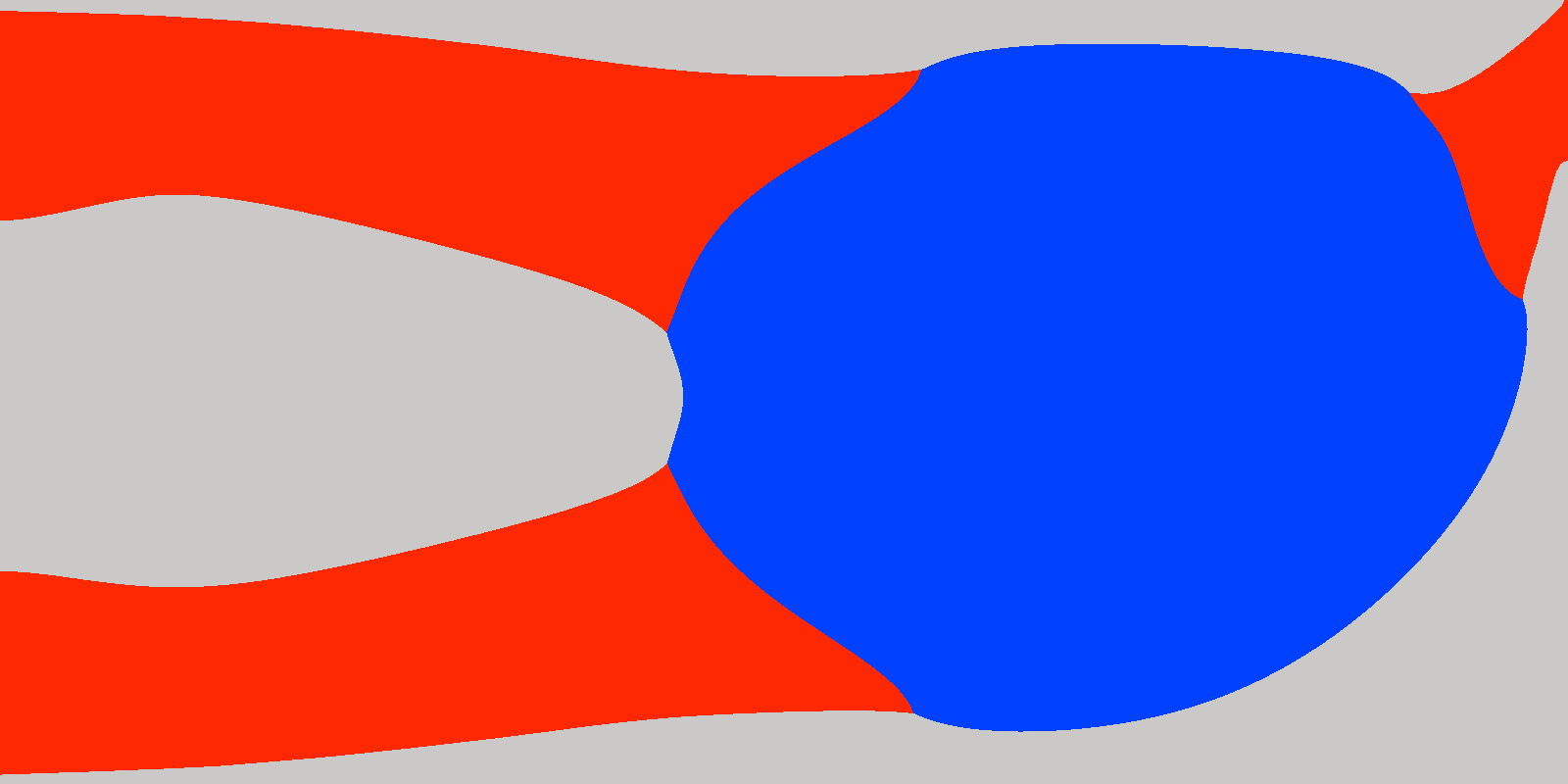}
		\subcaption{Case 15, step 20}\label{fig:init3 20}
	\end{minipage}
	\begin{minipage}[t]{0.24\linewidth}
		\centering
		\includegraphics[width=\linewidth]{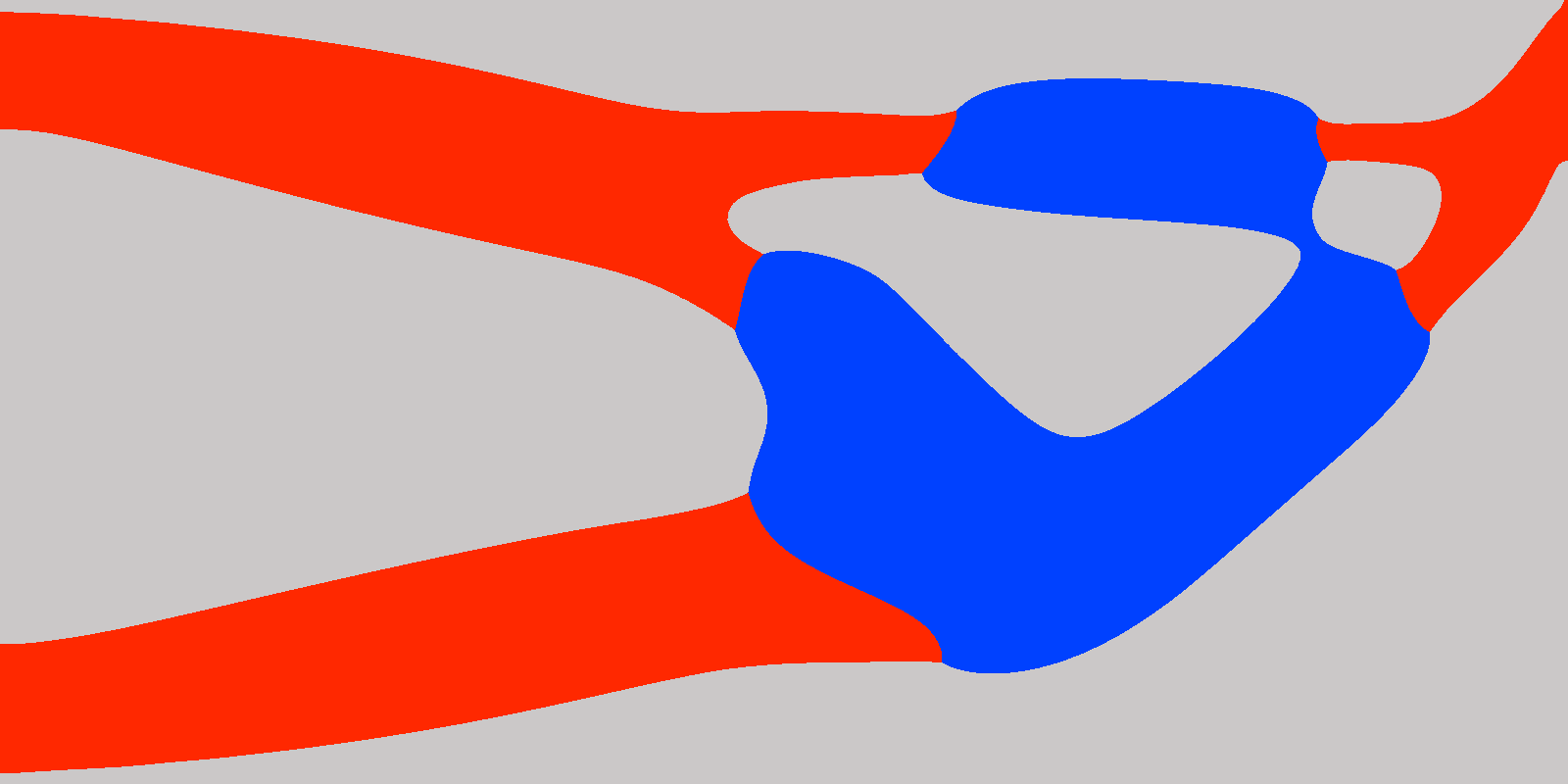}
		\subcaption{Case 15, step 50}\label{fig:int3 50}
	\end{minipage}
	\begin{minipage}[t]{0.24\linewidth}
		\centering
		\includegraphics[width=\linewidth]{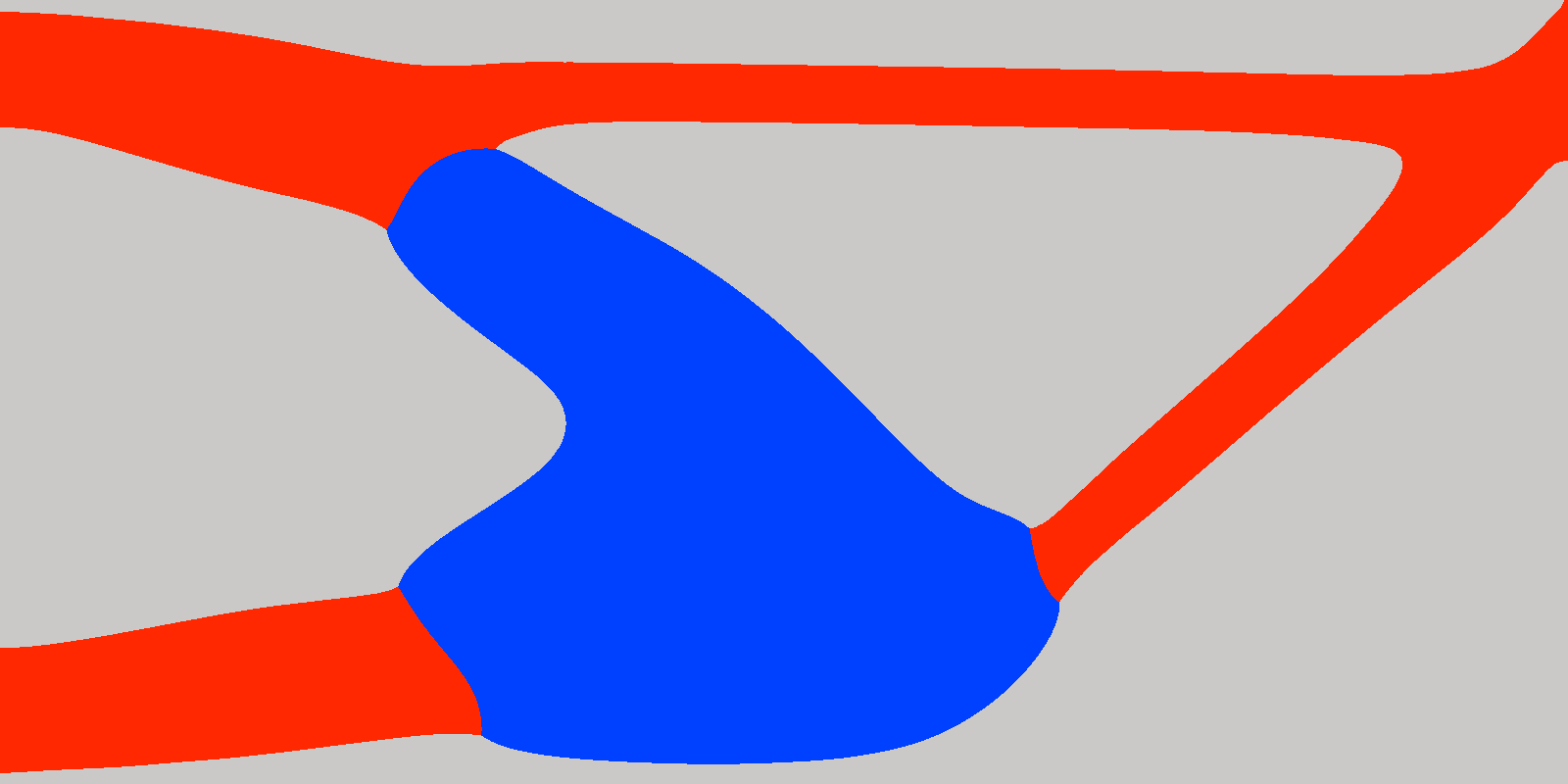}
		\subcaption{Case 15, optimal configration}\label{fig:init3 opt}
	\end{minipage}\\
	\begin{minipage}[t]{0.24\linewidth}
		\centering
		\includegraphics[width=\linewidth]{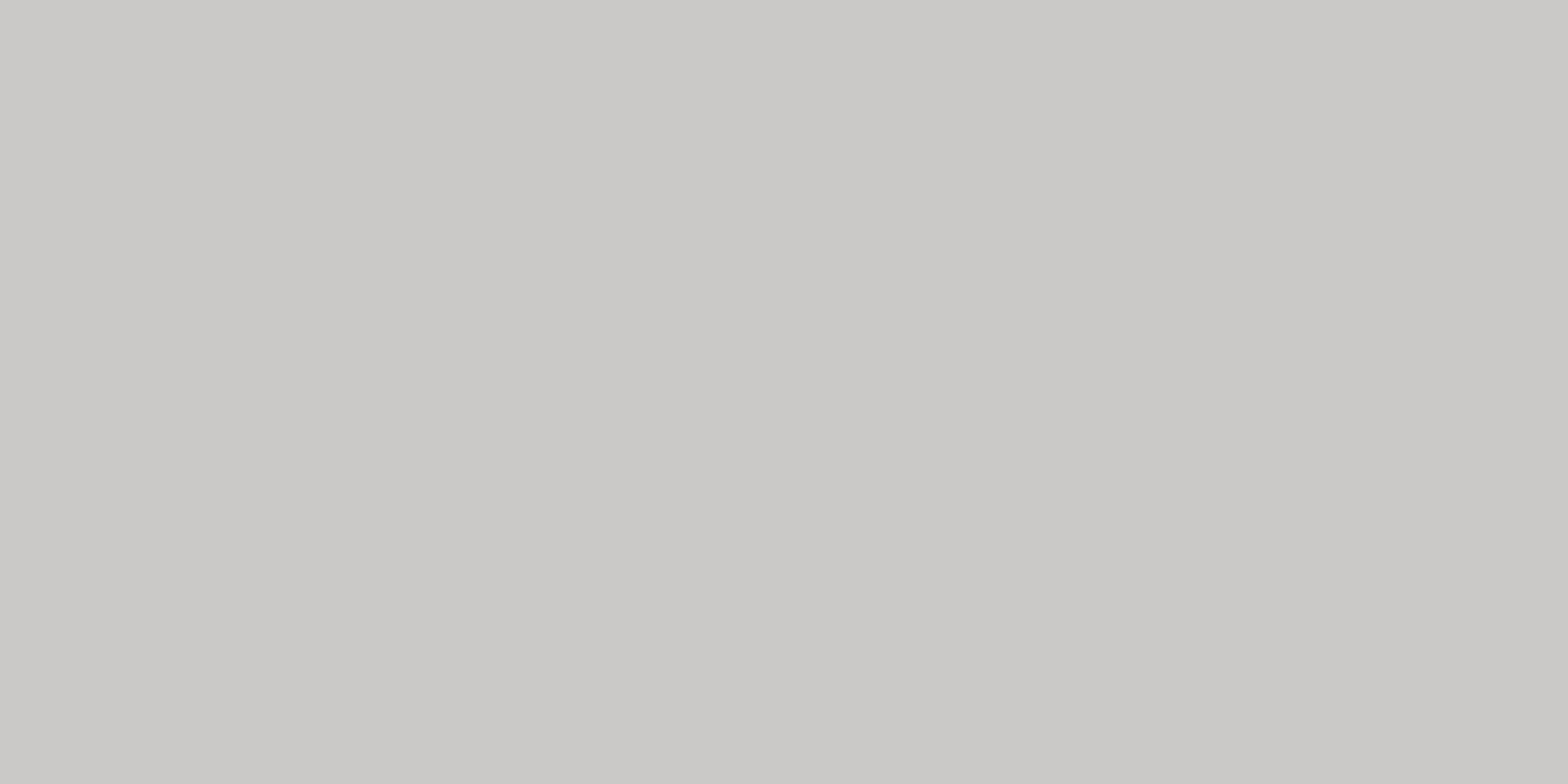}
		\subcaption{Case 16, initial configuration}\label{fig:init4 0}
	\end{minipage}
	\begin{minipage}[t]{0.24\linewidth}
		\centering
		\includegraphics[width=\linewidth]{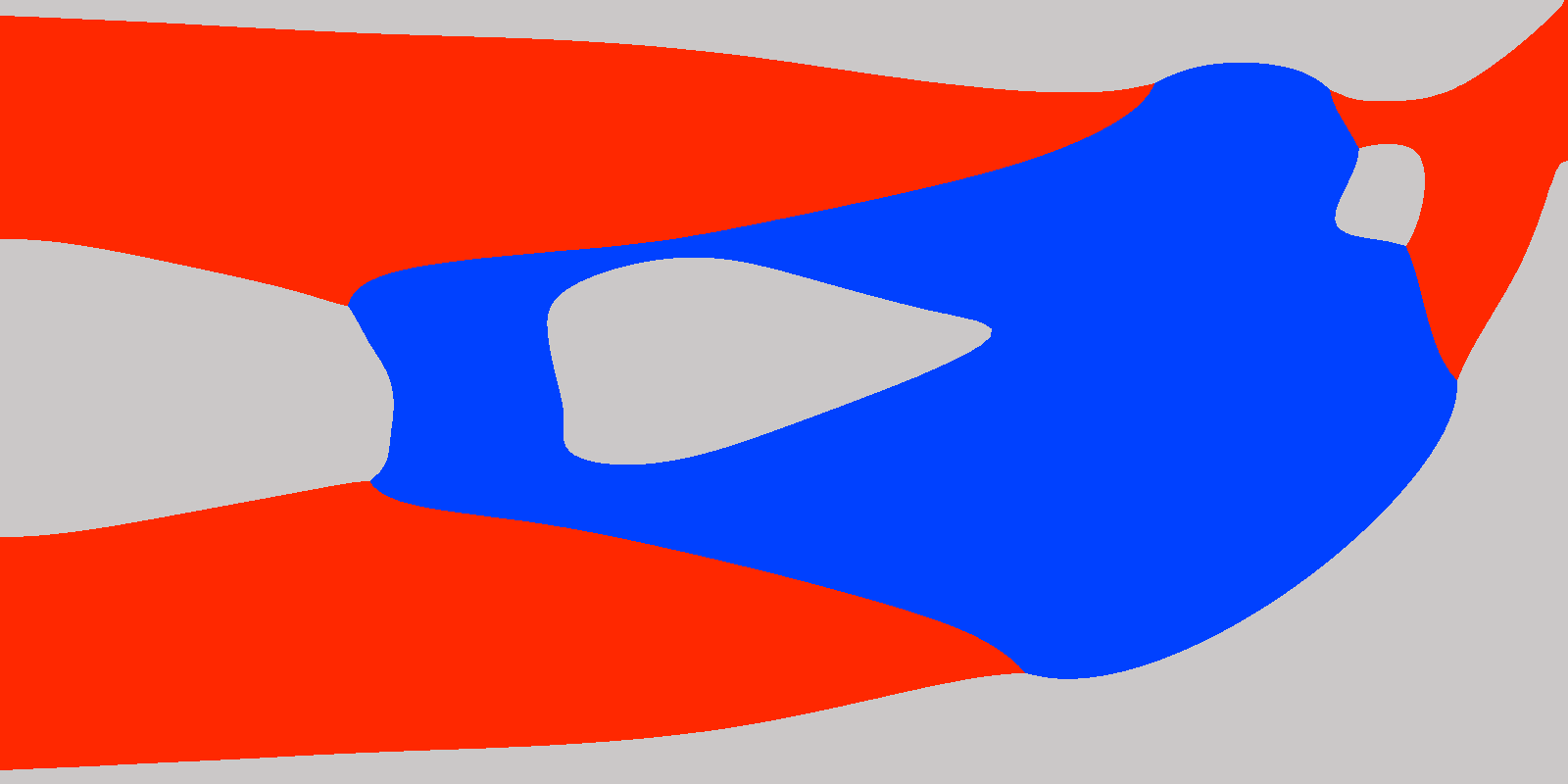}
		\subcaption{Case 16, step 20}\label{fig:init4 20}
	\end{minipage}
	\begin{minipage}[t]{0.24\linewidth}
		\centering
		\includegraphics[width=\linewidth]{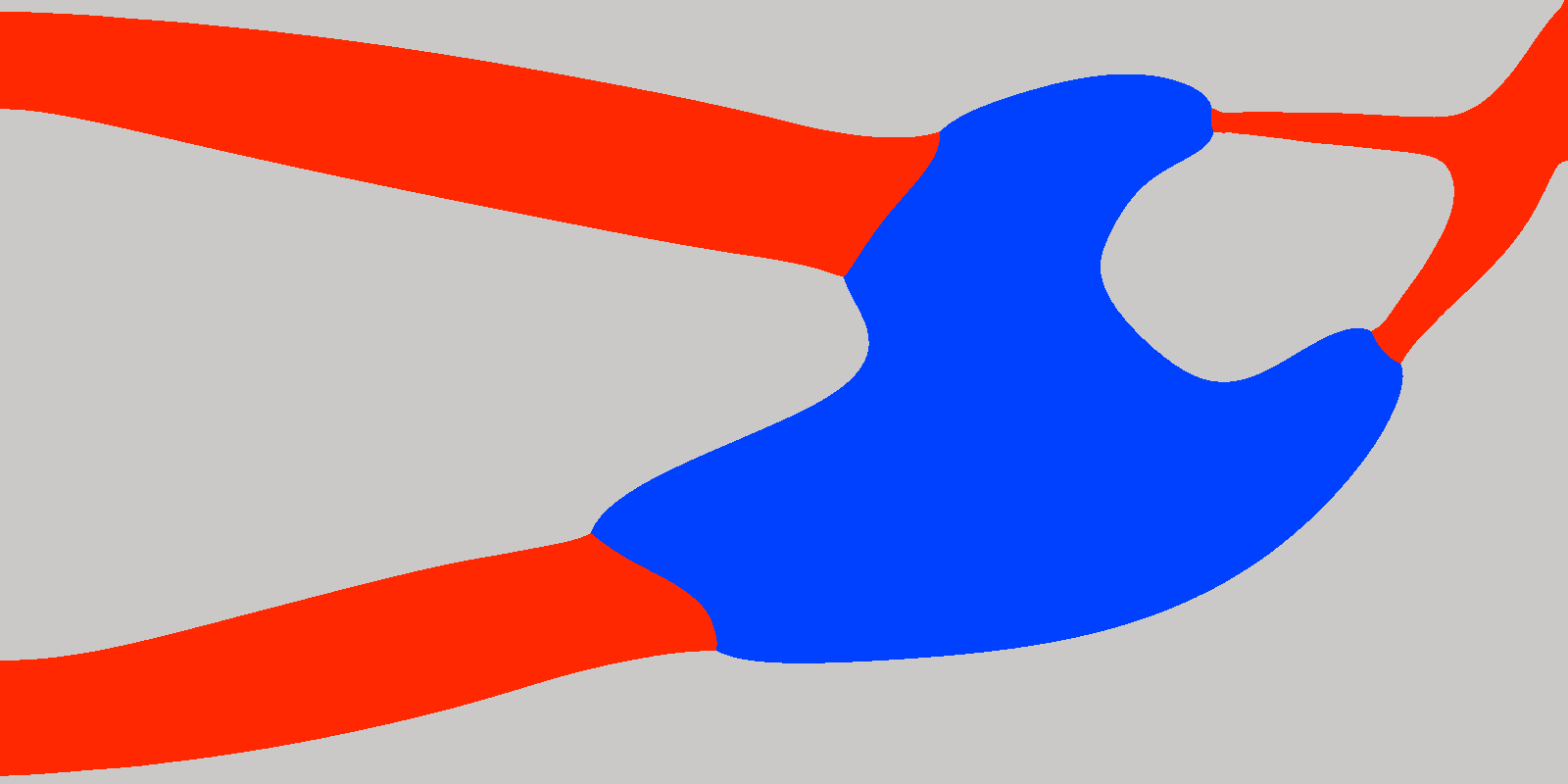}
		\subcaption{Case 16, step 50}\label{fig:int4 50}
	\end{minipage}
	\begin{minipage}[t]{0.24\linewidth}
		\centering
		\includegraphics[width=\linewidth]{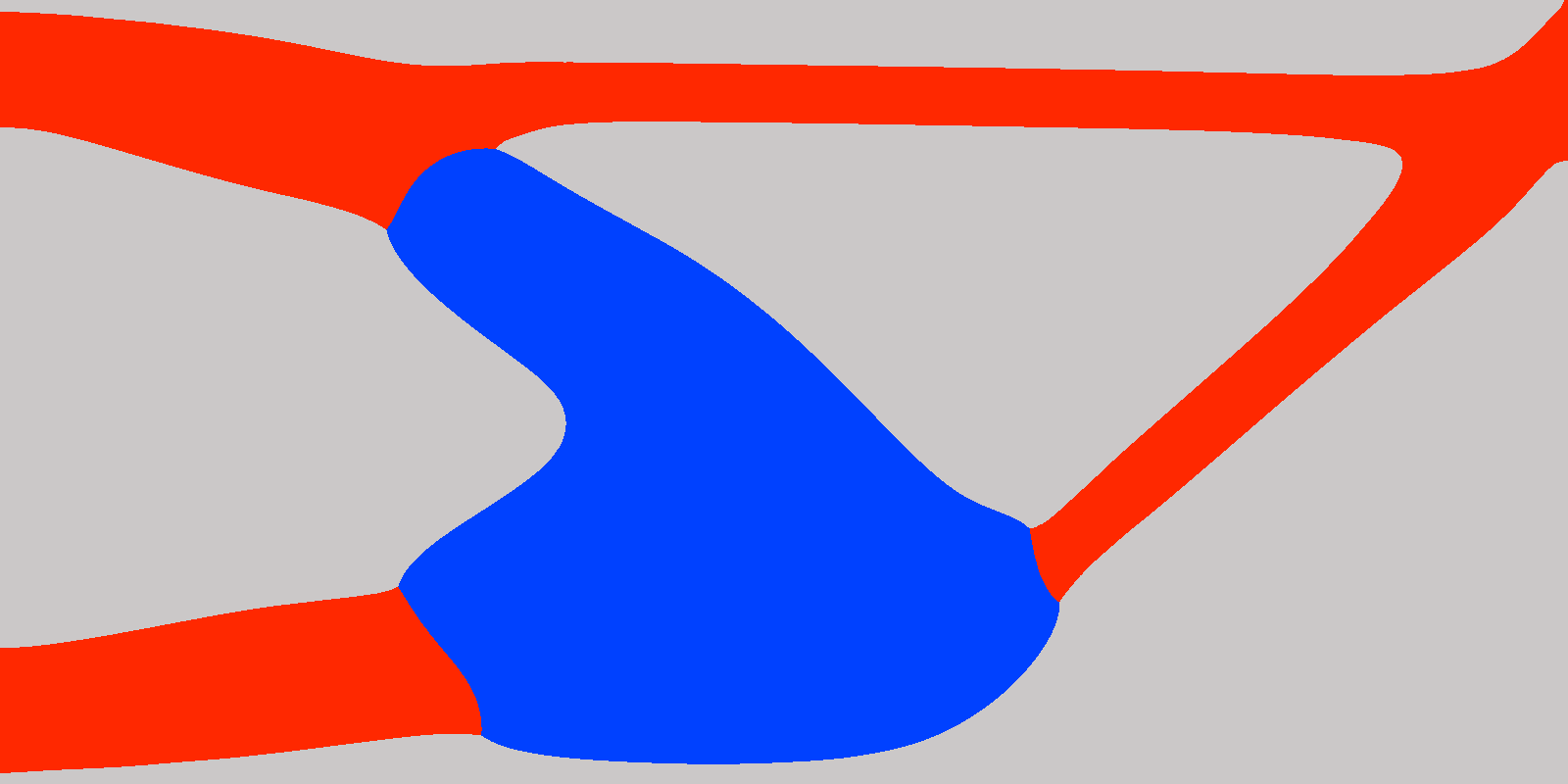}
		\subcaption{Case 16, optimal Configration}\label{fig:init4 opt}
	\end{minipage}
	\caption{Optimal configurations emerging from various initial configurations, showing that the proposed method is insensitive to the initial configuration.}\label{fig:init}
\end{figure}

Finally, we examined the effect of different initial configurations on the optimal configurations. The regularization parameter $\tau_{ij}$ was set to $1 \times 10^{-2}$. Fig. \ref{fig:init} shows the optimization results in Cases 13--16 with different initial configurations. Each initial configuration was set to a topologically different structure composed of material 0 or 1.
The upper volume constraints in Cases 13--16 were mostly satisfied (volume constraint functions $<10^{-10}$). The different initial configurations converged to almost identical optimal configurations (c.f. panels \subref{fig:init1 opt}, \subref{fig:init2 opt}, \subref{fig:init3 opt}, and \subref{fig:init4 opt} of Fig. \ref{fig:init}) and yielded the same value of the objective functions ($2.324896\times 10^{-11}$), confirming that all initial configurations led to the same appropriate optimal configuration. This result confirms the low dependency of the obtained optimal configurations on the initial configuration.

\subsection{Two-dimensional compliant mechanism optimization problems}
In this subsection, the proposed optimization method is applied to two-dimensional compliant mechanism optimization problems.

\begin{figure}[H]
	\centering
	\includegraphics[width=10cm]{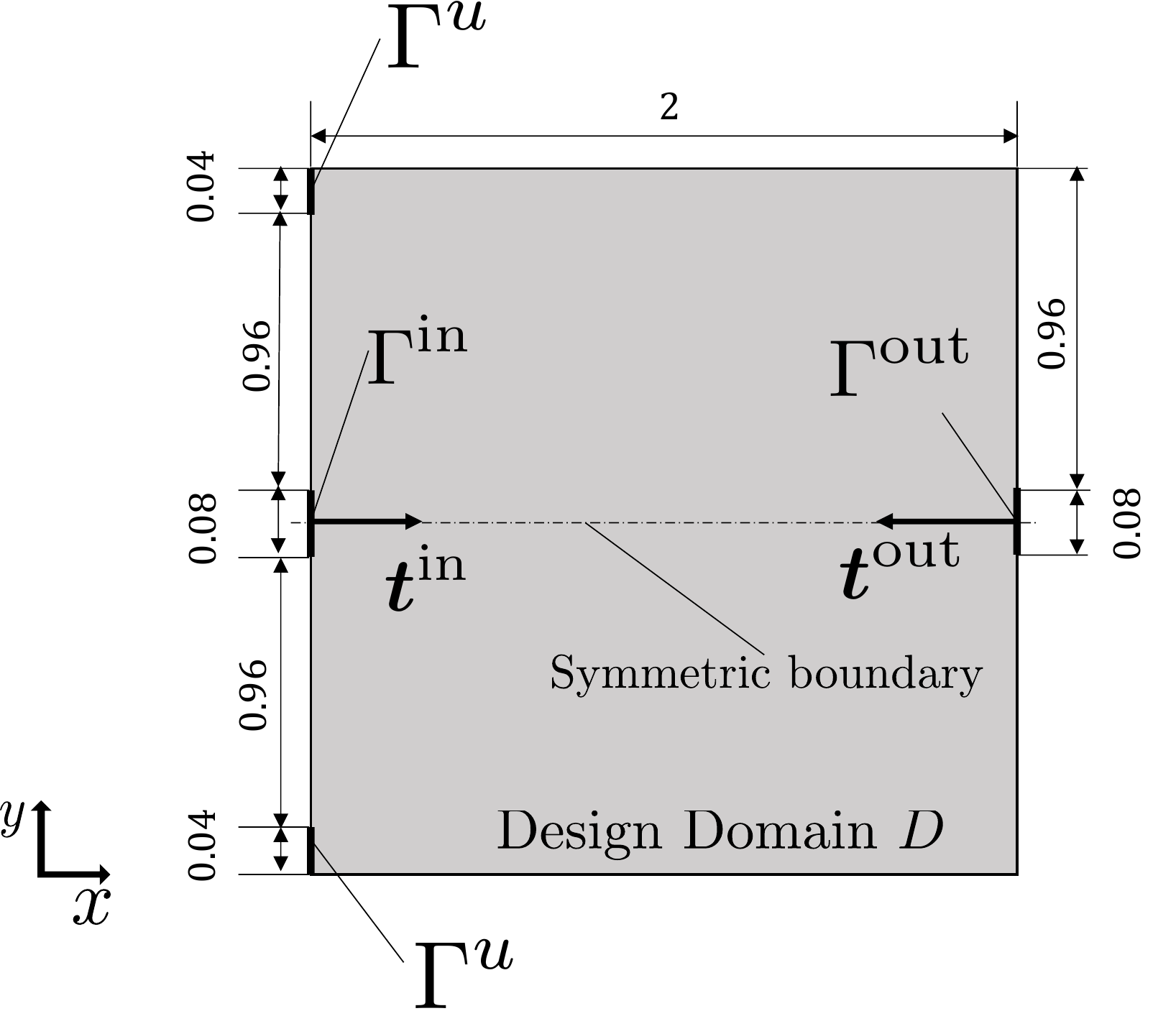}
	\caption{Setup of the two-dimensional compliant mechanism optimization problem}
	\label{fig: compmech problem 2d}
\end{figure}

\begin{figure}[H]
	\begin{minipage}[b]{0.24\linewidth}
		\centering
		\includegraphics[width=\linewidth]{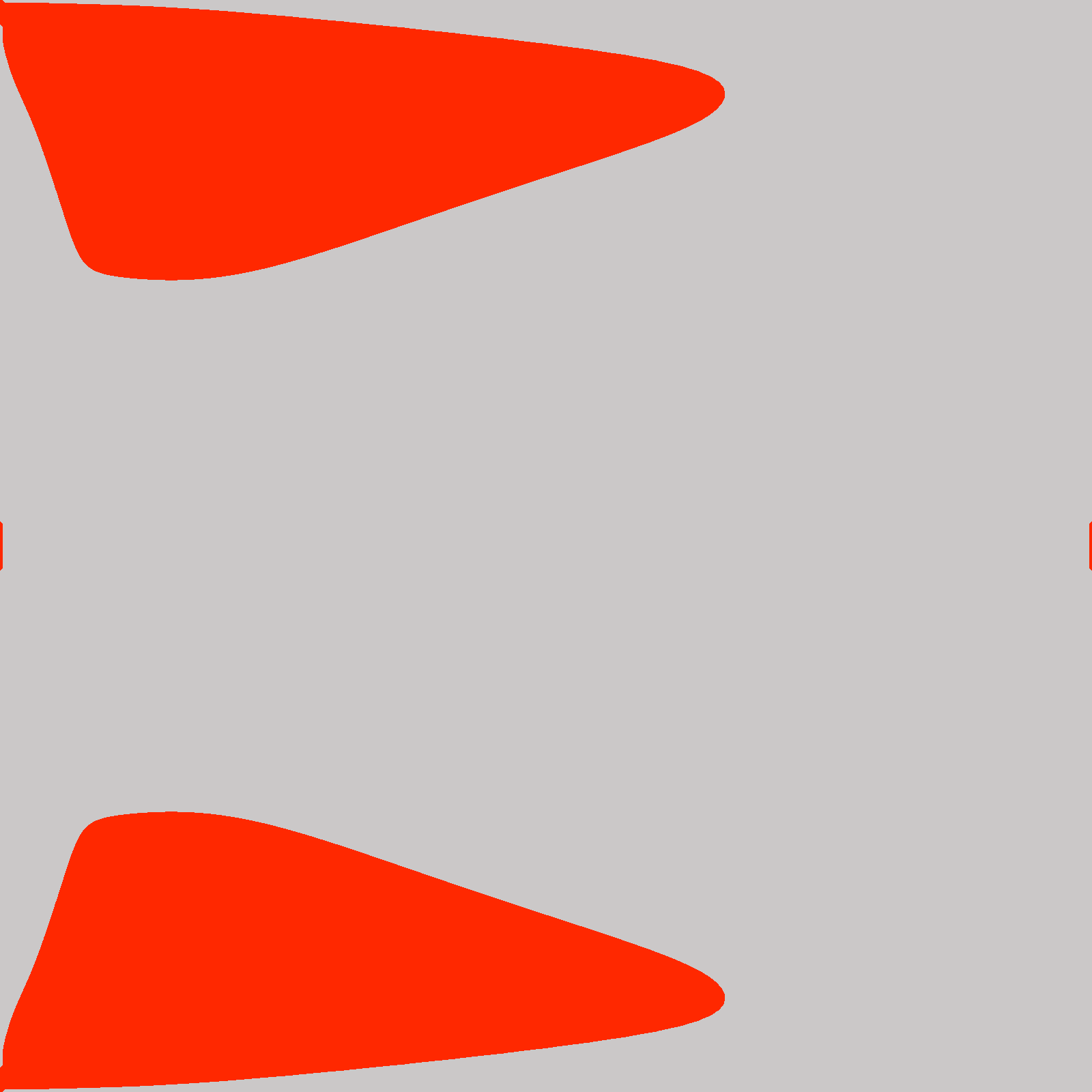}
		\subcaption{Step 10}\label{fig:compmech M2 10}
	\end{minipage}
	\begin{minipage}[b]{0.24\linewidth}
		\centering
		\includegraphics[width=\linewidth]{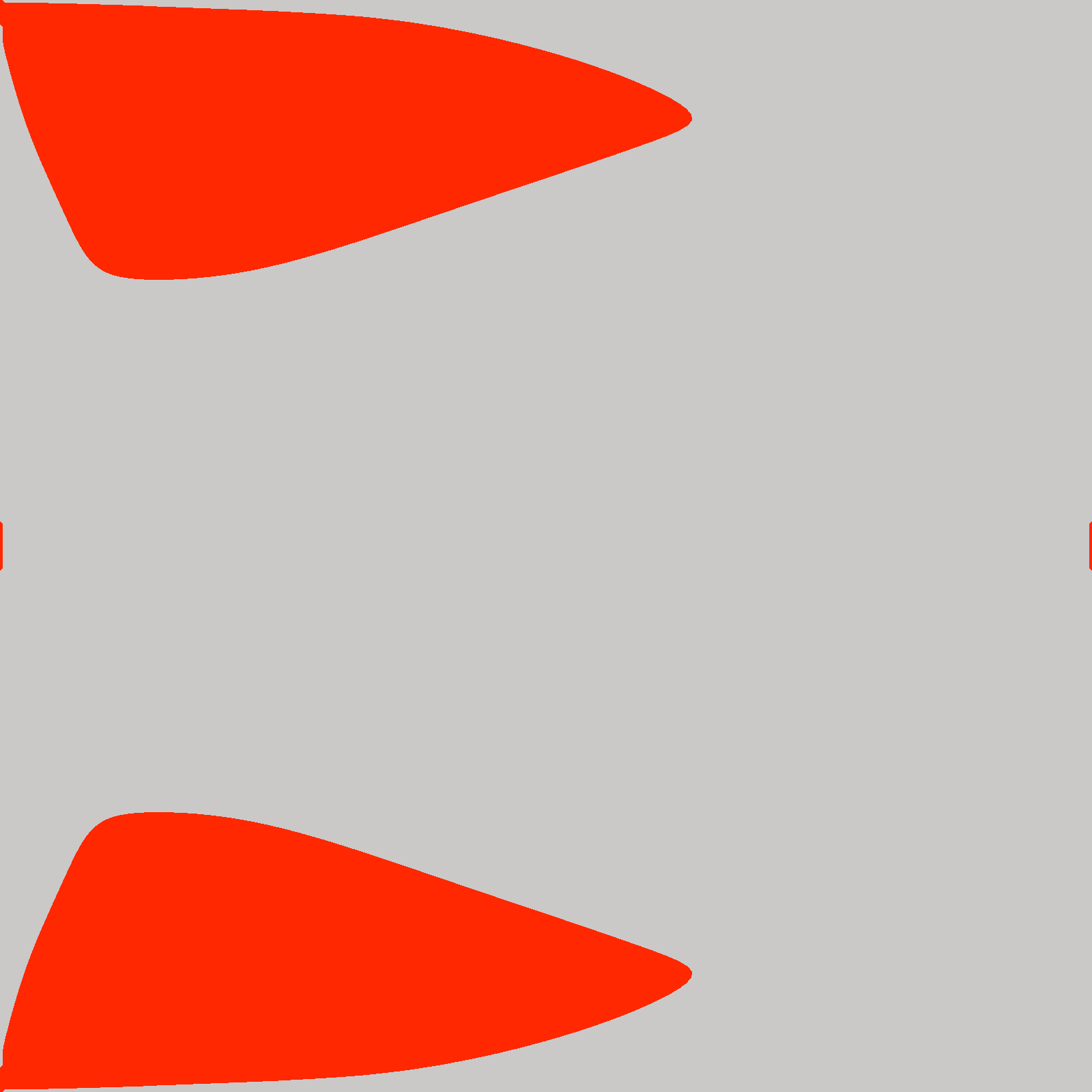}
		\subcaption{Step 50}\label{fig:compmech M2 50}
	\end{minipage}
	\begin{minipage}[b]{0.24\linewidth}
		\centering
		\includegraphics[width=\linewidth]{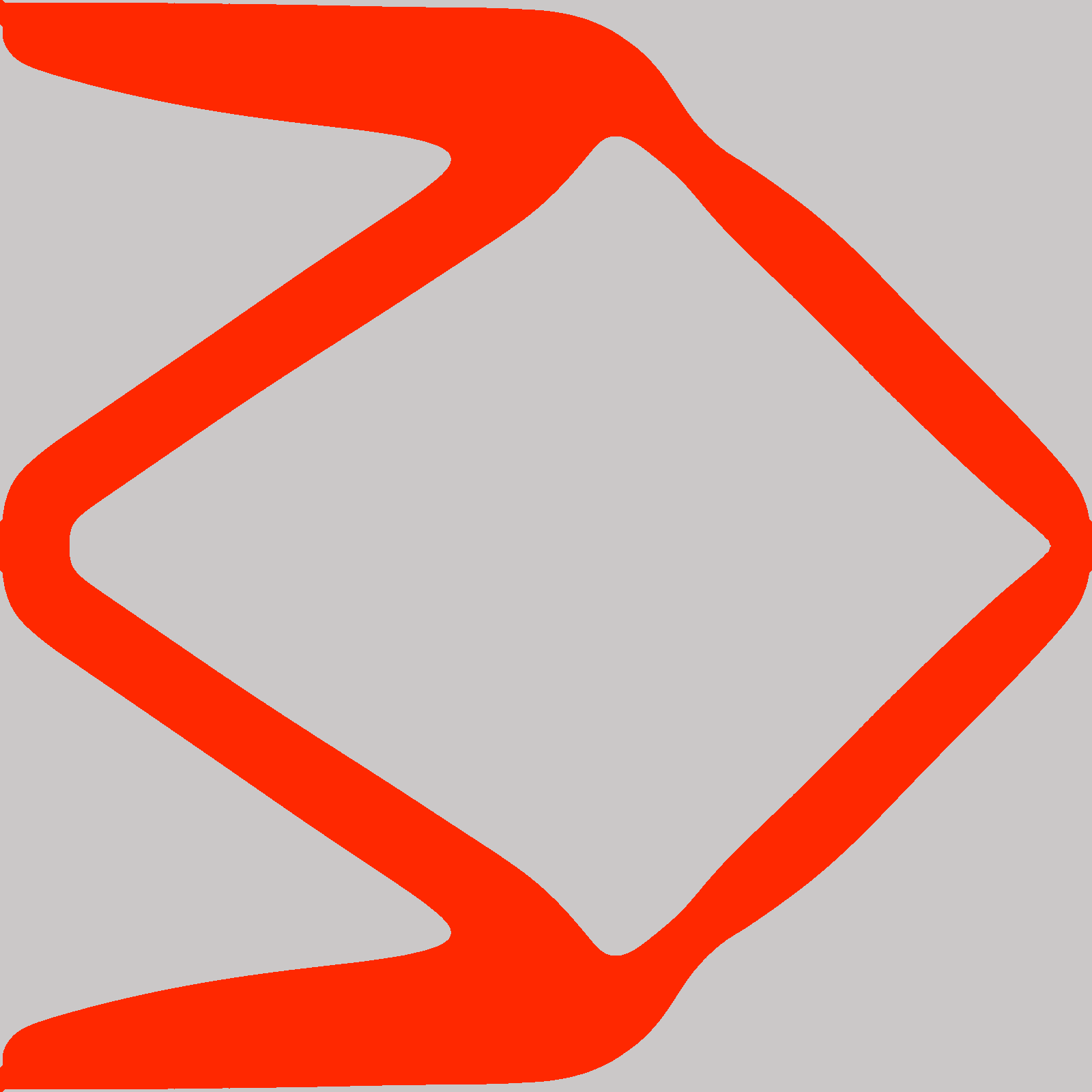}
		\subcaption{Step 200}\label{fig:compmech M2 200}
	\end{minipage}
	\begin{minipage}[b]{0.24\linewidth}
		\centering
		\includegraphics[width=\linewidth]{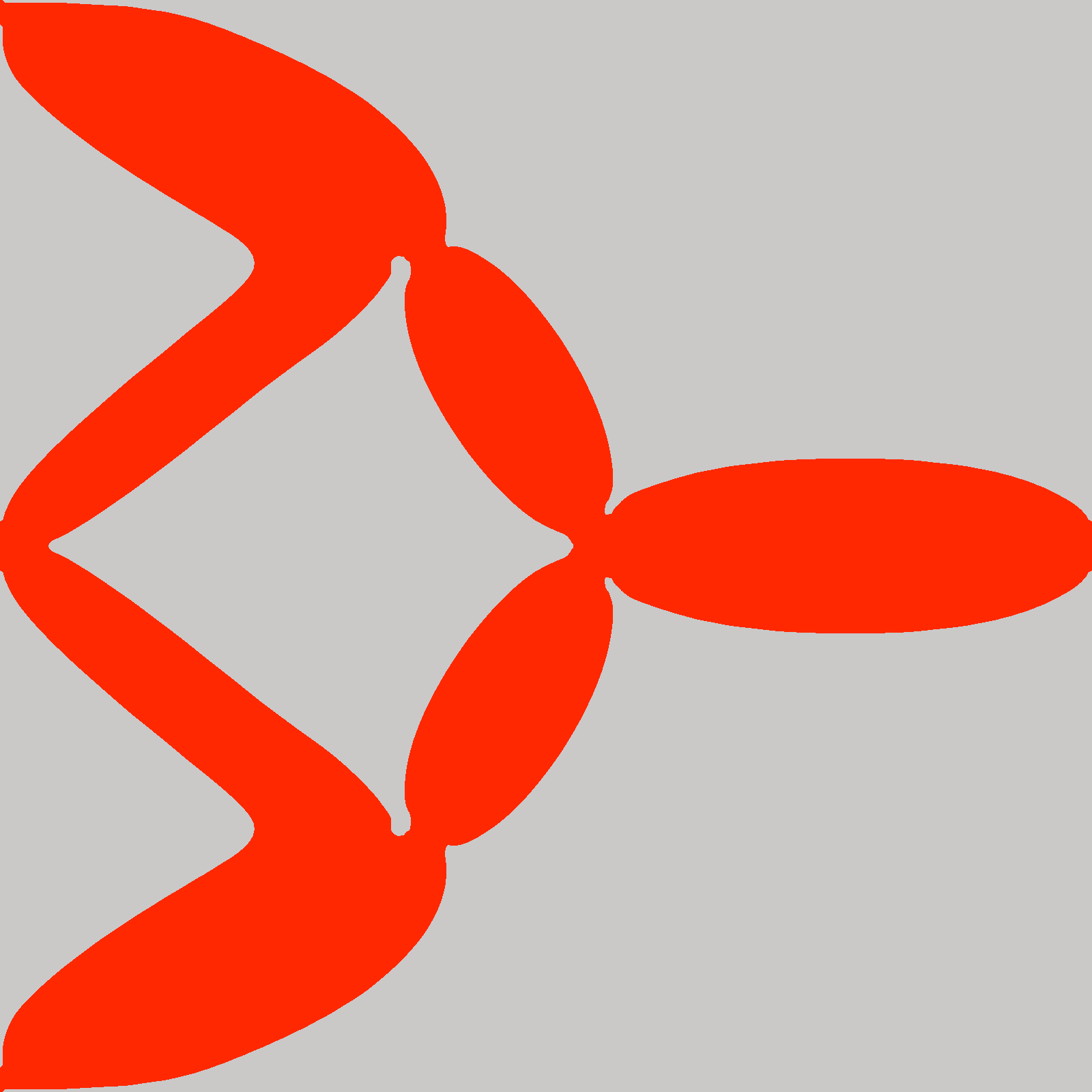}
		\subcaption{Optimal Configration}\label{fig:compmech M2 opt}
	\end{minipage}
	\caption{Intermediate results and optimal configuration of two materials (Case 17)}\label{fig:compmech M2}
\end{figure}

\begin{figure}[H]
	\begin{minipage}[b]{0.24\linewidth}
		\centering
		\includegraphics[width=\linewidth]{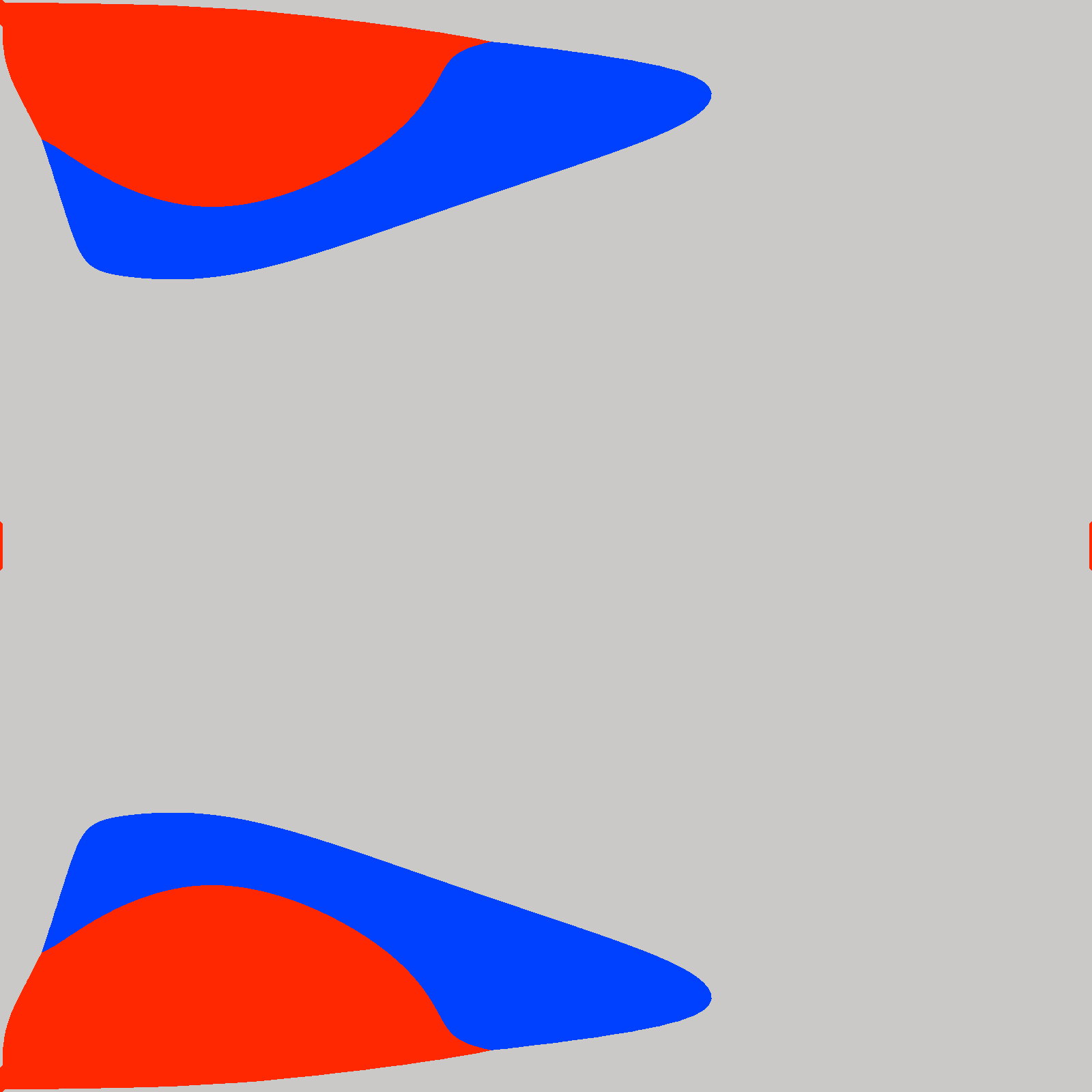}
		\subcaption{Step 10}\label{fig:compmech M3 10}
	\end{minipage}
	\begin{minipage}[b]{0.24\linewidth}
		\centering
		\includegraphics[width=\linewidth]{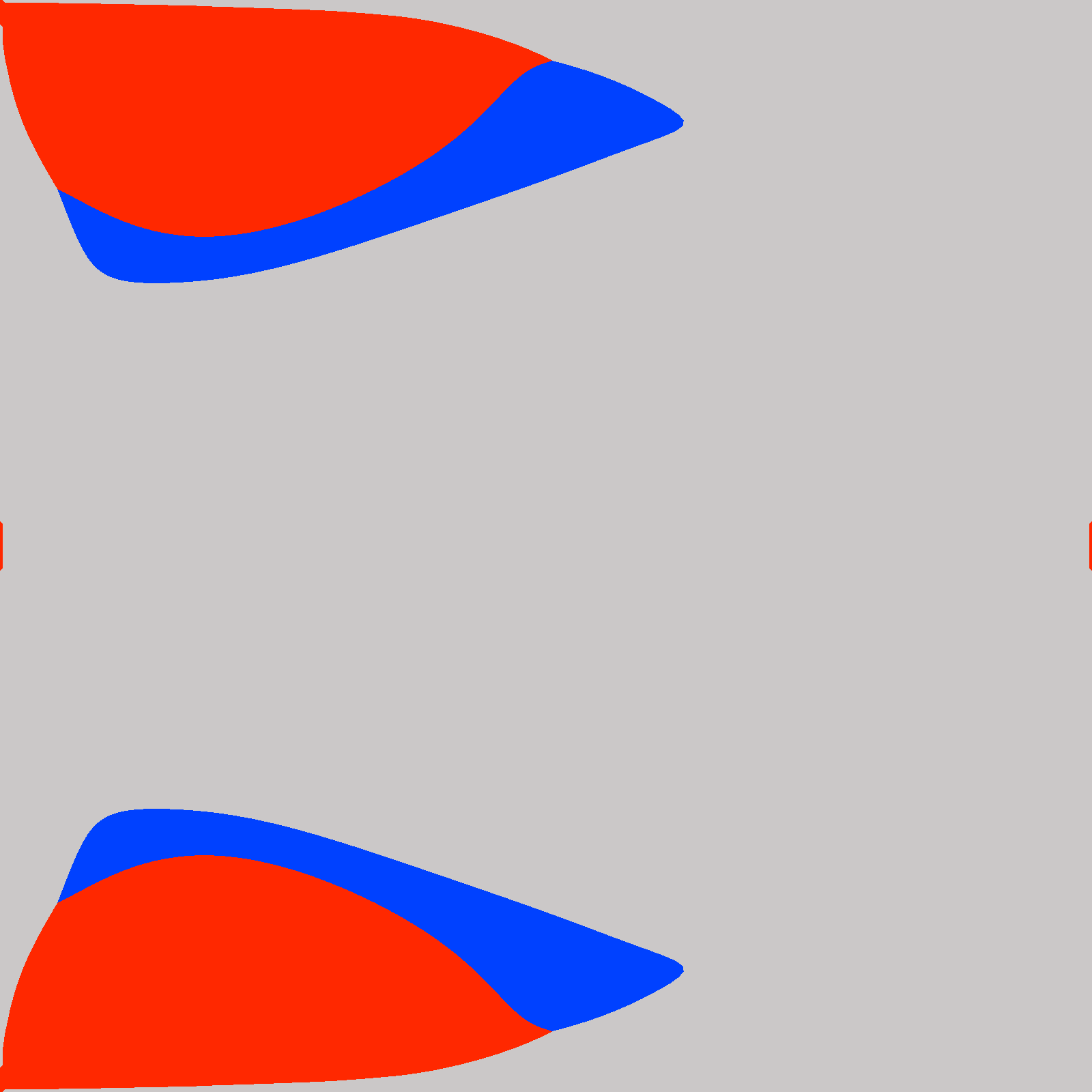}
		\subcaption{Step 50}\label{fig:compmech M3 50}
	\end{minipage}
	\begin{minipage}[b]{0.24\linewidth}
		\centering
		\includegraphics[width=\linewidth]{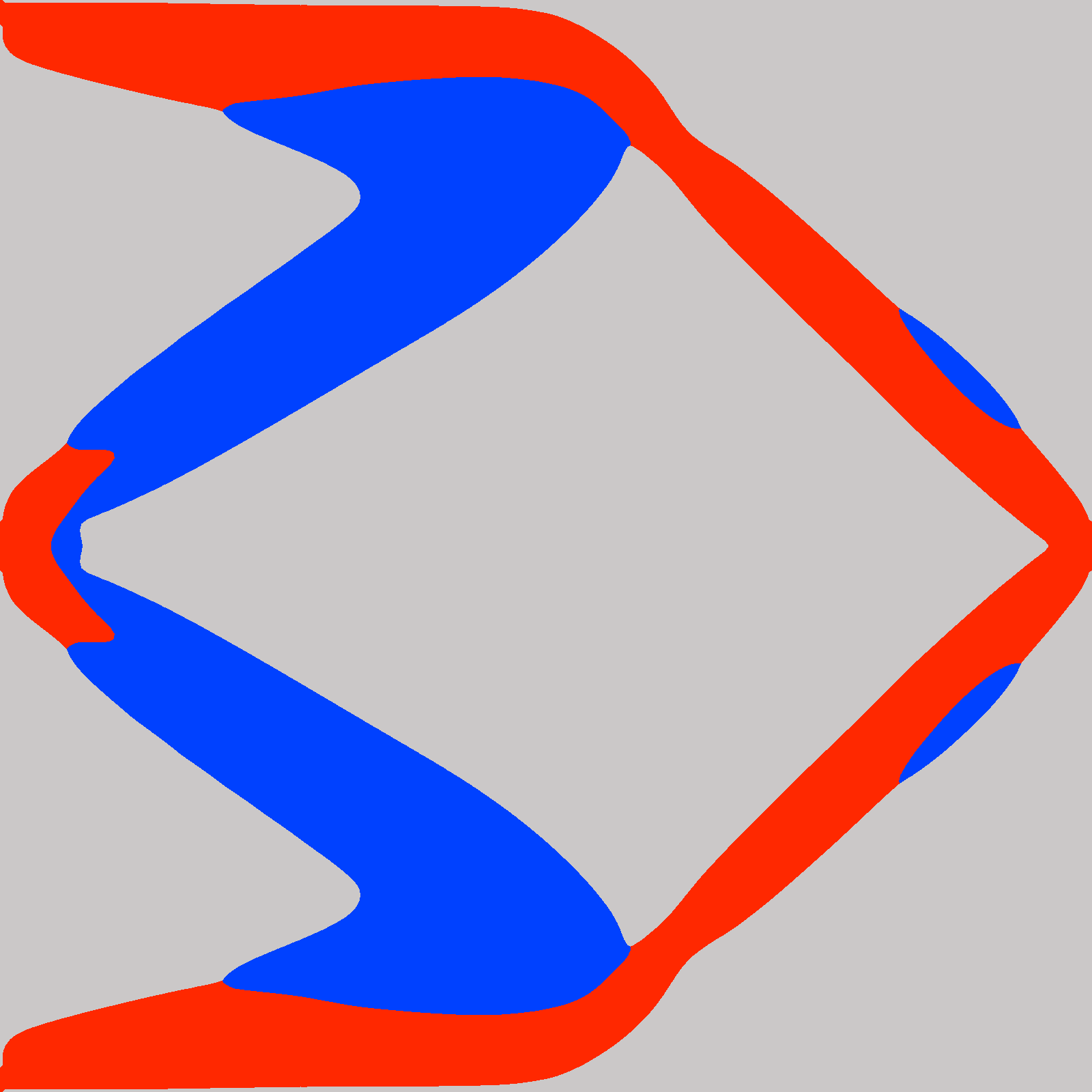}
		\subcaption{Step 200}\label{fig:compmech M3 200}
	\end{minipage}
	\begin{minipage}[b]{0.24\linewidth}
		\centering
		\includegraphics[width=\linewidth]{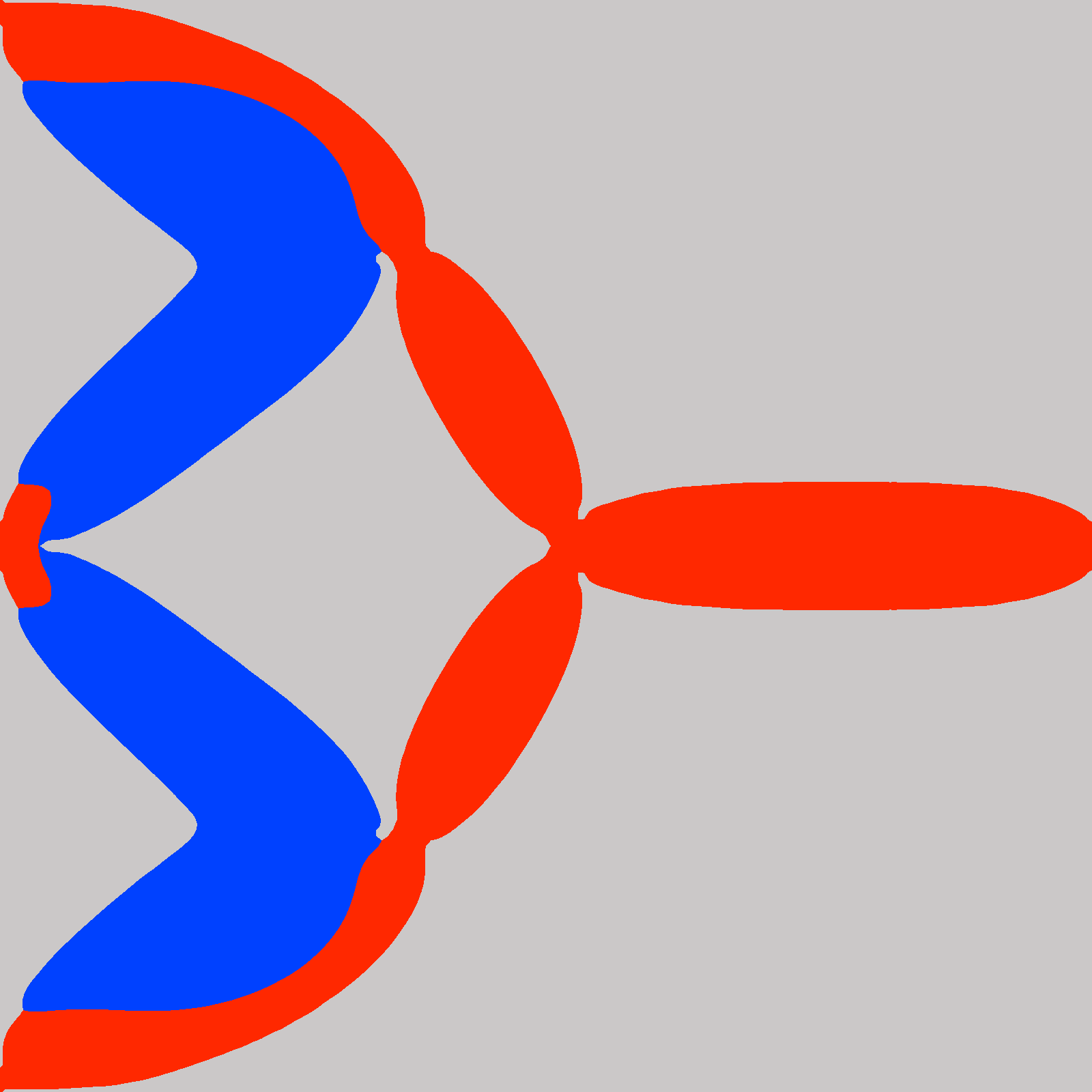}
		\subcaption{Optimal Configration}\label{fig:compmech M3 opt}
	\end{minipage}
	\caption{Intermediate results and optimal configuration of three materials (Case 18)}\label{fig:compmech M3}
\end{figure}

\begin{figure}[H]
	\begin{minipage}[b]{0.24\linewidth}
		\centering
		\includegraphics[width=\linewidth]{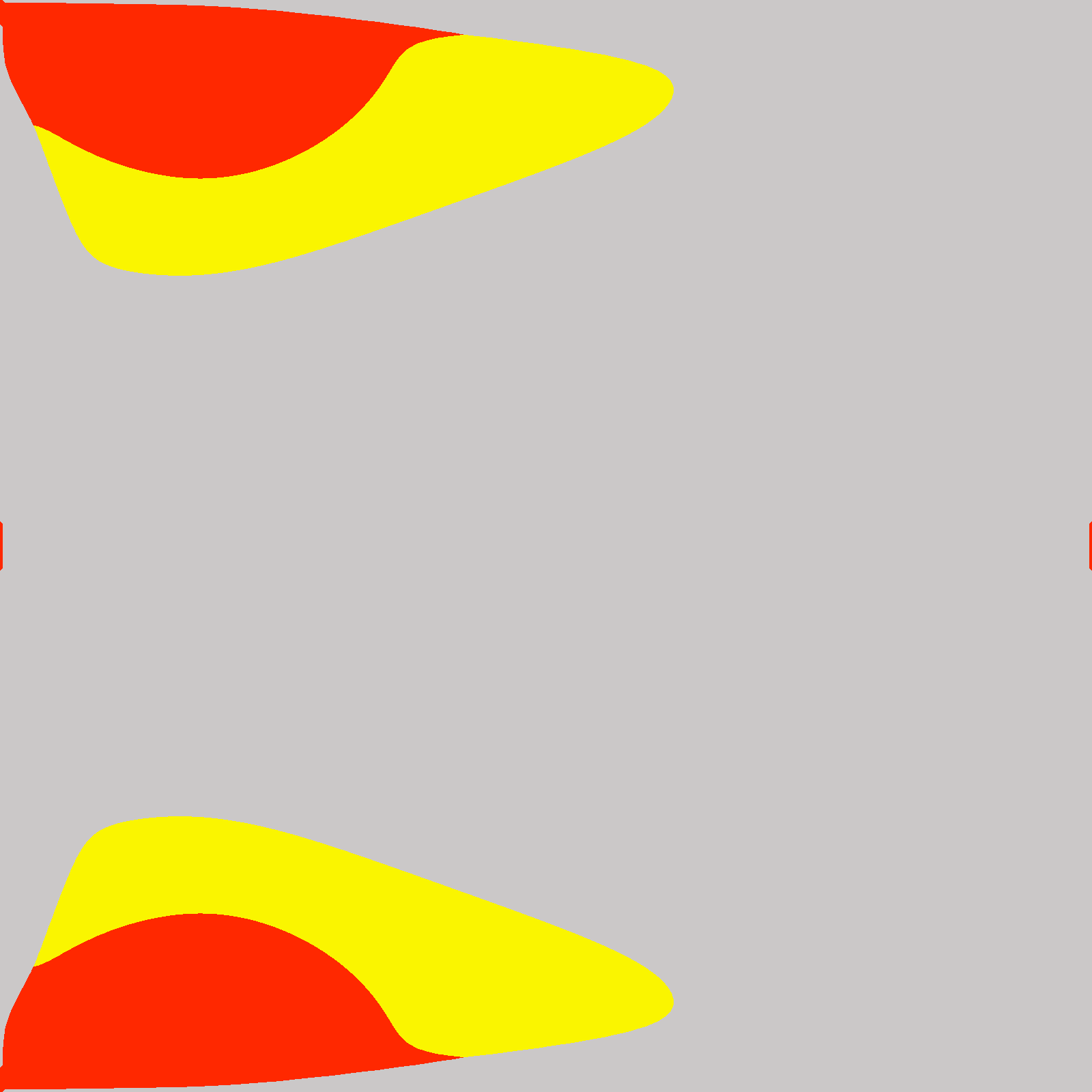}
		\subcaption{Step 10}\label{fig:compmech M4 10}
	\end{minipage}
	\begin{minipage}[b]{0.24\linewidth}
		\centering
		\includegraphics[width=\linewidth]{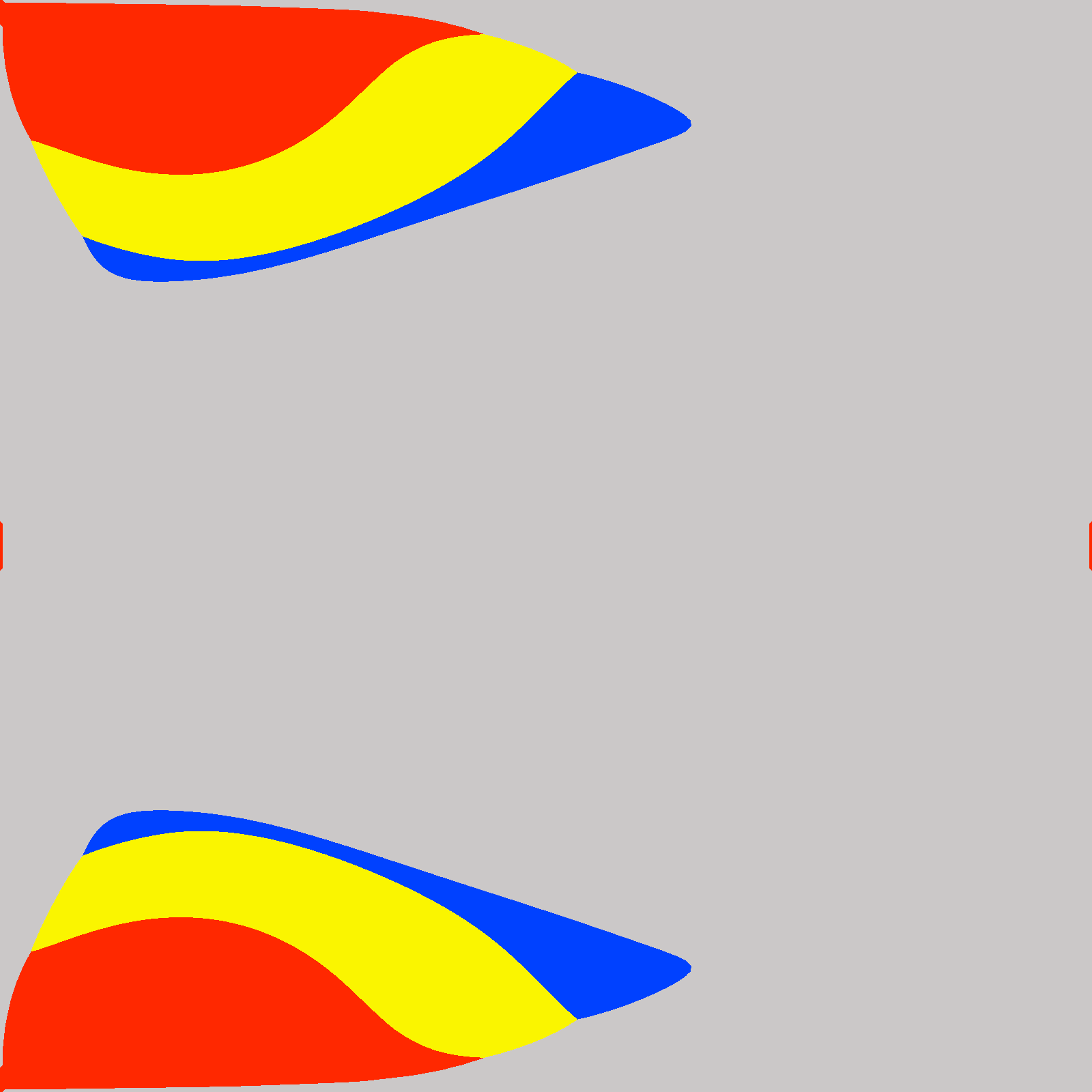}
		\subcaption{Step 50}\label{fig:compmech M4 50}
	\end{minipage}
	\begin{minipage}[b]{0.24\linewidth}
		\centering
		\includegraphics[width=\linewidth]{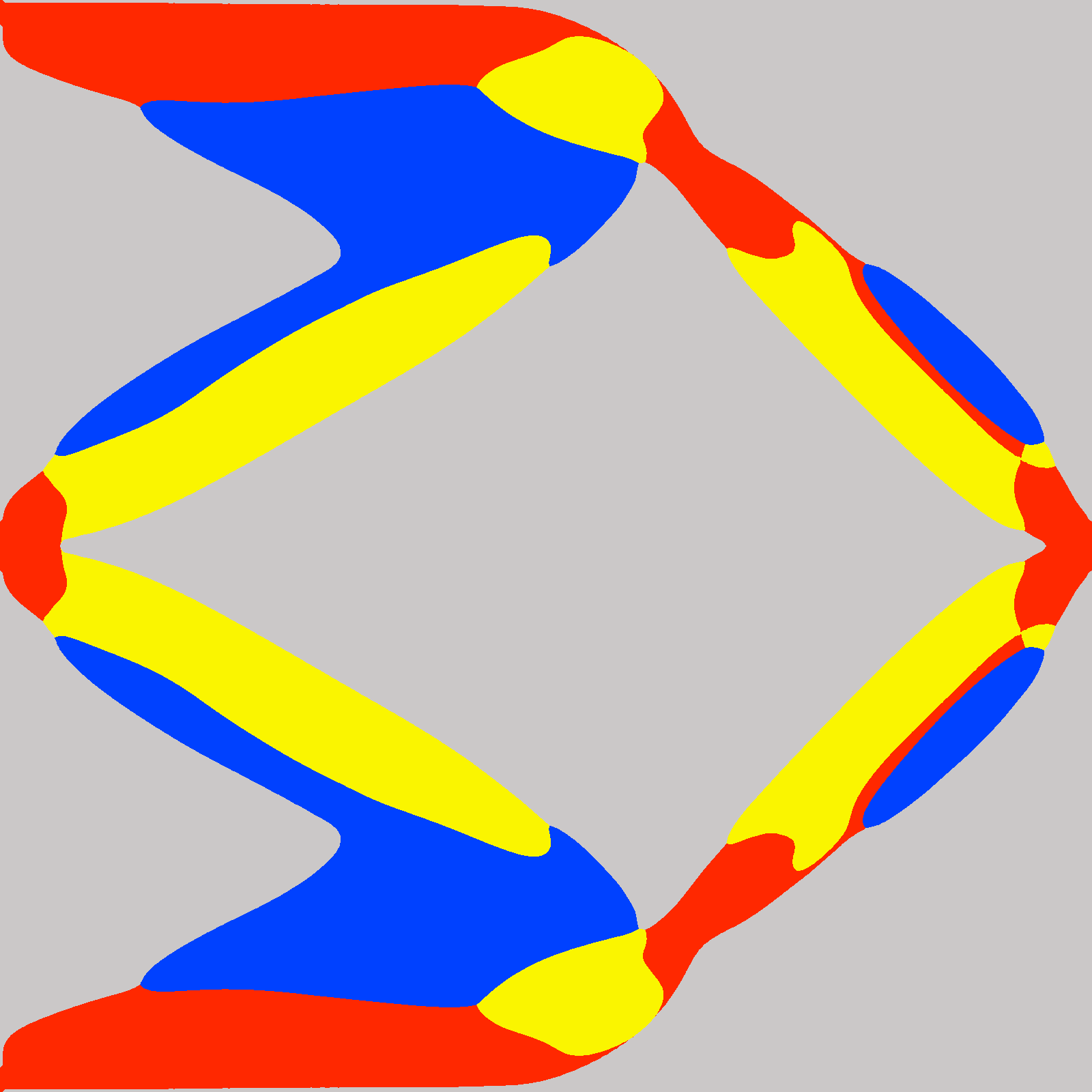}
		\subcaption{Step 200}\label{fig:compmech M4 200}
	\end{minipage}
	\begin{minipage}[b]{0.24\linewidth}
		\centering
		\includegraphics[width=\linewidth]{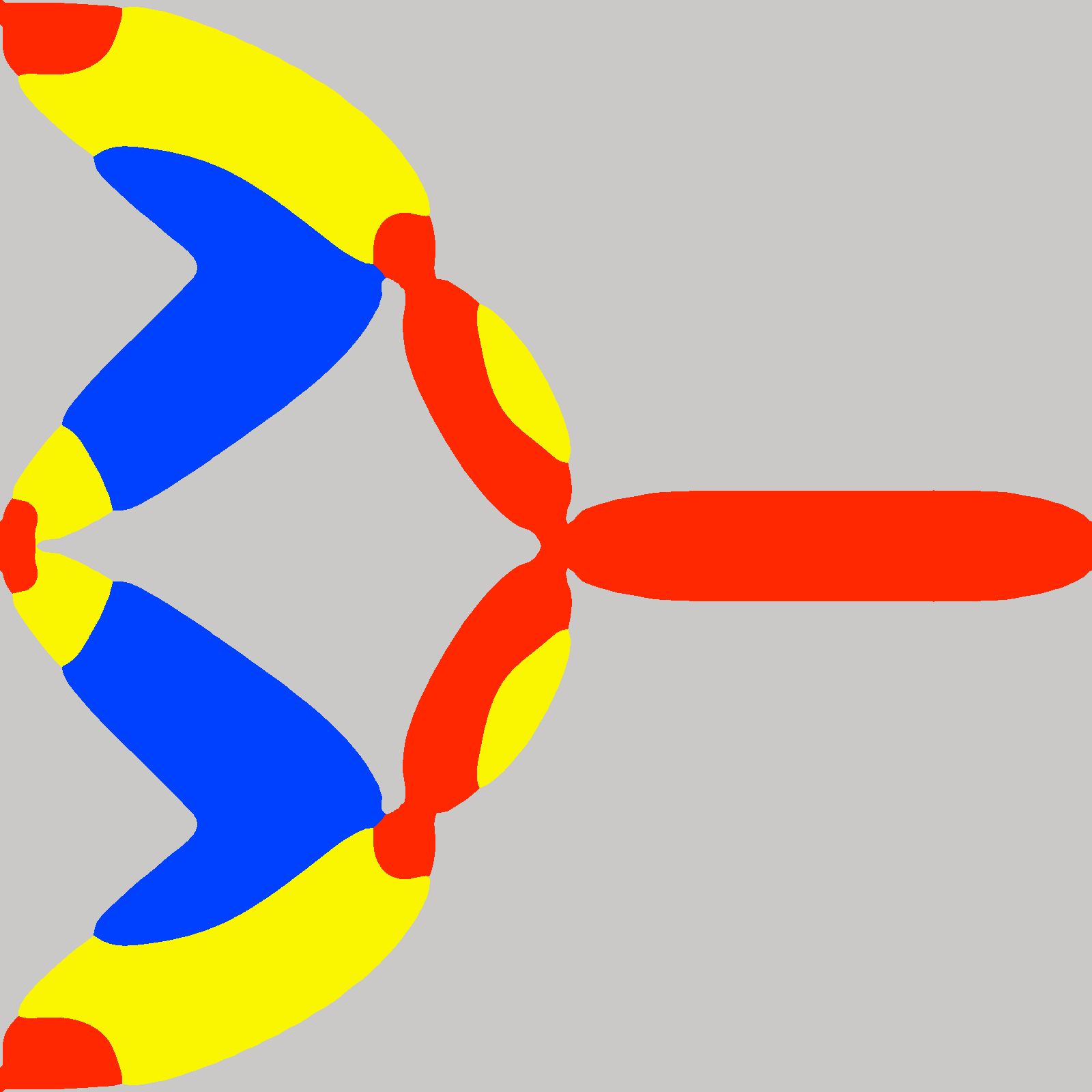}
		\subcaption{Optimal Configration}\label{fig:compmech M4 opt}
	\end{minipage}
	\caption{Intermediate results and optimal configuration of four materials (Case 19)}\label{fig:compmech M4}
\end{figure}

\begin{figure}[H]
	\begin{center}
		\begin{minipage}[b]{0.24\linewidth}
			\centering
			\includegraphics[width=\linewidth]{fig/2d_results/compmech/2D_v9_compmechM2_1620432074_ref_phi10000.png}
			\subcaption{Optimal configration, Case 17}\label{fig:compmech M2 opt 2}
		\end{minipage}
		\begin{minipage}[b]{0.24\linewidth}
			\centering
			\includegraphics[width=\linewidth]{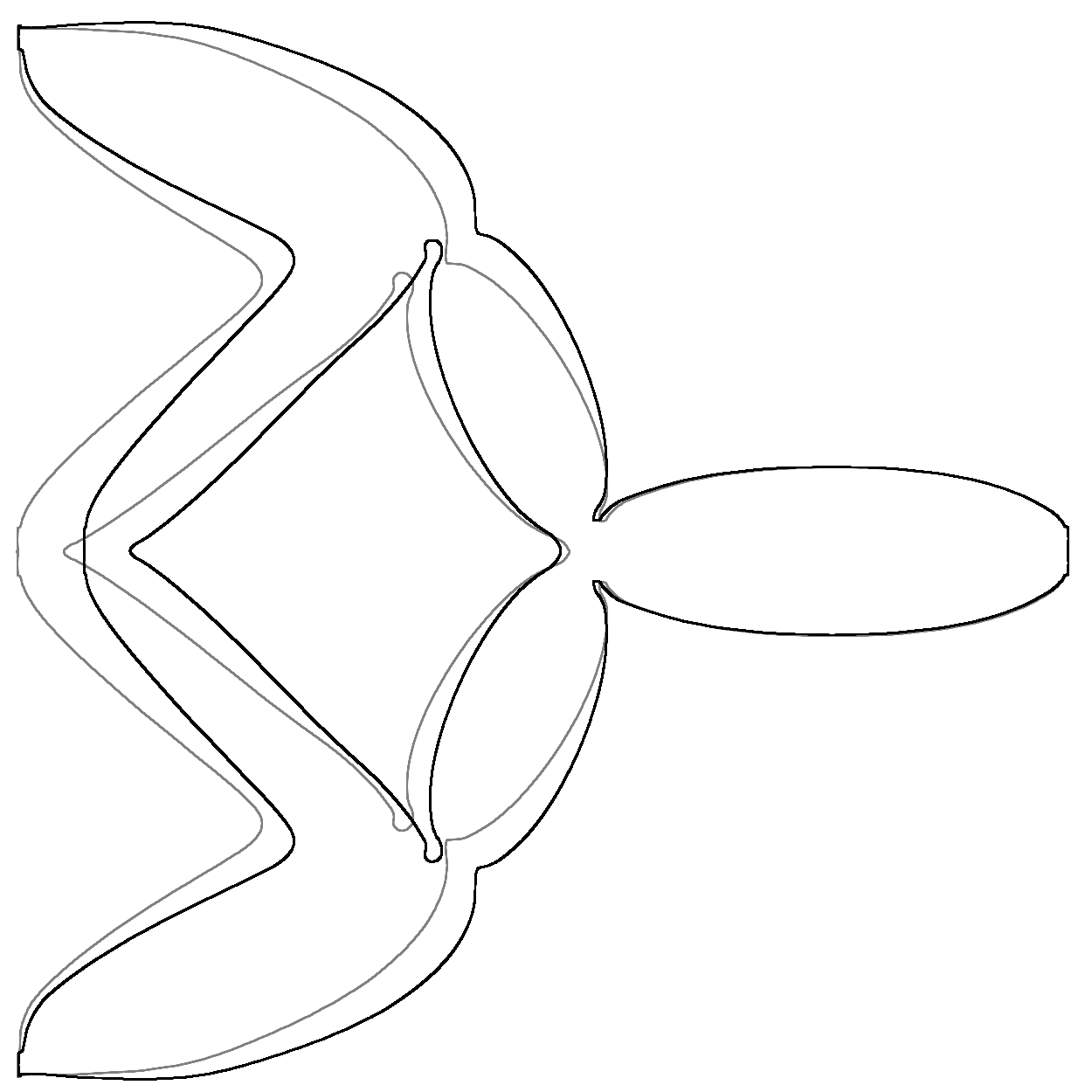}
			\subcaption{Deformation diagram with springs, Case 17}\label{fig:compmech M2 warp with spr}
		\end{minipage}
		\begin{minipage}[b]{0.24\linewidth}
			\centering
			\includegraphics[width=\linewidth]{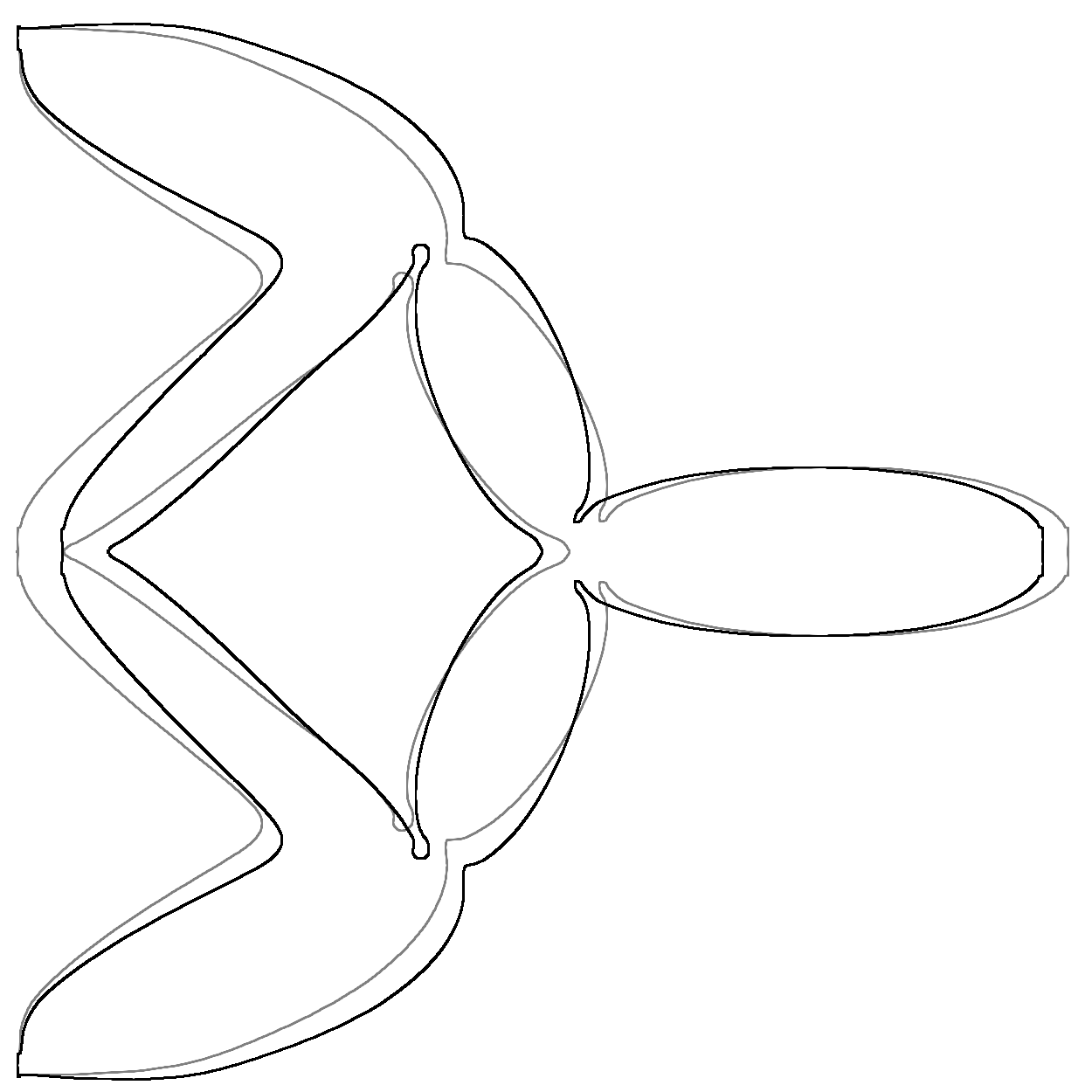}
			\subcaption{Deformation diagram without springs, Case 17}\label{fig:compmech M2 warp non spr}
		\end{minipage}\\
		\begin{minipage}[b]{0.24\linewidth}
			\centering
			\includegraphics[width=\linewidth]{fig/2d_results/compmech/2D_v9_M3compmech_1617978083_ref_phi10000.png}
			\subcaption{Optimal configration, Case 18}\label{fig:compmech M3 opt 2}
		\end{minipage}
		\begin{minipage}[b]{0.24\linewidth}
			\centering
			\includegraphics[width=\linewidth]{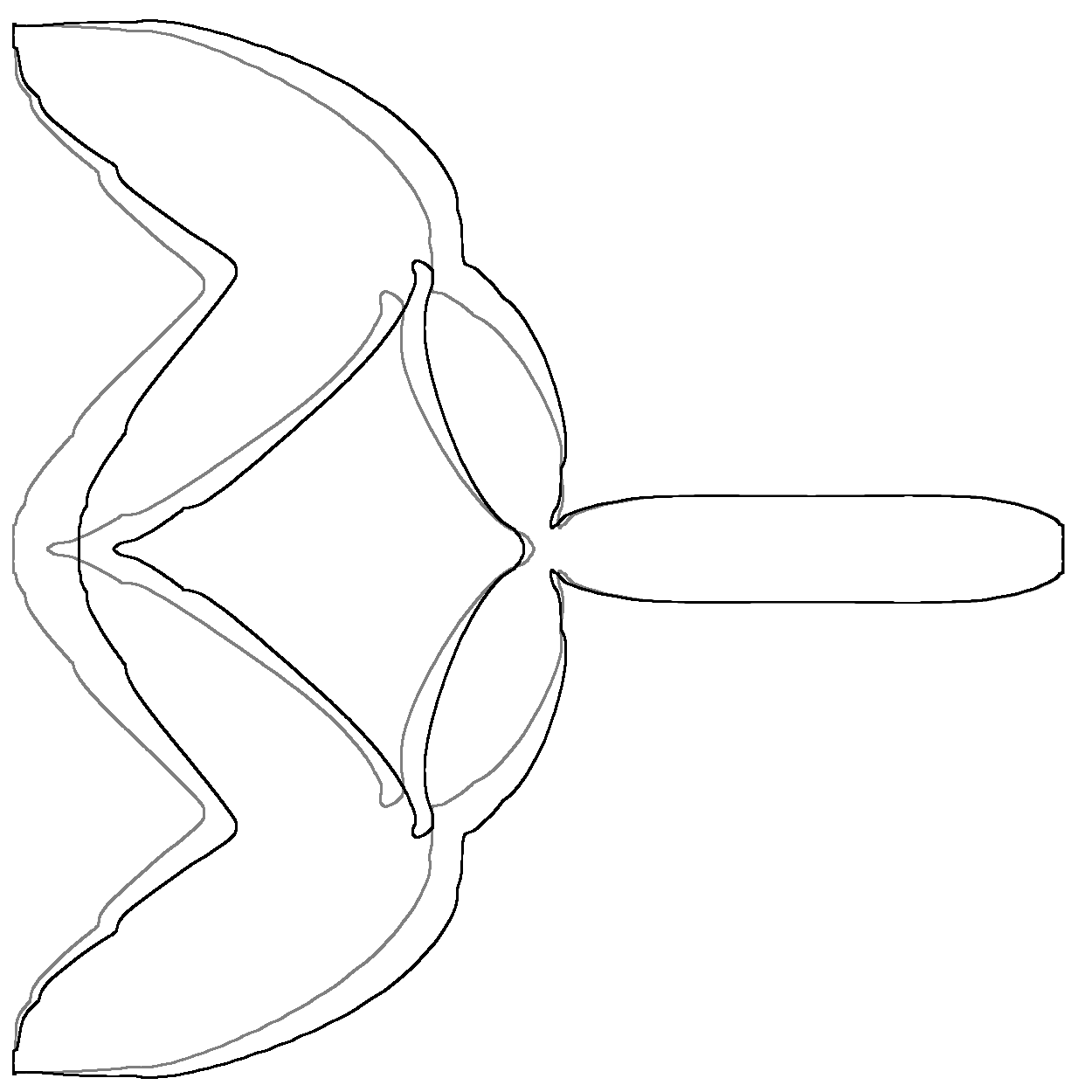}
			\subcaption{Deformation diagram with springs, Case 18}\label{fig:compmech M3 warp with spr}
		\end{minipage}
		\begin{minipage}[b]{0.24\linewidth}
			\centering
			\includegraphics[width=\linewidth]{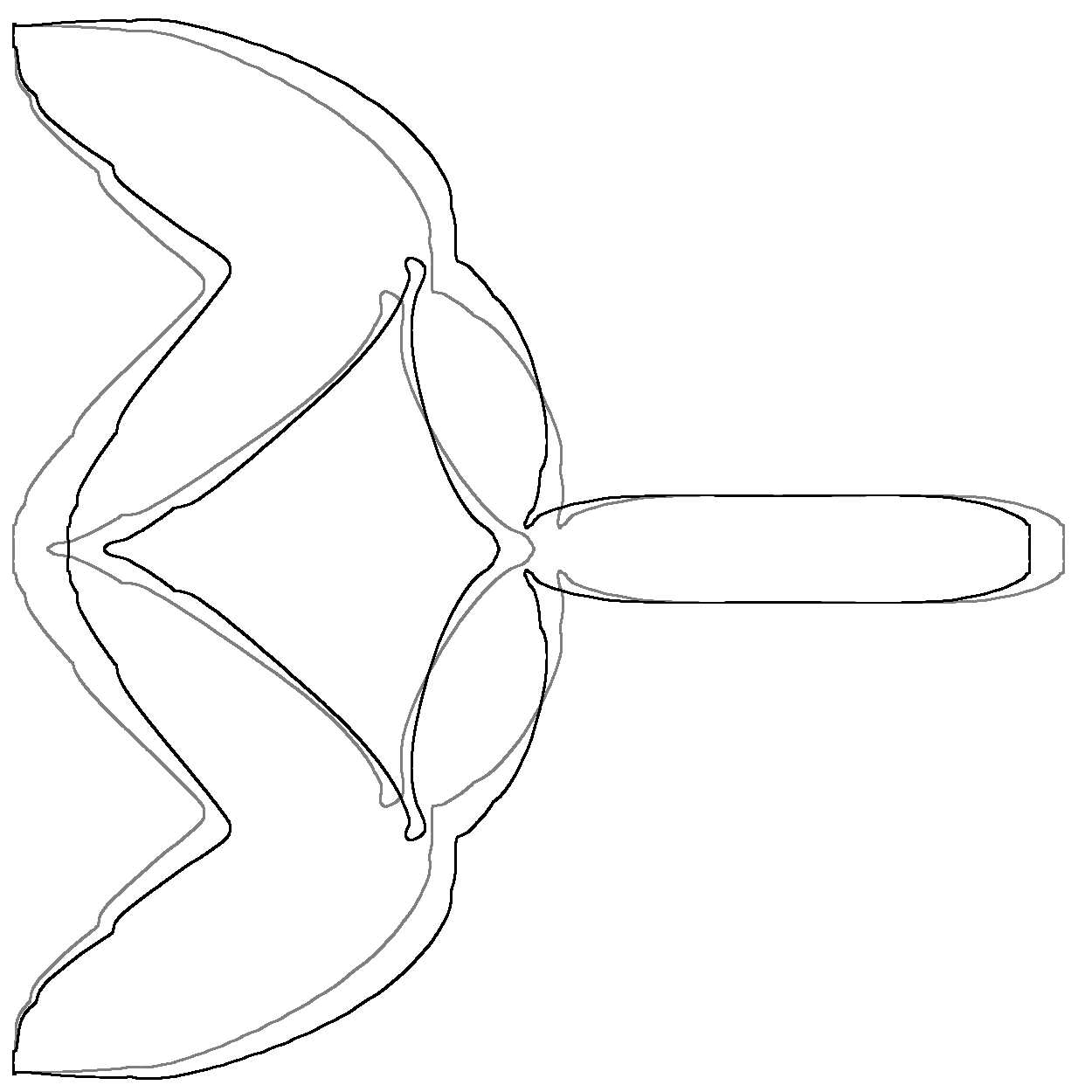}
			\subcaption{Deformation diagram without springs, Case 18}\label{fig:compmech M3 warp non spr}
		\end{minipage}\\
		\begin{minipage}[b]{0.24\linewidth}
			\centering
			\includegraphics[width=\linewidth]{fig/2d_results/compmech/2D_v9_M4compmech_1617978108_ref_phi10000.png}
			\subcaption{Optimal configration, Case 19}\label{fig:compmech M4 opt 2}
		\end{minipage}
		\begin{minipage}[b]{0.24\linewidth}
			\centering
			\includegraphics[width=\linewidth]{fig/2d_results/compmech/CompMechM4dispWithSpring.png}
			\subcaption{Deformation diagram with springs, Case 19}\label{fig:compmech M4 warp with spr}
		\end{minipage}
		\begin{minipage}[b]{0.24\linewidth}
			\centering
			\includegraphics[width=\linewidth]{fig/2d_results/compmech/CompMechM4dispNonSpring.png}
			\subcaption{Deformation diagram without springs, Case 19}\label{fig:compmech M4 warp non spr}
		\end{minipage}
	\end{center}
	\caption{Optimal designs of compliant mechanism and the deformation diagrams. Gray lines represent the structural boundaries between the void and other domains. Black lines outline the deformed structural boundaries (exaggerated by a factor of $5\times10^{14}$ in the middle panels and $5\times10^9$ in the right panels) }\label{fig:compmech warp}
\end{figure}

The stiffness values of the input and output springs were set to $({k^{\text{in}}}_{xx}, \\
{k^{\text{in}}}_{xy},{k^{\text{in}}}_{yx},{k^{\text{in}}}_{yy})=(4000\times10^{12},0,0,0)$ and $({k^{\text{out}}}_{xx},{k^{\text{out}}}_{xy},{k^{\text{out}}}_{yx},{k^{\text{out}}}_{yy})=(10\times10^{12},0,0,0)$, respectively. The characteristic length $L$ was set to 1 m. Fig. \ref{fig: compmech problem 2d} shows the fixed design domain and boundary conditions in this test. 
Material 1 was assumed at boundaries $\Gamma^u,\Gamma^\text{in}\text{~and~}\Gamma^\text{out}$, and material 0 was specified at the other boundaries. The boundary conditions of the level set functions were those of Eq. (\ref{eq:reaction-diffusion}). The two-dimensional compliant mechanism optimization problem was solved for 2, 3 and 4 materials in Cases 17, 18, and 19, respectively. The maximum volume ratios ${V^\text{max}}_m$ in each case are listed in Table \ref{tab:volumes 2}. The initial values of the X-LS functions $\phi_{ij}$ were set to 0, the regularization parameter $\tau_{ij}$ was set to $1\times10^{-3}$, and the fictitious time filtering coefficient ${K^\text{T}}'$ was set to 0.03.
\begin{table}[h]
	\centering
	\caption{Maximum volume ratios ${V^\text{max}}_m$[\%]}
	\begin{tabular}{lrrrr}\hline
		Number of &\multicolumn{4}{c}{Material number $m$} \\
		materials $M$& 
		~~0  & ~~1 &~~2 &~~3 \\ \hline
		2(Case 17)&100&30&-&- \\
		3(Case 18)&100&15&15&-  \\
		4(Case 19)&100&10&10&10 \\ \hline
	\end{tabular}
	\label{tab:volumes 2}
\end{table}
The intermediate results and optimal configurations in Cases 17, 18, and 19 are shown in Figs. \ref{fig:compmech M2}--\ref{fig:compmech M4}, respectively, and Fig. \ref{fig:compmech warp}  shows the obtained optimal structures and deformation diagrams in each case. As the virtual springs had large spring constants, the movements in the middle panels of Fig. \ref{fig:compmech warp} are barely noticeable. The right panels of Fig. \ref{fig:compmech warp} are the deformation diagrams without the virtual springs, i.e., with stiffness values of
$({k^{\text{in}}}_{xx},{k^{\text{in}}}_{xy},{k^{\text{in}}}_{yx},{k^{\text{in}}}_{yy})=(0,0,0,0)$ and $({k^{\text{out}}}_{xx},{k^{\text{out}}}_{xy},{k^{\text{out}}}_{yx},{k^{\text{out}}}_{yy})=(0,0,0,0)$. 
In all cases, the optimal configurations were smooth and clear. The non-cavity regions and regions of other structural materials were similarly shaped for different numbers of materials ($M$ = 2, 3, 4). 
The regions occupied by the structural materials formed 5 components; 2 “$>$” shaped parts in the left, a beam connected to the output port, and 2 rods in the center which connect the beam and “$>$” shaped parts.
These elements are connected by a constricted shape, and there is no hinge-like structure that connects the elements at a single point.
The values of the objective functions in Cases 17, 18, and 19 were $-6.46\times 10^{-20}, -5.31\times 10^{-20}$ and $ -5.19\times 10^{-11}$, respectively. The upper volume constraints were mostly satisfied (volume constraint functions $< 10^{-3}$) and the values of the objective functions were negative, indicating successful optimization with a pulling force at the left-side output port. 
As the virtual springs were arranged at the input and output ports to ensure sufficient strength of the structures, the output parts were hardly deformed in reality (Figs. \ref{fig:compmech warp}\subref{fig:compmech M2 warp with spr}, \subref{fig:compmech M3 warp with spr}, and \subref{fig:compmech M4 warp with spr}). When the springs were removed and the inputs were applied, the output parts were deformed to the left, as intended (Figs. \ref{fig:compmech warp}\subref{fig:compmech M2 warp non spr}, \subref{fig:compmech M3 warp non spr}, and \subref{fig:compmech M4 warp non spr}).

\subsection{Two-dimensional mean compliance and moment of inertia minimization problems}
\begin{figure}[H]
	\centering
		\includegraphics[width=10cm]{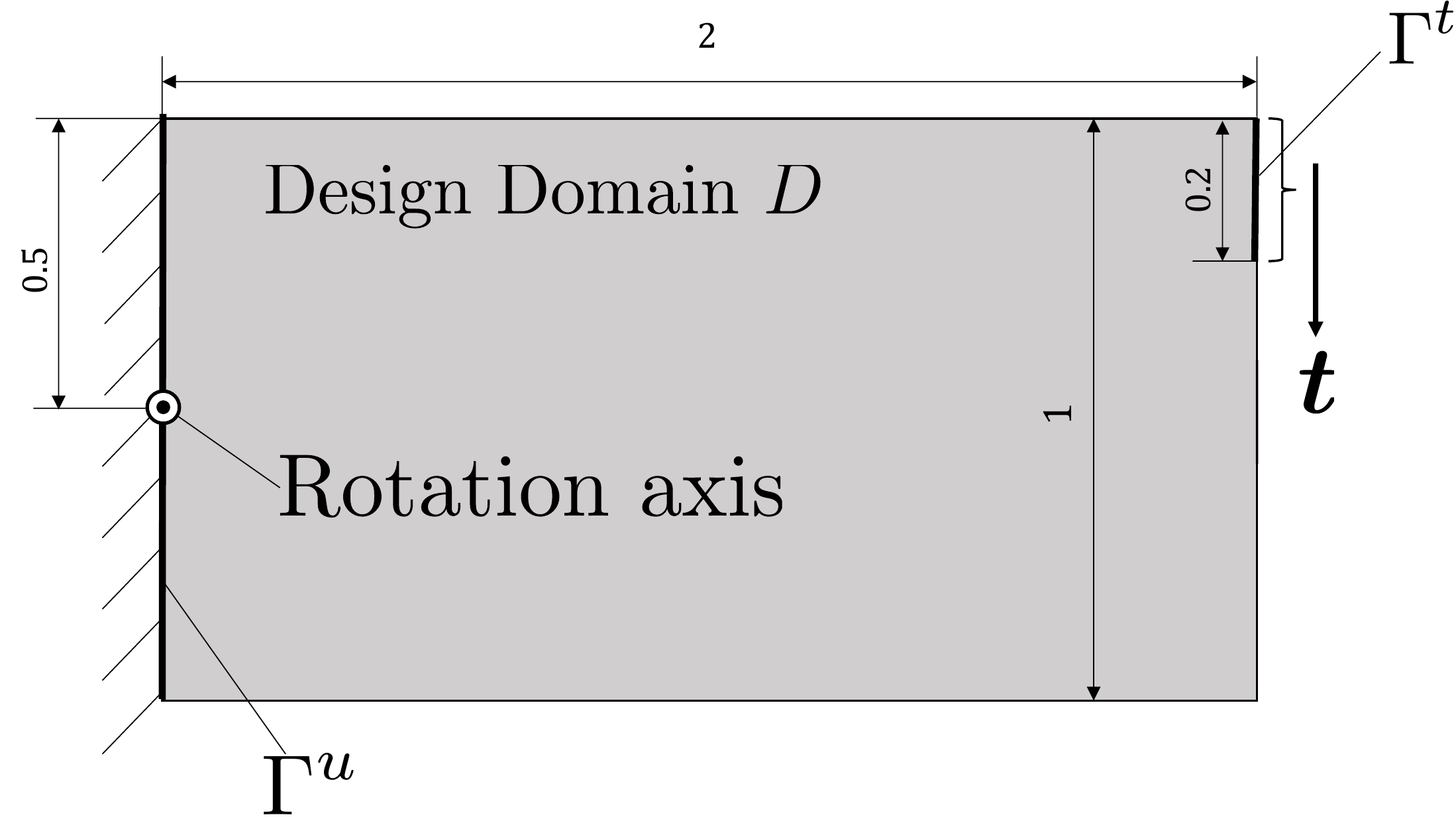}
	\caption{Two-dimensional problem settings for moment of inertia and mean-compliance minimization.}
	\label{fig: inertia problem 2d}
\end{figure}

In this subsection, the proposed optimization method is applied to the two-dimensional mean compliance and moment of inertia minimization problem. Fig. \ref{fig: inertia problem 2d} shows the fixed design domain, rotation axis, and boundary conditions in the simulation. The characteristic length $L$ was set to 1 m.
No material was assumed at the fixed displacement boundary $\Gamma^u$, material 1 was assumed at the boundary $\Gamma^t$ subjected to the traction force $\bm t$, and material 0 was assumed at the other boundaries. 
% the boundary conditions of the level set functions were set based on the Eq. (\ref{eq:reaction-diffusion}).

\begin{figure}[H]
	\begin{minipage}[b]{0.32\linewidth}
		\centering
		\includegraphics[width=\linewidth]{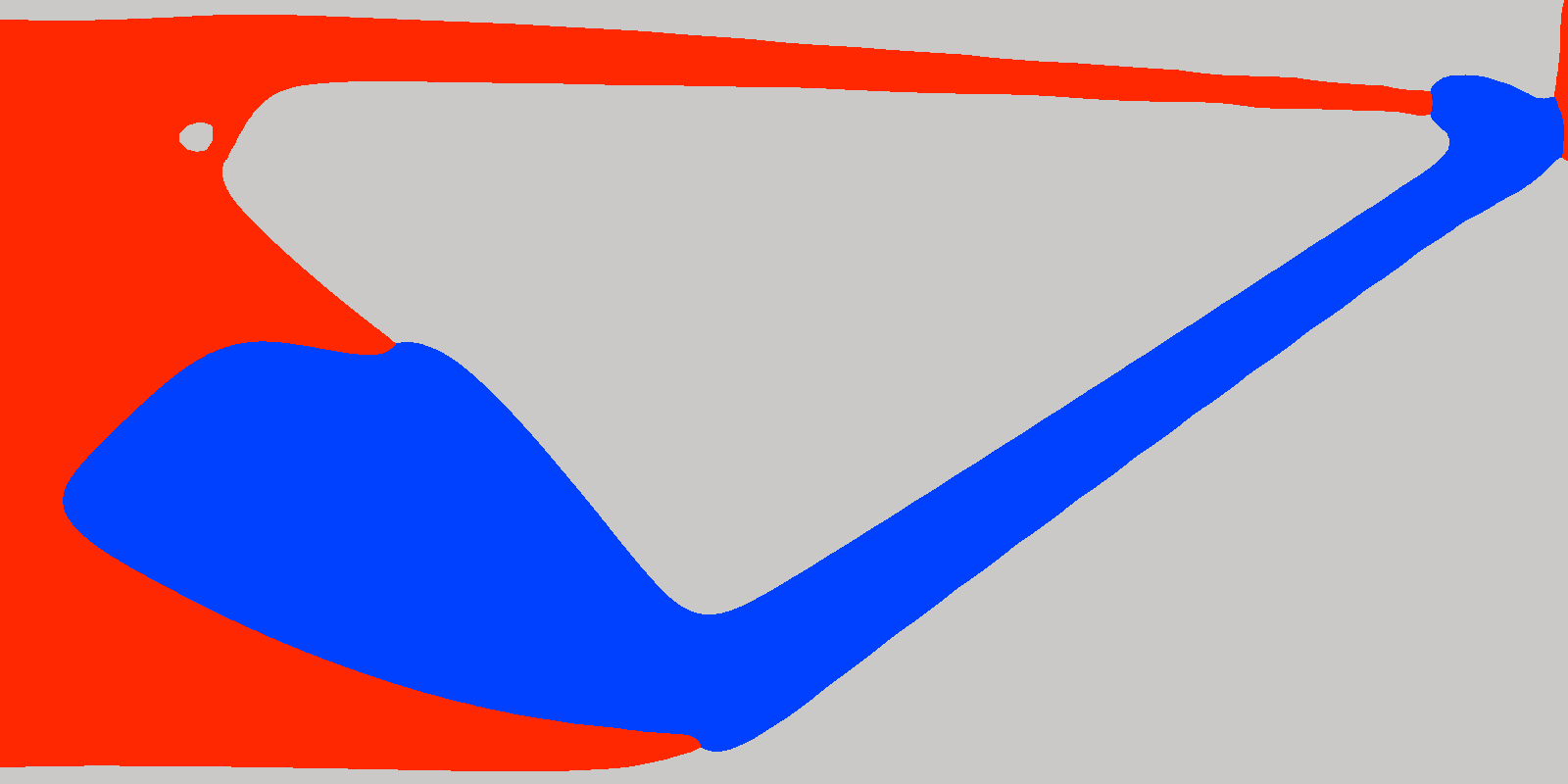}
		\subcaption{Weighting factor $w$ is $5\times10^{-11}$. \\(Case 20)}\label{fig:inertia 1}
	\end{minipage}
	\begin{minipage}[b]{0.32\linewidth}
		\centering
		\includegraphics[width=\linewidth]{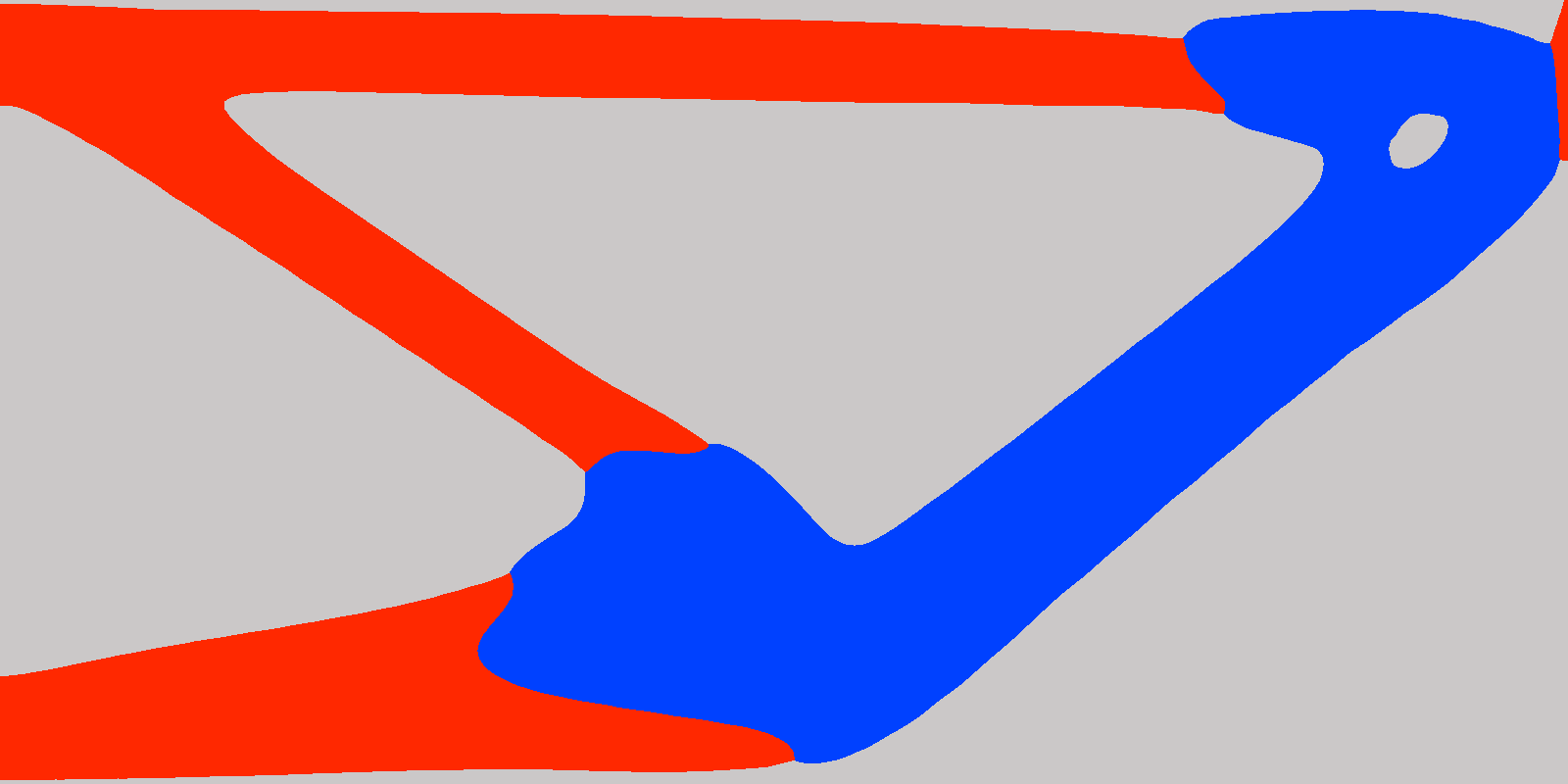}
		\subcaption{Weighting factor $w$ is $5\times10^{-12}$. \\(Case 21)}\label{fig:inertia 2}
	\end{minipage}
	\begin{minipage}[b]{0.32\linewidth}
		\centering
		\includegraphics[width=\linewidth]{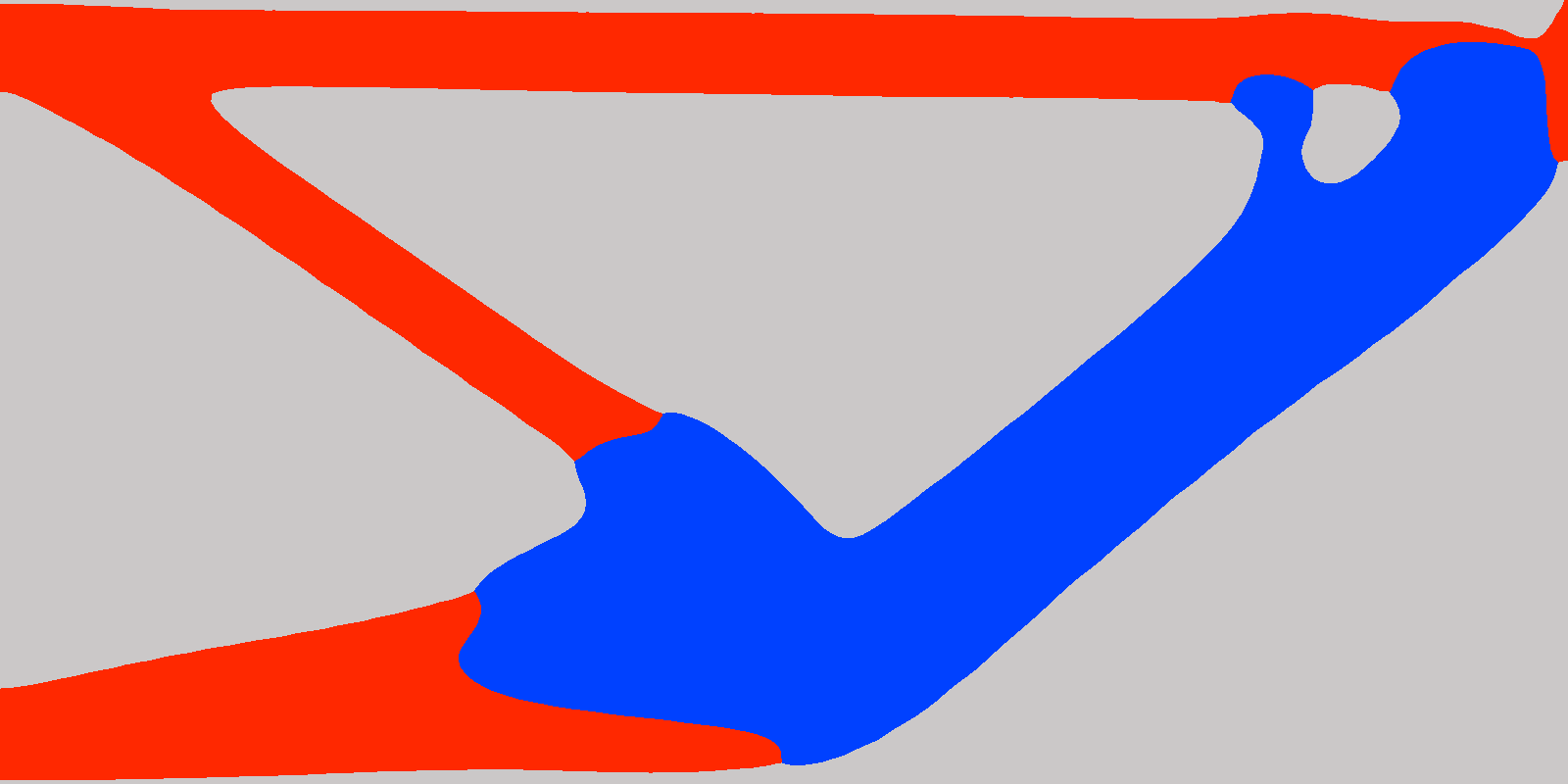}
		\subcaption{Weighting factor $w$ is $5\times10^{-13}$. \\(Case 22)}\label{fig:inertia 3}
	\end{minipage}
	\caption{Optimal configurations for inertia moment and mean-compliance minimization}\label{fig:inertia}
\end{figure}

In these examples, material 2 was assigned half the density of material 1 and material 0 had negligible mass density; that is, we set $( \rho_0,\rho_1, \rho_2)= (0, 2, 1)$. The upper limits of the volume constraint $({V^\text{max}}_0,{V^\text{max}}_1,{V^\text{max}}_2)$ were set to (100\%, 20\%, 20\%) of the volume of the fixed design domain. The initial values of the X-LS functions $\phi_{ij}$ were set to 0, and the regularization parameter $\tau_{ij}$ was set to $1 \times 10^{-3}$. We varied the weighting factor $w$ in Eq. (\ref{eq: J inertia}), which modulates the effects of minimizing the moment of inertia and the compliance, and compared the resulting optimal configurations. The weighting factors were set to $5\times10^{-13},5\times10^{-14}$ and $5\times10^{-15}$ in Cases 20, 21, and 22, respectively. The optimal configurations are shown in Fig. \ref{fig:inertia}.
In Cases 20, 21, and 22, the values of the objective functions were $3.61\times 10^{-11}, 2.13\times 10^{-11}$, and $2.08\times 10^{-11}$, respectively, and the values of the objective functions of moment of inertia $J_\text{I}$ defined in Eq. (\ref{eq: J inertia}) were $7.49\times 10^{1}, 1.30\times 10^{2}$, and $1.42\times 10^{2}$, respectively. The upper volume constraints were mostly satisfied ($<10^{-10}$ in Case 20 and 21 and $2\times10^{-3}$ in Case 22). The optimal structure exhibited a smooth boundary. As the weighting coefficient increased, the dense material moved closer to the axis of rotation, thus reducing the moment of inertia $J_\text{I}$. This result was deemed physically reasonable.

\subsection{Three-dimensional mean compliance minimization problems}
\begin{figure}[H]
	\centering
	\includegraphics[width=10cm]{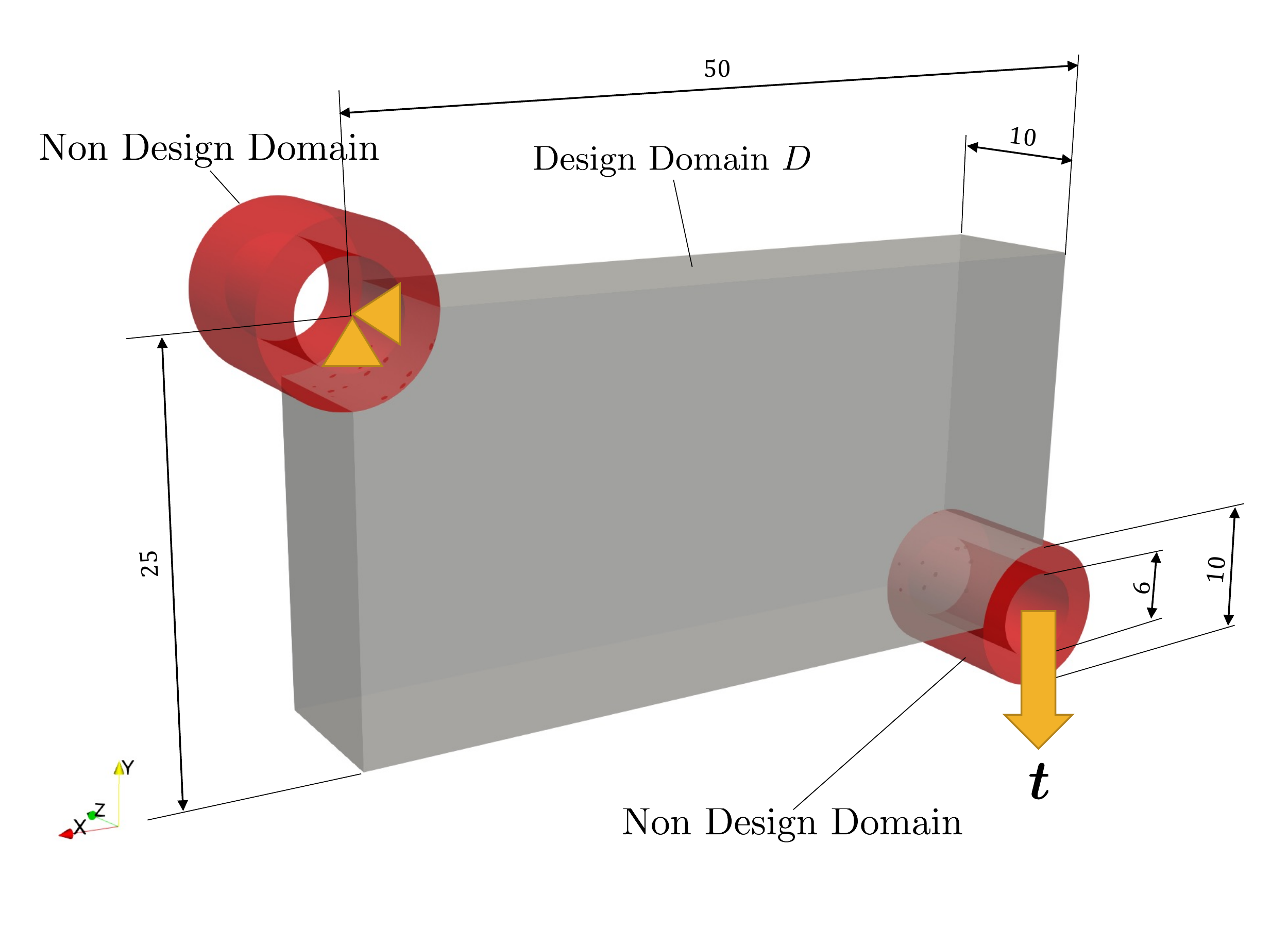}
	\caption{Three-dimensional setting of the mean-compliance minimization problem}
	\label{fig: problem 3d}
\end{figure}

The practicality of the method was evaluated on the mean-compliance minimization problem of a three-dimensional mechanical component. Fig. \ref{fig: problem 3d} shows the fixed design domain $D$ (gray region), the nondesign domain (red areas), and the boundary conditions. The characteristic length $L$ was set to 25 mm.
Also in these examples,  the isotropic linear elastic materials has Young's modulus shown in Table \ref{tab:materials} and Poisson's ratio $0.3$. Material 1 is assigned to the non-design domain. In the following figures of the optimization results, each material region is represented by a color as shown in the Table \ref{tab:materials}.
Except for the Case 25 bellow, among the boundaries of the design domain,  Material 1 is specified at the boundary with the non-design domain and material 0 is specified at the other boundaries. For the Case 25, among the boundaries of the design domain,  Material 1 is specified at the boundary with the non-design domain, no material is specified at the z-minimum and z-maximum plane and material 0 is specified at the other boundaries. The boundary conditions of the level set functions are set based on the Eq. (\ref{eq:reaction-diffusion}). 
Since the problem settings are symmetrical about the central plane in the z-axis direction, we analyzed half of the region using the symmetry condition.
Initial values of X-LS functions $\phi_{ij}$ is set to 0.

The Young's moduli of the isotropic linear elastic materials in these examples are listed in Table \ref{tab:materials} and the Poisson's ratio was 0.3. The nondesign domain was composed of material 1. The following figures show the optimization results. The material colors in these results are those assigned in Table \ref{tab:materials}. In all cases, material 1 was assumed at the boundary with the nondesign domain and material 0 was assumed at the other boundaries. In Case 25 alone, no material was assumed at the z-minimum and z-maximum plane. 
%The boundary conditions of the level set functions are those of Eq. (\ref{eq:reaction-diffusion}). 
As the problem settings were symmetric about the central plane in the z-axis direction, only half of the region was analyzed. The X-LS functions $\phi_{ij}$ were initialized to 0.

\subsubsection{Examples for $M=3,4$}
\begin{figure}[H]
	\begin{minipage}[t]{0.49\linewidth}
		\centering
		\includegraphics[width=\linewidth]{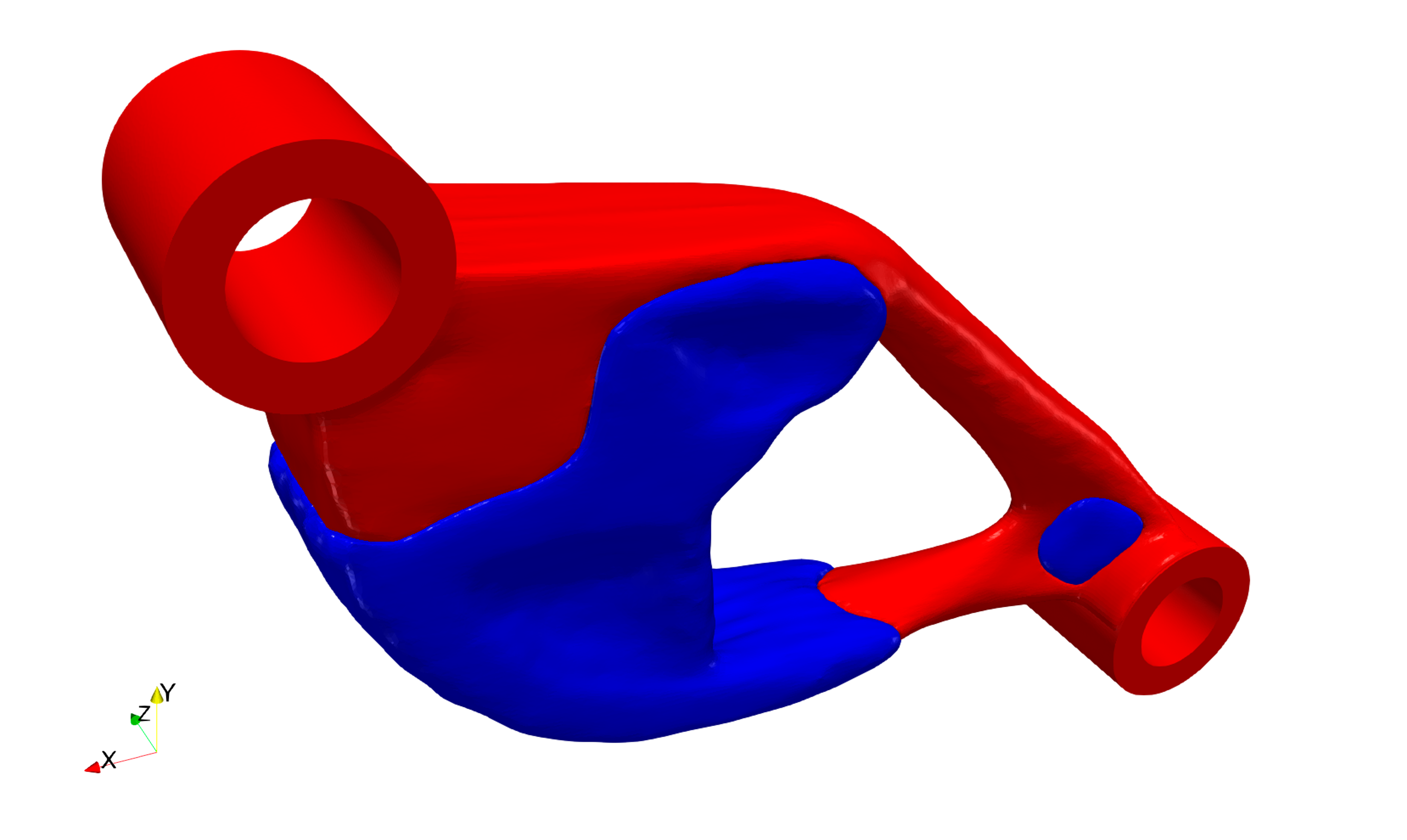}
		\subcaption{}\label{fig:M3front}
	\end{minipage}
	\begin{minipage}[t]{0.49\linewidth}
		\centering
		\includegraphics[width=\linewidth]{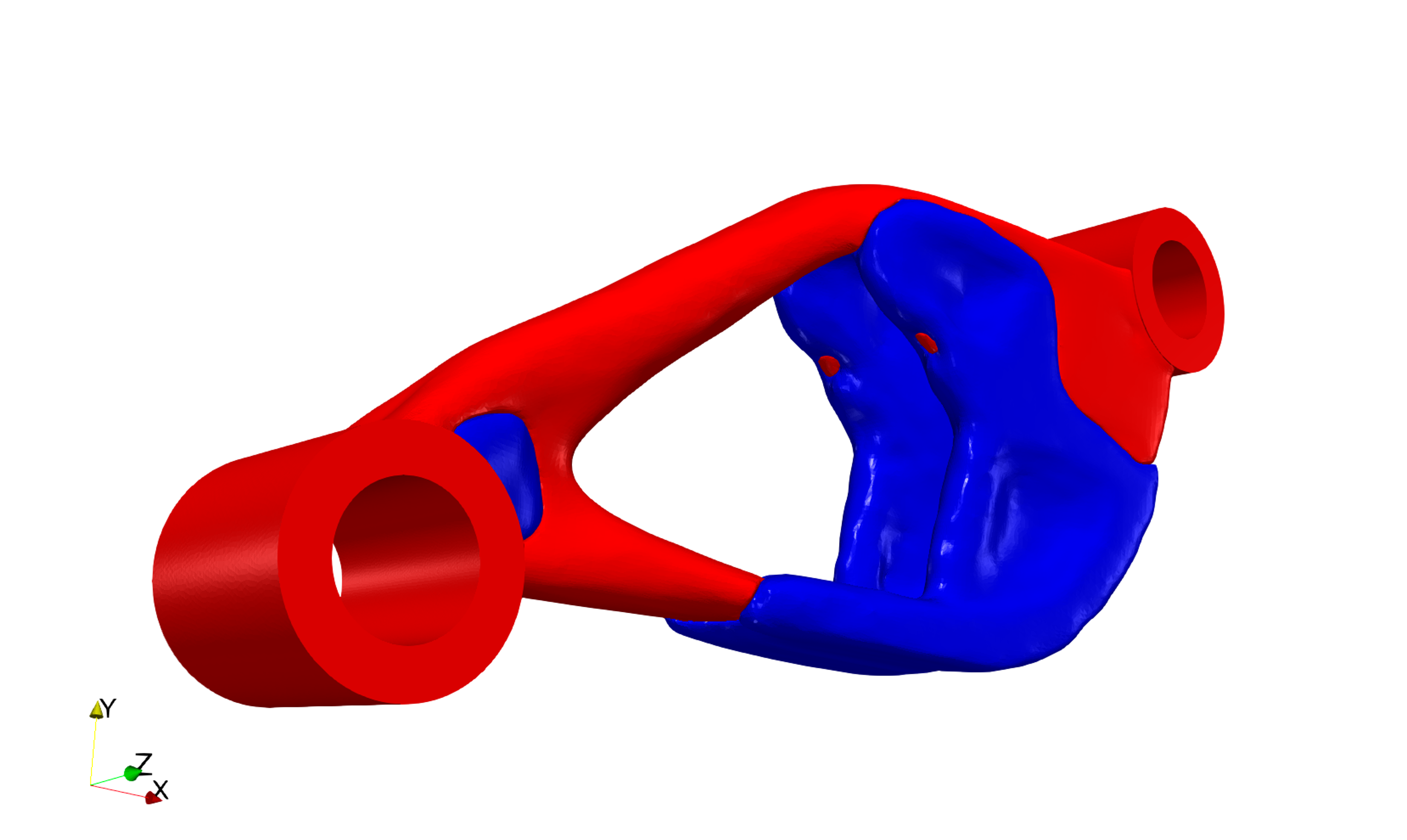}
		\subcaption{}\label{fig:M3back}
	\end{minipage}
	\caption{Optimal configuration in three-dimensional mean-compliance minimization of Case 23 (with three materials)
}\label{fig:3dM3 result}
\end{figure}

\begin{figure}[H]
	\begin{minipage}[b]{0.49\linewidth}
		\centering
		\includegraphics[width=\linewidth]{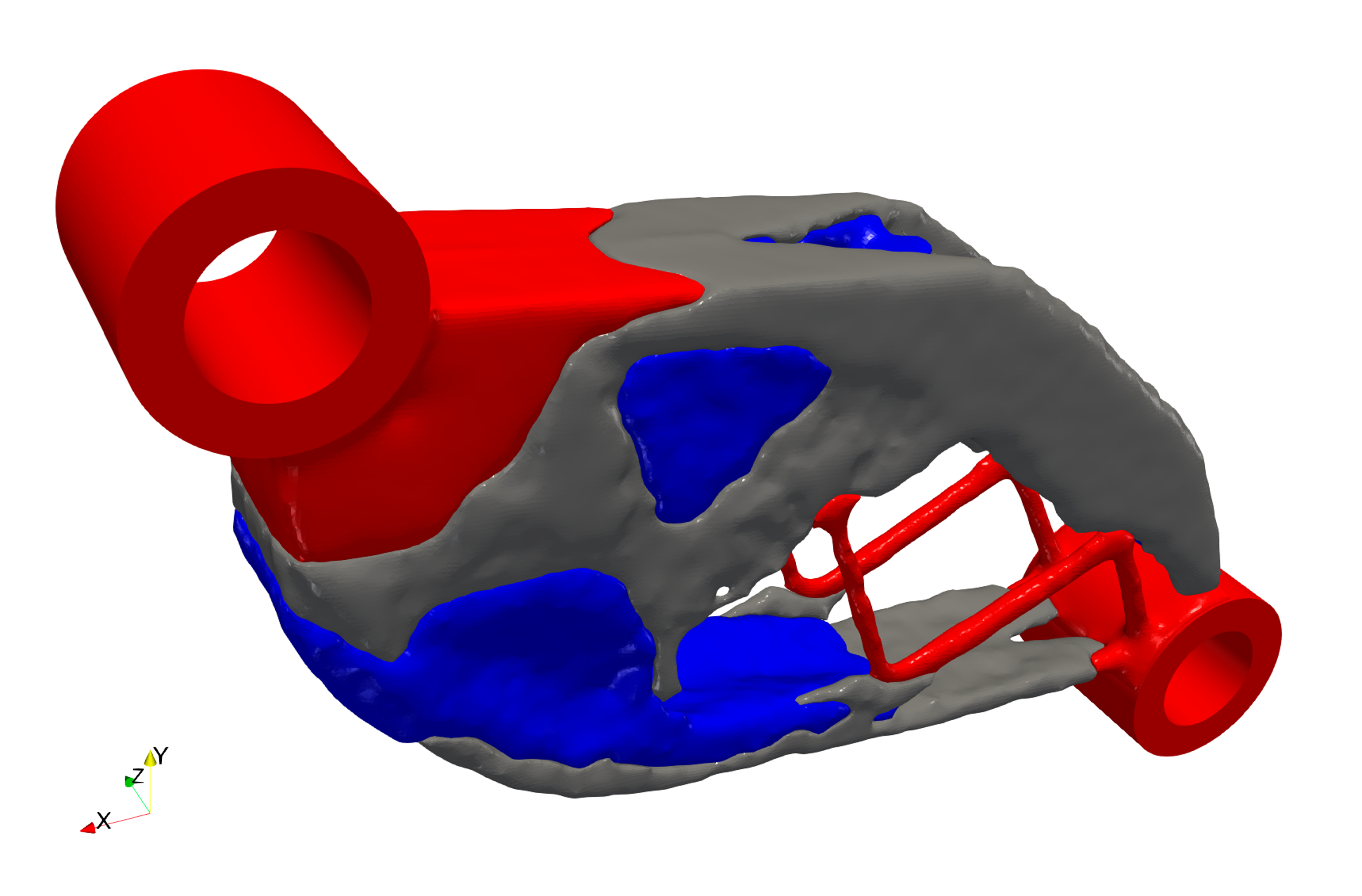}
		\subcaption{}\label{fig:M4front}
	\end{minipage}
	\begin{minipage}[b]{0.49\linewidth}
		\centering
		\includegraphics[width=\linewidth]{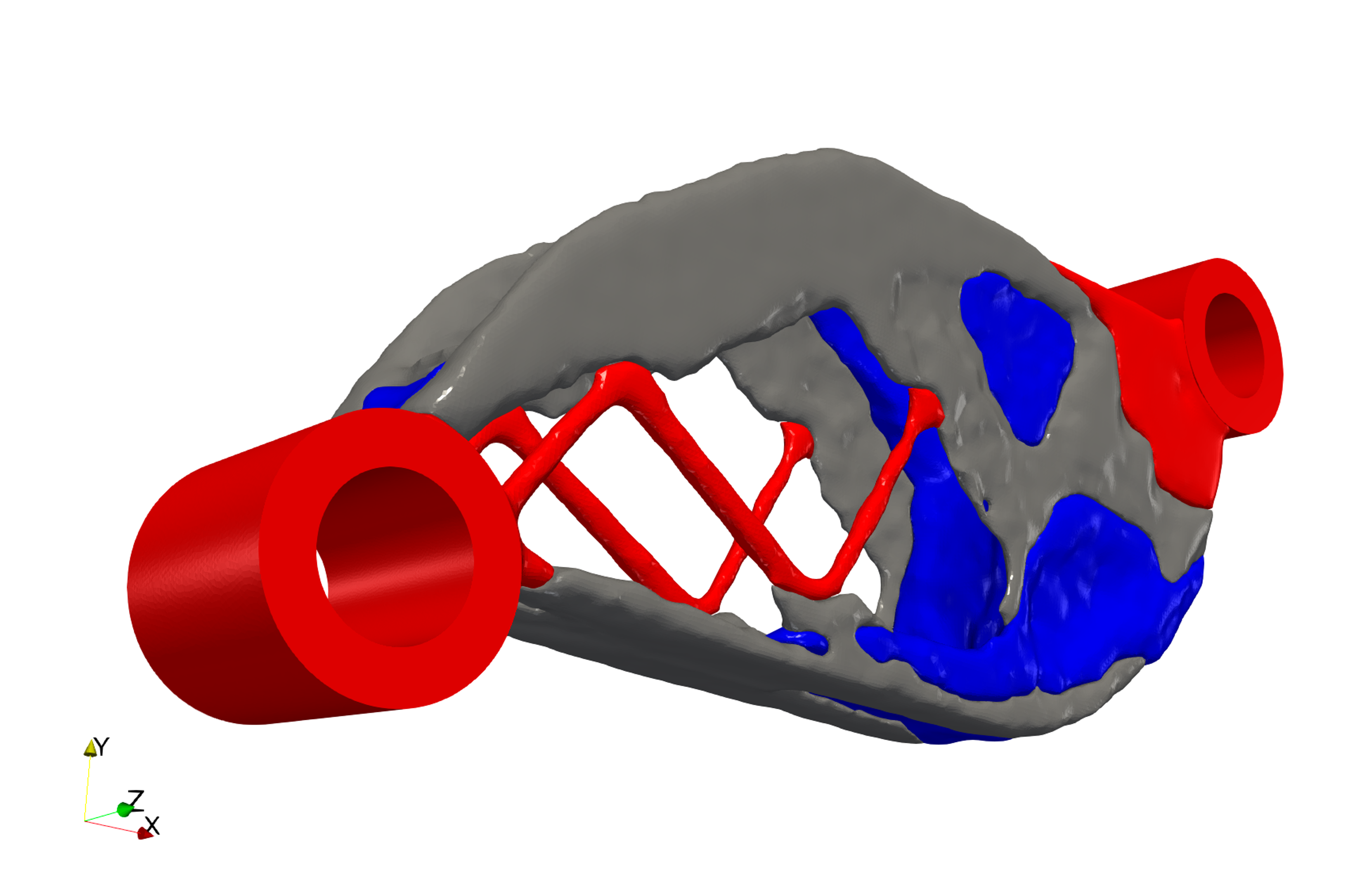}
		\subcaption{}\label{fig:M4back}
	\end{minipage}
	\caption{Optimal configuration in three-dimensional mean-compliance minimization of Case 24 (with four materials)}\label{fig:3dM4 result}
\end{figure}

We first optimized the structures in the cases of three and four materials (Cases 23 and 24, respectively). Case 23 included materials 0, 1, and 2 and Case 24 included Materials 0, 1, 2, and 3 (see Table \ref{tab:materials}). The maximum volume ratios were set to $({V^\text{max}}_0,{V^\text{max}}_1,{V^\text{max}}_2)$ = (100\%, 20\%, 20\%) in Case 23 and $({V^\text{max}}_0,{V^\text{max}}_1,{V^\text{max}}_2,{V^\text{max}}_3)$ = (100\%, 13.3\%, 13.3\%, 13.3\%) in Case 24. The regularization parameter $\tau_{ij}$ was set to $1\times10^{-4}$. The optimal configurations in Cases 23 and 24 are presented in Figs. \ref{fig:3dM4 result} and \ref{fig:3dM3 result}, respectively.

The values of the objective functions in Cases 23 and 24 were $0.190$. The upper volume constraints were mostly satisfied (volume constraint function $<10^{-5}$ in Case 23 and $<10^{-4}$ in Case 24). In the two optimal structures, the strong material was located near the fixed displacement boundary; the upper right, lower left, and central regions, where the stiffness contribution was low, were hollow. This configuration was compatible with structural mechanics. In addition, the lack of geometrically complex structures, such as checkerboard patterns, indicated that the regularization was properly applied.

\subsubsection{Uniform cross-section surface and piecewise linear surface constraints}
\begin{figure}[H]
	\begin{minipage}[b]{0.32\linewidth}
		\centering
		\includegraphics[width=\linewidth]{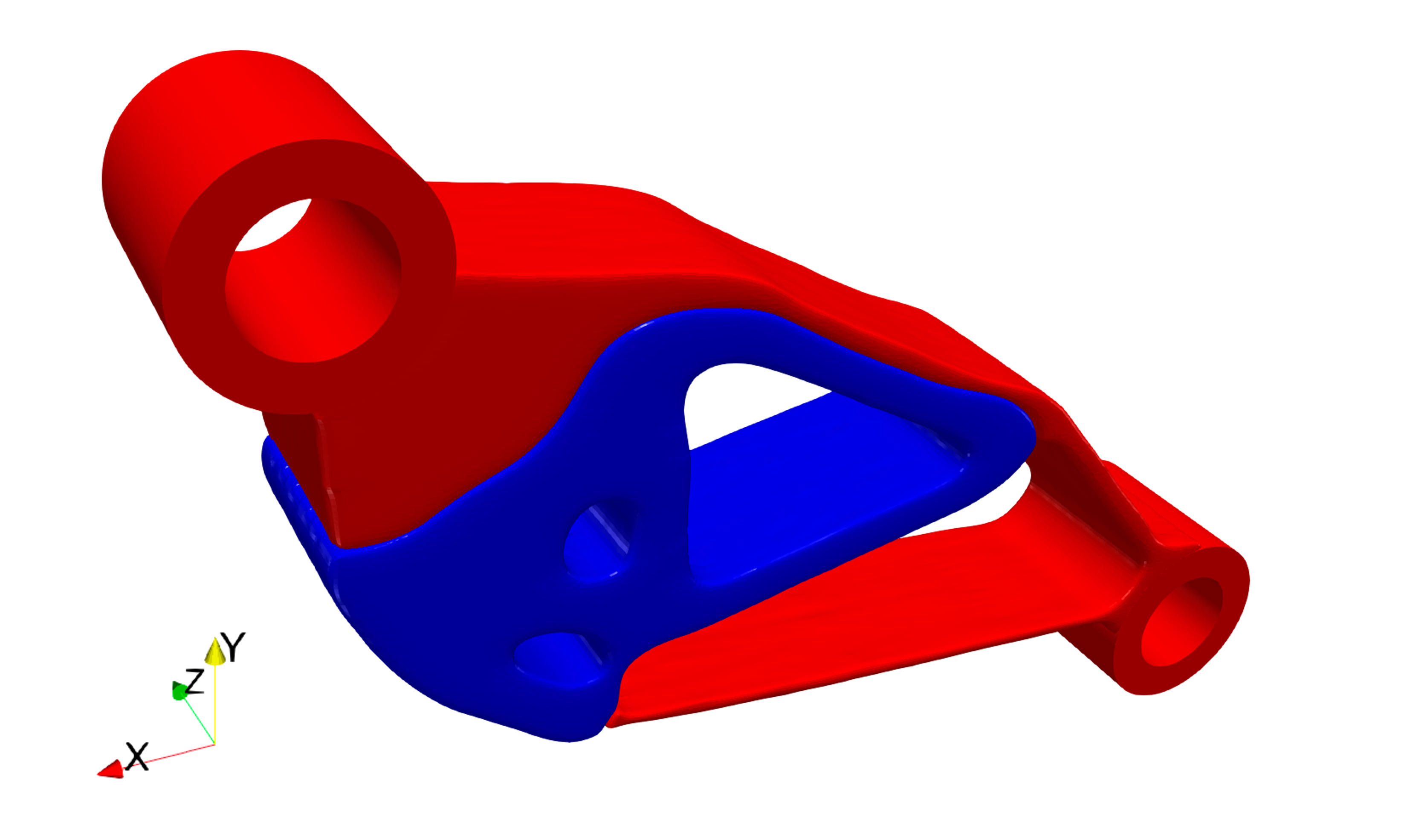}
		\subcaption{}\label{fig:ucsszallfront}
	\end{minipage}
	\begin{minipage}[b]{0.32\linewidth}
		\centering
		\includegraphics[width=\linewidth]{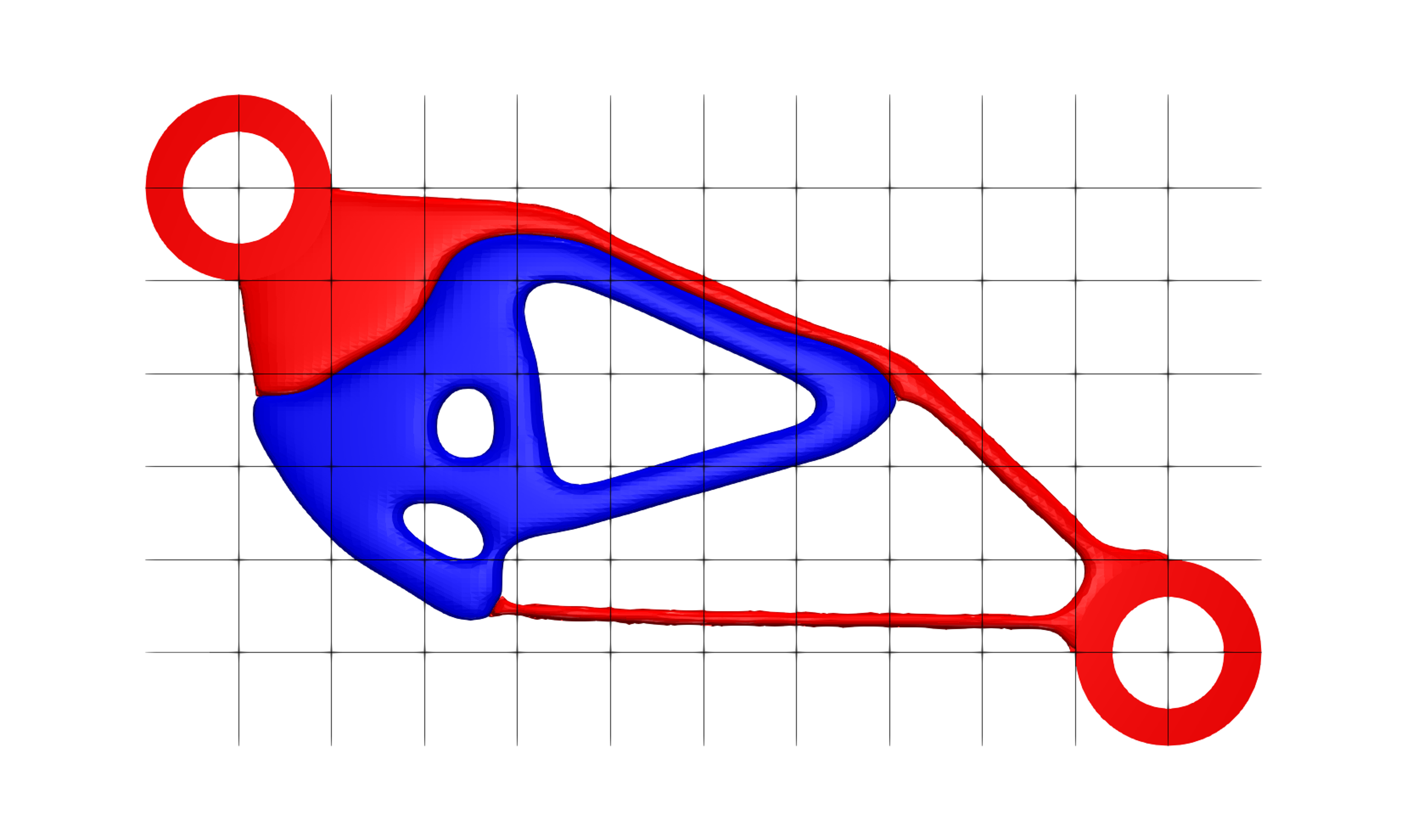}
		\subcaption{}\label{fig:ucsszallZView}
	\end{minipage}
	\begin{minipage}[b]{0.32\linewidth}
		\centering
		\includegraphics[width=\linewidth]{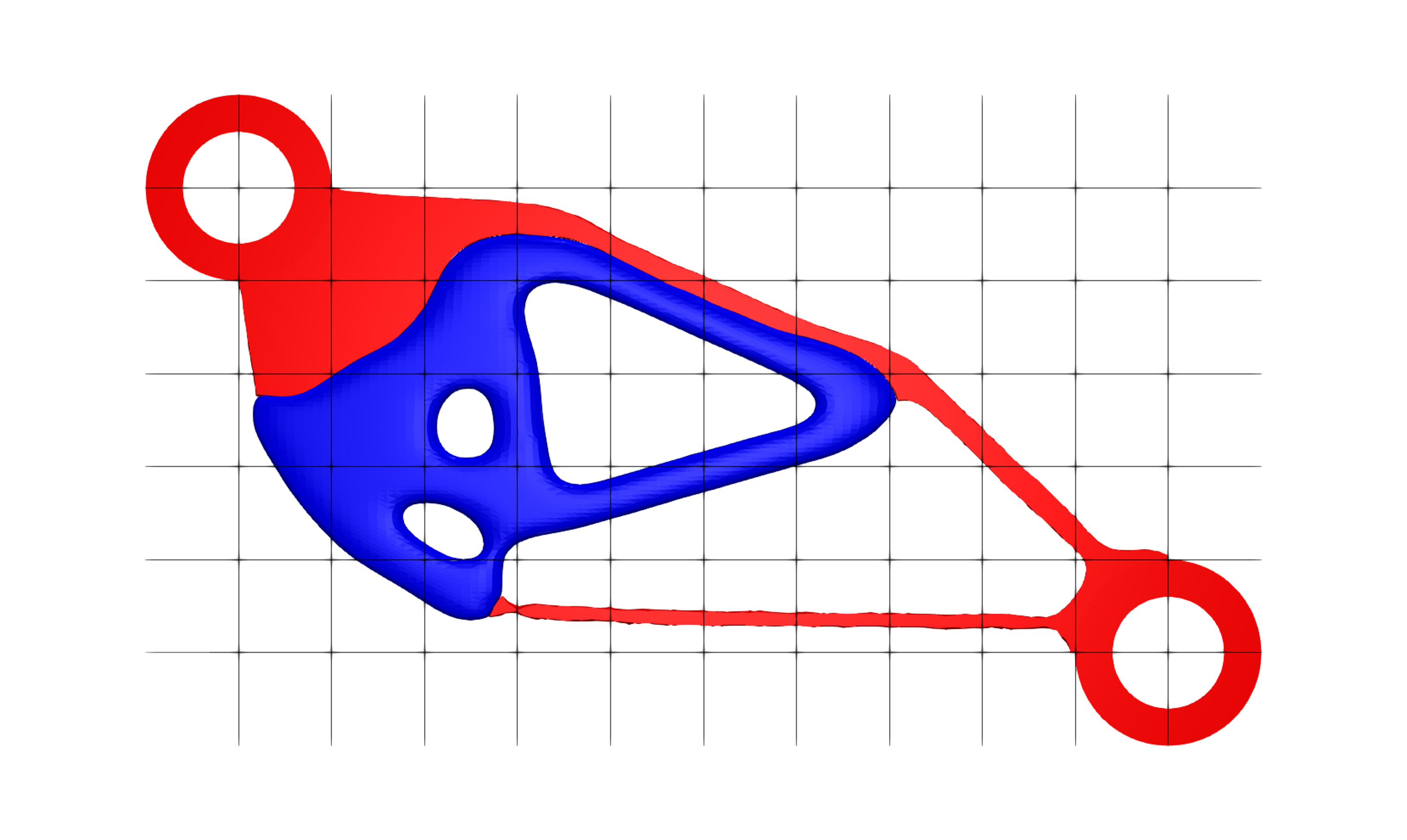}
		\subcaption{}\label{fig:ucsszallClip}
	\end{minipage}
	\caption{
		 Optimal configurations of the three-dimensional mean-compliance minimization with a uniform cross-section surface constraint between two materials (Case 25). \subref{fig:ucsszallClip} shows the sectional view. As shown in \subref{fig:ucsszallfront}, the boundary surfaces between two materials are vertical to the front surface.
		 }\label{fig:ucsszall result}
\end{figure}

\begin{figure}[H]
	\begin{minipage}[b]{0.32\linewidth}
		\centering
		\includegraphics[width=\linewidth]{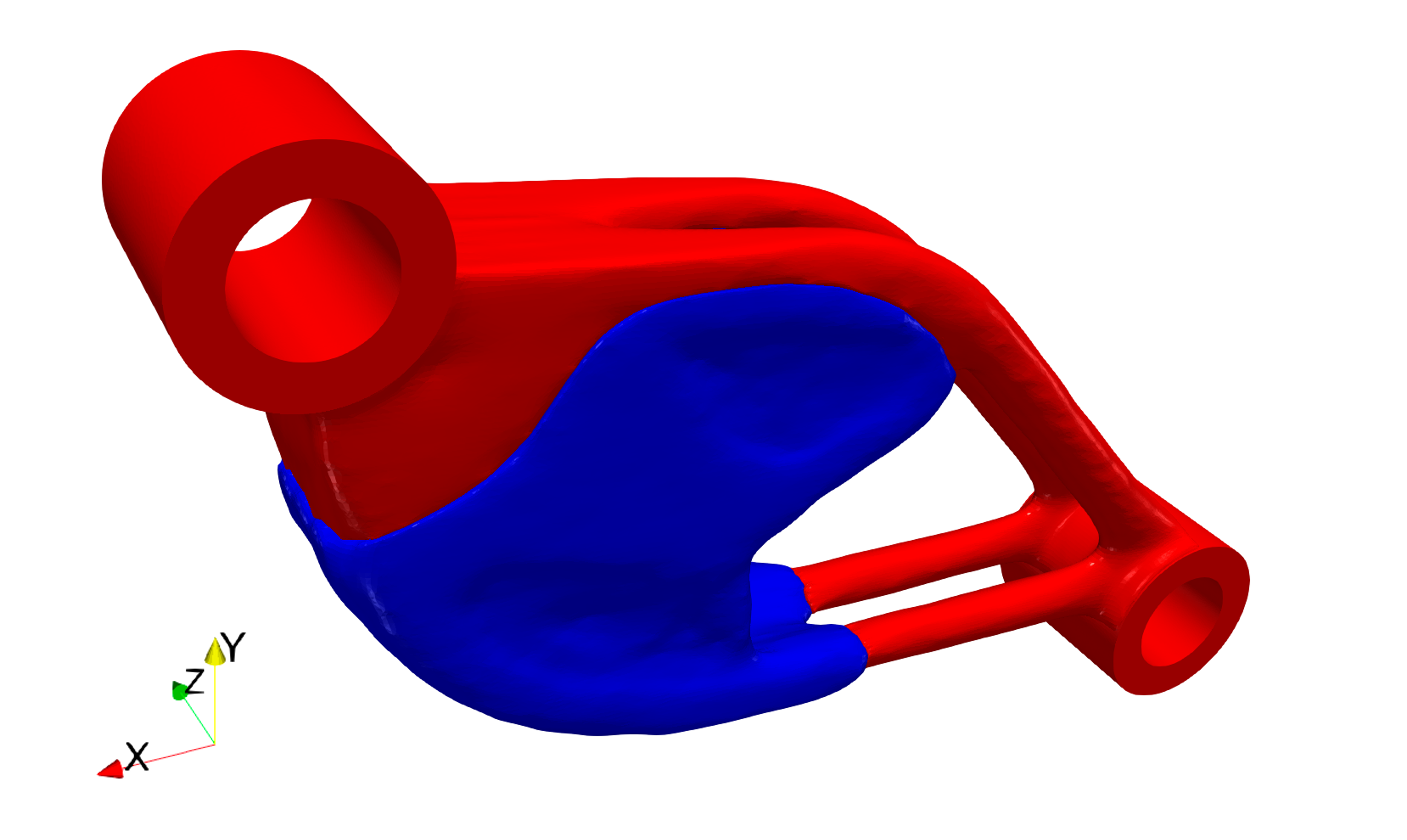}
		\subcaption{}\label{fig:ucssz12front}
	\end{minipage}
	\begin{minipage}[b]{0.32\linewidth}
		\centering
		\includegraphics[width=\linewidth]{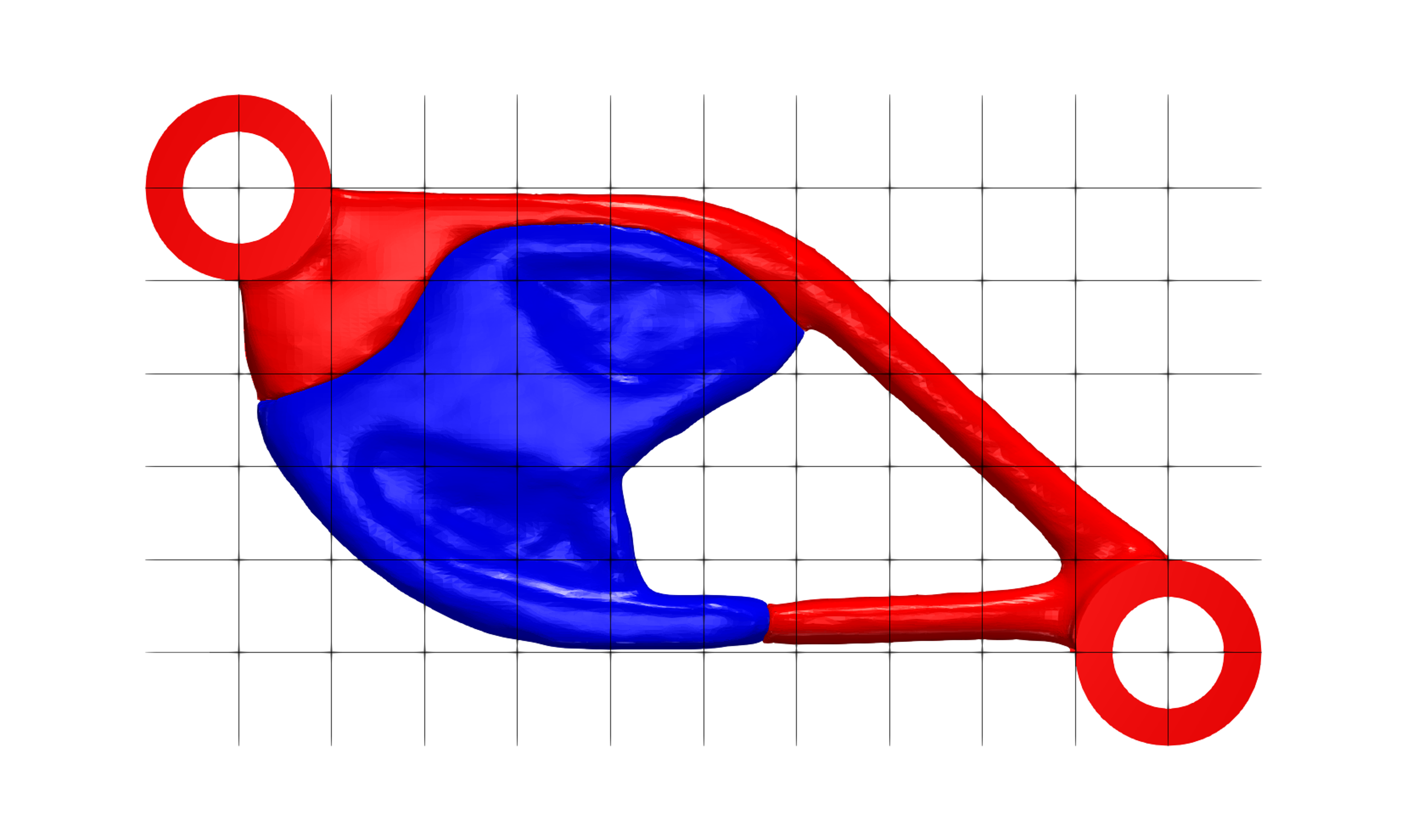}
		\subcaption{}\label{fig:ucssz12ZView}
	\end{minipage}
	\begin{minipage}[b]{0.32\linewidth}
		\centering
		\includegraphics[width=\linewidth]{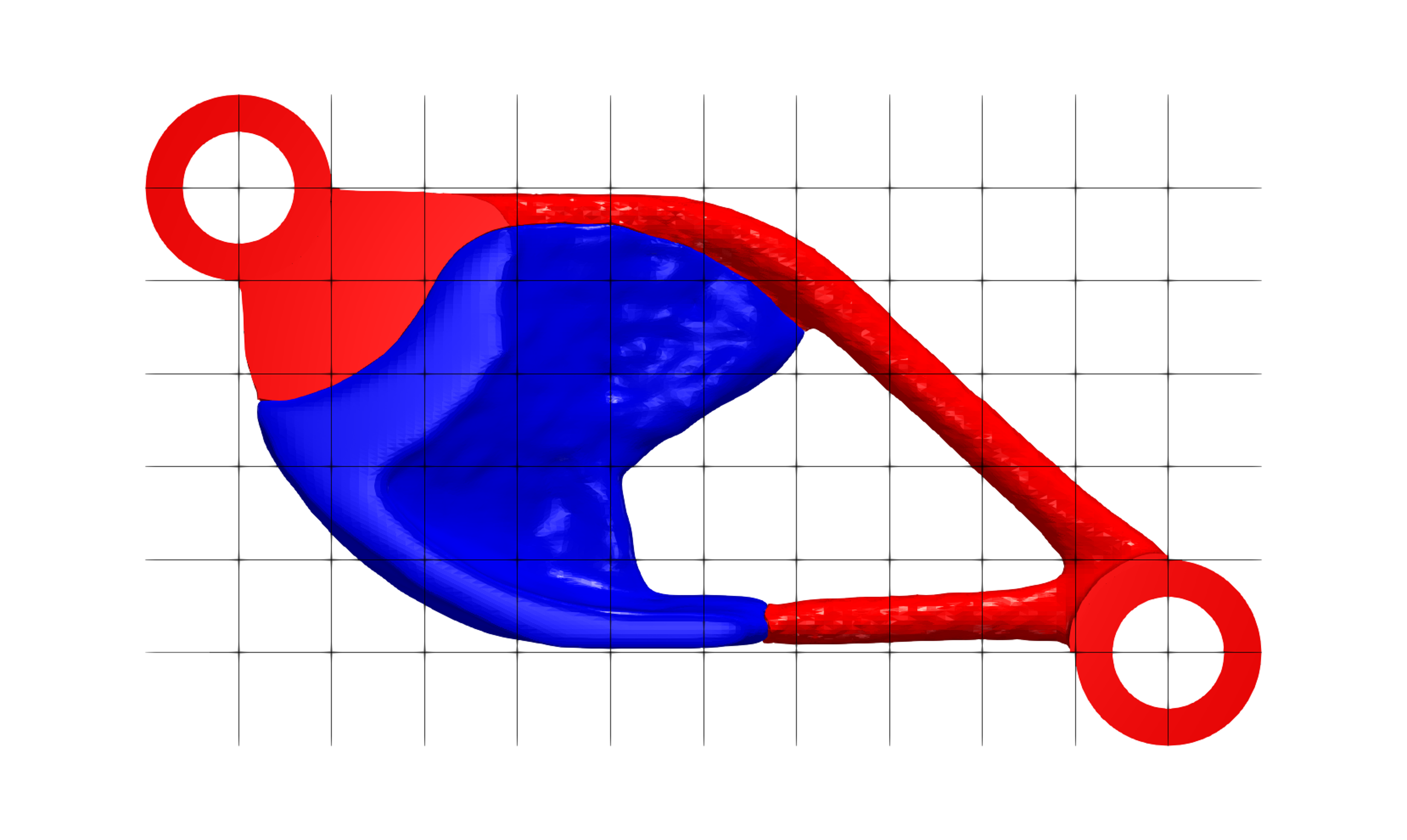}
		\subcaption{}\label{fig:ucssz12Clip}
	\end{minipage}
	\caption{
		Optimal configurations of the three-dimensional mean-compliance minimization with a uniform cross-section surface constraint between materials 1 and 2 (Case 26). \subref{fig:ucssz12Clip} shows the sectional view. Comparing \subref{fig:ucssz12ZView} and \subref{fig:ucssz12Clip}, we observe that the boundary surface between materials 1 and 2 is a congruent shape.}
	\label{fig:ucssz12 result}
\end{figure}

\begin{figure}[H]
	\begin{minipage}[b]{0.32\linewidth}
		\centering
		\includegraphics[width=\linewidth]{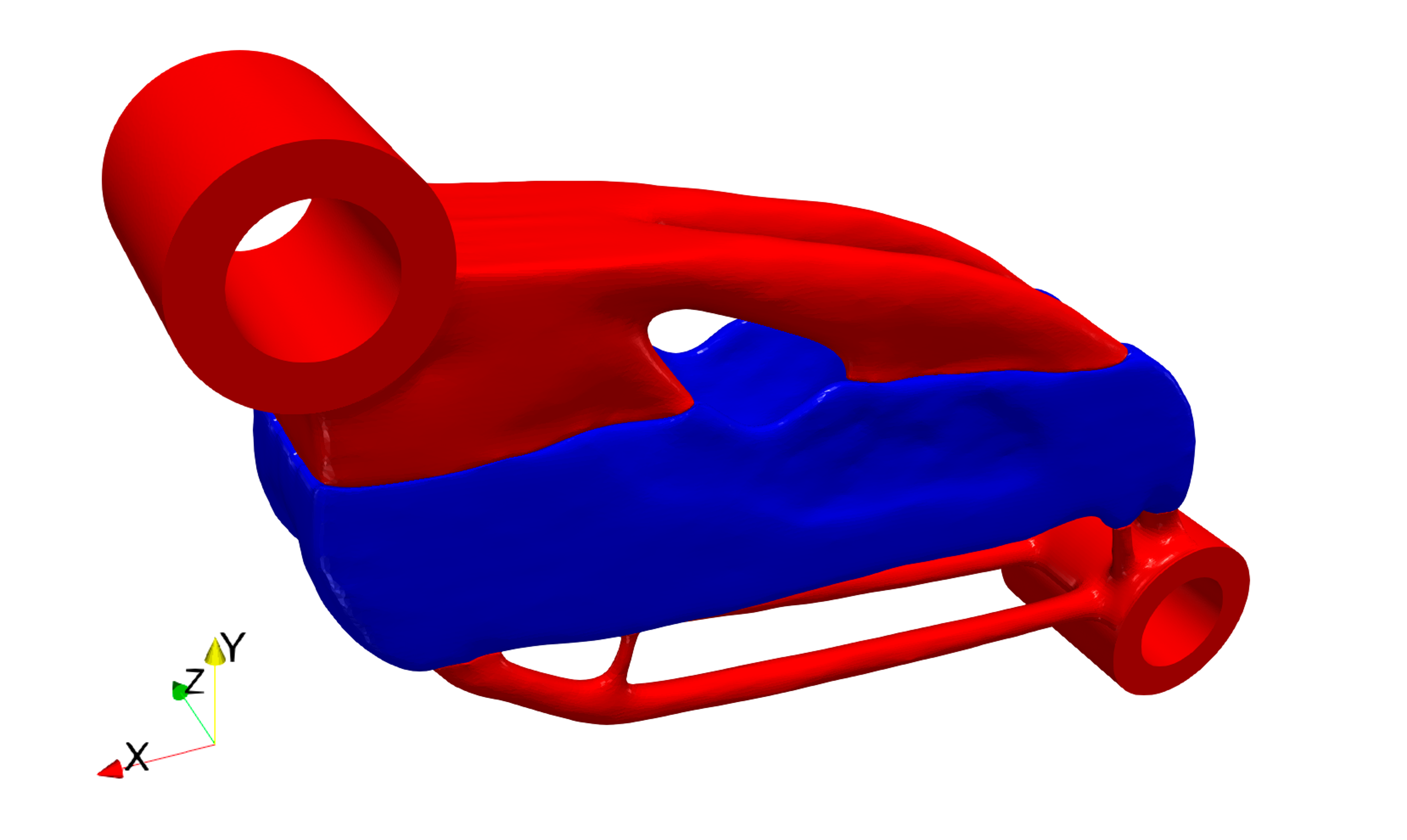}
		\subcaption{}\label{fig:ucssxzfront}
	\end{minipage}
	\begin{minipage}[b]{0.32\linewidth}
		\centering
		\includegraphics[width=\linewidth]{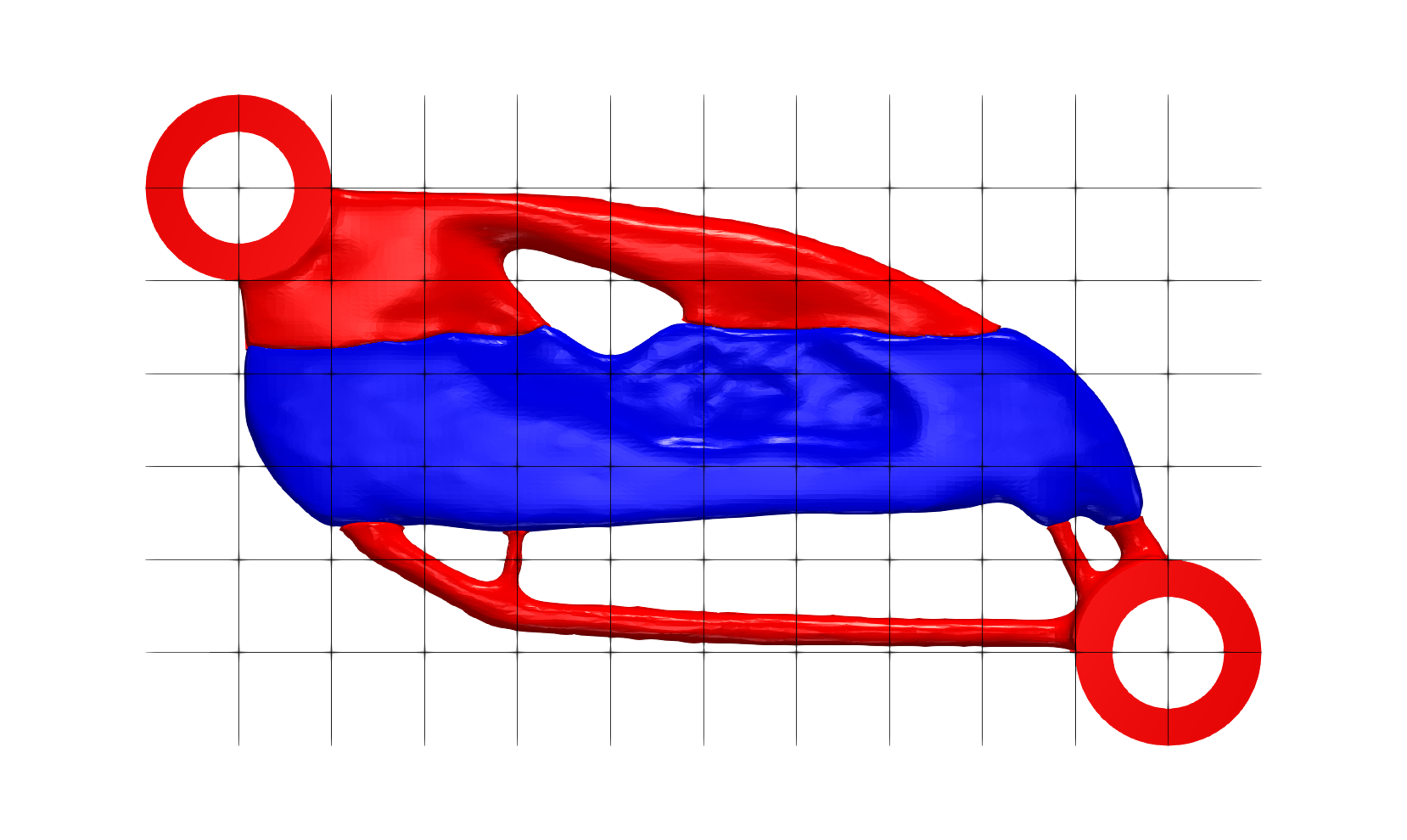}
		\subcaption{}\label{fig:ucssxzZView}
	\end{minipage}
	\begin{minipage}[b]{0.32\linewidth}
		\centering
		\includegraphics[width=\linewidth]{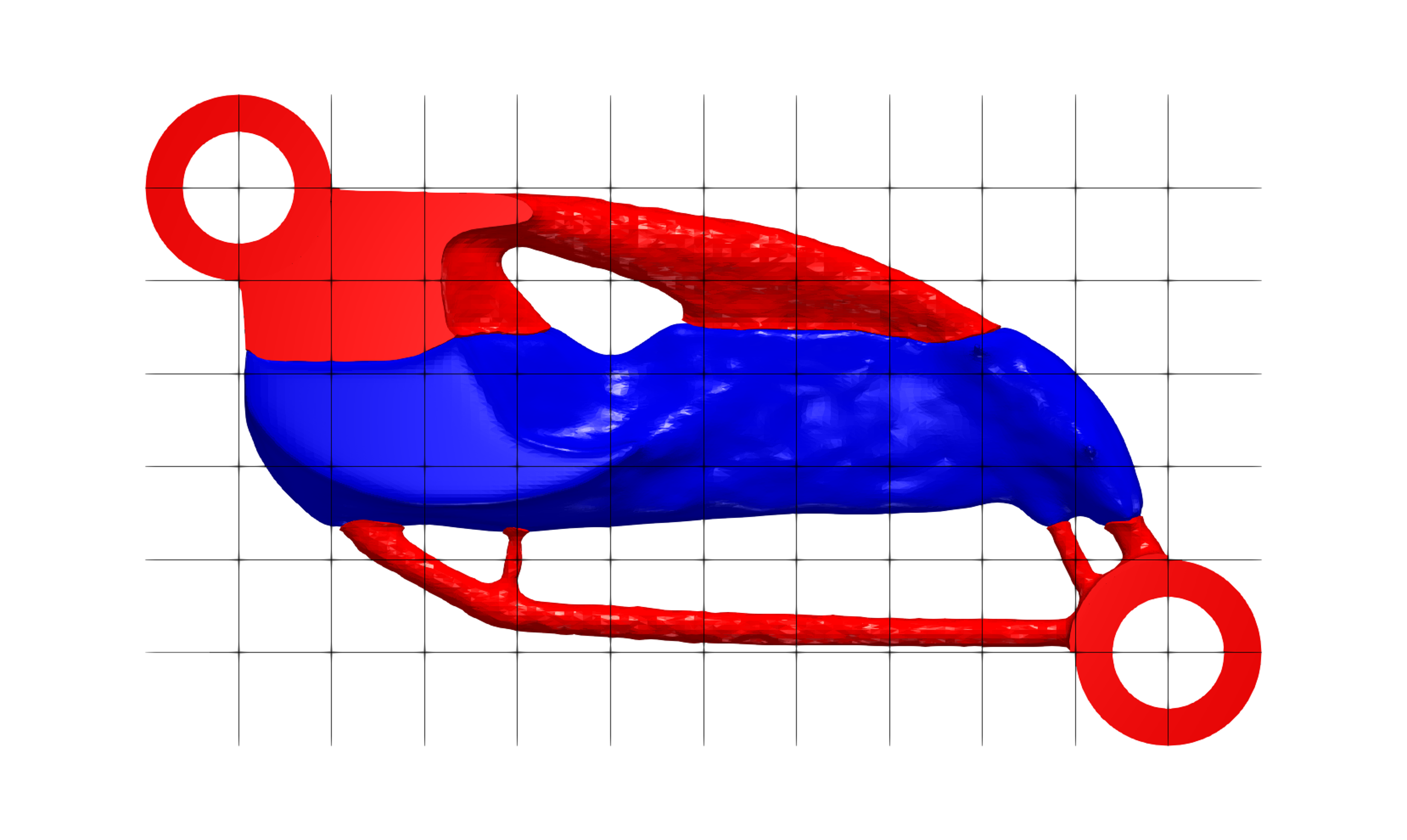}
		\subcaption{}\label{fig:ucssxzClip}
	\end{minipage}
	\caption{Optimal configurations of the three-dimensional mean-compliance minimization with a uniform cross-section surface constraint between materials 1 and 2 (Case 27). The uniform cross-section surface constraint is imposed in the directions parallel to the x- and z-axes. \subref{fig:ucssxzClip} shows the sectional view. The boundary surface between materials 1 and 2 became parallel to the x-z plane.}\label{fig:ucssxz result}
\end{figure}

\begin{figure}[H]
	\begin{minipage}[b]{0.32\linewidth}
		\centering
		\includegraphics[width=\linewidth]{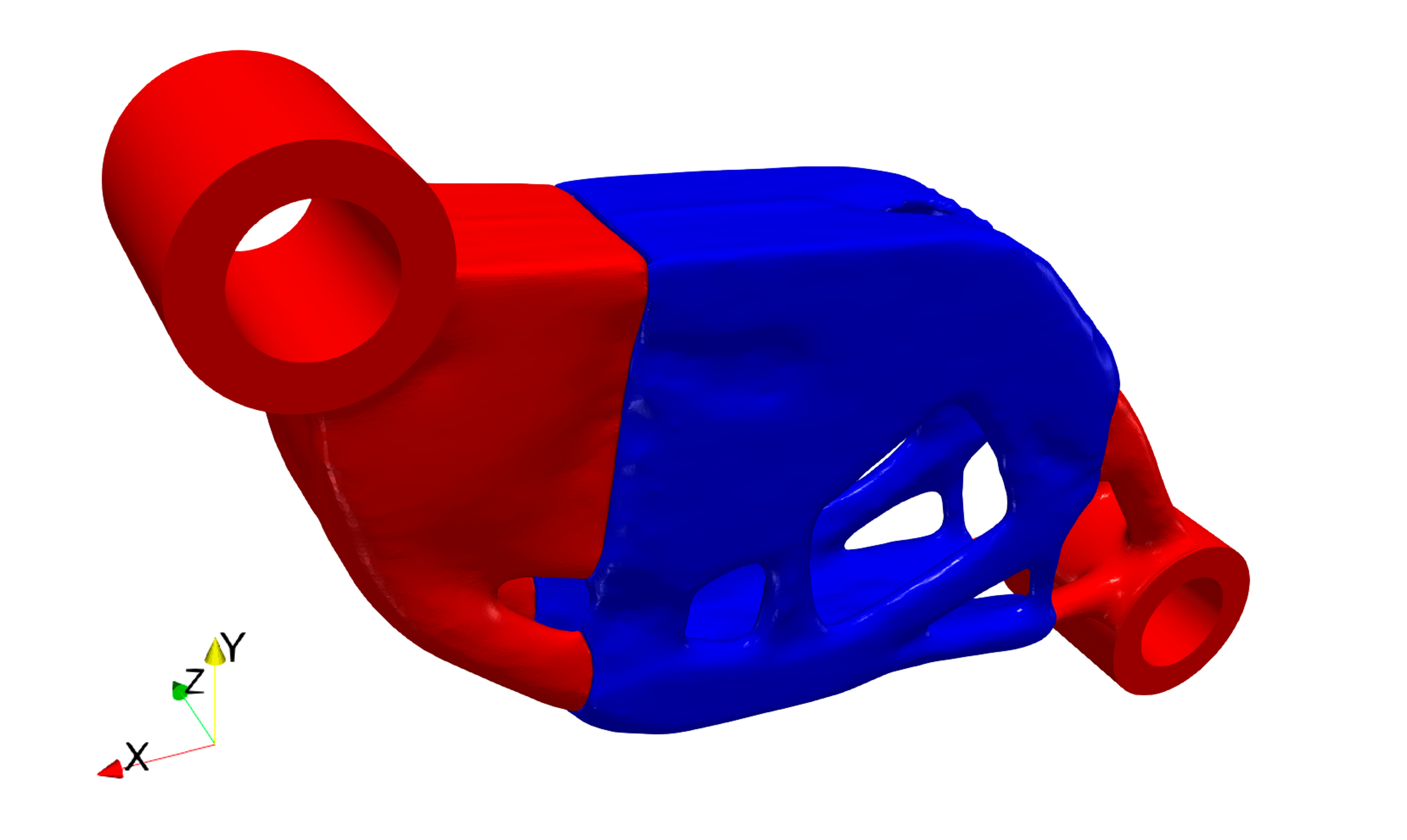}
		\subcaption{}\label{fig:ucssyzfront}
	\end{minipage}
%	\begin{minipage}[b]{\linewidth}
%		\centering
%		\includegraphics[width=\linewidth]{fig/3d_results/ucssyzback.png}
%		\subcaption{}\label{fig:ucssyzback}
%	\end{minipage}
	\begin{minipage}[b]{0.32\linewidth}
		\centering
		\includegraphics[width=\linewidth]{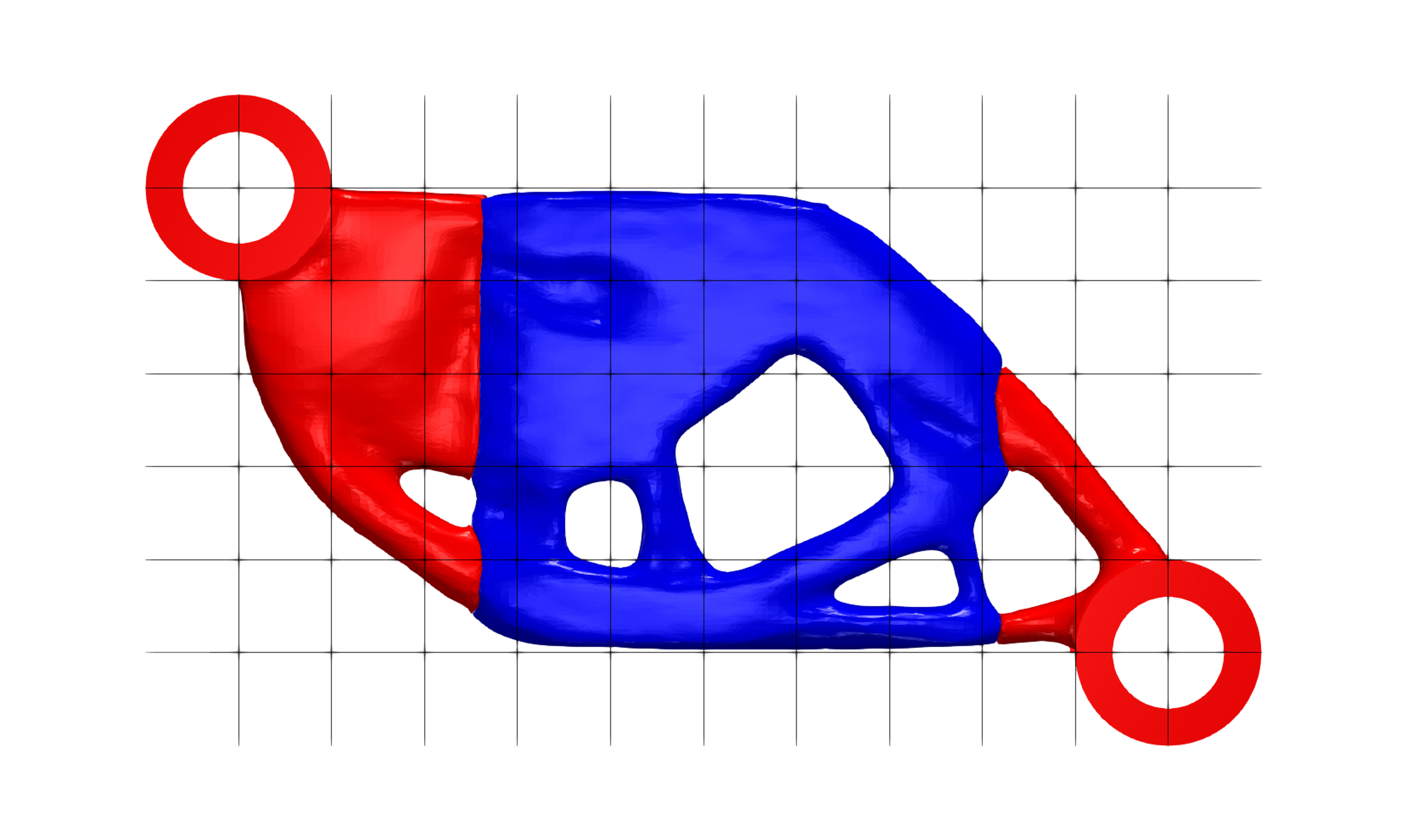}
		\subcaption{}\label{fig:ucssyzZView}
	\end{minipage}
	\begin{minipage}[b]{0.32\linewidth}
		\centering
		\includegraphics[width=\linewidth]{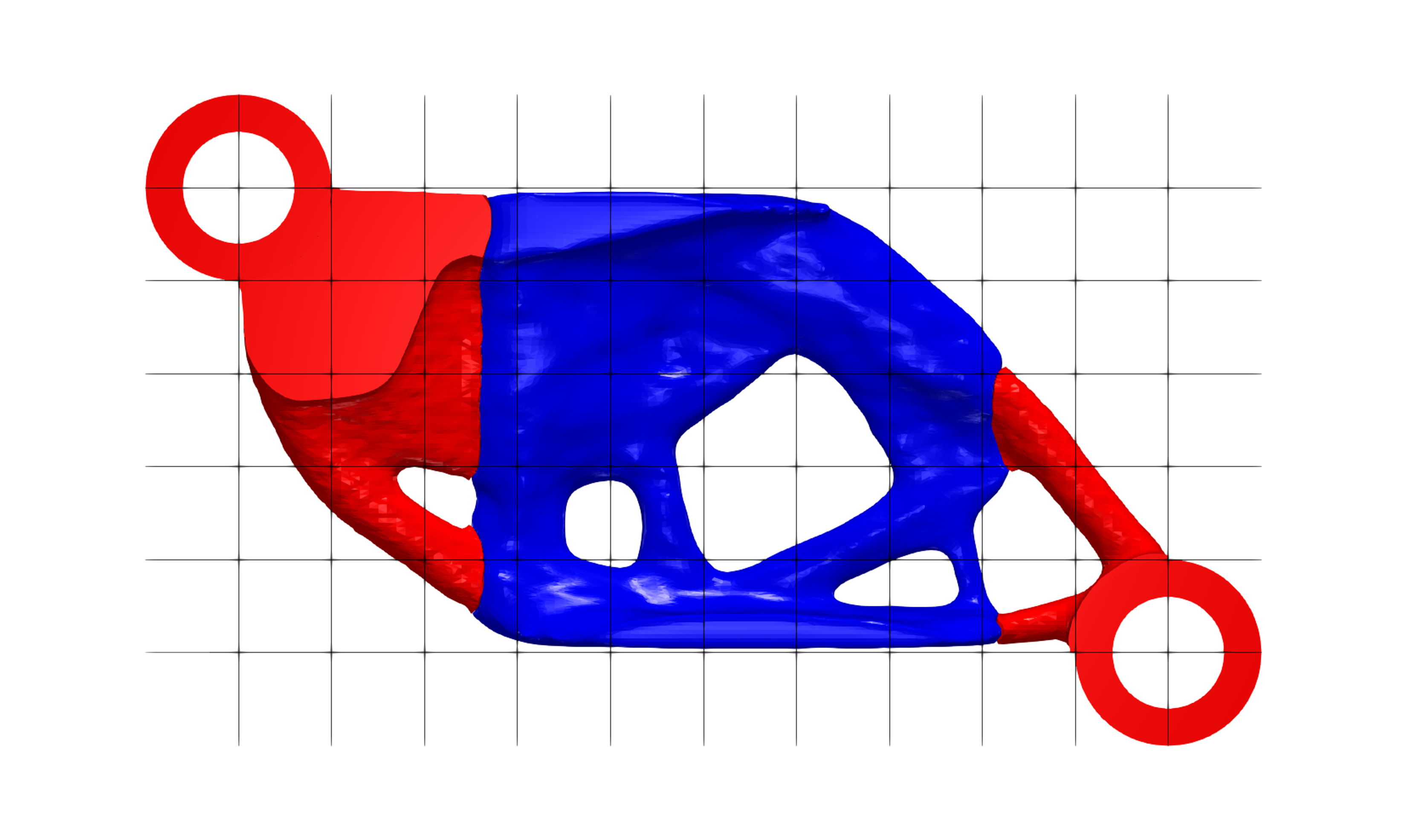}
		\subcaption{}\label{fig:ucssyzClip}
	\end{minipage}
	\caption{Optimal configurations of the three-dimensional mean-compliance minimization, with a uniform cross-section surface constraint between materials 1 and 2 (Case 28). The uniform cross-section surface constraint is imposed in the directions parallel to the y- and z-axes. \subref{fig:ucssyzClip} shows the sectional view. The boundary surface between materials 1
 and 2 became parallel to the y-z plane.
}\label{fig:ucssyz result}
\end{figure}

Similar to the two-dimensional compliance minimization problem, we next imposed uniform cross-section surface constraints and piecewise-linear surface constraints on configurations with three materials (materials 0, 1, and 2 in Table \ref{tab:materials}). The regularization parameter $\tau_{ij}$ was set to $1\times10^{-4}$. In Cases 25--28, we imposed uniform cross-section surface constraints and set the anisotropic regularization parameters $\tilde\tau_{ijk}$ as presented in Table \ref{tab:anisotropic regularization parameters for 3d}. 
\begin{table}[h]
\centering
\caption{Anisotropic regularization parameters $\tilde\tau_{ijk}$ in cases 25--28}
\begin{tabular}{lccccccccc}\hline
	Case &
	$\tilde\tau_{01x}$  & $\tilde\tau_{01y}$ &$\tilde\tau_{01z}$ &
	$\tilde\tau_{02x}$  & $\tilde\tau_{02y}$ &$\tilde\tau_{02z}$ &
	$\tilde\tau_{12x}$  & $\tilde\tau_{12y}$ &$\tilde\tau_{12z}$\\ \hline
	23&1&1&1&1 &1&1&1&1&1 \\
	25&1&1&$10^{3}$&1 &1&$10^{3}$ &1&1&$10^{3}$\\
	26&1&1&1&1 &1&1&1&1&$10^{3}$\\
	27&1&1&1&1 &1&1&$10^6$&1&$10^3$\\
	28&1&1&1&1 &1&1&1&$10^5$&$10^5$\\ \hline
\end{tabular}
\label{tab:anisotropic regularization parameters for 3d}
\end{table}
The normalization coefficients were set to $({K^\text{ucss}}_{01},{K^\text{ucss}}_{02},{K^\text{ucss}}_{12})=(1,1,1)$ in Cases 25 and 26, and $(1,1,10)$ in Cases 27 and 28.
In Case 25, the cross-section surface was assumed constant along the z-axis, like cookie cutters. In this problem setting, if material 0 is assumed on the front and back boundary surfaces of the design domain, the entire design domain will be material 0. Therefore, in Case 25 (but not in the other three-dimensional cases), the level set function was set to the Neumann condition $\bm n \cdot \nabla\phi_{ij}=0$ at the front and back boundaries of the design domain. In Case 26, the interface between the regions of materials 1 and 2 was constrained to be constant along the z-axis. In Cases 27 and 28, the surface was constrained to be flat, and the uniform cross-section surface constraint was imposed in the direction parallel to the x- and z-axes and parallel to the y- and z-axes, respectively. The optimal configurations in Cases 25--28 are displayed in Figs. \ref{fig:ucsszall result} -- \ref{fig:ucssyz result}.
\begin{figure}[H]
	\begin{minipage}[t]{0.32\linewidth}
		\centering
		\includegraphics[width=\linewidth]{fig/3d_results/M3front.png}
		\subcaption{}\label{fig:noucssfront}
	\end{minipage}
	\begin{minipage}[t]{0.32\linewidth}
		\centering
		\includegraphics[width=\linewidth]{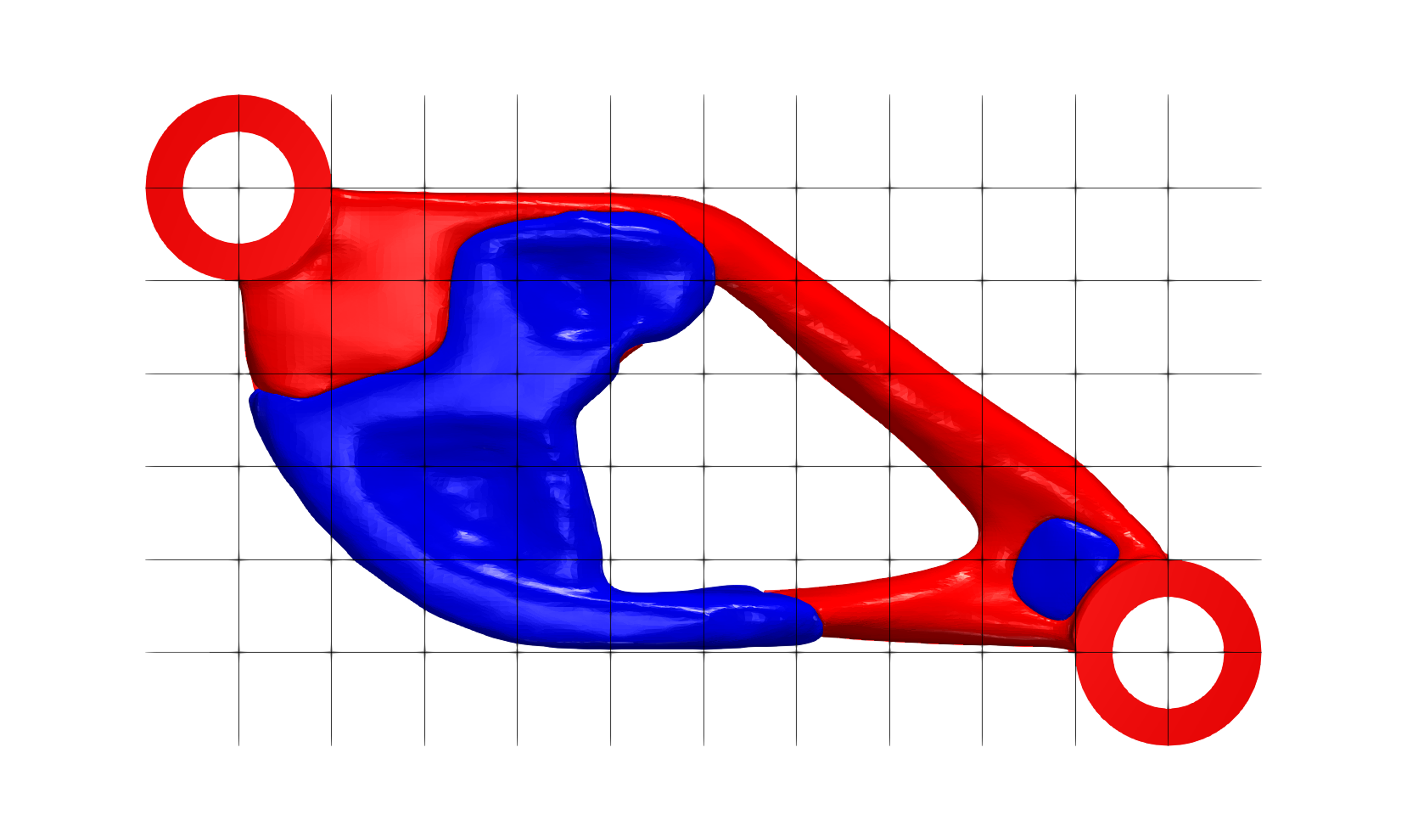}
		\subcaption{}\label{fig:noucssZView}
	\end{minipage}
	\begin{minipage}[t]{0.32\linewidth}
		\centering
		\includegraphics[width=\linewidth]{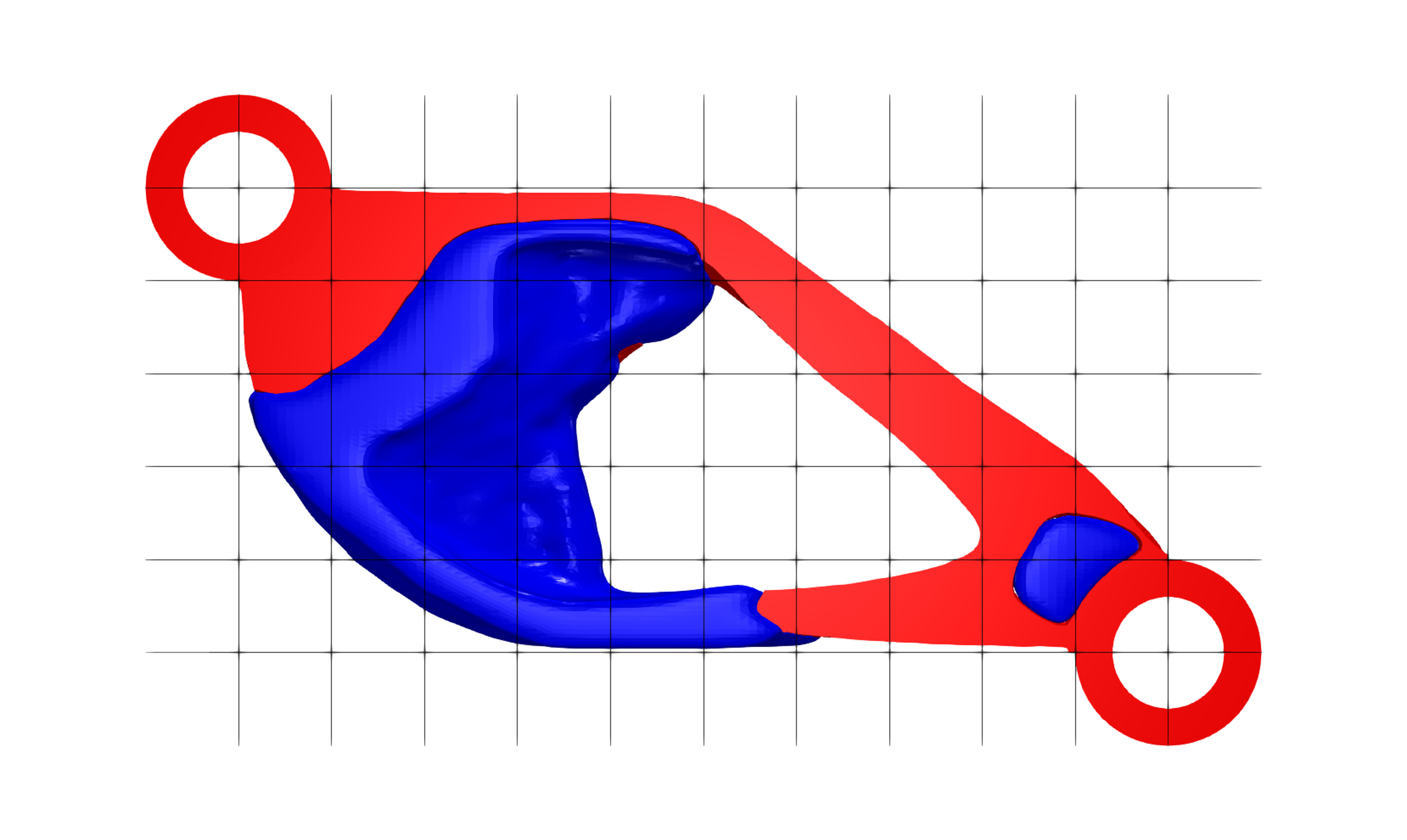}
		\subcaption{}\label{fig:noucssClip}
	\end{minipage}
	\caption{Optimal configurations of the three-dimensional mean-compliance minimization without the uniform cross-section surface constraint (Case 23). \subref{fig:noucssClip} shows the sectional view.}\label{fig:noucss result}
\end{figure}

The values of the objective functions in Cases 25, 26, 27, and 28 were 0.192, 0.190, 0.191, and 0.190, respectively. The upper volume constraints were mostly satisfied (volume constraint functions $<10^{-4}$ in Cases 25, 27, 28 and $<10^{-8}$ in Case 26). In Case 25 (see Fig. \ref{fig:ucsszall result}\subref{fig:ucsszallfront}), the uniform cross-section surface constraint was satisfied both at the exterior surface (the boundary between the cavity and other structural material regions) and at the boundary between materials 1 and 2, as evidenced by the coincidence of the front and cross-sectional views in panels \subref{fig:ucsszallfront} and \subref{fig:ucsszallClip} of Fig. \ref{fig:ucsszall result}, respectively. In Case 26, the uniform cross-section surface constraint was not satisfied at the exterior surface (boundary between the cavity and the other structural material regions; see Fig. \ref{fig:ucssz12 result}\subref{fig:ucssz12front}) but was satisfied at the boundary between materials 1 and 2, as evidenced by the coincidental front view in panel \subref{fig:ucssz12ZView} and sectional view in panel \subref{fig:ucssz12Clip} of Fig. \ref{fig:ucssz12 result}. In Cases 27 and 28, we confirmed that our method can successfully impose uniform cross-section surface constraints in two directions. In particular, the red and blue regions corresponding to materials 1 and 2, respectively, were divided by planes spanning the entire design domain parallel to the x-z plane in Case 27 (Fig. \ref{fig:ucssxz result}) and to the y-z plane in Case 28 (Fig. \ref{fig:ucssyz result}). These results indicate that the proposed method can selectively constrain the cross-sectional surface between two materials to be uniform in the three-dimensional case.

\begin{figure}[H]
	\begin{minipage}[b]{0.32\linewidth}
		\centering
		\includegraphics[width=\linewidth]{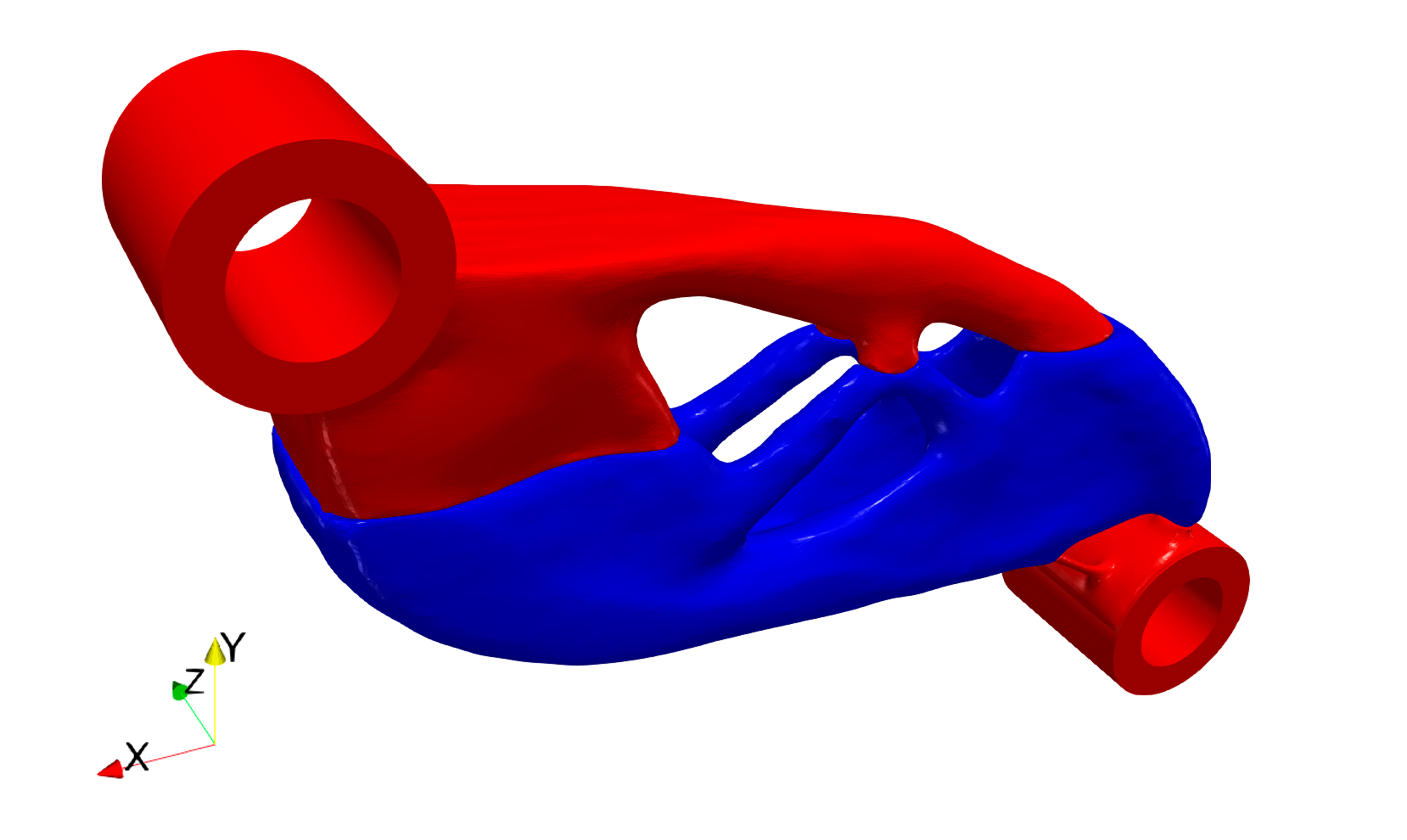}
		\subcaption{}\label{fig:lucssxzfront}
	\end{minipage}
	\begin{minipage}[b]{0.32\linewidth}
		\centering
		\includegraphics[width=\linewidth]{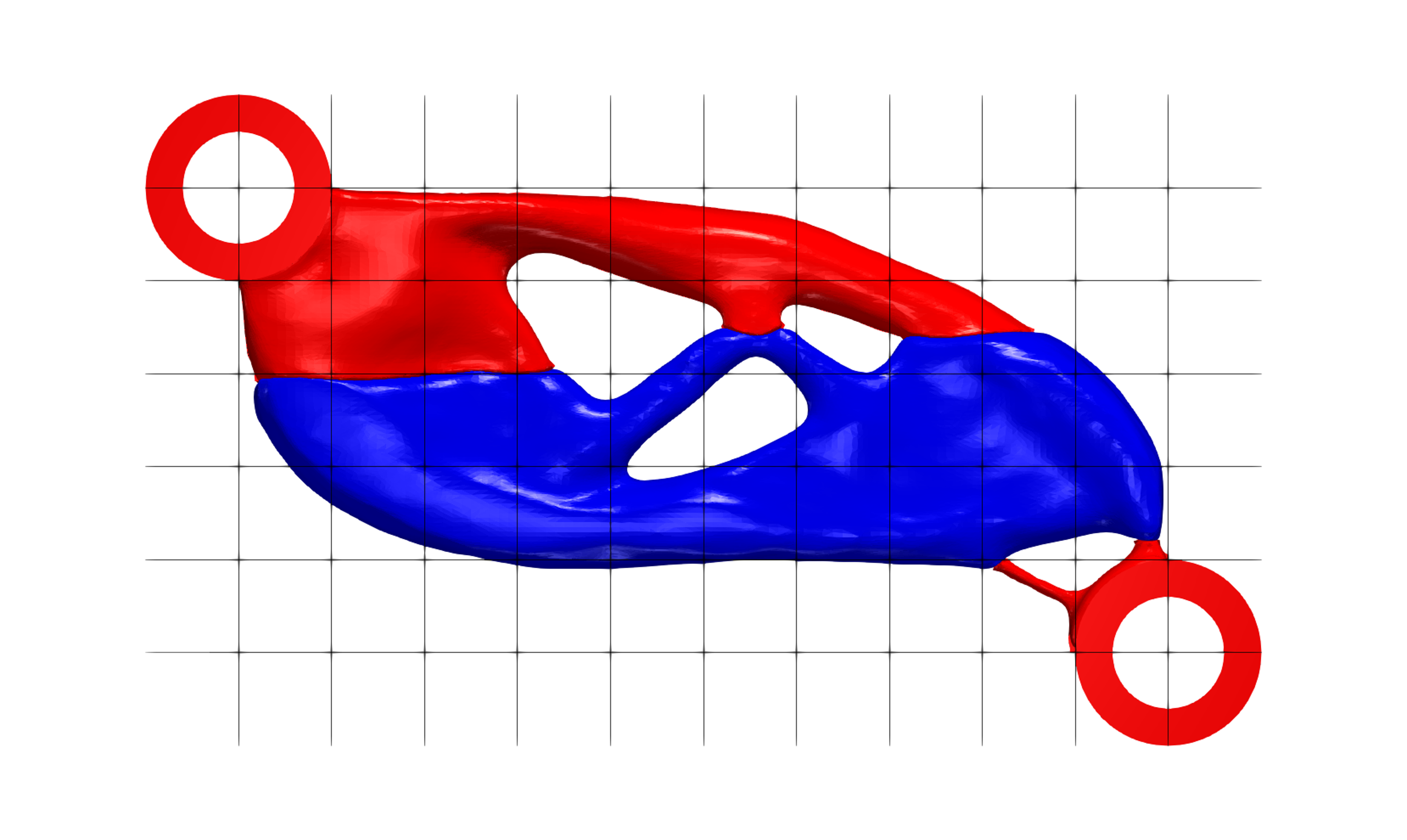}
		\subcaption{}\label{fig:lucssxzZView}
	\end{minipage}
	\begin{minipage}[b]{0.32\linewidth}
		\centering
		\includegraphics[width=\linewidth]{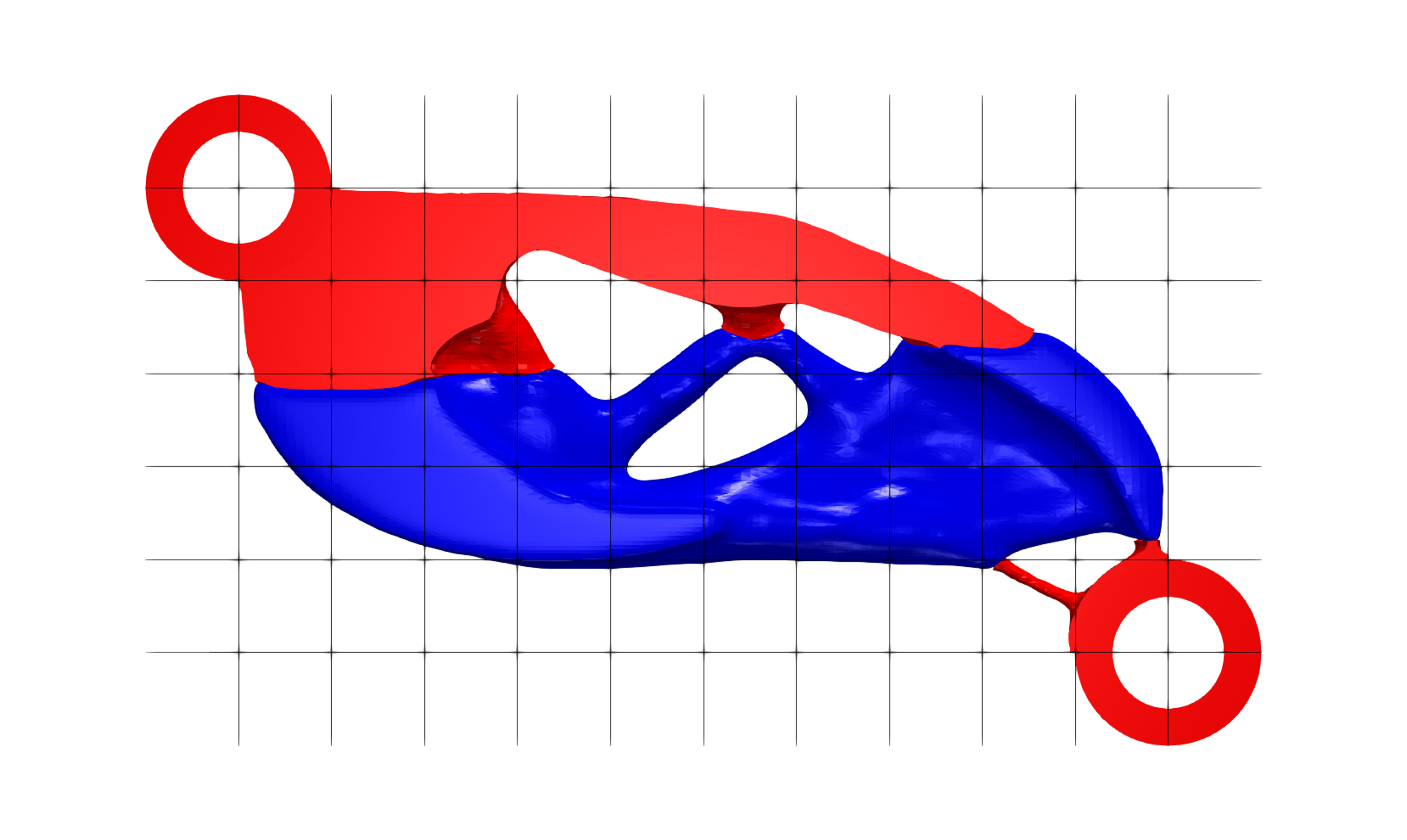}
		\subcaption{}\label{fig:lucssxzClip}
	\end{minipage}
	\caption{
		Optimal configurations in three-dimensional mean-compliance minimization with the boundaries between materials 1 and 2 constrained to be parallel to the x- and z-axes (Case 29). \subref{fig:lucssxzClip} shows the sectional view. The boundary surface between materials 1 and 2 became parallel to the x-z plane.}\label{fig:lucssxz result}
\end{figure}

\begin{figure}[H]
	\begin{minipage}[b]{0.32\linewidth}
		\centering
		\includegraphics[width=\linewidth]{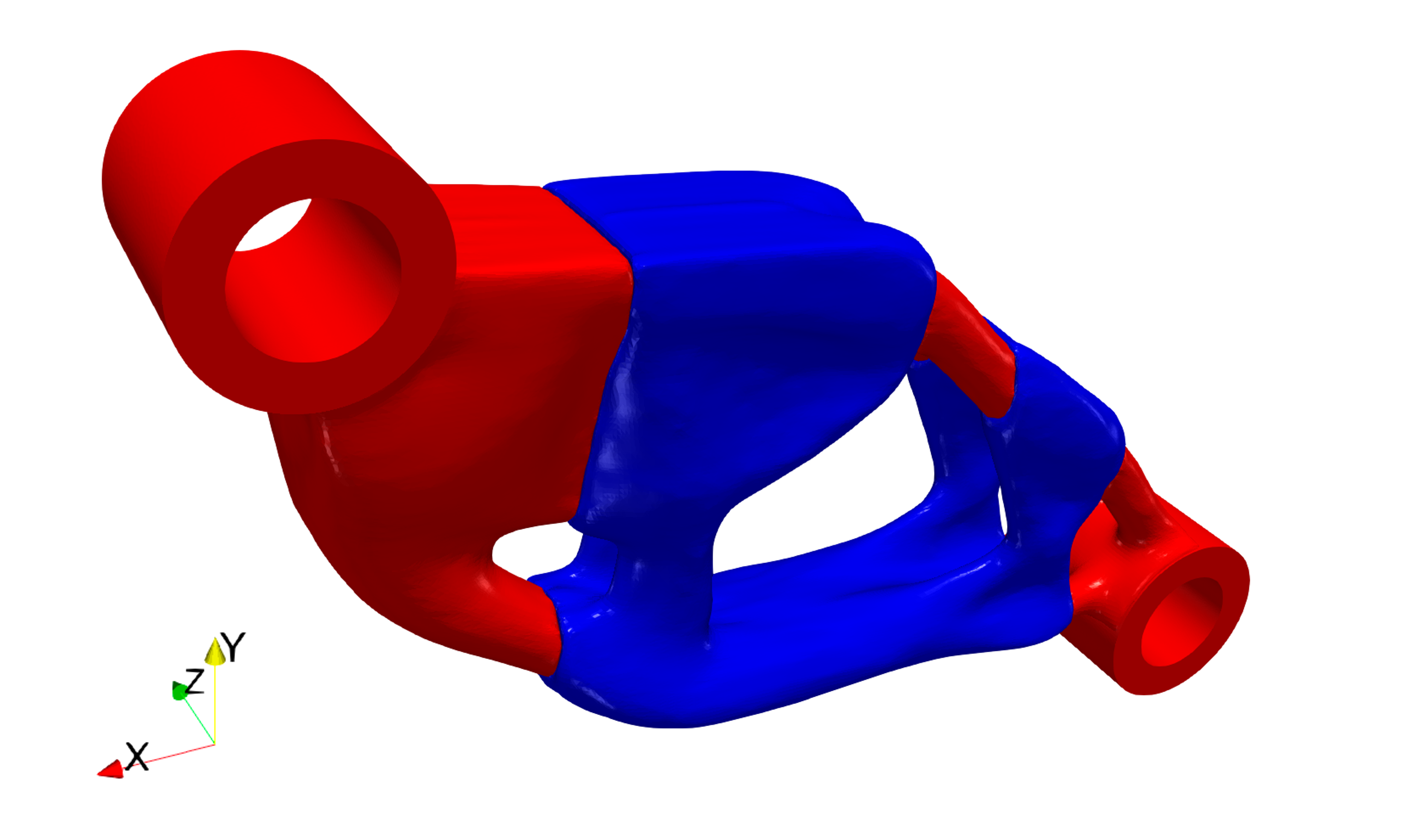}
		\subcaption{}\label{fig:lucssyzfront}
	\end{minipage}
	\begin{minipage}[b]{0.32\linewidth}
		\centering
		\includegraphics[width=\linewidth]{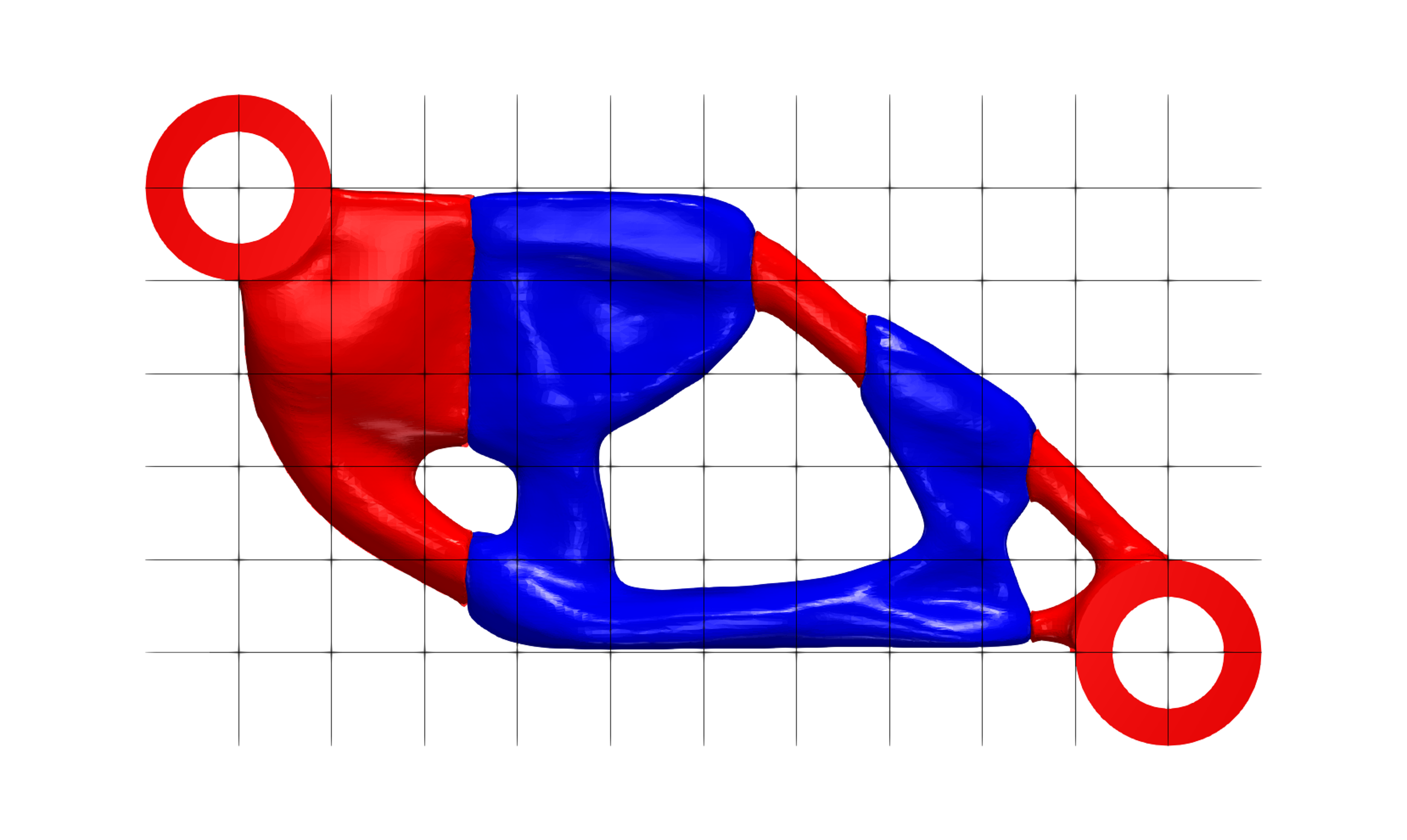}
		\subcaption{}\label{fig:lucssyzZView}
	\end{minipage}
	\begin{minipage}[b]{0.32\linewidth}
		\centering
		\includegraphics[width=\linewidth]{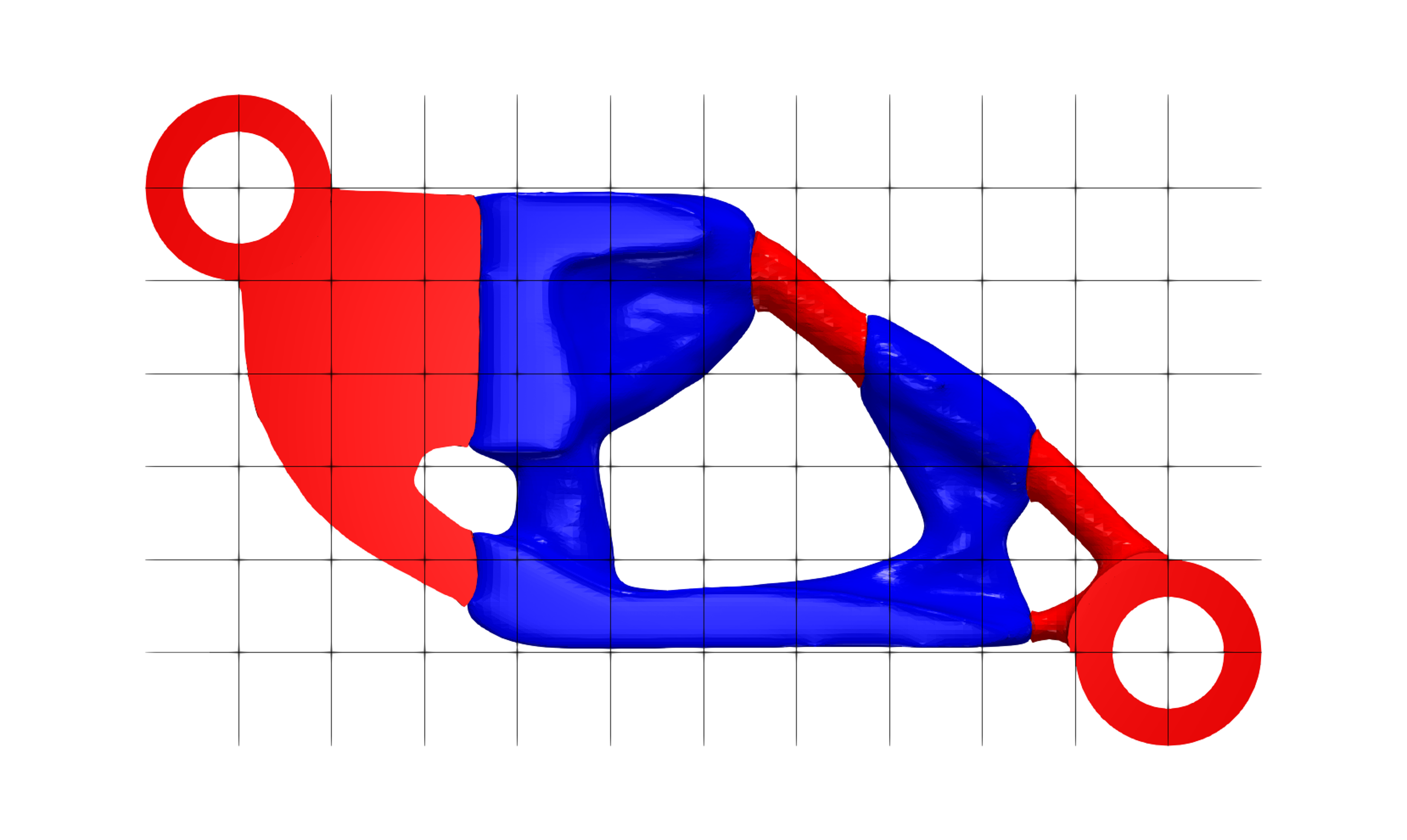}
		\subcaption{}\label{fig:lucssyzClip}
	\end{minipage}
	\caption{
	Optimal configurations in three-dimensional mean-compliance minimization with the boundaries between materials 1 and 2 constrained to be parallel to the y- and z-axes (Case 30). \subref{fig:lucssxzClip} shows the sectional view. The boundary surface between materials 1 and 2 became parallel to the y-z plane. 
}\label{fig:lucssyz result}
\end{figure}

%\subsubsection{Piecewise linear surface constraints}
In Cases 29 and 30, we imposed the piecewise-linear boundary constraints in two directions, that is, we imposed piecewise-linear surface constraints. The piecewise anisotropic regularization parameters $\tilde\tau_{ijk}^{'}$ were set as shown in Table \ref{tab: piecewiseanisotropic regularization parameters for 3d}. 
\begin{table}[h]
	\centering
	\caption{Piecewise anisotropic regularization parameters $\tilde\tau_{ijk}^{'}$ in Cases 29 and 30}
	\begin{tabular}{lccccccccc}\hline
		Case &
		$\tilde\tau_{01x}^{'}$  & $\tilde\tau_{01y}^{'}$ &$\tilde\tau_{01z}^{'}$ &
		$\tilde\tau_{02x}^{'}$  & $\tilde\tau_{02y}^{'}$ &$\tilde\tau_{02z}^{'}$ &
		$\tilde\tau_{12x}^{'}$  & $\tilde\tau_{12y}^{'}$ &$\tilde\tau_{12z}^{'}$\\ \hline
		29&1&1&1&1 &1&1&$10^6$&1&$10^3$\\
		30&1&1&1&1 &1&1&1&$10^5$&$10^5$\\ \hline
	\end{tabular}
	\label{tab: piecewiseanisotropic regularization parameters for 3d}
\end{table}
The normalization coefficients were set to
$\left( {K^\text{ucss}}_{01}, {K^\text{ucss}}_{02}, {K^\text{ucss}}_{12}\right)  \\=(1,1,10) $. In Cases 29 and 30, the piecewise-linear boundary constraint was imposed parallel to the x- and z-axes and to the y- and z-axes, respectively. The optimal configurations in Cases 29 and 30 are presented in Figs. \ref{fig:lucssxz result} and \ref{fig:lucssyz result}, respectively.
The values of the objective functions in Cases 29 and 30 were 0.192 and 0.190, respectively. The upper volume constraints were mostly satisfied (volume constraint functions $<5\times10^{-3}$ in Case 29 and $10^{-4}$ in Case 30). We confirmed that our method successfully imposed linear boundary constraints; that is, the boundary surfaces between materials 1 and 2 were parallel to the x-z plane in Case 29 (see Fig. \ref{fig:lucssxz result}) and to the y-z plane in Case 30 (Fig. \ref{fig:lucssyz result}). Under the uniform cross-section surface constraint, materials 1 and 2 were divided by planes over the entire design domain. The piecewise-linear boundary constraints allow boundaries to be displaced or lost if their cross sections cross the cavity region. In fact, in the optimal structure of Case 29, the two boundaries between the red and blue regions on the right side were higher than the boundary on the left side (see Fig. \ref{fig:lucssxz result}\subref{fig:lucssxzZView}). In Case 30, a red region appeared in the upper center but not in the lower center (Fig. \ref{fig:lucssyz result}\subref{fig:lucssyzZView}). These results indicate that the proposed method can impose a piecewise-linear boundary constraint in the three-dimensional case.

\section{Conclusion}\label{sec: Conclusion}
In this paper, we extended the level set method to multiple phases and applied the proposed method to the multi-material topology optimization problem. The study results are summarized below.
	\begin{enumerate}
		\item 
			We proposed an extended level set method for multiphase representation. For $M$ materials, the proposed method defines $M(M-1)$ level set functions. The boundary between materials $i$ and $j$ is expressed as the zero level set of the level set function $\phi_{ij}$ $(i,j\in \{0,1,\ldots,M-1\},~j\ne i)$.
		\item
			Based on the extended level set method, we formulated a multi-material topology optimization problem. The proposed method provides a high degree of freedom during the optimization. The topological derivatives are expressed in a simple form.
		\item
		 	Optimization procedures for the extended level set method were provided and implemented numerically. To apply the ersatz material approach to multiple materials, the characteristic functions of multi-materials were approximated and smoothed.
		\item 
			Several numerical examples were provided. We first applied the multi-material optimization method to two-dimensional compliance minimization, compliant mechanism optimization, and moment-of-inertia minimization problems using 2--9 different materials. In all numerical examples, the obtained optimal solutions were considered mechanically reasonable, validating the proposed method. We then introduced regularization parameters and examined their effects. It was shown that the proposed method can control the geometric complexity of the boundary between two materials for each combination of two materials. Finally the proposed method was validated using three-dimensional problems.
	\end{enumerate}

The shortcomings of the proposed method are described below.
\begin{enumerate}
	\item
		Due to the large number of level set functions, which are design variables, compared to other multi-material representation methods, a large amount of computation time is required to update them and calculate the characteristic functions. It could be improved by parallelization or by omitting the calculation of the part where the material is fixed.
		
	\item
		In the proposed method, there are degrees of freedom where the three boundaries should meet at a single point, but they are displaced. Therefore, there is a problem that some regions are not in either phase. In this paper, we formulate an approximate assignment of such a region to one of the phases. This approximation is only heuristic, and its validity is not guaranteed. 
		We cannot exclude the possibility that the approximation may not hold in the process of calculation. The vector-valued level set method can be seen as a solution to this problem by placing constraints between the design variables. However, these constraints are not necessarily essential constraints.
		One possible solution is to perform optimization with necessary and sufficient constraints.
\end{enumerate}

%%\textcolor{blue}{できれば，今後の展望についても述べてください．}
%The future prospects are described below.
%\begin{enumerate}
%	
%	\item
%		We should assess the applicability of the proposed method to other physical-field optimization problems such as heat transfer, vibration, electromagnetic, acoustic, and fluid flow problems. Optimization problems involving a combination of multiple physical fields are a particularly enticing prospect.
%	
%	\item
%		We should assess whether the method can optimize mechanical systems with multiple components and mechanisms.
%	
%	\item
%		The method should be applied to practical design problems, considering the material boundary characteristics and manufacturability.
%\end{enumerate}

\section*{Acknowledgment}
Funding: 
This work was supported by JST FOREST Program (Grant Number JPMJFR202J, Japan).

The authors would like to thank Enago (www.enago.jp) for the English language review.

\appendix

\section{Extended Level Set Method as a Generalization of Conventional Methods}\label{sec:generalization}
In this section, we explain that some of the existing multi-phase representation methods based on the level set method are special cases of the X-LS method. For each method, we first describe the concept of the methods and formulate the multiphase representation. We then give some constraint equations for the level set function in the X-LS method. Finally, we rewrite the multiphase representation of the X-LS method using the constraint equations and show that the constrained X-LS representation is equivalent to multiphase representation in the existing method.

\subsection{Color level set method}
\begin{figure}[H]
	\centering
	\includegraphics[width=10cm]{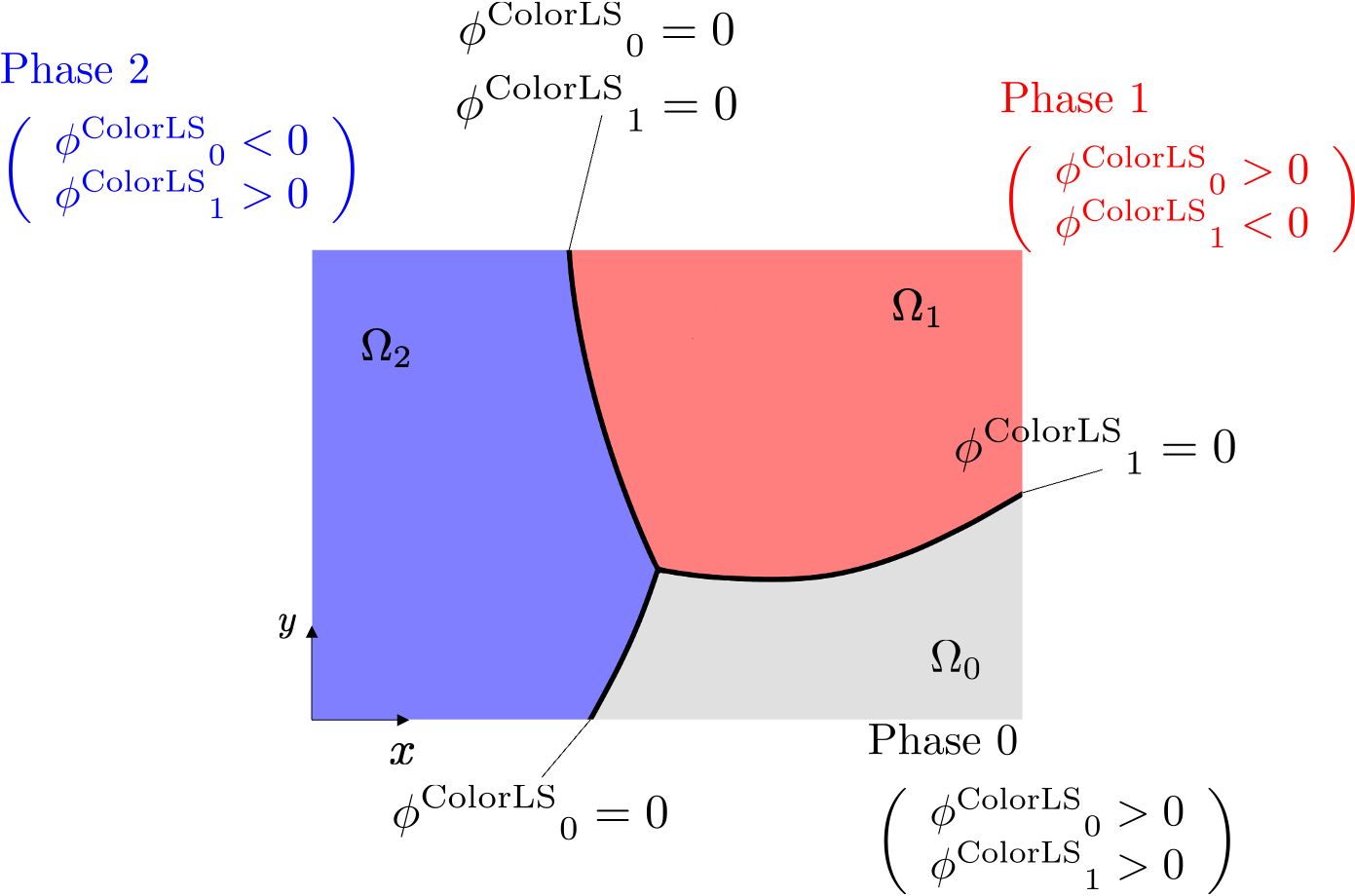}
	\caption{
	Concept of ColorLS. Regions $\Omega_0, \Omega_1$ and $\Omega_2$ (Phases 0, 1, and 2, respectively) are shaded gray, red, and blue, respectively. The values of the level set functions ${\phi^\text{ColorLS}}_0$ and ${\phi^\text{ColorLS}}_1$ are ${\phi^\text{ColorLS}}_0>0,{\phi^\text{ColorLS}}_1>0$ in Phase 0, ${\phi^\text{ColorLS}}_0>0,{\phi^\text{ColorLS}}_1<0$ in Phase 1, and ${\phi^\text{ColorLS}}_0<0,{\phi^\text{ColorLS}}_1>0$ in Phase 2.
	}
	\label{fig: colorls concept}       % Give a unique label
\end{figure}

ColorLS method is a multiphase representation method, which represents $2^n$ phases with $n$ level set functions, 
in a principle similar to combining colors from the three primary colors.
Fig. \ref{fig: colorls concept} shows an example of multiphase representation using the ColorLS method. For simplicity, we consider the case with $M=2^2$. To express each phase, we introduce $n=2$ level set functions, ${\phi^\text{ColorLS}}_0$ and ${\phi^\text{ColorLS}}_1$. We consider that  ${\phi^\text{ColorLS}}_0$ is positive in the gray and red regions (Phases 0 and 1, respectively) and negative in the blue region (Phase 2). Meanwhile,  ${\phi^\text{ColorLS}}_1$ is positive in the gray and blue regions (Phases 0 and 2, respectively) and negative in the red region (Phase 1). The boundary between Phases 0 and 1 is the zero isosurface of the level set function ${\phi^\text{ColorLS}}_1=0$. Similarly, the boundary between Phases 0 and 2 is the isosurface of ${\phi^\text{ColorLS}}_0=0$ and that between Phases 1 and 2 is the isosurface of ${\phi^\text{ColorLS}}_0=0$ and ${\phi^\text{ColorLS}}_1=0$.

In terms of these level set functions, the characteristic functions are expressed as follows:
\begin{align}
	&\psi_0 = H({\phi^\text{ColorLS}}_0)H({\phi^\text{ColorLS}}_1),\\
	&\psi_1 = H({\phi^\text{ColorLS}}_0)(1-H({\phi^\text{ColorLS}}_1)),\\
	&\psi_2 = (1-H({\phi^\text{ColorLS}}_0))H({\phi^\text{ColorLS}}_1),\\
	&\psi_3 = (1-H({\phi^\text{ColorLS}}_0))(1-H({\phi^\text{ColorLS}}_1)).
\end{align}

In the X-LS framework, this material representation is covered by imposing the following constraints on the X-LS function $\phi_{ij}$:
\begin{align}
	\phi_{20}=\phi_{31}=\phi_{21}={\phi^\text{ColorLS}}_0,\\
	\phi_{10}=\phi_{32}=\phi_{30}={\phi^\text{ColorLS}}_1.\label{eq:const colorls}
\end{align}
Material representation in the X-LS method is then transformed as 
\begin{align}
	\psi_0 &=\prod_{i\ne 0}H(\phi_{i0})\nonumber\\
	&= H({\phi^\text{ColorLS}}_1)H({\phi^\text{ColorLS}}_0)H({\phi^\text{ColorLS}}_1)\nonumber\\
	&= H({\phi^\text{ColorLS}}_0)H({\phi^\text{ColorLS}}_1),\\
	\psi_1 &=\prod_{i\ne 1}H(\phi_{i1})\nonumber\\
	&= H(-{\phi^\text{ColorLS}}_1)H({\phi^\text{ColorLS}}_0)H({\phi^\text{ColorLS}}_0)\nonumber\\
	&= H({\phi^\text{ColorLS}}_0)(1-H({\phi^\text{ColorLS}}_1)),\\
	\psi_2 &=\prod_{i\ne 2}H(\phi_{i2})\nonumber\\
	&= H(-{\phi^\text{ColorLS}}_0)H(-{\phi^\text{ColorLS}}_0)H({\phi^\text{ColorLS}}_1)\nonumber\\
	&= (1-H({\phi^\text{ColorLS}}_0))H({\phi^\text{ColorLS}}_1),\\
	\psi_3 &=\prod_{i\ne 3}H(\phi_{i3})\nonumber\\
	&= H(-{\phi^\text{ColorLS}}_1)H(-{\phi^\text{ColorLS}}_0)H(-{\phi^\text{ColorLS}}_1)\nonumber\\
	&= (1-H({\phi^\text{ColorLS}}_0))(1-H({\phi^\text{ColorLS}}_1)).
\end{align}
Thus, the representation of the characteristic function by X-LS under the constraints given by Eq. (\ref{eq:const colorls}) is equivalent to material representation by the ColorLS method. Therefore, the ColorLS method is a special case of the X-LS method when $M=4$. The same is true for $M\geq 5$, although the demonstration is omitted because the formula is very complex.

\subsection{Piecewise-constant level set method}
\begin{figure}[H]
	\centering
	\includegraphics[width=9cm]{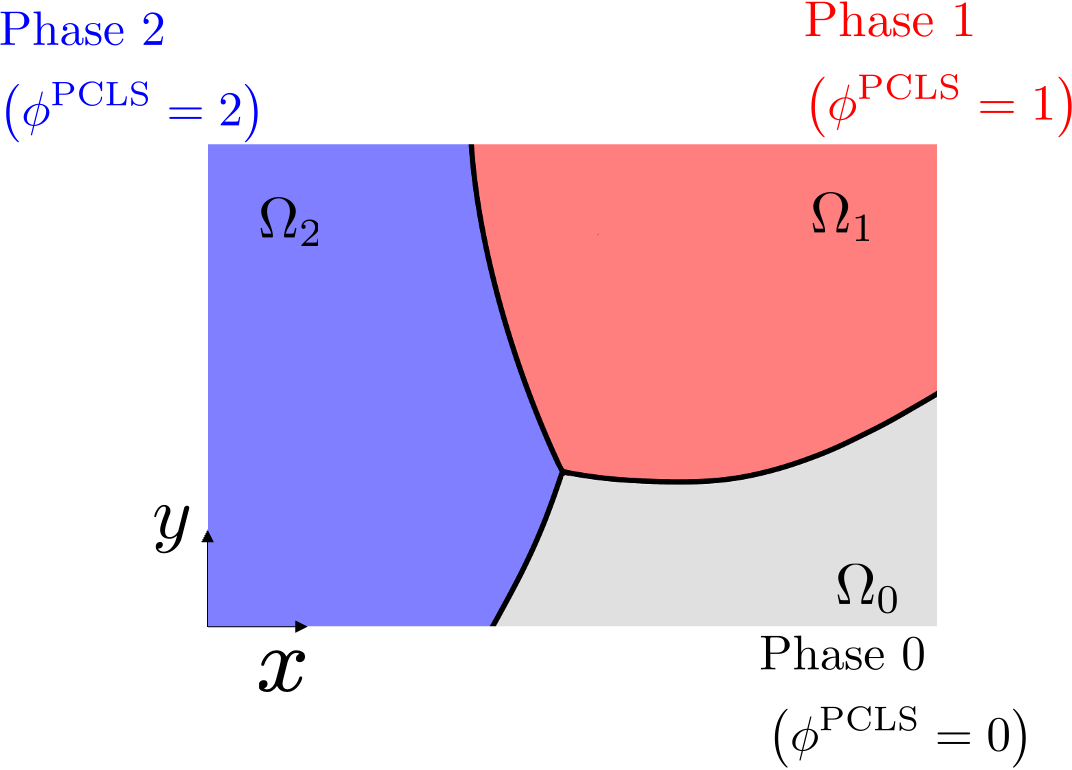}
	\caption{
		Concept of the PCLS method. The gray, red, and blue regions $\Omega_0,\Omega_1$, and $\Omega_2$ are domains containing Phase 0, 1, and 2, respectively, in which the PCLS function $\phi^\text{PCLS}$ has values of 0, 1, and 2, respectively. The boundaries (black lines) are described by discontinuities of the PCLS function $\phi^\text{PCLS}$.
	}
	\label{fig: pcls concept}       % Give a unique label
\end{figure}

In the PCLS approach \cite{lie2006variant}, each material phase is represented as the corresponding integer values of the PCLS function $\phi^\text{PCLS}$, and the boundaries are described as discontinuities in the PCLS function. Fig. \ref{fig: pcls concept} shows the concept of PCLS. The gray, red, and blue regions are the domains of Phases 0, 1, and 2, respectively, in which the piecewise-constant values of the PCLS function $\phi^\text{PCLS}$ are 0, 1, and 2, respectively.
The PCLS representation of $M$ material phases is given as
\begin{align}
	\phi^\text{PCLS}=k \quad \text{in} \quad\Omega_k, \quad k=0,1,\ldots,M-1
\end{align}
where $\Omega_k$ is the domain of Phase $k$. 

In terms of the PCLS function, the characteristic function $\psi_m$ can be expressed as follows:
\begin{align}
	&\psi_k = H(\phi^\text{PCLS}+0.5-k)\{1-H(\phi^\text{PCLS}-0.5-k)\},
\end{align}
with piecewise constant constraints;
\begin{align}
	\prod_{k=0}^{M-1}(\phi^\text{PCLS}-k)=0,
\end{align}
where $H$ is the Heaviside function defined by Eq. (\ref{eq: heaviside def}).

Here, we constrain the X-LS functions by the following equations.
\begin{align}
	\phi_{ij}+i=\phi^\text{PCLS}-0.5\quad \text{if}\quad j>i,\\
	\prod_{k=0}^{M-1}(\phi_{01}+0.5-k)=0.\label{eq:const pcls}
\end{align}
The phase representation by the X-LS method then transforms as
\begin{align}
	\psi_k&=\prod_{i\ne k}H(\phi_{ik})\nonumber\\
	&=\left(\prod_{i< k}H(\phi_{ik})\right)\left(\prod_{i>k}H(-\phi_{ki})\right)\nonumber\\
	&=\left(\prod_{i< k}H(\phi^\text{PCLS}-i-0.5)\right)\left(\prod_{i>k}H(-(\phi^\text{PCLS}-k-0.5))\right)\nonumber\\
	&=H(\phi^\text{PCLS}-(k-1)-0.5)\{1-H(\phi^\text{PCLS}-k-0.5)\}\nonumber\\
	&=H(\phi^\text{PCLS}-k+0.5)\{1-H(\phi^\text{PCLS}-k-0.5)\}.
\end{align}
Thus, the representation of the characteristic functions using the X-LS method under constraints Eq. (\ref{eq:const pcls}) is equivalent to the representation of the characteristic functions using the PCLS method. Therefore, the PCLS method is a special case of the X-LS method.

\subsection{Multi-material level set method}
\begin{figure}[H]
	\centering
	\includegraphics[width=12cm]{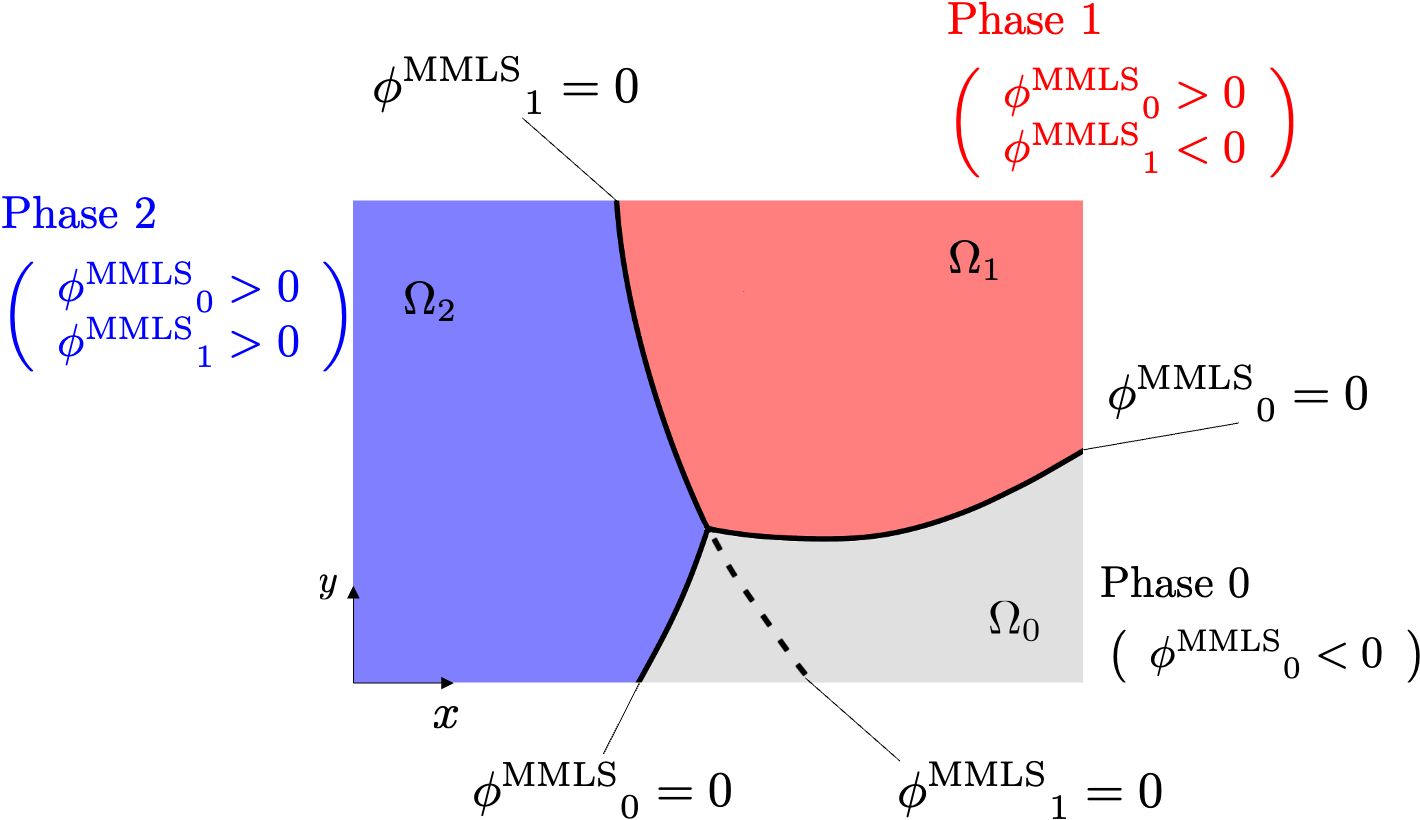}
	\caption{
		Concept of MMLS. In the gray, red, and blue domains, the level set functions are valued as ${\phi^\text{MMLS}}_0<0$ (Phase 0), ${\phi^\text{MMLS}}_0>0,{\phi^\text{MMLS}}_1<0$ (Phase 1), and ${\phi^\text{MMLS}}_0>0,{\phi^\text{MMLS}}_1>0$ (Phase 2), respectively.
	}
	\label{fig: mmls concept}       % Give a unique label
\end{figure}

The MMLS method \cite{wang2015multi,cui2016level,kishimoto2017optimal} sequentially represents $M$ phases by multiplying $M-1$ sets of level set functions. The MMLS method is conceptualized in Fig. \ref{fig: mmls concept}. The gray, red, and blue regions in this figure are assigned to Phases 0, 1, and 2, respectively. In the Phase 0 region, the level set function ${\phi^\text{MMLS}}_0$ is negative and the level set function ${\phi^\text{MMLS}}_1$ takes any value. In the Phase 1 and 2 regions, the level set function ${\phi^\text{MMLS}}_0$ is positive and the level set function ${\phi^\text{MMLS}}_1$ is negative and positive, respectively. The dashed and solid lines in Fig. \ref{fig: mmls concept} represent the zero isosurfaces coincident and not coincident with the actual phase boundaries, respectively. The boundary between Phase 0 and the other phases is the zero isosurface of the level set function ${\phi^\text{MMLS}}_0=0$. Meanwhile, the boundary between Phases 1 and 
2 is the zero isosurface of the level set function ${\phi^\text{MMLS}}_1=0$.
In the MMLS method, the characteristic function $\psi_m$ of each material $m$ can be expressed as follows:
\begin{align}
	&\psi_0 = 1-H({\phi^\text{MMLS}}_1),\\
	&\psi_k = (1-H({\phi^\text{MMLS}}_k))\prod^{k-1}_{i=0} H({\phi^\text{MMLS}}_i)\quad(k=1,\ldots,M-1),\\
	&\psi_M = \prod^{M-1}_{i=0} H({\phi^\text{MMLS}}_i).
\end{align}
Here, we impose the following constraints on the level set function in the X-LS method:
\begin{align}
	\phi_{i(i+1)}=\phi_{i(i+2)}=\ldots=\phi_{iM}={\phi^\text{MMLS}}_i \quad(i=0,1,\ldots,M-1).\label{eq:const mmls}
\end{align}
The material representation using X-LS then transforms as follows:
\begin{align}
	\psi_0&=\prod_{i\ne 0}H(\phi_{i0})\nonumber\\
	&=\prod_{i\ne 0}H(-{\phi^\text{MMLS}}_0)\nonumber\\
	&=(H(-{\phi^\text{MMLS}}_0))^{M-1}\nonumber\\
	&=H(-{\phi^\text{MMLS}}_0)\nonumber\\
	&=1-H({\phi^\text{MMLS}}_0),\\
	\psi_k&=\prod_{i\ne k}H(\phi_{ik})\nonumber\\
	&=(\prod_{i=0}^{k-1}H({\phi^\text{MMLS}}_i))(\prod_{i=k+1}^{M}H(-{\phi^\text{MMLS}}_k))\nonumber\\
	&=H(-{\phi^\text{MMLS}}_k)\prod_{i=0}^{k-1}H({\phi^\text{MMLS}}_i)\nonumber\\
	&=(1-H({\phi^\text{MMLS}}_k))\prod_{i=0}^{k-1}H({\phi^\text{MMLS}}_i),\\
	\psi_M&=\prod_{i\ne M}H(\phi_{iM})\nonumber\\
	&=\prod_{i=0}^{M-1}H({\phi^\text{MMLS}}_i).
\end{align}
Thus, the representation of the characteristic function using the X-LS method under constraints Eq. (\ref{eq:const mmls}) is equivalent to the phase representation using MMLS. Therefore, MMLS is a special case of X-LS.

\subsection{Vector-valued level set method}
The VVLS method defines an $M-1$ dimensional vector-valued function in a domain $D$ and an $M-1$ dimensional level set vector space divided into $M$ subregions corresponding to the $M$ phases. 
The phase of each point $\bm x \in D$ corresponds to the subregion of the level set vector space, in which the vector value of the VVLS function of that point $\bm\phi^\text{VVLS}(\bm x)$ exists.

\begin{figure}[H]
	\centering
	\begin{minipage}[t]{0.9\linewidth}
		\centering
		\includegraphics[width=\linewidth]{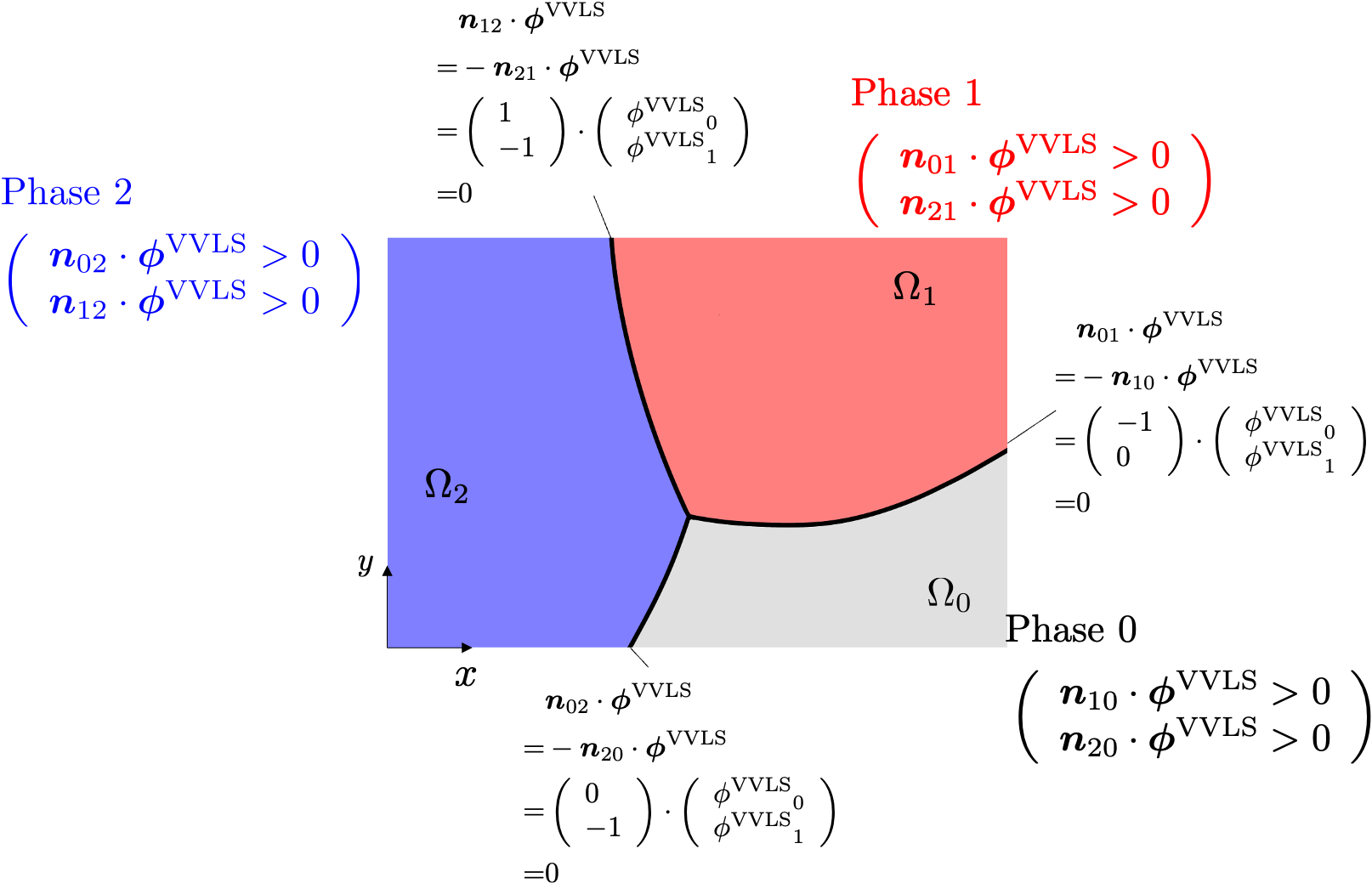}
	\end{minipage}\\
	\begin{minipage}[t]{0.9\linewidth}
		\centering
		\includegraphics[width=6cm]{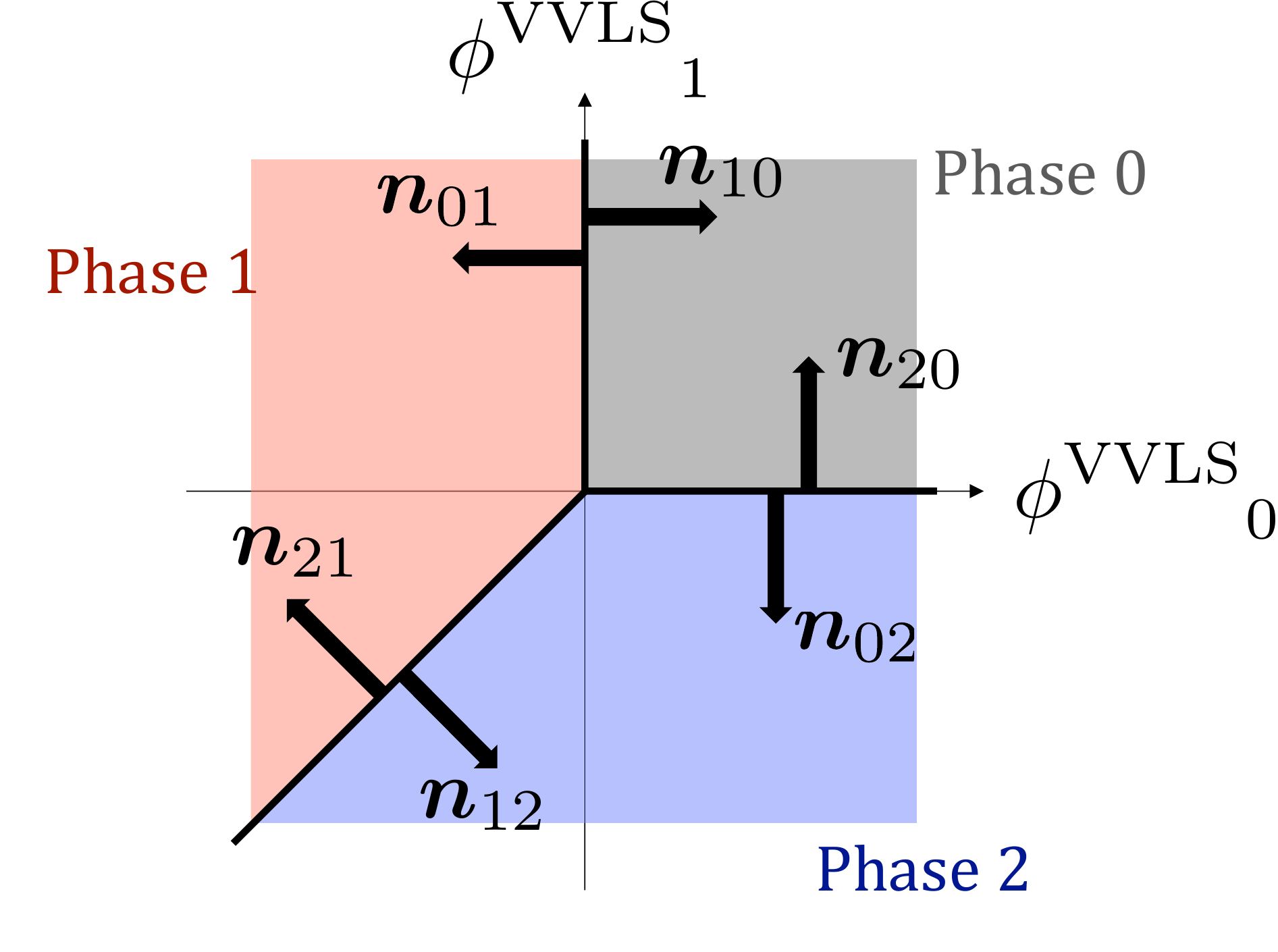}
	\end{minipage}
	\caption{
		Concept of VVLS, showing the material domains and VVLS functions (upper) and the VVLS functional space (lower). At each point $\x$ in the design domain $D$ in the upper figure, if the value of the level set function $\bm\phi^\text{VVLS}(\x)$ at that point belongs to the gray region in the level set function space (lower), then that point $\x$ is shown in gray. Similarly, if $\bm\phi^\text{VVLS}(\x)$ belongs to the red or blue region in the level set function vector space, then $\x$ is shown in red or blue, respectively, in the design domain.
	}
	\label{fig: vvls concept}       % Give a unique label
\end{figure}

Fig. \ref{fig: vvls concept} illustrates the VVLS method for $M=3$. The gray, red, and blue regions in the upper panel represent Phases 0, 1, and 2, respectively, and the lower panel shows the $M-1=2$-dimensional vector space. The gray, red, and blue regions are the subregions of the level set vector space corresponding to Phases 0, 1 and 2, respectively. The regions are separated by straight lines that intersect at the origin of the vector space. The vector $\bm n_{ij}$ is the normal vector of the boundary between the regions corresponding to phases $i$ and $j$. As these region boundaries are arbitrarily set, we set their normal vectors as 
$\bm n_{01}=-\bm n_{10}=(-1,0)^T,\bm n_{02}=-\bm n_{20}=(0,-1)^T,\bm n_{12}=-\bm n_{21}=(1,-1)^T$. Thus, in the upper panel of Fig. \ref{fig: vvls concept}, the value of the inner products of the VVLS function and the normal vectors are as follows:
\begin{align}
	\begin{cases}
		\bm n_{10}\cdot\bm\phi^\text{VVLS}>0\\
		\bm n_{20}\cdot\bm\phi^\text{VVLS}>0\\
	\end{cases}
	\quad \text{in} \quad \Omega_0,\\
	\begin{cases}
		\bm n_{01}\cdot\bm\phi^\text{VVLS}>0\\
		\bm n_{21}\cdot\bm\phi^\text{VVLS}>0\\
	\end{cases}
	\quad \text{in} \quad \Omega_1,\\
	\begin{cases}
		\bm n_{02}\cdot\bm\phi^\text{VVLS}>0\\
		\bm n_{12}\cdot\bm\phi^\text{VVLS}>0\\
	\end{cases}
	\quad \text{in} \quad \Omega_2.
\end{align}

In the upper panel of Fig. \ref{fig: vvls concept}, the boundary between the domains of Phases 0 and 1 is the zero isosurface of the first element of the VVLS function $\phi^0_\text{VVLS}=0$. Meanwhile, the boundary between the domains of Phases 0 and 2 is the zero isosurface of the second element of the VVLS function $\phi^1_\text{VVLS}=0$ and the boundary between Phases 1 and 2 is the zero isosurface of the inner product of the normal vector and level set function, i.e., $\bm n_{12}\cdot\bm\phi^\text{VVLS}=0$.
In the VVLS method, the characteristic function $\psi_k$ is expressed as follows:
\begin{align}
	&\psi_k = \prod_{i\ne k} H(\bm n_{ik}\cdot\bm\phi^\text{VVLS})\quad(k=0,\ldots,M-1).
\end{align}

Here, we consider using phases number is $M=3$ and giving the following equations as constraints, in which ${\phi^\text{VVLS}}_l$ are considered as intervening variables, for the level set function in the X-LS method.

Assuming three phases (i.e., $M=3$), we impose the following constraint equations on the X-LS method, where ${\phi^\text{VVLS}}_l$ are considered as intervening variables:
\begin{align}
	\phi_{ij}=\bm n_{ij}\cdot\bm\phi^\text{VVLS},
	\label{eq:const vvls}
\end{align}
where the vectors $\bm n_{ij},\bm\phi^\text{VVLS}$ are defined as follows:
\begin{align}
	\bm n_{01}=-\bm n_{10}=(-1,0)^T,\\
	\bm n_{02}=-\bm n_{20}=(0,-1)^T,\\
	\bm n_{12}=-\bm n_{21}=(1,-1)^T,\\
	\bm\phi^\text{VVLS}=({\phi^\text{VVLS}}_0,{\phi^\text{VVLS}}_1)^T.
\end{align}
The material representation by X-LS then transforms as
\begin{align}
	\psi_k&=\prod_{i\ne k}H(\phi_{ik})\nonumber\\
	&=\prod_{i\ne k}H(\bm n_{ik}\cdot\bm\phi^\text{VVLS}).
\end{align}
Thus, the representation of the characteristic functions using the X-LS method under constraints given by Eq. (\ref{eq:const vvls}) is equivalent to the material representation using VVLS. Therefore, VVLS is a special case of X-LS when $M=3$. The same holds for $M\geq 4$ but the demonstration is omitted because the formula is very complex.

In conclusion, the proposed method is the most general extension of the level set method for multiphase representation.

\clearpage
\bibliographystyle{elsarticle-num} 
%\section*{References}\label{sec:references}
%\bibliography{reference.bib}

%\begin{thebibliography}{00}
%
%	\bibitem{label}
%\end{thebibliography}
\end{document}